\documentclass[epsfig,12pt,onecolumn]{article}
\usepackage{amsfonts}
\usepackage{amssymb}
\usepackage{multicol}
\usepackage{graphicx}
\usepackage{float}
\usepackage{caption}
\usepackage{xcolor}
\usepackage[utf8]{inputenc}
\usepackage{amsmath}

\makeatletter \@addtoreset{equation}{section}
\renewcommand{\theequation}{\thesection.\arabic{equation}}
\def \be{\begin{equation}}
\def \ee{\end{equation}}
\def \bea{\begin{eqnarray}}
\def \eea{\end{eqnarray}}

\newcommand{\nc}{\newcommand}
\nc{\al}{\alpha} \nc{\bib}{\bibitem} \nc{\la}{\lambda}
\nc{\C}{\mbox{\hspace{1.24mm}\rule{0.2mm}{2.5mm}\hspace{-2.7mm} C}}
\nc{\R}{\mbox{\hspace{.04mm}\rule{0.2mm}{2.8mm}\hspace{-1.5mm} R}}
\input{tcilatex}
\begin{document}

\title{\rightline{\mbox{\small {LPHE-MS-sept-2022}} \vspace
{-0,2cm}} \textbf{From orthosymplectic structure to super topological matter}%
}
\author{L.B Drissi$^{1,2,3}$, E.H Saidi$^{1,3}$ \\
{\small 1. LPHE-MS, Science Faculty}, {\small Mohammed V University in
Rabat, Morocco.}\\
{\small 2.\ Peter Gr\"{u}nberg Institut and Institute for Advanced
Simulation, }\\
{\small Forschungszentrum J\"{u}lich \& JARA, D-52425 J\"{u}lich, Germany.}\\
{\small 3. Centre of Physics and Mathematics, CPM- Morocco.}}
\maketitle

\begin{abstract}
Topological supermatter is given by ordinary topological matter constrained
by supersymmetry or graded supergroups such as OSP(2N\TEXTsymbol{\vert}2N).
Using results on super oscillators and lattice QFT$_{d}$, we construct a
super tight binding model on hypercubic super lattice with supercharge $%
\boldsymbol{Q}=\sum_{\mathbf{k}}\hat{F}_{\mathbf{k}}.\boldsymbol{q}_{\mathbf{%
k}}.\hat{B}_{\mathbf{k}}$. We first show that the algebraic triplet $(\Omega
,G,J)$ of super oscillators can be derived from the $OSp(2N|2N)$ supergroup
containing the symplectic $Sp(2N)$ and the orthogonal $SO(2N)$ as even
subgroups. Then, we apply the obtained result on super oscillating matter to
super bands and investigate its topological obstructions protected by TPC
symmetries. We also give a classification of the Bose/Fermi coupling matrix $%
\boldsymbol{q}_{\mathbf{k}}$ in terms of subgroups of $OSp(2N|2N)$\ and show
that there are $2P_{N}$ (partition of $N$) classes $\boldsymbol{q}_{\mathbf{k%
}}$ given by unitary subgroups of $U\left( 2\right) \times U\left( N\right) $%
. Other features are also given.\newline
\textbf{Keywords:} Bosonic/fermionic oscillators, orthosymplectic structure,
topological super matter on super lattice, super tight binding model.
\end{abstract}


\section{Introduction}

Supersymmetry in Lorentzian 4D space time and higher dimensions plays a
crucial role in the study of relativistic superfield theory \textrm{\cite{1A}%
} and superstrings \textrm{\cite{1B}}. This Bose/Fermi symmetry is not an
exact symmetry at our energy scale; if it existed in early universe, it is
now broken. For that, several efforts have been made to find footprints of
supersymmetry in particle physics at high energies; but still without direct
experimental evidence. \newline
Supersymmetry in non relativistic physics has also received some attention
\textrm{\cite{1C}-\cite{1CD}; supersymmetric models have been constructed in
}numerous fields such as condensed matter \cite{2C1}, statistical and
quantum mechanics \cite{2CA,2CB}, optics \cite{2CD,2CE} and cosmology \cite%
{2CF}.\newline
Recently, a special attention has been given to supersymmetric topological
phases of matter \textrm{\cite{1D}-\cite{1DD}} and supersymmetric
entanglement \textrm{\cite{1E}-\cite{1EC}. There, }super models \textrm{\cite%
{1ED,1EE}} were constructed by using fermionic $\hat{f}^{\alpha }/\hat{f}%
_{\alpha }^{\dagger }$ and bosonic $\hat{b}^{\alpha }/\hat{b}_{\beta
}^{\dagger }$ oscillators coupled as%
\begin{equation}
Q=\sum_{\alpha ,\beta =1}^{N}\hat{f}_{\alpha }^{\dagger }\left[ A_{\beta
}^{\alpha }\right] \hat{b}^{\beta },\qquad ,\qquad Q^{\dagger }=\sum_{\alpha
,\beta =1}^{N}\hat{b}_{\beta }^{\dagger }\left[ \left( A^{\dagger }\right)
_{\alpha }^{\beta }\right] \hat{f}^{\alpha }  \label{11}
\end{equation}%
A simple coupling matrix $A_{\beta }^{\alpha }$ is given by the diagonal
frequency matrix $\delta _{\beta }^{\alpha }\sqrt{\nu }$ leading to the
familiar quantum supercharge $Q=\sqrt{\nu }\hat{f}_{\alpha }^{\dagger }\hat{b%
}^{\alpha }$. The Hamiltonian $H$ of this super oscillator is given by the
quadratic composite $QQ^{\dagger }+Q^{\dagger }Q$ generating four
contributions $H_{ff}+H_{bb}+H_{fb}+H_{bf}$ namely: \newline
$\left( \mathbf{i}\right) $ a quadratic fermionic $H_{ff}=\hat{f}_{\alpha
}^{\dagger }\left( h_{f}\right) _{\beta }^{\alpha }\hat{f}^{\beta }$ with
coupling matrix $h_{f}=AA^{\dagger }.$ \newline
$\left( \mathbf{ii}\right) $ a quadratic bosonic $H_{bb}=\hat{b}_{\alpha
}^{\dagger }\left( h_{b}\right) _{\beta }^{\alpha }\hat{b}^{\beta }$ with
coupling matrix $h_{b}=A^{\dagger }A$; and \newline
$\left( \mathbf{iii}\right) $ two \emph{quartic} terms $H_{fb}$ and $H_{bf},$
compensating each other, given by the coupling $\pm \hat{f}^{\dagger }\hat{b}%
^{\dagger }\left[ A\otimes A^{\dagger }\right] \hat{f}\hat{b}.$\newline
On the other hand, it is quite well known that, besides supersymmetry, the
bosonic and fermionic oscillators share several common features; but show
also different behaviours due to their different quantum statistics. In this
regard, the bosonic oscillator operators combined like $\hat{B}^{A}=(\hat{b}%
^{\alpha },\hat{b}_{\alpha }^{\dagger })$ are described by the symplectic
symmetry $Sp(2N)$ of the phase space $\mathcal{E}_{ph}$ of statistical
physics \textrm{\cite{1F}-\cite{2F}}; they are characterised by the
symplectic structure $\Omega ^{AB}$ given by the commutator $\hat{B}^{A}\hat{%
B}^{B}-\hat{B}^{B}\hat{B}^{A}$. However, the properties of the fermionic
oscillator operators, combined as $\hat{F}^{A}=(\hat{f}^{\alpha },\hat{f}%
_{\alpha }^{\dagger }),$ are described by the orthogonal symmetry $SO(2N)$
and are characterised by the metric $G^{AB}$ given by the anti-commutator $%
\hat{F}^{A}\hat{F}^{B}+\hat{F}^{B}\hat{F}^{A}$ \textrm{\cite{1G}-\cite{1GB}}%
. Because of the Bose/Fermi symmetry of the oscillating $\hat{B}/\hat{F}$
system, \textrm{it is legitimate} to ask on: $\left( i\right) $ the
relationship between supersymmetry and the combination of the orthogonal and
the symplectic structures, (termed below as the orthosymplectic structure). $%
\left( ii\right) $ the seek of lattice super QFTs governed by the
orthosymplectic property for probing supersymmetric effect in non
relativistic settings; and $\left( iii\right) $ the use of the
orthosymplectic idea to approach super topological matter while imposing
discrete TPC symmetries in the same spirit as in the derivation of the
periodic table classifying the topological insulators and superconductors in
tenfold way classes \cite{1GC}.\newline
\textrm{In this paper}, we contribute to super topological matter and super
band theory from the view of the orthosymplectic structure represented by
the triplet $\left( \Omega ,G,J\right) $ with $\Omega $ and $G$ as above;
the $J$ is a complex structure found to be given by the intersection $\Omega
.G^{-1}$ and described by the intersection $U\left( N\right) $ symmetry of $%
Sp(2N)$ and $SO(2N)$. We show that the algebraic properties of the super
oscillators are nicely described by the super group $OSp\left( 2N|2N\right)
; $ which when combined with lattice super QFT methods and TPC symmetries,
allow to approach the topological properties of the super bands. Among our
results, we cite the following:

\begin{description}
\item[$\left( \mathbf{1}\right) $] the construction of a super tight binding
model (TBM) for lattice supermatter described by a supercharge $\boldsymbol{Q%
}$ quadratic in local oscillator field operators $\hat{f}^{\alpha }\left(
\mathbf{r}_{i}\right) $, $\hat{b}^{\beta }\left( \mathbf{r}_{i}\right) $ as
follows%
\begin{equation}
\begin{tabular}{lll}
$\boldsymbol{Q}$ & $=$ & $\dsum\limits_{i,j}\left[ \hat{f}_{\alpha
}^{\dagger }\left( \mathbf{r}_{i}\right) \left( \kappa _{ij}\right) _{\beta
}^{\alpha }\hat{b}^{\beta }\left( \mathbf{r}_{j}\right) +\hat{f}^{\alpha
}\left( \mathbf{r}_{i}\right) \left( \varkappa _{ij}\right) _{\alpha \beta }%
\hat{b}^{\beta }\left( \mathbf{r}_{j}\right) \right] +$ \\
&  & $\dsum\limits_{i,j}\left[ \hat{b}_{\alpha }^{\dagger }\left( \mathbf{r}%
_{i}\right) \left( \kappa _{ij}^{\prime }\right) _{\beta }^{\alpha }\hat{f}%
^{\beta }\left( \mathbf{r}_{j}\right) +\hat{b}_{\alpha }^{\dagger }\left(
\mathbf{r}_{i}\right) \left( \varkappa _{ij}^{\prime }\right) ^{\alpha \beta
}\hat{f}_{\beta }^{\dagger }\left( \mathbf{r}_{j}\right) \right] $%
\end{tabular}
\label{SQ}
\end{equation}%
with coupling matrices $\kappa \left( \mathbf{r}_{i}-\mathbf{r}_{j}\right) $
and $\varkappa \left( \mathbf{r}_{i}-\mathbf{r}_{j}\right) $ assumed
translation invariant. This supercharge\textrm{\ can be presented shortly as}
$\hat{F}_{\mathbf{r}_{i}}\boldsymbol{J}_{ij}\hat{B}_{\mathbf{r}_{j}}$; its
square generates the super hamiltonian $H_{\text{super}}$ given by $\left(
i\right) $ the sum $\hat{F}_{\mathbf{r}_{i}}\left( h_{f}\right) _{ij}\hat{F}%
_{\mathbf{r}_{j}}+\hat{B}_{\mathbf{r}_{i}}\left( h_{b}\right) _{ij}\hat{B}_{%
\mathbf{r}_{j}},$ with coupling $\left( h_{f}\right) _{ij}$ and $\left(
h_{b}\right) _{ij}$ quadratic in $\boldsymbol{J}_{ij},$ and $\left(
ii\right) $ two extra terms that kill each other (Bose/Fermi compensation
effect).

\item[$\left( \mathbf{2}\right) $] the determination of the intrinsic
properties of the super $H_{\text{super}}$ and its topological distortions.
By using the Fourier transform, the supercharge operator expands as $\sum Q_{%
\mathbf{k}}$ with Fourier modes $Q_{\mathbf{k}}$ given by the quadratic form
$\hat{F}_{\mathbf{k}}.\boldsymbol{q}_{\mathbf{k}}.\hat{B}_{\mathbf{k}}.$ The
local $\boldsymbol{q}_{\mathbf{k}}$ living on the Brillouin torus $\mathbb{T}%
^{d}$ is a $2N\times 2N$ coupling matrix valued in the bi-fundamental
representation of $Sp(2N)\times SO(2N)$ with sub-bloc matrices $N\times N$
given by the Fourier transforms of $\mathbf{\kappa }_{ij},\mathbf{\kappa }%
_{ij}^{\prime },$ $\mathbf{\varkappa }_{ij},$ $\mathbf{\varkappa }%
_{ij}^{\prime }$. Here, we show that the super Hamiltonian $H_{\mathbf{k}}$
has two main contributions: $\left( \mathbf{i}\right) $ A fermion dependent $%
\left( H_{\mathbf{k}}\right) _{f}$ given by $\hat{F}_{\mathbf{k}}.%
\boldsymbol{h}_{f}.\hat{F}_{\mathbf{k}}$ with coupling matrix as $%
\boldsymbol{h}_{f}=\boldsymbol{q}_{\mathbf{k}}Z_{f}\boldsymbol{q}_{\mathbf{k}%
}^{\dagger }$ and $Z_{f}=\sigma _{z}\otimes I_{N}$. It can carry non-trivial
topological distortions as for non super Altland-Zirnbauer (AZ) matter. $%
\left( \mathbf{ii}\right) $ A boson dependent $H_{b}$ given by $\hat{B}_{%
\mathbf{k}}.\boldsymbol{h}_{b}.\hat{B}_{\mathbf{k}}$ with coupling matrix $%
\boldsymbol{h}_{b}=\boldsymbol{q}_{\mathbf{k}}^{\dagger }\boldsymbol{q}_{%
\mathbf{k}}$; it has a trivial topology.

\item[$\left( \mathbf{3}\right) $] the derivation of constraints on the
coupling matrix $\boldsymbol{q}_{\mathbf{k}}$\ that are required by the
embedding of $\mathcal{N}=2$ supersymmetric quantum mechanical matter ($%
\mathcal{N}=2$ super QM) within tight binding models based on the
orthosymplectic osp(2N\TEXTsymbol{\vert}2N). In this regards, we distinguish
two supermatter systems: orthosymplectic ({\small ORTIC}) and supersymmetric
({\small SUSY}). We also give solutions for $\boldsymbol{q}_{\mathbf{k}}$
which turn out to be classified by $\left( i\right) $\ the complex structure
$J$ represented by the diagonal $U\left( N\right) $\ subsymmetry of the
OSP(2N\TEXTsymbol{\vert}2N) supergroup\textrm{; and }$\left( ii\right) $ the
TPC symmetries of AZ matter.
\end{description}

The presentation is as follows: \textrm{In section 2}, we study the
structure triplet $\left( \Omega ,G,J\right) \equiv \mathbf{\tau }$ of
supersymmetric oscillator operators from the view of symplectic $Sp(2N)$ and
orthogonal $SO(2N)$ symmetries. \textrm{In section 3}, we give the relation
of the triplet $\mathbf{\tau }$ with the orthosymplectic $OSp(2N2|N)$ super
group; and its link with the graded algebra of quantum super oscillator.
\textrm{In section 4}, we study the link between the orthosymplectic osp(2N%
\TEXTsymbol{\vert}2M) Lie superalgebra and $\mathcal{N}=2$ super QM. Using
properties of osp(2N\TEXTsymbol{\vert}2M), we build out of $\boldsymbol{Q}$
several observables including two interesting ones termed as orthosymplectic
(ORTIC) Hamiltonian $H_{ortic}$ and supersymmetric (SUSY) Hamiltonian $%
H_{susy}$. \textrm{In section 5}, we develop the study of super tight
binding model by using the orthosymplectic supergroup representations and
methods of lattice super QFT$_{d}$. \textrm{In section 6}, we investigate
the topological properties of the super TBM. \textrm{In section 7}, we give
a conclusion and make comments. Last section is devoted to two appendices:
In appendix A, we deepen the study on the link between: $\left( i\right) $ $%
\mathcal{N}=2$ supersymmetry on world line, $\left( ii\right) $
orthosymplectic osp(2\TEXTsymbol{\vert}2) superalgebra and $\left(
iii\right) $ $\mathcal{N}=2$ $U\left( 1\right) $ superconformal invariance.
In appendix B, we develop the oscillator realisation of the osp(2N%
\TEXTsymbol{\vert}2M) used in our tight binding modeling and in the
construction of the supersymmetric fivefold ways of \textrm{\cite{1D2}}.

\section{Orthosymplectic structure}

\textrm{In this section, we study two building bloc structures }$\mathbf{%
\tau }$ and $\mathbf{\tilde{\tau}}$\textrm{\ of "orthosymplectic" manifolds}
$\mathcal{K}$\textrm{\ \cite{1H0} }with the aim of using them later to
investigate topological supermatter. These building blocs are given by the
following triplets,
\begin{equation}
\mathbf{\tau }=(g,\omega ,J)\quad ,\quad \mathbf{\tilde{\tau}}=(G,\Omega
,J^{T})  \label{tr}
\end{equation}%
They have an interpretation in the quantized phase space of free
supersymmetric oscillators. The $g$ and $G$ will be associated with fermions
$\digamma ,$ the $\omega $ and $\Omega $ with bosons $\xi ;$ and the $J$ and
$J^{T}$ give the link between them.\newline
First, we introduce the above triplets as recently formulated in \textrm{%
\cite{1H}}. Then, we investigate their useful properties. We show that they
constitute main pillars in dealing with the superalgebra of supersymmetric
quantum oscillators and its representations; thus offering a new way to
think about (\ref{tr}).

\subsection{Revisiting the $(g,\protect\omega ,J)$ and $(G,\Omega ,J^{T})$
triplets}

Following \textrm{\cite{1H}}, the triplet $\mathbf{\tau }=(g,\omega ,J)$ and
its dual $\mathbf{\tilde{\tau}}=(G,\Omega ,J^{T})$ play an important role in
the construction of the bosonic and fermionic Gaussian states. These are
algebraic structures that have representations in terms of non degenerate
real rank 2 tensors $\mathcal{T}$ in its three variants: contravariant $%
\mathcal{T}^{AB}$, covariant $\mathcal{T}_{AB}$ and mixed $\mathcal{T}%
_{B}^{A}$ as follows%
\begin{equation}
\begin{tabular}{lllll}
$g_{AB}$ & $\quad ,\quad $ & $\omega _{AB}$ & $\quad ,\quad $ & $J_{A}^{%
\text{ \ }B}$ \\
$G^{AB}$ & $\quad ,\quad $ & $\Omega ^{AB}$ & $\quad ,\quad $ & $J_{\text{ \
}B}^{A}$%
\end{tabular}
\label{matrix}
\end{equation}%
with $(J_{A}^{\text{ \ }B})^{T}=J_{\text{ \ }B}^{A\text{ }}.$ For later use,
we give below those interesting properties regarding the tensor realisation
of the triplet $(g,\omega ,J).$ Similar things can be written down for $%
(G,\Omega ,J^{T})$. \newline
The real $g_{AB}$ is symmetric ($g^{T}=g$) and positive defined ($\det g>0$%
). It acts as an orthogonal metric, it maps $SO\left( 2N,\mathbb{R}\right) $
contravariant vectors $\digamma ^{B}$ (fermions) to the covariant ones like
\begin{equation}
\digamma _{A}=g_{AB}\digamma ^{B}\qquad ,\qquad \digamma ^{B}=G^{BC}\digamma
_{C}\qquad ,\qquad G^{BC}g_{CA}=\delta _{A}^{B}
\end{equation}%
This metric is invariant under orthogonal transformation; i.e: $\Lambda
g\Lambda ^{T}=g$ with transformation matrix $\Lambda $ belonging to the
orthogonal group $SO\left( 2N,\mathbb{R}\right) .$ \newline
The $\omega _{AB}$ is the usual antisymmetric symplectic structure ($\omega
^{T}=-\omega ,\det \omega \neq 0$). It acts as a symplectic metric on
contravariant vectors $\xi ^{B}$ (bosons) to map them to covariant ones like
\begin{equation}
\xi _{A}=\omega _{AB}\xi ^{B}\qquad ,\qquad \xi ^{B}=\Omega ^{BC}\xi
_{C}\qquad ,\qquad \Omega ^{BC}\omega _{CA}=\delta _{A}^{B}
\end{equation}%
This symplectic metric is invariant under symplectic transformations; i.e: $%
\mathcal{S}\omega \mathcal{S}^{T}=\omega $ with matrix $\mathcal{S}$
belonging to the group $SP\left( 2N,\mathbb{R}\right) $ \textrm{\cite{1HA}}.
\newline
The mixed tensor $J$ is the complex structure living on the Kahler space $%
\mathcal{K}$ with square $J^{2}=-I_{id}$ showing that $J$ behaves as the
pure imaginary number unit $i$ ($i^{2}=-1)$. An interesting definition of $%
J_{B}^{A}$ is given by
\begin{equation}
J_{\text{ }B}^{A}=\Omega ^{AC}g_{CB}  \label{jmg}
\end{equation}%
from which we learn that it can act on both the orthogonal vectors $\mathbf{%
\digamma }$ (fermions) and the symplectic vectors $\mathbf{\xi }$ (bosons).
For example, we have%
\begin{equation}
\begin{tabular}{lll}
$J_{B}^{A}\digamma ^{B}$ & $=$ & $\Omega ^{AC}\digamma _{C}$ \\
$\xi _{A}J_{B}^{A}$ & $=$ & $-\xi ^{C}g_{CB}$%
\end{tabular}
\label{vp}
\end{equation}%
and%
\begin{equation}
\begin{tabular}{lll}
$\xi _{A}J_{B}^{A}\digamma ^{B}$ & $=$ & $\Omega ^{AC}\xi _{A}\digamma _{C}$
\\
& $=$ & $-g_{CB}\xi ^{C}\digamma ^{B}$%
\end{tabular}%
\end{equation}%
as well as%
\begin{equation}
\begin{tabular}{lll}
$\digamma _{A}J_{B}^{A}\digamma ^{B}$ & $=$ & $\Omega ^{AC}\digamma
_{A}\digamma _{C}$ \\
$\xi _{A}J_{B}^{A}\xi ^{B}$ & $=$ & $-g_{CB}\xi ^{C}\xi ^{B}$%
\end{tabular}%
\end{equation}%
To distinguish the labels of the $SO\left( 2N\right) $ and $Sp(2N)$
representations, we use below dotted labels for $SO\left( 2N\right) ;$\ for
example $\digamma ^{A},$ $g_{AB}$ and $J_{B}^{A}=\Omega ^{AC}g_{CB}$ will be
replaced by $\digamma ^{\dot{A}},$ $g_{\dot{A}\dot{B}}$ and $J_{\text{ }\dot{%
B}}^{A}=\Omega ^{AC}g_{\dot{C}\dot{B}}.$ The last relation requires the
identification of the labels $C$ and $\dot{C};$ this feature will be
discussed later.

\subsubsection{Duality and constraint relations}

The rank 2 tensors (\ref{matrix}) realising the 3+3 components of the
triplets (\ref{tr}) are not all of them free; they are subject to
relationships and constraint equations that we describe here after:

\begin{itemize}
\item \textbf{\ Duality relations}\newline
As noticed before, the $(g,\omega ,J)$ and $(G,\Omega ,J^{T})$ are dual
between them. The duality relations are given by
\begin{equation}
\begin{tabular}{lllllll}
$g_{\dot{A}\dot{C}}G^{\dot{C}\dot{B}}$ & $=$ & $+\delta _{\dot{A}}^{\dot{B}}$
& ,\qquad & $J_{\text{ }\dot{B}}^{A}J_{\text{ }C}^{\dot{B}}$ & $=$ & $\delta
_{C}^{A}$ \\
$\omega _{AC}\Omega ^{CB}$ & $=$ & $+\delta _{A}^{B}$ & ,\qquad & $J_{\text{
}C}^{\dot{A}}J_{\text{ }\dot{B}}^{C}$ & $=$ & $\delta _{\dot{B}}^{\dot{A}}$%
\end{tabular}%
\end{equation}%
These relations read in a condensed way as follows
\begin{equation}
\begin{tabular}{lllllll}
$g.G$ & $=$ & $+I_{2N}$ & ,\qquad & $J.J^{T}$ & $=$ & $+I_{2N}$ \\
$\omega .\Omega $ & $=$ & $+I_{2N}$ & ,\qquad & $J^{T}.J$ & $=$ & $+I_{2N}$%
\end{tabular}
\label{eq1}
\end{equation}

\item \textbf{Constraint Equations}\emph{\ }\newline
Amongst the above mentioned 3 basis components of the two triplets, only two
of them which are really basic ones. For example the symplectic $\omega
_{AB} $ of the bosonic sector and the $g_{\dot{A}\dot{B}}$ of the fermionic
sector. The following relations express one element of the triplet as a
product of two others:
\begin{equation}
\begin{tabular}{lllllll}
$J$ & $=$ & $+\Omega .g$ & $\qquad ,\qquad $ & $J^{T}$ & $=$ & $-g.\Omega $
\\
$J$ & $=$ & $-G.\omega $ & $\qquad ,\qquad $ & $J^{T}$ & $=$ & $+\omega .G$%
\end{tabular}
\label{JJ}
\end{equation}%
and
\begin{equation}
\begin{tabular}{lllllll}
$\Omega $ & $=$ & $+J.G$ & $\qquad ,\qquad $ & $\Omega $ & $=$ & $-G.J^{T}$
\\
$G$ & $=$ & $-J.\Omega $ & $\qquad ,\qquad $ & $G$ & $=$ & $+\Omega .J^{T}$%
\end{tabular}
\label{GG}
\end{equation}%
Eq(\ref{JJ}) captures the feature that the complex structure $J$ and its
underlying unitary group $U\left( N\right) $ is given by the intersection of
the orthogonal and the symplectic structures as depicted by the Figure
\textbf{\ref{triad}}.
\begin{figure}[tbph]
\begin{center}
\includegraphics[width=16cm]{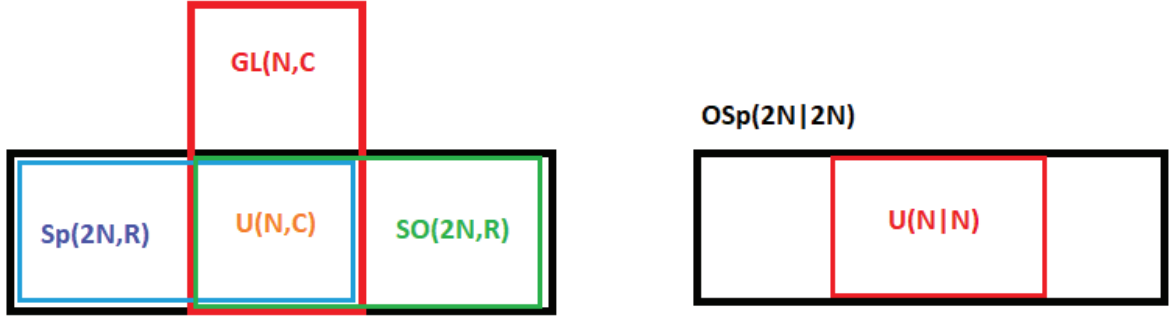}
\end{center}
\par
\vspace{-0.5cm}
\caption{On the left, the unitary group $U\left( N\right) $ as the
Intersection of three groups: the $GL\left( N,\mathbb{C}\right) $ in green
color, the $SP\left( 2N,\mathbb{R}\right) $ in blue; and the $SO(2N,\mathbb{R%
})$ in red. On the right the $U\left( N|N\right) $ supergroup as a graded
subsymmetry of $OSP\left( 2N|2N\right) .$ }
\label{triad}
\end{figure}

\item \textbf{Refining the constraints}\newline
From eqs(\ref{JJ}-\ref{GG}), we deduce remarkable relationships; in
particular the two following: \newline
$\left( \mathbf{i}\right) $ the triple intersection relation
\begin{equation}
\omega .J.G=I_{id}  \label{tri}
\end{equation}%
which can be also expressed in different, but equivalent, ways.\newline
$\left( \mathbf{ii}\right) $ the transformations
\begin{equation}
\begin{tabular}{lllllll}
$J.G.J^{T}$ & $=$ & $G$ & $\qquad ,\qquad $ & $J.\Omega .J^{T}$ & $=$ & $%
\Omega $ \\
$J^{T}.g.J$ & $=$ & $g$ & $\qquad ,\qquad $ & $J^{T}\omega .J$ & $=$ & $%
\omega $%
\end{tabular}
\label{eq4}
\end{equation}%
which are useful for explicit calculations.
\end{itemize}

\subsubsection{Canonical representation of the triplets}

If thinking about the symmetric matrices $g$ and $G$ as given by the
identity matrix $I_{2N}$, then the complex structure $J$ coincides with the
symplectic structure as shown by substituting in (\ref{tri}) $g=G=I_{2N};$
thus leading to%
\begin{equation}
\omega .J=I_{id}
\end{equation}%
For this canonical choice, we have $J=\Omega $ and $J^{T}=\omega .$ In
matrix notation \textrm{\cite{1E}},%
\begin{equation}
G=\left(
\begin{array}{cc}
I_{N} & 0 \\
0 & I_{N}%
\end{array}%
\right) \qquad ,\qquad J=\Omega =\left(
\begin{array}{cc}
0 & I_{N} \\
-I_{N} & 0%
\end{array}%
\right)  \label{eq5}
\end{equation}%
Because of the property $J^{2}=-I_{2N}$, one can define two interesting
quantities:

\begin{itemize}
\item \textrm{The two projectors}
\begin{equation}
P_{\pm }=\frac{1}{2}\left( I_{id}\pm iJ\right)
\end{equation}%
with
\begin{equation}
\left( P_{\pm }\right) ^{2}=P_{\pm }\quad ,\quad P_{+}+P_{-}=I_{id}
\end{equation}

\item Given a real contravariant (resp. covariant) vector $\xi ^{A}$ (resp. $%
\xi _{A}$) with 2N components, one can construct two chiral vectors $\xi
_{\pm }^{A}$ (resp. $\xi _{A}^{\pm }$) given by%
\begin{equation}
\begin{tabular}{lll}
$\xi _{+}^{A}$ & $=$ & $\xi ^{A}+iJ_{B}^{A}\xi ^{B}$ \\
$\xi _{-}^{A}$ & $=$ & $\xi ^{A}-iJ_{B}^{A}\xi ^{B}$%
\end{tabular}
\label{xj}
\end{equation}%
and
\begin{equation}
\begin{tabular}{lll}
$\xi _{B}^{+}$ & $=$ & $\xi _{B}+iJ_{B}^{A}\xi _{A}$ \\
$\xi _{B}^{-}$ & $=$ & $\xi _{B}-iJ_{B}^{A}\xi _{A}$%
\end{tabular}%
\end{equation}
\end{itemize}

\subsection{Supersymmetric phase space\emph{\ }$\mathcal{E}_{ph}^{2N|2N}$}

\begin{figure}[tbph]
\begin{center}
\includegraphics[width=12cm]{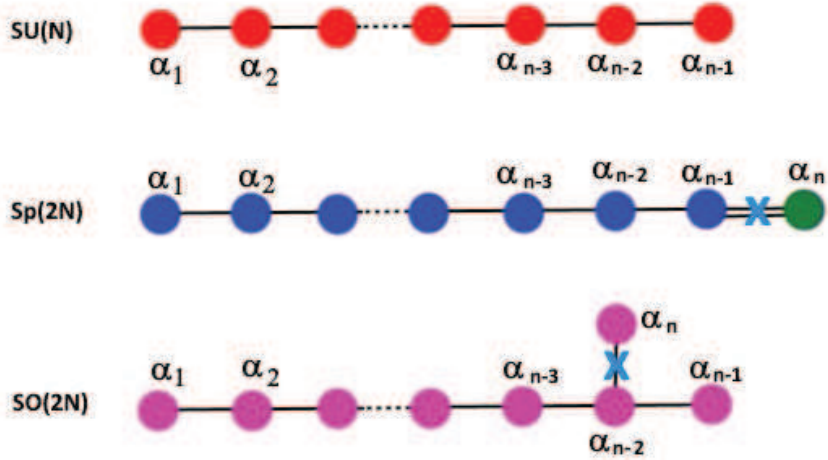}
\end{center}
\caption{The Dynkin diagrams of the unitary $SU\left( N\right) $, the $%
SP\left( 2N\right) $ and the $SO(2N)$ groups. By cutting the simple root $%
\protect\alpha _{N}$ in $SP\left( 2N\right) $ and the $SO(2N),$ we obtain
the diagram of $SU\left( N\right) $. }
\label{AC}
\end{figure}

So far we have used variables and structures carrying charges of the
symmetry groups $Sp\left( 2N\right) ,$ $SO\left( 2N\right) $ and $SU\left(
N\right) .$ The unitary group $SU\left( N\right) $ is a subgroup of $%
Sp\left( 2N\right) $ and of $SO\left( 2N\right) $ as schematized by the
Figure \textbf{\ref{triad}}. This feature can be also exhibited by using the
Dynkin diagrams of the Lie algebras of these Lie groups as given by the
Figure \textbf{\ref{AC}}. The diagram of $SU\left( N\right) $ has $\left(
N-1\right) $ nodes while those of $Sp\left( 2N\right) $ and $SO\left(
2N\right) $ have $N$ nodes \textrm{\cite{YS}}; by cutting the node $\alpha
_{N}$ of $Sp\left( 2N\right) $ and $SO\left( 2N\right) ,$ one obtains the
graph of $SU\left( N\right) .$
This unitary group $SU\left( N\right) $ plays an important role in our
present study; we put it aside for the moment; we will take it up later.
Moreover, we will think of the symplectic (bosons) and the orthogonal
(fermions) groups as
\begin{equation}
G_{\bar{0}}=Sp\left( 2N\right) \times SO\left( 2N\right)
\end{equation}%
that is the even part of the orthosymplectic group \textrm{\cite{sp}}
\begin{equation}
OSp\left( 2N|2N\right) =G_{\bar{0}}\ltimes G_{\bar{1}}
\end{equation}%
with $G_{\bar{1}}$ given by $(2N,2\dot{N}\mathbf{)},$ the bi-fundamental
representation of $G_{\bar{0}}$ with $2\mathbf{N}$ standing for the
fundamental representation of $Sp\left( 2N\right) $ and the $2\mathbf{\dot{N}%
}$ referring to the fundamental representation of $SO\left( 2N\right) .$
\newline
Among the relevant features of $OSp\left( 2N|2N\right) $ and the realisation
of the orthosymplectic manifold $\mathcal{K}$ for our present study, we cite
the supersymmetric phase space $\mathcal{E}_{ph}^{2N|2N}$ with graded
coordinates like%
\begin{equation}
Z^{\underline{A}}=\left(
\begin{array}{c}
\xi ^{A} \\
\lambda ^{\dot{A}}%
\end{array}%
\right) \qquad \leftrightarrow \qquad \left(
\begin{array}{c}
bosons \\
femions%
\end{array}%
\right)
\end{equation}%
and super label $\underline{A}=(A,\dot{A})$. The bosonic $\xi ^{A}$ is a
vector of $Sp\left( 2N\right) $ with label taking values as $A=1,...,2N$; it
parameterises $\mathcal{E}_{ph}^{2N|0}$. The fermionic $\lambda ^{\dot{A}}$
is a vector of $SO\left( 2N\right) $ with label $\dot{A}=1,...,2\dot{N}$; it
parameterises $\mathcal{E}_{ph}^{0|2N}.$\newline
Because of the $\mathbb{Z}_{2}$- grading of supersymmetry, the
orthosymplectic space $\mathcal{K}$ has two kinds of variables: bosonic
phase space variables $\xi ^{A}$ and fermionic ones $\lambda ^{\dot{A}}$
splitting as follows%
\begin{equation}
\xi ^{A}=\left(
\begin{array}{c}
x^{I} \\
p_{I}%
\end{array}%
\right) \quad ,\quad \lambda ^{\dot{A}}=\left(
\begin{array}{c}
\mathrm{\gamma }_{x}^{\dot{I}} \\
\mathrm{\gamma }_{p}^{\dot{I}}%
\end{array}%
\right)
\end{equation}%
where $x^{I}$ and $p_{I}$ refer to the position variables and their
momentums. The $\mathrm{\gamma }_{x}^{\dot{I}}$ and $\mathrm{\gamma }_{p}^{%
\dot{I}}$ are real fermionic variables; they are the super partners of $%
x^{I} $ and $p_{I}$.

\subsubsection{Bosonic sector of $\mathcal{K}$}

The $\mathcal{K}$ contains the bosonic phase space $\mathcal{E}_{ph}^{2N|0}$
of Hamiltonian systems as a subspace where live the $Sp\left( 2N\right) $
symplectic symmetry. The $\mathcal{E}_{ph}^{2N|0}$ is coordinated by the $%
N+N $ Darboux variables $x^{I},p_{I}$ with label $I=1,...,N$. These real
symplectic variables can be combined into two interesting ways: $\left(
i\right) $ $Sp\left( 2N,\mathbb{R}\right) $ vectors; or $\left( ii\right) $ $%
U\left( N,\mathbb{C}\right) $ complex vectors.

\begin{description}
\item[$\left( \mathbf{1}\right) $] As $Sp\left( 2N,\mathbb{R}\right) $
vectors $\xi ^{A}$ and $\xi _{A}$ given by
\begin{equation}
\xi ^{A}=\left(
\begin{array}{c}
x^{I} \\
p_{I}%
\end{array}%
\right) \qquad ,\qquad \xi _{A}=\left( -p_{I},x^{I}\right)  \label{x1}
\end{equation}%
with $x^{I}=\xi ^{I}$ and $p_{I}=\xi ^{I+N}$. The contravariant $\xi ^{A}$
and the covariant $\xi _{A}$ vectors has 2N components and are related to
each other by the metric as
\begin{equation}
\xi _{A}=\omega _{AB}\xi ^{B}\qquad ,\qquad \xi ^{A}=\Omega ^{AB}\xi _{B}
\end{equation}%
with antisymmetric $\omega _{AB}$ and its inverse $\Omega ^{AB}$ thought of
in terms of their canonical form as follows
\begin{equation}
\omega _{AB}=\left(
\begin{array}{cc}
0 & -I_{N} \\
I_{N} & 0%
\end{array}%
\right) \qquad ,\qquad \Omega ^{AB}=\left(
\begin{array}{cc}
0 & I_{N} \\
-I_{N} & 0%
\end{array}%
\right)
\end{equation}%
Using two vectors $\xi ^{A}$ and $\xi ^{\prime B}$, their symplectic
invariant is given by $\xi _{A}\xi ^{\prime A}=\omega _{AB}\xi ^{B}\xi
^{\prime A}.$ By substituting (\ref{x1}), we get
\begin{equation}
\xi _{A}\xi ^{\prime A}=x^{I}p_{I}^{\prime }-p_{I}x^{\prime I}
\end{equation}%
capturing manifestly the property $\omega _{AB}\xi ^{A}\xi ^{B}=0$ because
of the commutativity property of the bosonic variable namely $\xi ^{A}\xi
^{B}=\xi ^{B}\xi ^{A}$ and anti-commutativity of the symplectic metric $%
\omega _{AB}=-\omega _{BA}.$

\item[$\left( \mathbf{2}\right) $] As $U\left( N,\mathbb{C}\right) $ vectors
by using complex coordinates like
\begin{equation}
z^{I}=x^{I}+ip_{I}\qquad ,\qquad z_{I}^{\ast }=x^{I}-ip_{I}  \label{x2}
\end{equation}%
in term of which the symplectic invariant $\xi _{A}\xi ^{\prime A}$ is given
by the imaginary part of $z_{I}^{\ast }z^{\prime I}$,
\begin{equation}
\xi _{A}\xi ^{\prime A}=\func{Im}\left( z_{I}^{\ast }z^{\prime I}\right)
\label{met}
\end{equation}%
So, the real 2N dimensional bosonic phase space $\mathcal{E}_{ph}^{2N|0}$ is
homomorphic to a N-dimensional complex space that we imagine it here as $%
\mathbb{C}^{N}$ but with metric (\ref{met}). Notice that the complex
variables can be also combined as
\begin{equation}
\zeta ^{A}=\left(
\begin{array}{c}
z^{I} \\
z_{I}^{\ast }%
\end{array}%
\right) \qquad ,\qquad \zeta _{A}=\left( -z_{I}^{\ast },z^{I}\right)
\label{zeta}
\end{equation}%
with reality condition $\left( \zeta ^{A}\right) ^{\ast }=\zeta _{A}$ given
by $\zeta _{A}=\omega _{AB}\zeta ^{B}.$ The passage from the $\zeta $%
-coordinate basis to the $\xi $-coordinate is given by
\begin{equation}
\zeta ^{A}=\mathcal{P}_{B}^{A}\xi ^{B}
\end{equation}%
with%
\begin{equation}
\mathcal{P}_{B}^{A}=\left(
\begin{array}{cc}
\delta _{J}^{I} & i\delta _{J}^{I} \\
\delta _{J}^{I} & -i\delta _{J}^{I}%
\end{array}%
\right)  \label{mp}
\end{equation}
\end{description}

\subsubsection{Fermionic sector of $\mathcal{K}$}

In addition to the bosonic $\mathcal{E}_{ph}^{2N|0}$, the space $\mathcal{K}$
contains also a super extension $\mathcal{E}_{ph}^{0|2N}$. This is an odd
part which is coordinated by the $N+N$ fermionic variables given by ($%
\mathrm{\gamma }_{x}^{I},\mathrm{\gamma }_{pI}).$ For the particular case $%
N=1$, the odd space $\mathcal{E}_{ph}^{0|2}$ has two real variables $(%
\mathrm{\gamma }_{x},\mathrm{\gamma }_{p})$ while the full space $\mathcal{E}%
_{ph}^{2|2}$\ has four real variables: two bosonic $\left( x,p\right) $ and
the two fermionic $(\mathrm{\gamma }_{x},\mathrm{\gamma }_{p})$; that is
super coordinates as%
\begin{equation}
\tilde{Z}=\left( x,p;\mathrm{\gamma }_{x},\mathrm{\gamma }_{p}\right)
\end{equation}%
As far as the generic odd sub-superspace $\mathcal{E}_{ph}^{0|2N}$ is
concerned, notice that its coordinate variables can be also combined into
two ways, the same as for the bosonic $\mathcal{E}_{ph}^{2N|0}$:

\begin{description}
\item[$\left( \mathbf{1}\right) $] As real \textrm{orthogonal} vectors $%
\lambda ^{\dot{A}}$ and $\lambda _{\dot{A}}$\ like
\begin{equation}
\lambda ^{\dot{A}}=\left(
\begin{array}{c}
\mathrm{\gamma }_{x}^{\dot{I}} \\
\mathrm{\gamma }_{p\dot{I}}%
\end{array}%
\right) \qquad ,\qquad \lambda _{\dot{A}}=\left( \mathrm{\gamma }_{x\dot{I}},%
\mathrm{\gamma }_{p}^{\dot{I}}\right)  \label{y1}
\end{equation}%
where we have used dotted labels $\dot{A},\dot{I}$ in order to distinguish
them from the labels $A,I$ of eq(\ref{x1}). The contravariant $\lambda ^{%
\dot{A}}$ and the covariant $\lambda _{\dot{A}}$ vectors are 2N dimensional;
they are linked by
\begin{equation}
\lambda _{\dot{A}}=g_{\dot{A}\dot{B}}\lambda ^{\dot{B}}\qquad ,\qquad
\lambda ^{\dot{A}}=G^{\dot{A}\dot{B}}\lambda _{\dot{B}}
\end{equation}%
with symmetric metric $g_{\dot{A}\dot{B}}$ and its inverse $G^{\dot{A}\dot{B}%
}$ whose canonical forms are as
\begin{equation}
g_{\dot{A}\dot{B}}=\left(
\begin{array}{cc}
I_{N} & 0 \\
0 & I_{N}%
\end{array}%
\right) \qquad ,\qquad G^{\dot{A}\dot{B}}=\left(
\begin{array}{cc}
I_{N} & 0 \\
0 & I_{N}%
\end{array}%
\right)
\end{equation}%
In terms of these orthogonal vectors, we can calculate interesting
quantities like
\begin{equation}
g_{\dot{A}\dot{B}}\lambda ^{\dot{B}}\lambda ^{\prime \dot{A}}=\mathrm{\gamma
}_{x\dot{I}}\mathrm{\gamma }_{x}^{\prime \dot{I}}+\mathrm{\gamma }_{p}^{\dot{%
I}}\mathrm{\gamma }_{p\dot{I}}^{\prime }
\end{equation}%
from which we learn the property $g_{\dot{A}\dot{B}}\lambda ^{\dot{A}%
}\lambda ^{\dot{B}}=0$ because of the anti-commutativity property fermionic
coordinates ($\lambda ^{\dot{A}}\lambda ^{\dot{B}}=-\lambda ^{\dot{B}%
}\lambda ^{\dot{A}}$) and the symmetry $g_{\dot{A}\dot{B}}=g_{\dot{B}\dot{A}%
}.$

\item[$\left( \mathbf{2}\right) $] By using complex fermionic variables
\begin{equation}
\mathrm{\beta }^{\dot{I}}=\mathrm{\gamma }_{x}^{\dot{I}}+i\mathrm{\gamma }_{p%
\dot{I}}\qquad ,\qquad \mathrm{\beta }_{\dot{I}}^{\ast }=\mathrm{\gamma }%
_{x}^{\dot{I}}-i\mathrm{\gamma }_{p\dot{I}}  \label{y2}
\end{equation}%
in term of which we can also define%
\begin{equation}
\mathrm{\chi }^{\dot{A}}=\left(
\begin{array}{c}
\mathrm{\beta }^{\dot{I}} \\
\mathrm{\beta }_{\dot{I}}^{\ast }%
\end{array}%
\right) \qquad ,\qquad \mathrm{\chi }_{\dot{A}}=\left( \mathrm{\beta }_{\dot{%
I}},\mathrm{\beta }^{\ast \dot{I}}\right)  \label{xi}
\end{equation}%
and
\begin{equation}
g_{\dot{A}\dot{B}}\mathrm{\chi }^{\dot{B}}\mathrm{\tilde{\chi}}^{\dot{A}}=%
\mathrm{\beta }_{\dot{I}}\mathrm{\tilde{\beta}}^{\dot{I}}+\mathrm{\beta }%
^{\ast \dot{I}}\mathrm{\tilde{\beta}}_{\dot{I}}^{\ast }
\end{equation}%
with the property $g_{\dot{A}\dot{B}}\mathrm{\chi }^{\dot{B}}\mathrm{\chi }^{%
\dot{A}}=0$.
\end{description}

\section{Supersymmetric oscillators and observables}

In this section, we give the relationship between the triplets (\ref{tr})
and the free supersymmetric oscillator algebra and its observables. We also
construct the super oscillator realisation of the orthosymplectic symmetry
and derive a family of observables\textrm{\ }$\mathcal{O}_{\eta }(\hat{\xi},%
\hat{\lambda})$ which includes the supercharge $Q=Q(\hat{\xi},\hat{\lambda})$
operator and the supersymmetric Hamiltonian\textrm{\ }%
\begin{equation}
Q^{2}=H\qquad ,\qquad H=H_{b}+H_{f}
\end{equation}%
For that we begin by studying the algebra of the $N+N$ quantum super
oscillators. We refer to this supersymmetric algebra as the generalised
super Heisenberg algebra (for short \textsc{SHA}$^{2N|2N}$).

\subsection{The super algebra: \textsc{SHA}$^{2N|2N}$}

The supersymmetric oscillator algebra is an extension of the usual
Heisenberg algebra of quantum bosonic oscillator $\hat{b}=(\hat{x}+i\hat{p})/%
\sqrt{2}$ by harmonic fermionic operators $\hat{c}=(\mathrm{\hat{\gamma}}%
_{x}+i\mathrm{\hat{\gamma}}_{p})/\sqrt{2}.$ Here, the $\hat{b}/\hat{b}%
^{\dagger }$ and the $\hat{c}/\hat{c}^{\dagger }$ satisfy graded
commutations relations; in particular $[\hat{b},\hat{b}^{\dagger }]=1$ and $%
\{\hat{c},\hat{c}^{\dagger }\}=1$ describing \textsc{SHA}$^{2|2}.$ For the
generators of \textsc{SHA}$^{2N|2N}$, we have $2N+2N$ operators
\begin{equation}
\hat{x}^{I},\quad \hat{p}_{I};\quad \mathrm{\hat{\gamma}}_{x}^{\dot{I}%
},\quad \mathrm{\hat{\gamma}}_{p\dot{I}}\qquad \Leftrightarrow \qquad \hat{b}%
,\quad \hat{b}^{\dagger },\quad \hat{c},\quad \hat{c}^{\dagger }
\end{equation}%
in one to one with the even coordinates $x^{I},p_{I}$ ($\hat{b},\hat{b}%
^{\dagger }$); and the odd $\mathrm{\gamma }_{x}^{\dot{I}},\mathrm{\gamma }%
_{p\dot{I}}$ ($\hat{c},\hat{c}^{\dagger }$). These operators obey the
following graded commutators \textrm{\cite{SA},}
\begin{equation}
\begin{tabular}{lllllll}
$\left[ \hat{x}^{I},\hat{p}_{J}\right] $ & $=$ & $i\delta _{J}^{I}$ & $%
\qquad ,\qquad $ & $\{\mathrm{\hat{\gamma}}_{x}^{\dot{I}},\mathrm{\hat{\gamma%
}}_{x}^{\dot{J}}\}$ & $=$ & $\delta ^{\dot{I}\dot{J}}$ \\
$\left[ \hat{x}^{I},\hat{x}^{J}\right] $ & $=$ & $0$ & $\qquad ,\qquad $ & $%
\{\mathrm{\hat{\gamma}}_{\dot{I}p},\mathrm{\hat{\gamma}}_{p\dot{J}}\}$ & $=$
& $\delta _{\dot{I}\dot{J}}$ \\
$\left[ \hat{p}_{I},\hat{p}_{J}\right] $ & $=$ & $0$ & $\qquad ,\qquad $ & $%
\{\mathrm{\hat{\gamma}}_{x}^{\dot{I}},\mathrm{\hat{\gamma}}_{p\dot{J}}\}$ & $%
=$ & $0$%
\end{tabular}
\label{xp}
\end{equation}%
and vanishing crossed relations. For the particular case $N=1,$ the four
generators of \textsc{SHA}$^{2|2}$ are given by $\hat{x},\hat{p},\hat{\gamma}%
_{x},\hat{\gamma}_{p}$; they obey the following non trivial relations
\begin{eqnarray}
\left[ \hat{x},\hat{p}\right] &=&i  \notag \\
\left( \hat{\gamma}_{x}\right) ^{2} &=&\left( \hat{\gamma}_{p}\right) ^{2}=%
\frac{1}{2}I_{id}  \label{gx} \\
\hat{\gamma}_{x}\hat{\gamma}_{p} &=&-\hat{\gamma}_{p}\hat{\gamma}_{x}  \notag
\end{eqnarray}%
A typical realisation of these relations is given by
\begin{equation}
\begin{tabular}{lllllll}
$\hat{x}$ & $=$ & $x$ & $\qquad ,\qquad $ & $\hat{\gamma}_{x}$ & $=$ & $%
\frac{1}{\sqrt{2}}\mathbf{\sigma }_{1}$ \\
$\hat{p}$ & $=$ & $i\frac{\partial }{\partial x}$ & $\qquad ,\qquad $ & $%
\hat{\gamma}_{p}$ & $=$ & $\frac{1}{\sqrt{2}}\mathbf{\sigma }_{2}$%
\end{tabular}%
\end{equation}%
where $\mathbf{\sigma }_{1}=\mathbf{\sigma }_{x}$ and $\mathbf{\sigma }_{2}=%
\mathbf{\sigma }_{y}$ are Pauli matrices. For the generalised super
Heisenberg algebra \textsc{SHA}$^{2N|2N},$ the above realisation extends as
follow%
\begin{equation}
\hat{x}^{I}=x^{I}\qquad ,\qquad \hat{p}_{I}=i\frac{\partial }{\partial x^{I}}
\end{equation}%
and%
\begin{equation}
\hat{\gamma}_{x}^{\dot{I}}=\frac{1}{\sqrt{2}}\Gamma ^{2\dot{I}-1}\qquad
,\qquad \hat{\gamma}_{p\dot{I}}=\frac{1}{\sqrt{2}}\Gamma ^{2\dot{I}}
\end{equation}%
where $\Gamma ^{\dot{A}}=\left( \Gamma ^{2\dot{I}-1},\Gamma ^{2\dot{I}%
}\right) $ are 2N dimensional Clifford algebra%
\begin{equation}
\Gamma ^{\dot{A}}\Gamma ^{\dot{B}}+\Gamma ^{\dot{B}}\Gamma ^{\dot{A}%
}=2\delta ^{\dot{A}\dot{B}}
\end{equation}%
with $\delta ^{\dot{A}\dot{B}}$ thought of as the canonical form of the
orthogonal metric $G^{\dot{A}\dot{B}}$. In this regard, notice that
\begin{equation}
\Sigma ^{\dot{A}\dot{B}}=\frac{1}{2i}(\Gamma ^{\dot{A}}\Gamma ^{\dot{B}%
}-\Gamma ^{\dot{B}}\Gamma ^{\dot{A}})
\end{equation}%
are the generators of the spinor representation of $SO(2N,\mathbb{R}).$

\subsection{The \textsc{SHA}$^{2N|2N}$ and the triplets (\protect\ref{tr})}

We study two aspects of the \textsc{SHA}$^{2N|2N}$ superalgebra. \textrm{%
First}, we show that the graded commutation relations (\ref{xp}) defining
the \textsc{SHA}$^{2N|2N}$ are intimately related with the triplets (\ref{tr}%
) which we recombine into three pairs as follows.
\begin{equation}
\left. \left( \mathbf{\tau ,\tilde{\tau}}\right) \right. =\left( \left.
\begin{array}{c}
\Omega ^{AB} \\
G^{\dot{A}\dot{B}}%
\end{array}%
\right. ,\qquad \left.
\begin{array}{c}
\omega _{AB} \\
g_{\dot{A}\dot{B}}%
\end{array}%
\right. ,\qquad \left.
\begin{array}{c}
J_{B}^{\dot{A}} \\
J_{\dot{A}}^{B}%
\end{array}%
\right. \right)  \label{23}
\end{equation}%
This link indicates that eqs(\ref{xp}) can be expressed in three different,
but equivalent, ways depending on the algebraic structure we want to
exhibit. \textrm{Second}, we construct the orthosymplectic symmetry
underlying the quantum super oscillators and the super Hamiltonian
underlying their dynamics.

\subsubsection{Three bases for \textsc{SHA}$^{2N|2N}$}

Below, we give the three bases we can use to write down the graded
commutators of the \textsc{SHA}$^{2N|2N}$ Lie superalgebra. These three
bases are distinguished by the tensorial properties of the generators namely
$\left( i\right) $ contravariant, $\left( ii\right) $ covariant and $\left(
iii\right) $ mixed.

\begin{description}
\item[$\left( \mathbf{1}\right) $ ] \textbf{Contravariant basis}\emph{\ }$\{%
\hat{\xi}^{A},\hat{\lambda}^{\dot{A}}\}$ \newline
Using the contravariant supersymmetric oscillators operators $\hat{\xi}^{A}$
and $\hat{\lambda}^{\dot{A}},$ associated with super coordinate variables $%
\xi ^{A}$ (\ref{x1}) and $\lambda ^{\dot{A}}$ (\ref{y1}), the graded
commutation relations defining the \textsc{SHA}$^{2N|2N}$ read as follows%
\begin{equation}
\begin{tabular}{lll}
$\lbrack \hat{\xi}^{A},\hat{\xi}^{B}]$ & $=$ & $i\Omega ^{AB}$ \\
$\{\hat{\lambda}^{\dot{A}},\hat{\lambda}^{\dot{B}}\}$ & $=$ & $G^{\dot{A}%
\dot{B}}$%
\end{tabular}
\label{X1}
\end{equation}%
and vanishing others. In this definition, the right hand side of (\ref{X1})
are given by the contravariant symplectic $\Omega ^{AB}$ and orthogonal $G^{%
\dot{A}\dot{B}}$ structures.

\item[$\left( \mathbf{2}\right) $] \textbf{Covariant basis}\emph{\ }$\{\hat{%
\xi}_{A},\hat{\lambda}_{\dot{A}}\}$\newline
The super \textsc{SHA}$^{2N|2N}$ (\ref{xp}) can be also defined by using the
covariant operators $\hat{\xi}_{A}$ and $\hat{\lambda}_{\dot{A}},$
associated with the phase space super variables $\xi _{A}$ and $\lambda _{%
\dot{A}}$. The non vanishing graded commutation relations of $\hat{\xi}_{A}$
and $\hat{\lambda}_{\dot{A}}$ are given by%
\begin{equation}
\begin{tabular}{lll}
$\lbrack \hat{\xi}_{A},\hat{\xi}_{B}]$ & $=$ & $i\omega _{AB}$ \\
$\{\hat{\lambda}_{\dot{A}},\hat{\lambda}_{\dot{B}}\}$ & $=$ & $g_{\dot{A}%
\dot{B}}$%
\end{tabular}
\label{Y1}
\end{equation}%
In this definition, the right hand side of (\ref{Y1}) is given by the
covariant symplectic $\Omega _{AB}$ and the orthogonal $G_{\dot{A}\dot{B}}$
structures.

\item[$\left( \mathbf{3}\right) $] \textbf{Mixed basis} \newline
The superalgebra \textsc{SHA}$^{2N|2N}$ in the mixed basis is defined as
follows,%
\begin{equation}
\begin{tabular}{lll}
$\lbrack \hat{\xi}^{\dot{A}},\hat{\xi}_{B}]$ & $=$ & $-iJ_{B}^{\dot{A}}$ \\
$\{\hat{\lambda}_{\dot{A}},\hat{\lambda}^{B}\}$ & $=$ & $J_{\dot{A}}^{B}$%
\end{tabular}
\label{mixed}
\end{equation}%
where the right hand side is given by the complex structure $J_{B}^{\dot{A}}$
and its transpose $J_{\dot{A}}^{B}$. This complex structure is related to
the symplectic and the orthogonal structures by (\ref{JJ}) namely,%
\begin{equation}
J_{B}^{\dot{A}}=-\sum_{\dot{C}=C=1}^{2N}G^{\dot{A}\dot{C}}\omega _{CB}\qquad
,\qquad J_{\dot{A}}^{B}=\sum_{\dot{C}=C=1}^{2N}\Omega ^{BC}g_{\dot{C}\dot{A}}
\end{equation}%
Notice that the requirement of the condition $\dot{C}=C$ breaks the $%
Sp\left( 2N\right) \times SO\left( 2N\right) $ down to the diagonal $%
S[U\left( 2\right) \times U\left( N\right) ]$. The mixed basis can be
derived from (\ref{X1}-\ref{Y1}) by using eq(\ref{JJ}). Indeed, starting for
example from the superalgebra (\ref{Y1}); then multiplying both sides of the
super commutators by $G^{\dot{A}\dot{C}}$ and $\Omega ^{AC}$ like
\begin{equation}
\begin{tabular}{lll}
$G^{\dot{A}\dot{C}}[\hat{\xi}_{C},\hat{\xi}_{B}]$ & $=$ & $iG^{\dot{A}\dot{C}%
}\omega _{CB}$ \\
$\Omega ^{AC}\{\hat{\lambda}_{\dot{C}},\hat{\lambda}_{\dot{B}}\}$ & $=$ & $%
\Omega ^{AC}g_{\dot{C}\dot{B}}$%
\end{tabular}%
\end{equation}%
and setting
\begin{equation}
\hat{\xi}^{\dot{A}}=\sum_{\dot{C}=C=1}^{2N}G^{\dot{A}\dot{C}}\hat{\xi}%
_{C}\qquad ,\qquad \hat{\lambda}^{A}=\sum_{\dot{C}=C=1}^{2N}\Omega ^{AC}\hat{%
\lambda}_{\dot{C}}
\end{equation}%
we end up exactly with eq(\ref{mixed}).
\end{description}

\subsubsection{From \textsc{SHA}$^{2N|2N}$ to orthosymplectic $osp\left(
2N|2N\right) $}

\begin{figure}[tbph]
\begin{center}
\includegraphics[width=12cm]{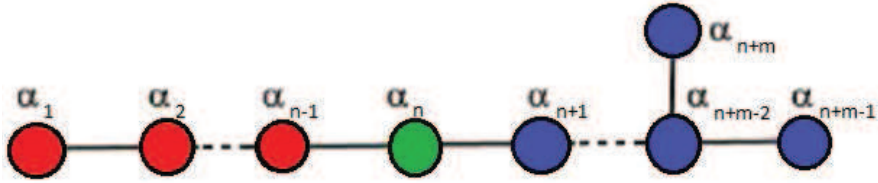}
\end{center}
\par
\vspace{-0.5cm}
\caption{Distinguished super Dynkin diagram of the orthosymplectic Lie
algebra $osp(2m|2n).$ The $\protect\alpha _{i}$'s on the nodes are the
simple roots of the Lie superalgebra. The green node is fermionic while the
(symplectic) red and (orthogonal) blue ones are bosonic.}
\label{osp}
\end{figure}

Given the super oscillators operators $\hat{\xi}^{A}$ and $\hat{\lambda}^{%
\dot{A}}$ obeying the \textsc{SHA}$^{2N|2N}$ (\ref{X1}), we can construct
quantum observables $\mathcal{O}(\hat{\xi},\hat{\lambda})$ given by
polynomials of $\hat{\xi}^{A}$ and $\hat{\lambda}^{\dot{A}}$ as%
\begin{equation}
\mathcal{O}_{\text{\textsc{n}}+\text{\textsc{m}}}=\sum_{k=1}^{\text{\textsc{n%
}}}\sum_{l=1}^{\text{\textsc{m}}}{\large a}_{A_{1}...A_{k}\dot{A}_{1}...\dot{%
A}_{l}}(\hat{\xi}^{A_{1}}...\hat{\xi}^{A_{k}})(\hat{\lambda}^{\dot{A}_{1}}...%
\hat{\lambda}^{\dot{A}_{l}})  \label{ob}
\end{equation}%
As illustrations, we have for \textsc{n}$+$\textsc{m}$=1$, the two basic $%
\mathcal{O}_{\left( \text{\textsc{1}},\text{\textsc{0}}\right) }=\hat{\xi}%
^{A}$ and $\mathcal{O}_{\left( \text{\textsc{0}},\text{\textsc{1}}\right) }=%
\hat{\lambda}^{\dot{A}}$. For \textsc{n}$+$\textsc{m}$=3,$ we have the four
following generators%
\begin{equation}
\begin{tabular}{lllllll}
$\mathcal{O}_{\left( \text{\textsc{3}},\text{\textsc{0}}\right) }$ & : & $%
\hat{\xi}^{A}\hat{\xi}^{B}\hat{\xi}^{C}$ & $\qquad ,\qquad $ & $\mathcal{O}%
_{\left( \text{\textsc{2}},\text{\textsc{1}}\right) }$ & : & $\hat{\xi}^{A}%
\hat{\xi}^{B}\hat{\lambda}^{\dot{C}}$ \\
$\mathcal{O}_{\left( \text{\textsc{1}},\text{\textsc{2}}\right) }$ & : & $%
\hat{\xi}^{A}\hat{\lambda}^{\dot{B}}\hat{\lambda}^{\dot{C}}$ & $\qquad
,\qquad $ & $\mathcal{O}_{\left( \text{\textsc{0}},\text{\textsc{3}}\right)
} $ & : & $\hat{\lambda}^{\dot{A}}\hat{\lambda}^{\dot{B}}\hat{\lambda}^{\dot{%
C}}$%
\end{tabular}%
\end{equation}%
In what follow, we study the interesting set of observables $\mathcal{O}%
_{\left( \text{\textsc{n}},\text{\textsc{m}}\right) }$ given by quadratic
monomials; that is observables $\mathcal{O}_{\left( \text{\textsc{n}},\text{%
\textsc{m}}\right) }$ with labels constrained like \textsc{n}$+$\textsc{m}$%
=2.$ This set is generated by%
\begin{equation}
\mathcal{O}^{AB}=\hat{\xi}^{A}\hat{\xi}^{B},\qquad \mathcal{O}^{A\dot{B}}=%
\hat{\xi}^{A}\hat{\lambda}^{\dot{B}},\qquad \mathcal{O}^{\dot{A}\dot{B}}=%
\hat{\lambda}^{\dot{A}}\hat{\lambda}^{\dot{B}}  \label{bo}
\end{equation}%
We show amongst others that these observables generate the $osp\left(
2N|2N\right) $ orthosymplectic Lie superalgebra. \newline
Recall that the $osp\left( 2N|2N\right) $ Lie superalgebra has two sectors: $%
\left( i\right) $ an even sector $\mathcal{G}_{\bar{0}}$ given by $sp\left(
2N\right) \oplus so\left( 2N\right) $ having $4N^{2}$ bosonic generators. $%
\left( ii\right) $ an odd sector $\mathcal{G}_{\bar{1}}$ given by the ($2N,2%
\dot{N})$ representation of $\mathcal{G}_{\bar{0}}$. This module $\mathcal{G}%
_{\bar{1}}$ has also $4N^{2}$ Fermionic generators. The distinguished Dynkin
super diagram of the orthosymplectic group $OSP\left( 2N|2N\right) $ is
given by the Figure \textbf{\ref{osp}.}

\ \ \ \

\textbf{A)} \textbf{Symplectic} $sp\left( 2N\right) $ \textbf{and orthogonal}
$so\left( 2N\right) $\newline
We begin by noticing that the quadratic monomials of the quantum bosonic
oscillators $\hat{\xi}^{A}$ give the so-called oscillator representation of
the symplectic Lie algebra $sp\left( 2N,\mathbb{R}\right) .$ Similarly, the
quadratic monomials of the fermionic oscillators $\hat{\lambda}^{A}$ give
the oscillator representation of the orthogonal Lie algebra $so\left( 2N,%
\mathbb{R}\right) $ Lie algebras. The $sp\left( 2N\right) $ and $so\left(
2N\right) $ are bosonic Lie algebras with respective generators $\mathcal{K}%
^{\left( AB\right) }$ and $\mathcal{J}^{\left[ AB\right] }$ realised in
terms of the super oscillator operators as given below%
\begin{equation}
\begin{tabular}{c|ccc|c}
\ {\small Lie algebra \ } & \multicolumn{3}{|c|}{\small bosonic generators}
& $\dim $ \\ \hline
$sp\left( 2N\right) $ & $\left.
\begin{array}{c}
\text{ } \\
\text{ }%
\end{array}%
\right. 2\mathcal{K}^{\left( AB\right) }$ & $=$ & $\hat{\xi}^{A}\hat{\xi}%
^{B}+\hat{\xi}^{B}\hat{\xi}^{A}$ & $N\left( 2N+1\right) $ \\ \hline
$so\left( 2N\right) $ & $\left.
\begin{array}{c}
\text{ } \\
\text{ }%
\end{array}%
\right. 2\mathcal{J}^{\left[ \dot{A}\dot{B}\right] }$ & $=$ & $\hat{\lambda}%
^{\dot{A}}\hat{\lambda}^{\dot{B}}-\hat{\lambda}^{\dot{B}}\hat{\lambda}^{\dot{%
A}}$ & $N\left( 2N-1\right) $ \\ \hline
\end{tabular}
\label{ORT}
\end{equation}%
These generators $\mathcal{K}^{\left( AB\right) }$ and $\mathcal{J}^{\left[
AB\right] }$ can be also presented in terms of the usual creation $\hat{b}%
^{\dagger }/\hat{c}^{\dagger }$ and annihilations $\hat{b}/\hat{c}$ as
follows%
\begin{equation}
\mathcal{K}^{AB}=\left(
\begin{array}{cc}
\mathcal{K}^{IJ} & \mathcal{K}_{J}^{I} \\
\mathcal{K}_{I}^{J} & \mathcal{K}_{IJ}%
\end{array}%
\right) \qquad ,\qquad \mathcal{J}^{\dot{A}\dot{B}}=\left(
\begin{array}{cc}
\mathcal{J}^{\dot{I}\dot{J}} & \mathcal{J}_{\dot{J}}^{\dot{I}} \\
\mathcal{J}_{\dot{I}}^{\dot{J}} & \mathcal{J}_{\dot{I}\dot{J}}%
\end{array}%
\right)
\end{equation}%
with%
\begin{equation}
\begin{tabular}{lllllll}
$\mathcal{K}^{IJ}$ & $=$ & $\hat{b}^{I}\hat{b}^{J}$ & $\qquad ,\qquad $ & $%
\mathcal{J}^{\dot{I}\dot{J}}$ & $=$ & $\hat{c}^{\dot{I}}\hat{c}^{\dot{J}}$
\\
$\mathcal{K}_{I}^{J}$ & $=$ & $\hat{b}_{I}^{\dagger }\hat{b}^{J}+\frac{1}{2}%
\delta _{I}^{J}$ & $\qquad ,\qquad $ & $\mathcal{J}_{\dot{I}}^{\dot{J}}$ & $%
= $ & $\hat{c}_{\dot{I}}^{\dagger }\hat{c}^{\dot{J}}-\frac{1}{2}\delta _{%
\dot{I}}^{\dot{J}}$ \\
$\mathcal{K}_{IJ}$ & $=$ & $\hat{b}_{I}^{\dagger }\hat{b}_{J}^{\dagger }$ & $%
\qquad ,\qquad $ & $\mathcal{J}_{\dot{I}\dot{J}}$ & $=$ & $\hat{c}_{\dot{I}%
}^{\dagger }\hat{c}_{\dot{J}}^{\dagger }$%
\end{tabular}
\label{obs}
\end{equation}%
Notice that the $\mathcal{K}_{I}^{J},$ $\mathcal{K}_{J}^{I}$ operators are
the generators of $U\left( N\right) \subset Sp(2N);$ and the $\mathcal{J}_{%
\dot{I}}^{\dot{J}},$ $\mathcal{J}_{\dot{J}}^{\dot{I}}$\ operators are the
generators of $U\left( N\right) ^{\prime }\subset SO(2N).$

\ \ \ \

\textbf{B)} \textbf{Fermionic operators} $\mathcal{F}^{A\dot{B}}$\newline
These operators $\mathcal{F}^{A\dot{B}}$ are given by crossing products of
the bosonic $\hat{\xi}^{A}$ and fermionic $\hat{\lambda}^{\dot{B}}$
oscillator operators like
\begin{equation}
\mathcal{F}^{A\dot{B}}=\hat{\xi}^{A}\hat{\lambda}^{\dot{B}}  \label{FRT}
\end{equation}%
They split in terms of $U\left( N\right) \times U\left( N\right) ^{\prime }$
labels as follows
\begin{equation}
\mathcal{F}^{A\dot{B}}=\left(
\begin{array}{cc}
\hat{b}^{I}\hat{c}^{\dot{J}} & \hat{b}^{I}\hat{c}_{\dot{J}}^{\dagger } \\
\hat{b}_{I}^{\dagger }\hat{c}^{\dot{J}} & \hat{b}_{I}^{\dagger }\hat{c}_{%
\dot{J}}^{\dagger }%
\end{array}%
\right) =\left(
\begin{array}{cc}
\boldsymbol{F}^{I\dot{J}} & \boldsymbol{G}_{\dot{J}}^{I} \\
\boldsymbol{\bar{G}}_{I}^{\dot{J}} & \boldsymbol{\bar{F}}_{I\dot{J}}%
\end{array}%
\right)  \label{FAB}
\end{equation}%
The interesting features of these operators are listed below:\newline
$\left( \mathbf{i}\right) $ they are fermionic generators. \newline
$\left( \mathbf{ii}\right) $ they relate bosons and fermions. We have
\begin{equation}
\{\mathcal{F}^{A\dot{B}},\hat{\lambda}^{\dot{C}}\}=G^{\dot{B}\dot{C}}\hat{\xi%
}^{A}\qquad ,\qquad \left[ \mathcal{F}^{A\dot{B}},\hat{\xi}^{C}\right]
=i\Omega ^{AC}\hat{\lambda}^{\dot{B}}
\end{equation}%
$\left( \mathbf{iii}\right) $ The set of the operators $\mathcal{K}^{AB},$ $%
\mathcal{J}^{\dot{A}\dot{B}}$ and $\mathcal{F}^{A\dot{B}}$ generate the osp$%
\left( 2N|2N\right) $ orthosymplectic Lie superalgebra; it contains $u\left(
N|N\right) $ as a sub- superalgebra. The even sector of these superalgebras
read as
\begin{equation}
\begin{tabular}{lll}
$osp\left( 2N|2N\right) _{\bar{0}}$ & $=$ & $sp\left( 2N\right) \oplus
so\left( 2N\right) $ \\
$u\left( N|N\right) _{\bar{0}}$ & $=$ & $u\left( N\right) \oplus u\left(
N\right) $%
\end{tabular}%
\end{equation}%
while the odd sectors are respectively given by the bi-fundamentals $\left(
2N,2\dot{N}\right) $ for $osp\left( 2N|2N\right) _{\bar{1}};$ and $(N,%
\overline{\dot{N}})$ and $\left( \bar{N},\dot{N}\right) $ for $u\left(
N|N\right) _{\bar{1}}$.

\section{From osp(2N\TEXTsymbol{\vert}2M) to super QM}

In this section, we use properties of the orthosymplectic osp(2N\TEXTsymbol{%
\vert}2M) Lie superalgebra to build Hamiltonians $H_{susy}$ modeling
supersymmetric quantum mechanical systems with two supersymmetric charges ($%
\mathcal{N}=2$ super QM)
\begin{equation}
Q=Q_{1}+iQ_{2}\qquad ,\qquad Q^{\dagger }=Q_{1}-iQ_{2}  \label{2Q}
\end{equation}%
First, we describe the osp(2\TEXTsymbol{\vert}2) orthosymplectic model as a
theory enveloping $\mathcal{N}=2$ super QM. Then we embed the $\mathcal{N}=2$
super QM into the larger osp(2N\TEXTsymbol{\vert}2M) theory with integers $%
M\geq N\geq 1.$\newline
To that purpose, we begin by studying osp(2\TEXTsymbol{\vert}2) theory
having four conserved fermionic charges obeying general graded commutation
relations to be given later. To avoid confusion between the osp(2\TEXTsymbol{%
\vert}2) orthosymplectic (ORTIC) superalgebra and supersymmetric (SUSY)
algebra, we refer to the osp(2\TEXTsymbol{\vert}2) supercharges as $%
Q_{osp_{2|2}}$ (sometimes also as $Q_{ortic}$) and to the supersymmetric
ones like $Q_{susy}$ because the latter obeys extra constraints. The $%
Q_{osp_{2|2}}$ characterising the osp(2\TEXTsymbol{\vert}2) theory is given
by a fermionic operator valued in the odd sector of osp(2\TEXTsymbol{\vert}%
2); by using (\ref{FAB}), we have $Q_{osp_{2|2}}=\sum \Lambda _{\dot{B}A}%
\mathcal{F}^{\dot{B}A}$ expanding like
\begin{equation}
Q_{osp_{2|2}}=t_{I\dot{J}}\boldsymbol{F}^{\dot{J}I}+r_{I}^{\dot{J}}%
\boldsymbol{G}_{\dot{J}}^{I}+w_{\dot{J}}^{I}\boldsymbol{\bar{G}}_{I}^{\dot{J}%
}+s^{I\dot{J}}\boldsymbol{\bar{F}}_{\dot{J}I}\in osp_{(2|2)_{\bar{1}}}
\end{equation}%
where $\Lambda _{\dot{B}A}$ are complex numbers. The Hamiltonian $%
H_{osp_{2|2}}$ of this theory is given by the anticommutator $%
\{Q_{osp_{2|2}},Q_{osp_{2|2}}^{\dagger }\}$; it is valued in the even sector
of osp(2\TEXTsymbol{\vert}2). The $Q_{osp_{2|2}}^{\dagger }$ is the adjoint
conjugate of $Q_{osp_{2|2}};$ in general it is different from $Q_{osp_{2|2}}$%
; but we may also have $Q_{osp_{2|2}}^{\dagger }=Q_{osp_{2|2}}$. So, given $%
Q_{osp_{2|2}}$, we can construct a family of observables in the osp(2%
\TEXTsymbol{\vert}2) theory by using the anticommutators as%
\begin{equation}
\begin{tabular}{lll}
$\{Q_{osp_{2|2}},Q_{osp_{2|2}}^{\dagger }\}$ & $=$ & $H_{osp_{2|2}}$ \\
$\{Q_{osp_{2|2}},Q_{osp2|2}\}$ & $=$ & $Z_{osp_{2|2}}$ \\
$\{Q_{osp_{2|2}}^{\dagger },Q_{osp_{2|2}}^{\dagger }\}$ & $=$ & $%
Z_{osp_{2|2}}^{\dagger }$%
\end{tabular}%
\end{equation}%
valued into $osp(2|2)_{\bar{0}}$. We also have commutators like%
\begin{equation}
\left[ H_{osp_{2|2}},Q_{osp_{2|2}}\right] =\tilde{Q}_{osp_{2|2}}
\end{equation}%
valued into $osp(2|2)_{\bar{1}}$. To engineer supersymmetric quantum
mechanical models (super QM) out of the osp(2\TEXTsymbol{\vert}2) theory, we
have to impose constraints required by supersymmetry (SUSY). A basic set of
such constraints is given by%
\begin{equation}
\left\{ Q_{osp_{2|2}},Q_{osp_{2|2}}\right\} =0\qquad ,\qquad \lbrack
H_{osp_{2|2}},Q_{osp_{2|2}}]=0
\end{equation}%
Under these SUSY constraints, the supercharge $Q_{osp_{2|2}}$ and $%
H_{osp_{2|2}}$ reduce respectively down to $Q_{susy}$ and $H_{susy}$; they
sit in particular subspaces of the odd and even sectors of osp(2\TEXTsymbol{%
\vert}2). Notice that for the case of one complex SUSY charge ($%
Q_{susy}^{\dagger }\neq Q_{susy})$, we talk about $\mathcal{N}=2$ super QM
while for hermitian $Q_{susy},$ we have $\mathcal{N}=1$ super QM; for
further details see appendix A. Notice also that the construction given in
this section can be also viewed as a front matter towards the building of
tight binding models for super AZ matter living in the Brillouin Zone; see
next sections. There, the $Q_{osp_{2|2}}$ and $H_{osp_{2|2}}$ (resp. $%
Q_{susy}$ and $H_{susy}$) should be read as $Q_{\mathbf{k}}^{osp_{2|2}}$ and
$H_{\mathbf{k}}^{osp_{2|2}}$ (resp. $Q_{\mathbf{k}}^{susy}$ and $H_{\mathbf{k%
}}^{susy}$) with $\mathbf{k}$ standing from the momentum variable in the
Brillouin torus.

\subsection{\textbf{The osp(2\TEXTsymbol{\vert}2)} model}

Here, we first construct the osp(2\TEXTsymbol{\vert}2) orthosymplectic model
based on $Q_{osp_{2|2}}$; then we derive the constraint relations towards $%
\mathcal{N}=2$ super QM resting on $Q_{susy}$. After that, we work out
typical solutions for Hamiltonians $H_{susy}$ descending from $Q_{susy}$.

\subsubsection{Oscillator realisation of t\textbf{he osp(2\TEXTsymbol{\vert}%
2) structure}}

The engineering of a simple orthosymplectic model that is invariant under
the graded OSP(2\TEXTsymbol{\vert}2) symmetry relies on using one fermionic $%
\hat{c}/\hat{c}^{\dagger }$ oscillator and one bosonic $\hat{b}/\hat{b}%
^{\dagger }$ describing super particle excitations super QM. In terms of
these quantum oscillators, the eight generators of the osp(2\TEXTsymbol{\vert%
}2) Lie superalgebra are realised as follows:

\begin{itemize}
\item \textbf{Generators of the even sector}\newline
The four bosonic operators generating the $SO(2)\times SP\left( 2\right) $
subsymmetry of OSP(2\TEXTsymbol{\vert}2) are given by%
\begin{equation}
\begin{tabular}{cccccc}
{\small bosonic generators} & : & $\boldsymbol{J}_{0}$ & $\boldsymbol{S}_{0}$
& $\boldsymbol{S}_{+}$ & $\boldsymbol{S}_{-}$ \\
{\small oscillator realisation} & : & $\frac{1}{4}\left( \hat{c}^{\dagger }%
\hat{c}-\hat{c}\hat{c}^{\dagger }\right) $ & $\frac{1}{4}(\hat{b}^{\dagger }%
\hat{b}+\hat{b}\hat{b}^{\dagger })$ & $\frac{1}{2}\hat{b}^{\dagger }\hat{b}%
^{\dagger }$ & $\frac{1}{2}\hat{b}\hat{b}$%
\end{tabular}
\label{JS}
\end{equation}%
with $2\boldsymbol{J}_{0}=(\mathcal{N}_{f}-1/2)$ generating $SO(2)$; and $2%
\boldsymbol{S}_{0}=(\mathcal{N}_{b}+1/2)$ being the Cartan charge operator
of $SP\left( 2,\mathbb{R}\right) .$ These four generators obey the usual
commutations relations of the $so(2)\oplus sp\left( 2\right) $ Lie algebra;
see appendix A.

\item \textbf{Generators of the odd sector}\newline
The four fermionic operators generating the odd sector of osp(2\TEXTsymbol{%
\vert}2) transform in the $\left( 2;2\right) $ representation of $%
SO(2)\times SP\left( 2\right) $; they are denoted like $\boldsymbol{F}%
_{p}^{q}$ with $q=\pm $ labelling the charges of $so(2)$ while $p=\pm $ the
charges of $sp(2).$ The oscillator realisation of the $\boldsymbol{F}%
_{p}^{q} $'s reads as follows%
\begin{equation}
\begin{tabular}{cccccc}
{\small fermionic} & : & $\boldsymbol{F}_{-}^{+}$ & $\boldsymbol{F}_{+}^{-}$
& $\boldsymbol{F}_{+}^{+}$ & $\boldsymbol{F}_{-}^{-}$ \\
{\small realisation} & : & $\hat{c}^{\dagger }\hat{b}$ & $\hat{c}\hat{b}%
^{\dagger }$ & $\hat{c}^{\dagger }\hat{b}^{\dagger }$ & $\hat{c}\hat{b}$%
\end{tabular}
\label{FF}
\end{equation}%
with $\left( \mathbf{i}\right) $ the adjoint conjugations $\left(
\boldsymbol{F}_{-}^{-}\right) ^{\dagger }=\boldsymbol{F}_{+}^{+}$ and $%
\left( \boldsymbol{F}_{-}^{+}\right) ^{\dagger }=\boldsymbol{F}_{+}^{-};$
and $\left( \mathbf{ii}\right) \ $the nilpotency $\left( \boldsymbol{F}%
_{p}^{q}\right) ^{2}=0$ due to $\hat{c}\hat{c}=0$. By using the notation $%
\hat{c}^{q}$ and $\hat{b}_{p},$ the above fermionic generators can be also
presented collectively like $\boldsymbol{F}_{p}^{q}=\hat{c}^{q}\hat{b}_{p}$
with $\hat{\upsilon}^{+}=\hat{\upsilon}^{\dagger }$ and $\hat{\upsilon}^{-}=%
\hat{\upsilon}.$
\end{itemize}

\textbf{A) Graded commutations}\newline
The anticommutation relations between the fermionic operators are given by%
\begin{equation}
\begin{tabular}{lll}
$\left\{ \boldsymbol{F}_{-}^{-},\boldsymbol{F}_{+}^{+}\right\} $ & $=$ & $%
2S_{0}-2J_{0}$ \\
$\left\{ \boldsymbol{F}_{-}^{-},\boldsymbol{F}_{-}^{+}\right\} $ & $=$ & $%
2S_{-}$ \\
$\left\{ \boldsymbol{F}_{-}^{-},\boldsymbol{F}_{+}^{-}\right\} $ & $=$ & $0$%
\end{tabular}%
\qquad ,\qquad
\begin{tabular}{lll}
$\left\{ \boldsymbol{F}_{-}^{+},\boldsymbol{F}_{+}^{-}\right\} $ & $=$ & $%
2S_{0}+2J_{0}$ \\
$\left\{ \boldsymbol{F}_{+}^{-},\boldsymbol{F}_{+}^{+}\right\} $ & $=$ & $%
2S_{+}$ \\
$\left\{ \boldsymbol{F}_{-}^{+},\boldsymbol{F}_{+}^{+}\right\} $ & $=$ & $0$%
\end{tabular}
\label{01}
\end{equation}%
where appear the observables $S_{0}\pm J_{0}$ and where $\boldsymbol{F}%
_{q}^{q}$ does not anticommute with $\boldsymbol{F}_{q}^{-q}$ because they
are equal to $S_{q}.$ The commutation relations between the four fermionic $%
\boldsymbol{F}_{p}^{q}$ and the bosonic step operators $S_{\pm }$ read as
follows
\begin{equation}
\begin{tabular}{lll}
$\left[ S_{-},\boldsymbol{F}_{-}^{q}\right] $ & $=$ & $0$ \\
$\left[ S_{+},\boldsymbol{F}_{+}^{q}\right] $ & $=$ & $0$%
\end{tabular}%
,\qquad
\begin{tabular}{lll}
$\left[ S_{+},\boldsymbol{F}_{-}^{q}\right] $ & $=$ & $-\boldsymbol{F}%
_{+}^{q}$ \\
$\left[ S_{-},\boldsymbol{F}_{+}^{q}\right] $ & $=$ & $+\boldsymbol{F}%
_{-}^{q}$%
\end{tabular}%
\end{equation}%
while the commutators of the $\boldsymbol{F}_{p}^{q}$'s with the Cartan
charge operators are given by $[J_{0},\boldsymbol{F}_{p}^{q}]=q\boldsymbol{F}%
_{p}^{q}/2$ and $[S_{0},\boldsymbol{F}_{p}^{q}]=-p\boldsymbol{F}_{p}^{q}/2$
or equivalently
\begin{equation}
\begin{tabular}{lll}
$\lbrack J_{0}+S_{0},\boldsymbol{F}_{p}^{q}]$ & $=$ & $\frac{q-p}{2}%
\boldsymbol{F}_{p}^{q}$ \\
$\lbrack J_{0}-S_{0},\boldsymbol{F}_{p}^{q}]$ & $=$ & $\frac{q+p}{2}%
\boldsymbol{F}_{p}^{q}$%
\end{tabular}
\label{03}
\end{equation}

\textbf{B) Consequences of eqs(\ref{01}-\ref{03})\newline
}We begin by noticing that in general the four fermionic $\boldsymbol{F}%
_{p}^{q}$ do not commute with the Cartan operators $J_{0}\pm S_{0};$ so they
cannot diagonalised in the same basis. This feature is read from (\ref{03});
for $p=q$, we find that $\boldsymbol{F}_{q}^{q}$ commutes with $J_{0}+S_{0}$%
; but does not commute with $J_{0}-S_{0}$ as shown below%
\begin{equation}
\begin{tabular}{lll}
$\lbrack J_{0}+S_{0},\boldsymbol{F}_{q}^{q}]$ & $=$ & $0$ \\
$\lbrack J_{0}-S_{0},\boldsymbol{F}_{q}^{q}]$ & $=$ & $q\boldsymbol{F}%
_{q}^{q}$%
\end{tabular}%
\end{equation}%
Similarly for $p=-q$, we have a vanishing $[J_{0}-S_{0},\boldsymbol{F}%
_{-q}^{q}]=0$ but a non vanishing commutator $[J_{0}+S_{0},\boldsymbol{F}%
_{-q}^{q}]=q\boldsymbol{F}_{-q}^{q}.$ This violation of the commutation
relation of $J_{0}\pm S_{0}$ with the four fermionic $\boldsymbol{F}_{p}^{q}$
is a feature of the orthosymplectic model. \newline
To find bosonic operators commuting with the four fermionic $\boldsymbol{F}%
_{\pm }^{\pm }$, we need to either reduce osp(2\TEXTsymbol{\vert}2) to
particular sub- superalgebras; or go to the enveloping algebra of osp(2%
\TEXTsymbol{\vert}2); for example by considering the Casimirs of osp(2%
\TEXTsymbol{\vert}2). A tricky way to engineer a quadratic operator $%
\mathcal{C}$ commuting with the four fermionic generators, that is obeying
\begin{equation}
\lbrack \mathcal{C},\boldsymbol{F}_{p}^{q}]=0
\end{equation}%
is by taking advantage of the structure of eq(\ref{03}). Thinking of $%
\mathcal{C}$ as given by the product of the Cartans $\left(
S_{0}+J_{0}\right) $ and $\left( S_{0}-J_{0}\right) $; we end up with $%
\mathcal{C}=S_{0}^{2}-J_{0}^{2}$; thanks to $S_{0}J_{0}=J_{0}S_{0}$. By
computing the commutator between $\boldsymbol{F}_{p}^{q}$ and $%
S_{0}^{2}-J_{0}^{2},$ we find that it vanishes identically; this is because
of the property $q^{2}=p^{2}=1.$

\subsubsection{From osp(2\TEXTsymbol{\vert}2) towards $\mathcal{N}=2$ super
QM}

The odd sector of osp(2\TEXTsymbol{\vert}2) has four fermionic charges given
by the two complex $\boldsymbol{F}_{+}^{+},$ $\boldsymbol{F}_{-}^{+},$ and
their adjoint conjugates $\boldsymbol{F}_{-}^{-},\boldsymbol{F}_{+}^{-}.$
This sector is suggestive for building $\left( \mathbf{i}\right) $
orthosymplectic models with diagonal observables proportional to $%
S_{0}+J_{0} $ and $S_{0}-J_{0}$; and $\left( \mathbf{ii}\right) $ deriving $%
\mathcal{N}=2 $ super \textrm{QM models }with supersymmetric hamiltonians $%
H_{susy}$ obtained by constraining the OSP(2\TEXTsymbol{\vert}2) invariance.
\newline
To build observables $\mathcal{O}_{osp_{2|2}}$ characterising the osp(2%
\TEXTsymbol{\vert}2) models, we consider a complex fermionic charge $%
Q_{osp_{2|2}}$ which is given by a generic linear combination of the osp(2%
\TEXTsymbol{\vert}2) fermionic generators as
\begin{equation}
Q_{osp_{2|2}}=\sum_{p=\pm }\sum_{q=\pm }X_{q}^{p}\boldsymbol{F}%
_{p}^{q}\qquad ,\qquad Q_{osp_{2|2}}^{\dagger }=\sum_{p=\pm }\sum_{q=\pm }%
\boldsymbol{\bar{F}}_{p}^{q}\bar{X}_{q}^{p}  \label{3Q}
\end{equation}%
with $\boldsymbol{F}_{p}^{q}$ as in (\ref{FF}), $\boldsymbol{\bar{F}}%
_{p}^{q}=(\boldsymbol{F}_{p}^{q})^{\dagger }$ and where the $X_{q}^{p}$ is a
complex 2$\times $2 matrix which in tight binding modeling is interpreted in
terms of hoppings. Using this fermionic charge and its adjoint $Q^{\dagger
}, $ we can construct the Hamiltonian $H_{osp_{2|2}}$ describing the OSP(2%
\TEXTsymbol{\vert}2) model. It is given by%
\begin{equation}
H_{osp_{2|2}}=Q_{osp_{2|2}}Q_{osp_{2|2}}^{\dagger }+Q_{osp_{2|2}}^{\dagger
}Q_{osp_{2|2}}  \label{osp22}
\end{equation}%
By substituting (\ref{3Q}) into (\ref{osp22}) while using the short notation
$H_{pr}^{qs}=\{\boldsymbol{F}_{p}^{q},\boldsymbol{F}_{r}^{s}\}$ with
anticommutators $\{\boldsymbol{F}_{p}^{q},\boldsymbol{F}_{r}^{s}\}$ valued
in $so\left( 2\right) \oplus sp\left( 2\right) $ as shown by eq(\ref{01}),
we can present the $H_{osp_{2|2}}$ as follows
\begin{equation}
H_{osp_{2|2}}=\sum X_{q}^{p}H_{pr}^{qs}\bar{X}_{s}^{r}
\end{equation}%
So the hamiltonian $H_{osp_{2|2}}$ is generally valued in $so\left( 2\right)
\oplus sp\left( 2\right) $; and as such it has the typical expansion $f_{0}%
\boldsymbol{S}_{0}+f_{+}\boldsymbol{S}_{-}+f_{-}\boldsymbol{S}_{+}+g_{0}%
\boldsymbol{J}_{0}$ with $f_{0,\pm },$ $g_{0}$ some coupling parameters that
can be read from $X_{q}^{p}\bar{X}_{s}^{r}$; and where $\boldsymbol{S}%
_{0,\pm },\boldsymbol{J}_{0}$ are the generators of $so\left( 2\right)
\oplus sp\left( 2\right) $ given by eq(\ref{JS}). Notice also that for osp(2%
\TEXTsymbol{\vert}2) models, the $Q_{osp_{2|2}}$ is in general not nilpotent
($Q_{osp_{2|2}}^{2}\neq 0$) and the Hamiltonian $H_{osp_{2|2}}$ does not
commute with $Q_{osp_{2|2}};$ i.e:%
\begin{equation}
H_{osp_{2|2}}Q_{osp_{2|2}}\neq Q_{osp_{2|2}}H_{osp_{2|2}}\qquad ,\qquad
Q_{osp_{2|2}}^{2}\neq 0
\end{equation}%
To construct $\mathcal{N}=2$ super QM models out of the osp(2\TEXTsymbol{%
\vert}2) orthosymplectic ones based on $Q_{osp_{(2|2)}}$, we impose the $%
\mathcal{N}=2$ supersymmetric algebra on 1D world line (1D $\mathcal{N}=2$)
which is defined by the following graded commutation relations%
\begin{equation}
\begin{tabular}{lll}
$\left\{ Q,Q^{\dagger }\right\} $ & $=$ & $H_{susy}$ \\
$\left\{ Q,Q\right\} $ & $=$ & $0$ \\
$\left[ H_{susy},Q\right] $ & $=$ & $0$%
\end{tabular}
\label{SSA}
\end{equation}%
In this regards, notice the following: $\left( \mathbf{a}\right) $ the
graded relations (\ref{SSA}) can be also interpreted in terms of the $%
\mathcal{N}=1$ supersymmetry in 2D world sheet generated by the 2D Majorana
operator $Q_{\alpha }=(Q_{1},Q_{2});$ for details \textrm{see appendix A}. $%
\left( \mathbf{b}\right) $ For $Q^{\dagger }=Q$, the relations (\ref{SSA})
reduce to $2Q^{2}=H_{susy}.$ \newline
A simple realisation of the above 1D $\mathcal{N}=2$ superalgebra (\ref{SSA}%
) in terms of the quantum oscillators $\hat{c}$ and $\hat{b}$ is given by
the following family of supercharges%
\begin{equation}
Q=\hat{c}^{\dagger }\hat{B},\qquad Q^{\dagger }=\hat{B}^{\dagger }\hat{c}
\label{ro}
\end{equation}%
where $Q$ has no dependence into $\hat{c}$ and where the bosonic operator $%
\hat{B}$ and its useful properties are given are as follows
\begin{equation}
\begin{tabular}{lll}
$\hat{B}$ & $=$ & $X\hat{b}+\hat{b}^{\dagger }Y$ \\
$\hat{B}^{\dagger }$ & $=$ & $\hat{b}^{\dagger }\bar{X}+\bar{Y}\hat{b}$%
\end{tabular}%
\qquad ,\qquad
\begin{tabular}{ccc}
$\lbrack \hat{B},\hat{B}^{\dagger }]$ & $=$ & $Z$ \\
$X\bar{X}-Y\bar{Y}$ & $=$ & $Z$%
\end{tabular}
\label{or}
\end{equation}%
Here, the $X$ and $Y$ are two complex parameters and $Z=X\bar{X}-Y\bar{Y}$
is the discriminant of the mapping $\hat{b}\rightarrow \hat{B}.$ The
quadratic $\hat{B}^{\dagger }\hat{B}$ and $\hat{B}\hat{B}^{\dagger }$
operators descending from (\ref{or}) read as follows%
\begin{equation}
\begin{tabular}{lll}
$\hat{B}\hat{B}^{\dagger }$ & $=$ & $X\bar{X}\hat{b}\hat{b}^{\dagger }+Y\bar{%
Y}\hat{b}^{\dagger }\hat{b}+X\bar{Y}\hat{b}\hat{b}+Y\bar{X}\hat{b}^{\dagger }%
\hat{b}^{\dagger }$ \\
$\hat{B}^{\dagger }\hat{B}$ & $=$ & $\bar{X}X\hat{b}^{\dagger }\hat{b}+Y\bar{%
Y}\hat{b}\hat{b}^{\dagger }+X\bar{Y}\hat{b}\hat{b}+Y\bar{X}\hat{b}^{\dagger }%
\hat{b}^{\dagger }$%
\end{tabular}
\label{BB}
\end{equation}%
Notice that the relations (\ref{BB}) are just linear combinations of the
generators $\boldsymbol{S}_{0,\pm }$ of the symplectic sp(2,R) subsymmetry
of osp(2\TEXTsymbol{\vert}2); they read as follows,%
\begin{equation}
\begin{tabular}{lll}
$\hat{B}\hat{B}^{\dagger }$ & $=$ & $2\left( X\bar{X}+Y\bar{Y}\right)
\boldsymbol{S}_{0}+2X\bar{Y}\boldsymbol{S}_{-}+2Y\bar{X}\boldsymbol{S}_{+}+%
\frac{1}{2}Z$ \\
$\hat{B}^{\dagger }\hat{B}$ & $=$ & $2\left( X\bar{X}+Y\bar{Y}\right)
\boldsymbol{S}_{0}+2X\bar{Y}\boldsymbol{S}_{-}+2Y\bar{X}\boldsymbol{S}_{+}-%
\frac{1}{2}Z$ \\
$\hat{B}^{\dagger }\hat{B}+\hat{B}\hat{B}^{\dagger }$ & $=$ & $4\left( X\bar{%
X}+Y\bar{Y}\right) \boldsymbol{S}_{0}+4X\bar{Y}\boldsymbol{S}_{-}+4Y\bar{X}%
\boldsymbol{S}_{+}$%
\end{tabular}%
\end{equation}%
where in addition to $\boldsymbol{S}_{0,\pm }$ given by (\ref{JS}) we have a
central charge $Z$ commuting with them. From the particular oscillator
realisation (\ref{ro}), we can perform several calculations and derive first
results of the embedding of 1D $\mathcal{N}=2$ supersymmetry into osp(2%
\TEXTsymbol{\vert}2). Particular results are as listed below:

\begin{itemize}
\item \emph{\ Algebra of }$\hat{B}$\emph{\ and} $\hat{B}^{\dagger }:$\newline
Using eq(\ref{or}), we calculate the useful commutation relations
\begin{equation}
\left[ \hat{B}^{\dagger }\hat{B},\hat{B}^{\dagger }\right] =\hat{B}^{\dagger
}Z\qquad ,\qquad \left[ \hat{B}^{\dagger }\hat{B},\hat{B}\right] =-Z\hat{B}
\end{equation}%
They reduce to the usual $[\hat{b}^{\dagger }\hat{b},\hat{b}^{\dagger }]=%
\hat{b}^{\dagger }$ and $[\hat{b}^{\dagger }\hat{b},\hat{b}]=-\hat{b}$ for
the case $Z=1$ corresponding to $X\bar{X}=1$ and $Y=0.$ From these
relations, we learn that $\pm Z$ are some how charges of the new bosonic
oscillator under the bosonic operator number $\hat{B}^{\dagger }\hat{B}.$

\item \emph{Nilpotency constraint equation} $\left\{ Q,Q\right\} =0:$\newline
We can also check that we have indeed the nilpotency condition $Q^{2}=0$
required by 1D $\mathcal{N}=2$ supersymmetry. This feature follows from the
nilpotency of the fermionic oscillator namely $\left( \hat{c}^{\dagger
}\right) ^{2}=0$ and the commutativity $XY=YX$. Though trivial in this
example, the last commutativity relation is required when embedding $%
\mathcal{N}=2$ super QM models into osp(2N\TEXTsymbol{\vert}2M).

\item \emph{Supersymmetric Hamiltonian }$H_{susy}:$\newline
The realisation of the supersymmetric Hamiltonian operator $H_{susy}$ in
terms of the oscillators $\hat{b}/\hat{b}^{\dagger }$ and $\hat{c}/\hat{c}%
^{\dagger }$ reads as follows
\begin{equation}
\begin{tabular}{lll}
$H_{susy}$ & $=$ & $\hat{B}^{\dagger }\hat{B}+\hat{c}^{\dagger }Z\hat{c}$ \\
& $=$ & $\frac{1}{2}\left( \hat{B}^{\dagger }\hat{B}+\hat{B}^{\dagger }\hat{B%
}\right) +\hat{c}^{\dagger }Z\hat{c}-\frac{1}{2}Z$%
\end{tabular}%
\end{equation}%
with $\hat{B}^{\dagger }\hat{B}$ given by (\ref{BB}). In terms of the $%
SO(2,R)\times SP(2,R)$ generators $\boldsymbol{J}_{0}$ and $\boldsymbol{S}%
_{0,\pm }$, the above supersymmetric Hamiltonian is given by the following
linear combination%
\begin{equation}
H_{susy}=2\left( X\bar{X}+Y\bar{Y}\right) \boldsymbol{S}_{0}+2X\bar{Y}%
\boldsymbol{S}_{-}+2Y\bar{X}\boldsymbol{S}_{+}+2Z\boldsymbol{J}_{0}
\end{equation}%
with $Z=X\bar{X}-Y\bar{Y}.$ From this supersymmetric Hamiltonian, we learn
the bosonic and the fermionic contributions to $H_{susy}$ namely
\begin{equation}
\begin{tabular}{lll}
$H_{bose}$ & $=$ & $2\left( X\bar{X}+Y\bar{Y}\right) \boldsymbol{S}_{0}+2X%
\bar{Y}\boldsymbol{S}_{-}+2Y\bar{X}\boldsymbol{S}_{+}$ \\
$H_{fermi}$ & $=$ & $2Z\boldsymbol{J}_{0}$%
\end{tabular}%
\end{equation}

\item \emph{The commutation }$H_{susy}Q=QH_{susy}:$\newline
Using the oscillator realisation (\ref{ro}-\ref{or}) of the supercharge $Q$
and the supersymmetric Hamiltonian $H_{susy}$, we calculate the commutator $%
[H_{susy},Q]$; we find that is equal to $\hat{c}^{\dagger }Z\hat{B}-\hat{c}%
^{\dagger }Z\hat{B}$ which vanishes identically. In this regard, notice the
two following: $\left( \mathbf{i}\right) $ Given a supersymmetric highest
weight state $\left\vert \mathrm{\phi }\right\rangle $ (ground state)
constrained as%
\begin{equation}
Q\left\vert \mathrm{\phi }\right\rangle =0,\qquad H\left\vert \mathrm{\phi }%
\right\rangle =\varepsilon _{\mathrm{\phi }}\left\vert \mathrm{\phi }%
\right\rangle  \label{ef}
\end{equation}%
its super partner $\left\vert \mathrm{\psi }\right\rangle $ is given by $%
Q^{\dagger }\left\vert \mathrm{\phi }\right\rangle $ with the property $%
H_{susy}\left\vert \mathrm{\psi }\right\rangle =\varepsilon _{\mathrm{\phi }%
}\left\vert \mathrm{\psi }\right\rangle ;$ thanks to the commutation $%
H_{susy}Q=QH_{susy}.$ Obviously such property does not hold for $%
Q_{osp_{2|2}}$ because $H_{osp_{2|2}}Q_{osp_{2|2}}$ differs from $%
Q_{osp_{2|2}}H_{osp_{2|2}}.$ $\left( \mathbf{ii}\right) $ By thinking about $%
H_{susy}$ in terms of its bosonic contribution $H_{bose}=\hat{B}^{\dagger }%
\hat{B}$ and the fermionic $H_{fermi}=\hat{c}^{\dagger }Z\hat{c}$ as well as
on free total energy $\varepsilon _{\mathrm{\phi }}$ as the sum $\varepsilon
_{\mathrm{\phi }}^{fermi}+\varepsilon _{\mathrm{\phi }}^{bose}$, we learn
that neither $H_{bose}$ commutes with $Q$ nor $H_{fermi}$ commutes with $Q$
since we have%
\begin{equation}
\left[ H_{bose},Q\right] =-ZQ\qquad ,\qquad \left[ H_{fermi},Q\right] =+ZQ
\end{equation}%
But, from these remarkable relations, we learn that
\begin{equation*}
\left[ H_{bose}^{2},Q\right] =\left[ H_{fermi}^{2},Q\right] =Z^{2}Q
\end{equation*}%
with $\left[ D^{2},Q\right] $ given by the adjoint action $\left[ D,\left[
D,Q\right] \right] $ and where here the operator $D$ stands for $H_{bose}$
and $H_{fermi}.$ Moreover, using (\ref{ef}), we end up with
\begin{equation}
\begin{tabular}{lll}
$H_{bose}\left\vert \mathrm{\psi }\right\rangle $ & $=$ & $\left(
\varepsilon _{\mathrm{\phi }}^{bose}-Z\right) \left\vert \mathrm{\psi }%
\right\rangle $ \\
$H_{fermi}\left\vert \mathrm{\psi }\right\rangle $ & $=$ & $\left(
\varepsilon _{\mathrm{\phi }}^{fermi}+Z\right) \left\vert \mathrm{\psi }%
\right\rangle $%
\end{tabular}%
\end{equation}%
So for bosonic ground states with $\varepsilon _{\mathrm{\phi }%
}^{bose}=\varepsilon _{\mathrm{\phi }}^{fermi}=0,$ their fermionic partners $%
\left\vert \mathrm{\psi }\right\rangle $ have also vanishing energy $%
\varepsilon _{\mathrm{\phi }}$ but with opposite contributions $\pm Z$ from
bosonic and fermionic excitations.
\end{itemize}

\subsection{$\mathcal{N}=2$ super QM within OSp(2N\TEXTsymbol{\vert}2M)}

A more involved realisation of orthosymplectic osp(2\TEXTsymbol{\vert}2)
models and consequently the $\mathcal{N}=2$ super QM ones is given by
embedding osp(2\TEXTsymbol{\vert}2) into OSp(2N\TEXTsymbol{\vert}2M) with $%
M\geq N\geq 1$. Here, the modeling relies on using N fermionic $\hat{c}^{i}/%
\hat{c}_{i}^{\dagger }$ oscillators and M bosonic $\hat{b}^{\alpha }/\hat{b}%
_{\alpha }^{\dagger }$ with label $i$ running from $1$ to $N$ and label $%
\alpha =1,...,M$. But later, we will restrict the study to $M=N.$

\subsubsection{Realisation of OSp(2N\TEXTsymbol{\vert}2M)}

In terms of these N+M quantum graded oscillators $\hat{c}^{i}$ and $\hat{b}%
^{\alpha }$, the generators of the osp(2N\TEXTsymbol{\vert}2M) Lie
superalgebra are realised as follows:

\textbf{1) Even sector of OSp(2N\TEXTsymbol{\vert}2M)}\newline
The even part of OSp(2N\TEXTsymbol{\vert}2M) is given by $SO(2N)\times
Sp(2M) $; the oscillator realisation of the orthogonal $SO(2N)$ and the
symplectic $Sp(2M)$ generators is given by
\begin{equation}
\begin{tabular}{ccc|ccc}
\multicolumn{3}{c|}{$SO(2N)$} & \multicolumn{3}{|c}{$Sp(2M)$} \\ \hline
$\mathcal{O}_{i}^{j}$ & $=$ & $\frac{1}{4}\left( \hat{c}_{i}^{\dagger }\hat{c%
}^{j}-\hat{c}^{j}\hat{c}_{i}^{\dagger }\right) $ & $\mathcal{S}_{\alpha
}^{\beta }$ & $=$ & $\frac{1}{4}\left( \hat{b}_{\alpha }^{\dagger }\hat{b}%
^{\beta }+\hat{b}^{\beta }\hat{b}_{\alpha }^{\dagger }\right) $ \\
$\mathcal{O}^{\left[ ij\right] }$ & $=$ & $\frac{1}{4}\left( \hat{c}^{i}\hat{%
c}^{j}-\hat{c}^{j}\hat{c}^{i}\right) $ & $\mathcal{S}^{\left[ \alpha \beta %
\right] }$ & $=$ & $\frac{1}{4}\left( \hat{b}^{\alpha }\hat{b}^{\beta }+\hat{%
b}^{\beta }\hat{b}^{\alpha }\right) $ \\
$\mathcal{O}_{\left[ ij\right] }^{\dagger }$ & $=$ & $\frac{1}{4}\left( \hat{%
c}_{i}^{\dagger }\hat{c}_{j}^{\dagger }-\hat{c}_{j}^{\dagger }\hat{c}%
_{i}^{\dagger }\right) $ & $\mathcal{S}_{\left[ \alpha \beta \right]
}^{\dagger }$ & $=$ & $\frac{1}{4}\left( \hat{b}_{\alpha }^{\dagger }\hat{b}%
_{\beta }^{\dagger }+\hat{b}_{\beta }^{\dagger }\hat{b}_{\alpha }^{\dagger
}\right) $ \\ \hline
\end{tabular}
\label{osp2n}
\end{equation}%
The $N+M$ commuting Cartan generators are as follows%
\begin{equation}
J_{i}=\frac{1}{4}\left( \hat{c}_{i}^{\dagger }\hat{c}^{i}-\hat{c}^{i}\hat{c}%
_{i}^{\dagger }\right) ,\qquad S_{\alpha }=\frac{1}{4}(\hat{b}_{\alpha
}^{\dagger }\hat{b}^{\alpha }+\hat{b}^{\alpha }\hat{b}_{\alpha }^{\dagger })
\label{osp2m}
\end{equation}%
with no summation on the labels.

\textbf{2) Odd sector of OSp(2N\TEXTsymbol{\vert}2M)}\newline
The 4NM fermionic generators of OSp(2N\TEXTsymbol{\vert}2M) are given by%
\begin{equation}
\begin{tabular}{lllllll}
$\boldsymbol{G}_{i}^{\alpha }$ & $=$ & $\hat{c}_{i}^{\dagger }\hat{b}%
^{\alpha }$ & $\qquad ,\qquad $ & $\boldsymbol{F}^{i\alpha }$ & $=$ & $\hat{c%
}^{i}\hat{b}^{\alpha }$ \\
$\boldsymbol{\bar{G}}_{\alpha }^{i}$ & $=$ & $\hat{c}^{i}\hat{b}_{\alpha
}^{\dagger }$ & $\qquad ,\qquad $ & $\boldsymbol{\bar{F}}_{i\alpha }$ & $=$
& $\hat{c}_{i}^{\dagger }\hat{b}_{\alpha }^{\dagger }$%
\end{tabular}%
\end{equation}%
they carry eigenvalue charges under the $J_{i}$ and the $S_{\alpha }$
operators generating the Cartan subsymmetry of OSp(2N\TEXTsymbol{\vert}2M).
For example, we have $[J_{i},\boldsymbol{G}_{l}^{\alpha }]=+\frac{1}{2}%
\delta _{il}\boldsymbol{G}_{i}^{\alpha }$\ and $[S_{\alpha },\boldsymbol{G}%
_{l}^{\beta }]=-\frac{1}{2}\delta _{\alpha \beta }\boldsymbol{G}_{i}^{\beta
}.$

\subsubsection{1D $\mathcal{N}=2$ supersymmetry}

A particular realisation of the $1D$ $\mathcal{N}=2$ supersymmetric charges $%
Q$ and $Q^{\dagger }$ is given by the extension of the representation (\ref%
{ro}) reading as follows
\begin{equation}
Q=\hat{c}_{i}^{\dagger }\hat{B}^{i}\qquad ,\qquad Q^{\dagger }=\hat{B}%
_{i}^{\dagger }\hat{c}^{i}  \label{cB}
\end{equation}%
constrained by\ the supersymmetric constraints $Q^{2}=0$ and $[H_{susy},Q]=0$
with hamiltonian. $H_{susy}=\left\{ Q,Q^{\dagger }\right\} .$ In these
expressions, the bosonic operators $\hat{B}^{i}$ and $\hat{B}_{i}^{\dagger }$
are defined by the following linear combinations%
\begin{equation}
\hat{B}^{i}=X_{\alpha }^{i}\hat{b}^{\alpha }+\hat{b}_{\alpha }^{\dagger
}Y^{\alpha i}\qquad ,\qquad \hat{B}_{i}^{\dagger }=\bar{Y}_{i\alpha }\hat{b}%
^{\alpha }+\hat{b}_{\alpha }^{\dagger }\bar{X}_{i}^{\alpha }  \label{Xb}
\end{equation}%
where the complex coupling tensors $X_{\alpha }^{i}$ and $Y^{\alpha i}$ are
respectively $N\times M$ and $M\times N$ rectangular matrices; they will be
constrained below by imposing the superalgebra (\ref{SSA}). From the
relations (\ref{cB}-\ref{Xb}), we can calculate useful quantities; in
particular the following ones:

\begin{itemize}
\item \emph{Algebra of }$\hat{B}^{i}$\emph{'s and }$\hat{B}_{j}^{\dagger }$%
\emph{'s}\newline
It is given by the following commutations
\begin{equation}
\lbrack \hat{B}^{i},\hat{B}_{j}^{\dagger }]=Z_{j}^{i}\qquad ,\qquad
Z_{j}^{i}=X_{\alpha }^{i}\bar{X}_{j}^{\alpha }-\bar{Y}_{j\alpha }Y^{\alpha i}
\label{BBB}
\end{equation}%
and%
\begin{equation}
\begin{tabular}{lllllll}
$\lbrack \hat{B}^{i},\hat{B}^{j}]$ & $=$ & $\Delta ^{\left[ ij\right] }$ & $%
\qquad ,\qquad $ & $\Delta ^{\left[ ij\right] }$ & $=$ & $X_{\alpha
}^{i}Y^{\alpha j}-X_{\alpha }^{j}Y^{\alpha i}$ \\
$\lbrack \hat{B}_{i}^{\dagger },\hat{B}_{j}^{\dagger }]$ & $=$ & $\bar{\Delta%
}_{ij}$ & $\qquad ,\qquad $ & $\bar{\Delta}_{ij}$ & $=$ & $\bar{Y}_{i\alpha }%
\bar{X}_{j}^{\alpha }-\bar{Y}_{j\alpha }\bar{X}_{i}^{\alpha }$%
\end{tabular}%
\end{equation}%
As far as these commutations are concerned, notice the two following: $%
\left( \mathbf{i}\right) $ the commutation $\hat{B}^{i}\hat{B}^{j}=\hat{B}%
^{j}\hat{B}^{i}$ requires $\Delta ^{\left[ ij\right] }=0$. $\left( \mathbf{ii%
}\right) $ By expressing this antisymmetric tensor like%
\begin{equation}
\Delta ^{\left[ ij\right] }=\left( X.Y\right) ^{ij}-\left( X.Y\right) ^{ji}
\end{equation}
with $\left( X.Y\right) ^{ij}=X_{\alpha }^{i}Y^{\alpha j}$, the vanishing
condition $\Delta ^{\left[ ij\right] }=0$ can be solved by taking $\left(
X.Y\right) ^{ij}=\eta _{XY}G^{ij}$ with $G^{ij}$ a symmetric tensor and $%
\eta _{XY}$ a complex parameter.

\item \emph{Supersymmetric Hamiltonian}\newline
The supersymmetric Hamiltonian is defined by $\left\{ Q,Q^{\dagger }\right\}
=H_{susy}$; by substituting (\ref{cB}-\ref{Xb}) and using the algebra of the
$\hat{B}^{i}$'s, we obtain%
\begin{equation}
H_{susy}=\hat{B}_{i}^{\dagger }\hat{B}^{i}+\hat{c}_{i}^{\dagger }Z_{j}^{i}%
\hat{c}^{j}
\end{equation}%
having the property $H_{bose}+H_{fermi}$ with%
\begin{equation}
\begin{tabular}{lll}
$H_{bose}$ & $=$ & $\hat{B}_{i}^{\dagger }\hat{B}^{i}$ \\
$H_{fermi}$ & $=$ & $\hat{c}_{i}^{\dagger }Z_{j}^{i}\hat{c}^{j}$%
\end{tabular}%
\end{equation}%
The bosonic Hamiltonian $\hat{B}_{i}^{\dagger }\hat{B}^{i}$ can be also
presented in other ways like: $\left( \mathbf{i}\right) $ in terms of $\hat{b%
}^{\alpha }$'s and $\hat{b}_{\beta }^{\dagger }$s as follows
\begin{equation}
\begin{tabular}{lll}
$\hat{B}_{i}^{\dagger }\hat{B}^{i}$ & $=$ & $\hat{b}_{\beta }^{\dagger }(%
\bar{X}_{i}^{\beta }X_{\alpha }^{i})\hat{b}^{\alpha }+\hat{b}_{\beta
}^{\dagger }\left( Y^{\beta i}\bar{X}_{i}^{\alpha }\right) \hat{b}_{\alpha
}^{\dagger }+$ \\
&  & $\hat{b}^{\beta }\left( X_{\beta }^{i}\bar{Y}_{i\alpha }\right) \hat{b}%
^{\alpha }+\hat{b}^{\beta }\left( \bar{Y}_{i\beta }Y^{\alpha i}\right) \hat{b%
}_{\alpha }^{\dagger }$%
\end{tabular}%
\end{equation}%
$\left( \mathbf{ii}\right) $ in the matrix language as%
\begin{equation}
H_{bose}=\left( \hat{b}_{\beta }^{\dagger },\hat{b}^{\beta }\right) \left(
\begin{array}{cc}
\bar{X}_{i}^{\beta }X_{\alpha }^{i} & Y^{\beta i}\bar{X}_{i}^{\alpha } \\
X_{\beta }^{i}\bar{Y}_{i\alpha } & \bar{Y}_{i\beta }Y^{\alpha i}%
\end{array}%
\right) \left(
\begin{array}{c}
\hat{b}^{\alpha } \\
\hat{b}_{\alpha }^{\dagger }%
\end{array}%
\right)
\end{equation}%
and $\left( \mathbf{iii}\right) $ by using the generators of $SO(2N)\times
SP\left( 2M\right) ,$
\begin{equation}
\begin{tabular}{lll}
$\hat{B}_{i}^{\dagger }\hat{B}^{i}$ & $=$ & $2\left( X_{\alpha }^{i}\bar{Y}%
_{i\beta }\right) \boldsymbol{S}^{\left( \alpha \beta \right) }+2Y^{\alpha i}%
\bar{X}_{i}^{\beta }\boldsymbol{\bar{S}}_{\left( \alpha \beta \right) }$ \\
&  & $+2\left( \bar{X}_{i}^{\alpha }X_{\beta }^{i}+Y^{\alpha i}\bar{Y}%
_{i\beta }\right) \boldsymbol{S}_{\alpha }^{\beta }-\frac{1}{2}Z$%
\end{tabular}%
\end{equation}%
with $Z=tr\left( Z_{j}^{i}\right) $ given by%
\begin{equation}
Z=\bar{X}_{i}^{\alpha }X_{\alpha }^{i}-Y^{\alpha i}\bar{Y}_{i\alpha }
\end{equation}

\item \emph{the conditions} $\left\{ Q,Q\right\} =\left[ H_{susy},Q\right]
=0 $\newline
First, the nilpotency property $Q^{2}=0$ of the supersymmetric algebra
follows from two things: $\left( \mathbf{i}\right) $ the anticommutations $%
\hat{c}_{i}^{\dagger }\hat{c}_{j}^{\dagger }=-\hat{c}_{j}^{\dagger }\hat{c}%
_{i}^{\dagger }$ which usually hold; and $\left( \mathbf{ii}\right) $ the
commutations $\hat{B}^{i}\hat{B}^{j}=\hat{B}^{j}\hat{B}^{i}$ which are
ensured by demanding $\Delta ^{\left[ ij\right] }=0$ solved by $\left(
X.Y\right) ^{ij}=\eta _{XY}G^{\left( ij\right) }.$ \newline
Regarding the vanishing of the commutation relation $\left[ H_{susy},Q\right]
,$ we use the splitting $H_{susy}=H_{bose}+H_{fermi}$; then calculate first $%
[H_{bose},Q],$ which by using the previous relationships, leads to $-\hat{c}%
_{l}^{\dagger }Z_{i}^{l}\hat{B}^{i}$ with $Z_{i}^{l}\ $as in (\ref{BBB}).
Doing the same thing for $[H_{fermi},Q],$ we end up with the value $\hat{c}%
_{i}^{\dagger }Z_{j}^{i}\hat{B}^{j}$ which cancels the previous contribution.
\end{itemize}

\subsection{Towards tight binding modeling}

In this subsection, we give a useful parameterisation of the fermionic and
the bosonic oscillators to be used in the construction of tight binding
modeling of super AZ matter.

\subsubsection{The $\hat{c}/\hat{c}^{\dagger }$ and $\hat{b}/\hat{b}%
^{\dagger }$ as local field operators}

Here, we will think about the fermionic $\hat{c}/\hat{c}^{\dagger }$ and the
bosonic $\hat{b}/\hat{b}^{\dagger }$ operators, used in the building of the
fermionic charges $Q,$ in terms of local field operators as $\hat{c}\left(
\mathbf{r}_{i}\right) /\hat{c}^{\dagger }\left( \mathbf{r}_{i}\right) $ and $%
\hat{b}\left( \mathbf{r}_{i}\right) /\hat{b}^{\dagger }\left( \mathbf{r}%
_{i}\right) .$ These local fields living on lattice with coordinates $%
\mathbf{r}_{i}$ will be used later for the study of the tight binding
modeling of super AZ matter.

\ \ \

\textbf{A) Local fermionic oscillators}\newline
The complex fermionic $\sqrt{2}\hat{c}^{\dot{I}}{\small (\mathbf{r})}=\hat{%
\gamma}^{\dot{I}}{\small (\mathbf{r})}+i\hat{\eta}_{\dot{I}}{\small (\mathbf{%
r})}$ and its adjoint $\sqrt{2}\hat{c}_{^{\dot{I}}}^{\dagger }{\small (%
\mathbf{r})}=\hat{\gamma}^{\dot{I}}{\small (\mathbf{r})}-i\hat{\eta}_{\dot{I}%
}{\small (\mathbf{r})}$ combine into the orthogonal $SO(2N)$ vector $\hat{%
\lambda}^{\dot{A}}{\small (\mathbf{r})}=(\hat{c}^{\dot{I}}{\small (\mathbf{r}%
)},\hat{c}_{^{\dot{I}}}^{\dagger }{\small (\mathbf{r})})$ with $\dot{I}%
=1,...,N$ and off diagonal metric $g_{\dot{A}\dot{B}}.$ In this
parametrisation, the $\hat{c}^{\dot{I}}{\small (\mathbf{r})}$ transforms in
the fundamental representation of the maximal unitary $U\left( N\right) $
contained in $SO(2N)$ and the $\hat{c}_{^{\dot{I}}}^{\dagger }{\small (%
\mathbf{r})}$ transforms in the anti-fundamental. Notice that, it is the
hermitian vector operator%
\begin{equation}
\hat{\digamma}^{\dot{A}}{\small (\mathbf{r})}=\left(
\begin{array}{c}
\hat{\gamma}^{\dot{I}}{\small (\mathbf{r})} \\
\hat{\eta}_{\dot{I}}{\small (\mathbf{r})}%
\end{array}%
\right)
\end{equation}%
made of the Majoranas that transforms under the vector representation of $%
SO(2N)$ with diagonal metric $\delta _{\dot{A}\dot{B}}$. The passage between
the two frames is given by $\hat{\lambda}^{\dot{A}}{\small (\mathbf{r})}=V_{%
\dot{B}}^{\dot{A}}\hat{\digamma}^{\dot{B}}{\small (\mathbf{r})}$ with
\begin{equation}
V_{\dot{B}}^{\dot{A}}=\frac{1}{\sqrt{2}}\left(
\begin{array}{cc}
\delta _{\dot{J}}^{\dot{I}} & i\delta _{\dot{I}}^{\dot{J}} \\
\delta _{\dot{J}}^{\dot{I}} & -i\delta _{\dot{J}}^{\dot{I}}%
\end{array}%
\right)
\end{equation}

\textbf{B) Local bosonic oscillators}\newline
The complex bosonic $\sqrt{2}\hat{b}^{I}{\small (\mathbf{r})}=\hat{X}^{I}%
{\small (\mathbf{r})}+i\hat{P}_{I}{\small (\mathbf{r})}$ and its adjoint $%
\sqrt{2}\hat{b}_{I}^{\dagger }{\small (\mathbf{r})}=\hat{X}^{I}{\small (%
\mathbf{r})}-i\hat{P}_{I}{\small (\mathbf{r})}$ form the symplectic $SP(2M)$
vector $\hat{\xi}^{A}{\small (\mathbf{r})}=(\hat{b}^{I}{\small (\mathbf{r})},%
\hat{b}_{I}^{\dagger }{\small (\mathbf{r})})$ with label $I$ running from $1$
to $M.$ In this parametrisation, the $\hat{b}^{I}{\small (\mathbf{r})}$
transforms in the fundamental representation of the maximal unitary $U\left(
M\right) $ contained in $SP(2M)$ and the $\hat{b}_{I}^{\dagger }{\small (%
\mathbf{r})}$ transforms in the anti-fundamental. Here also notice that, it
is the real vector%
\begin{equation}
\hat{\phi}^{A}\left( \mathbf{r}\right) =\left(
\begin{array}{c}
\hat{X}^{I}\left( \mathbf{r}\right) \\
\hat{P}_{I}\left( \mathbf{r}\right)%
\end{array}%
\right)
\end{equation}%
that transforms with the usual antisymmetric symplectic $\omega _{AB}.$ The
bridge between the two frames is given by $\hat{\xi}^{A}{\small (\mathbf{r})}%
=U_{B}^{A}\hat{\phi}^{B}{\small (\mathbf{r})}$ with
\begin{equation}
U_{B}^{A}=\frac{1}{\sqrt{2}}\left(
\begin{array}{cc}
\delta _{J}^{I} & i\delta _{I}^{J} \\
\delta _{J}^{I} & -i\delta _{I}^{J}%
\end{array}%
\right)
\end{equation}

\textbf{C) Oscillators on lattice} $\mathbb{L}$\newline
The graded symmetry of the above super oscillator system $(\hat{c}^{\dot{I}},%
\hat{b}^{I})$ is given by the orthosymplectic OSP(2N\TEXTsymbol{\vert}2M).
It contains the $SP(2N)\times SP(2M)$ invariance and the $U(N)\times U(M)$
group as bosonic subsymmetries. Moreover, to build tight binding super
models, we have to think about the bosonic $\hat{\xi}^{A}$ and the fermionic
$\hat{\lambda}^{\dot{A}}$ oscillators as local lattice QFT$_{d}$ operators
labeled like
\begin{equation}
\hat{\xi}^{A}\left( \mathbf{r}_{i}\right) =\left(
\begin{array}{c}
\hat{b}^{I}\left( \mathbf{r}_{i}\right) \\
\hat{b}_{I}^{\dagger }\left( \mathbf{r}_{i}\right)%
\end{array}%
\right) \qquad ,\qquad \hat{\lambda}^{\dot{A}}\left( \mathbf{r}_{i}\right)
=\left(
\begin{array}{c}
\hat{c}^{\dot{I}}\left( \mathbf{r}_{i}\right) \\
\hat{c}_{^{\dot{I}}}^{\dagger }\left( \mathbf{r}_{i}\right)%
\end{array}%
\right)  \label{pa}
\end{equation}%
and%
\begin{equation}
\hat{\xi}_{A}^{\dagger }\left( \mathbf{r}_{i}\right) =\left( \hat{b}%
_{I}^{\dagger }\left( \mathbf{r}_{i}\right) ,\hat{b}^{I}\left( \mathbf{r}%
_{i}\right) \right) \qquad ,\qquad \hat{\lambda}_{\dot{A}}^{\dagger }\left(
\mathbf{r}_{i}\right) =\left( \hat{c}_{^{\dot{I}}}^{\dagger }\left( \mathbf{r%
}_{i}\right) ,\hat{c}^{\dot{I}}\left( \mathbf{r}_{i}\right) \right)
\label{ap}
\end{equation}%
with variables $\mathbf{r}_{i}$ referring to the oscillators' positions in
the hyper cubic lattice $\mathbb{L}$ as in the Figure \textbf{\ref{D2}}.
\textrm{By using the hermitian phase space operators }$(\hat{X}^{I}{\small (%
\mathbf{r}_{i})},\hat{P}_{I}{\small (\mathbf{r}_{i})})$\textrm{\ and the
Majoranas }$(\hat{\gamma}^{\dot{I}}{\small (\mathbf{r}_{i})},\hat{\eta}_{%
\dot{I}}{\small (\mathbf{r}_{i})}$\textrm{) that make the local complex
bosonic }$\hat{b}^{I}{\small (\mathbf{r}_{i})}$ and \textrm{the local
complex fermionic} $\hat{c}^{\dot{I}}{\small (\mathbf{r}_{i})}$, we can also
express the above $\hat{\xi}^{A}{\small (\mathbf{r}_{i})}$ and $\hat{\lambda}%
^{\dot{A}}{\small (\mathbf{r}_{i})}$ in terms of the hermitian $sp(2N,%
\mathbb{R})$ field $\hat{\phi}^{A}{\small (\mathbf{r}_{i})}$ and the $so(2N,%
\mathbb{R})$ Majorana $\hat{\digamma}^{\dot{A}}{\small (\mathbf{r}_{i})}$
given by
\begin{equation}
\hat{\phi}^{A}{\small (\mathbf{r}_{i})}=\left(
\begin{array}{c}
\hat{X}^{I}{\small (\mathbf{r}_{i})} \\
\hat{P}_{I}{\small (\mathbf{r}_{i})}%
\end{array}%
\right) \qquad ,\qquad \hat{\digamma}^{\dot{A}}{\small (\mathbf{r}_{i})}%
=\left(
\begin{array}{c}
\hat{\gamma}^{\dot{I}}{\small (\mathbf{r}_{i})} \\
\hat{\eta}_{\dot{I}}{\small (\mathbf{r}_{i})}%
\end{array}%
\right)
\end{equation}

\subsubsection{Restriction to OSp(2N\TEXTsymbol{\vert}2N)}

In the analysis given below, we restrict the orthosymplectic symmetric OSP(2N%
\TEXTsymbol{\vert}2M) down to the particular case where $M=N.$ This
constraint has been motivated by the modeling of supermatter living on the
hypercubic super lattice of the Figure \textbf{\ref{D2} }for which the $%
\mathbb{L}_{fermi}$ is isomorphic to $\mathbb{L}_{bose}$. However, our
analysis can be also used for super lattices $\mathbb{L}_{fermi}^{{\small (N)%
}}/\mathbb{L}_{bose}^{{\small (M)}}$ with $M\neq N$. Indeed, following
\textrm{\cite{1D2}}, there exist several super lattice constructions
involving different numbers of fermionic and bosonic oscillators ($M\neq N$%
). To fix the ideas, we cite here after four examples of such lattice pairs,%
\begin{equation}
\begin{tabular}{c|c}
$\mathbb{L}_{fermi}^{2D}$ & $\mathbb{L}_{bose}^{2D}$ \\ \hline
{\small Honeycomb} & {\small Kagome} \\
{\small Square-octagon} & {\small squagome}%
\end{tabular}%
\qquad ;\qquad
\begin{tabular}{c|c}
$\mathbb{L}_{fermi}^{3D}$ & $\mathbb{L}_{bose}^{3D}$ \\ \hline
{\small Hyper-honeycomb} & {\small Hyperkagome} \\
{\small Diamond } & {\small Pyrochlore}%
\end{tabular}
\label{tab}
\end{equation}%
\begin{equation*}
\text{ \ \ }
\end{equation*}%
For explicit details regarding the super bands associated with the super
lattices in eq(\ref{tab}), we refer to \textrm{\cite{1D2}}. Moreover it is
interesting to notice that the quantity $\nu =M-N$ defining the
super-dimension of the graded space $\mathbb{R}^{N|M}$ has interesting
interpretations; in particular: $\left( \mathbf{i}\right) $ as the so-called
Maxwell-Callading index of topological mechanics given by eq(5) of the work
\textrm{\cite{1D2}}. $\left( \mathbf{ii}\right) $ like zero modes of the
supersymmetric Hamiltonian (flat bands); and $\left( \mathbf{iii}\right) $
as values of non vanishing Witten index Tr$\left( -\right) ^{N_{F}}$ \textrm{%
\cite{witten}}. \newline
From the above description, it follows that models with orthosymplectic
OSP(2N\TEXTsymbol{\vert}2N) has a vanishing index $\nu =0$ and then no flat
bands. Within this picture, we demand the two following conditions for our
OSP(2N\TEXTsymbol{\vert}2N) model:

\begin{description}
\item[$\left( \mathbf{1}\right) $] \textbf{Local OSP(2N\TEXTsymbol{\vert}2N)
invariance on Lattice}: \newline
The bosonic $\hat{\xi}\left( \mathbf{r}_{i}\right) $ and the fermionic $\hat{%
\lambda}\left( \mathbf{r}_{i}^{\prime }\right) $ ---or equivalently the
symplectic $\hat{\phi}^{A}\left( \mathbf{r}_{i}\right) $ and the orthogonal $%
\hat{\digamma}^{\dot{A}}\left( \mathbf{r}_{i}^{\prime }\right) $--- live on
identical hypercubic lattices $\Theta _{\xi }$ and $\Theta _{\lambda }$
isomorphic to $\mathbb{Z}^{d}$ with size $\left\vert \Theta \right\vert
=L^{d}$. This feature insures that the number of degrees of freedom of the
bosonic $\hat{\xi}$'s is equal to the fermionic $\hat{\lambda}$'s; thus the
OSP(2N\TEXTsymbol{\vert}2N) invariance. The two $\Theta _{\xi }\equiv
\mathbb{L}_{Bose}$ and $\Theta _{\lambda }\equiv \mathbb{L}_{fermi}$ are as
illustrated in the Figure \textbf{\ref{D2} }for the example 2D square
lattice.
\begin{figure}[tbph]
\begin{center}
\includegraphics[width=8cm]{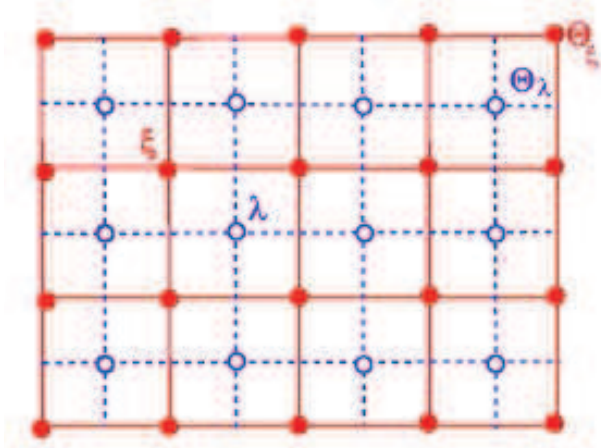}
\end{center}
\par
\vspace{-0.5cm}
\caption{2D lattice made of two identical sublattices: Bosonic operators
live on red sublattice $\Theta _{\protect\xi }$. Fermionic operators live on
the blue sublattice $\Theta _{\protect\lambda }$. }
\label{D2}
\end{figure}
\textrm{So, the super lattice }$\Theta _{\xi }/\Theta _{\lambda }$\textrm{\
has 2N local bosonic} ($\hat{b}^{I}\left( \mathbf{r}\right) ,\hat{b}%
_{I}^{\dagger }\left( \mathbf{r}\right) $) and 2N local fermionic ($\hat{c}^{%
\dot{I}}\left( \mathbf{r}\right) ,\hat{c}_{\dot{I}}^{\dagger }\left( \mathbf{%
r}\right) $) operators; thus inducing a local OSP(2N\TEXTsymbol{\vert}2N)
symmetry on the Brillouin Zone.

\item[$\left( \mathbf{2}\right) $] \textbf{Fourier modes of field operators}%
\emph{\ }$\hat{\phi}^{A}\left( \mathbf{r}\right) $ \textbf{and} $\hat{%
\digamma}^{\dot{A}}\left( \mathbf{r}\right) $\emph{\ }\newline
The 4N local hermitian oscillator operators $\hat{X}^{I}\left( \mathbf{r}%
\right) ,$ $\hat{P}_{I}\left( \mathbf{r}\right) ,$ $\hat{\gamma}^{\dot{I}%
}\left( \mathbf{r}\right) ,$ $\hat{\eta}_{\dot{I}}\left( \mathbf{r}\right) $
---or equivalently the complex $\hat{b}^{I}\left( \mathbf{r}\right) ,\hat{c}%
^{\dot{I}}\left( \mathbf{r}\right) $ and their adjoints $\hat{b}%
_{I}^{\dagger }\left( \mathbf{r}\right) ,\hat{c}_{\dot{I}}^{\dagger }\left(
\mathbf{r}\right) $ --- sit in a unit cell of the super lattice subject to a
periodic boundary condition. It has 2N bosonic ($\hat{b}^{I}\left( \mathbf{r}%
\right) ,\hat{b}_{I}^{\dagger }\left( \mathbf{r}\right) $) and 2N fermionic (%
$\hat{c}^{\dot{I}}\left( \mathbf{r}\right) ,\hat{c}_{\dot{I}}^{\dagger
}\left( \mathbf{r}\right) $) operators in agreement with the OSP(2N%
\TEXTsymbol{\vert}2N) symmetry requiring an equal number of degrees of
freedom.\newline
The momentum modes $\hat{\phi}_{\mathbf{k}}^{A}=(\hat{X}_{\mathbf{k}}^{I},%
\hat{P}_{I\mathbf{k}})$ and $\hat{\digamma}_{\mathbf{k}}^{\dot{A}}=(\hat{%
\gamma}_{\mathbf{k}}^{\dot{I}},\hat{\eta}_{\dot{I}\mathbf{k}})$ descending
from the Fourier transform of the $\hat{\phi}^{A}\left( \mathbf{r}\right) $
and $\hat{\digamma}^{\dot{A}}\left( \mathbf{r}\right) $ (equivalently $\hat{b%
}^{I}\left( \mathbf{r}\right) ,\hat{c}^{\dot{I}}\left( \mathbf{r}\right) $)
are given by the usual relation%
\begin{equation}
F\left( \mathbf{k}\right) =\frac{1}{\sqrt{\left\vert \Theta \right\vert }}%
\sum_{\mathbf{r\in \Theta }}e^{-i\mathbf{k.r}}F\left( \mathbf{r}\right)
\label{FR}
\end{equation}%
where $F\left( \mathbf{k}\right) $ stands for $\hat{\phi}^{A}\left( \mathbf{k%
}\right) \equiv \hat{\phi}_{\mathbf{k}}^{A}$ and for $\hat{\digamma}^{\dot{A}%
}\left( \mathbf{k}\right) \equiv \hat{\digamma}_{\mathbf{k}}^{\dot{A}}.$
Similar relations are valid for the $\hat{b}_{\mathbf{k}}^{I}$ and $\hat{c}_{%
\mathbf{k}}^{\dot{I}}$ descending from $\hat{b}^{I}\left( \mathbf{r}\right) $
and $\hat{c}^{\dot{I}}\left( \mathbf{r}\right) ;$ they read as $\hat{b}_{%
\mathbf{k}}^{I}=(\hat{X}_{\mathbf{k}}^{I}+i\hat{P}_{I\mathbf{k}})/\sqrt{2}$
and $\hat{c}_{\mathbf{k}}^{\dot{I}}=(\hat{\gamma}_{\mathbf{k}}^{\dot{I}}+i%
\hat{\eta}_{\dot{I}\mathbf{k}})/\sqrt{2}$. These Fourier modes will be used
below.
\end{description}

\section{ORTIC and SUSY tight binding models}

In this section, we use the orthosymplectic group properties (\ref{ORT}-\ref%
{FRT}) to develop the study of the osp(2N\TEXTsymbol{\vert}2N)
orthosymplectic (ORTIC) and the\textrm{\ }$\mathcal{N}$-\textrm{\ }%
supersymmetric (SUSY) tight binding models respectively based on the
fermionic $Q_{{\small orth}}$ and $Q_{{\small susy}}$ charges. The section
is organised in \textrm{three subsections}; the first subsection concerns
the ORTIC and the SUSY observables on lattice. The second regards the ORTIC
and the SUSY tight binding models. The third subsection deals with the
building of osp(2N\TEXTsymbol{\vert}2N) and $\mathcal{N}=2$ super TBMs.

\subsection{ORTIC and SUSY observables on lattice}

\begin{figure}[tbph]
\begin{center}
\includegraphics[width=7cm]{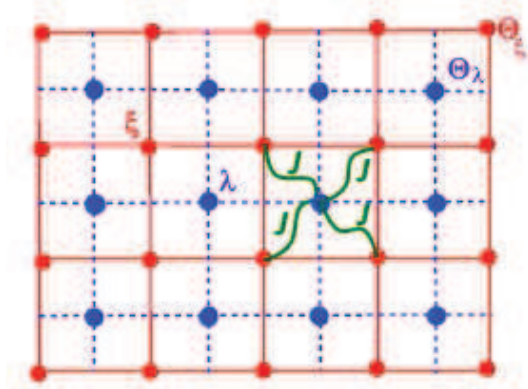}
\end{center}
\par
\vspace{-0.5cm}
\caption{Coupling of fermions and bosons. Here, the fermionic $\hat{\protect%
\lambda}$ (in blue) interacts with its four closed neighbors $\hat{\protect%
\xi}$ (in red). The coupling is given by $J$ (in green). Similar couplings
can be drawn for a bosonic site.}
\label{DD}
\end{figure}

Here, we extend the observables (\ref{ob}-\ref{bo}) of the orthosymplectic
algebra to band theory. We focus on two particular observables $Q_{\text{%
{\small ortho}}}$ and $H_{\text{{\small ortho}}}$ as well as $Q_{\text{%
{\small susy}}}$ and $H_{\text{{\small susy}}}$; they are quadratic in the
super oscillator operators and are related as described here below:

$\left( \mathbf{1}\right) $ \textbf{the orthosymplectic} $Q_{\text{{\small %
ortho}}}$\newline
This is a fermionic charge given by $\sum_{i,j}\hat{\lambda}_{\dot{A}}\left(
\mathbf{r}_{i}\right) [\boldsymbol{J}_{ij}]_{B}^{\dot{A}}\hat{\xi}^{B}\left(
\mathbf{r}_{j}\right) $ with local field operators $\hat{\lambda}_{\dot{A}}=(%
\hat{c}_{^{\dot{I}}}^{\dagger },\hat{c}^{\dot{I}})$ and $\hat{\xi}^{A}=(\hat{%
b}^{I},\hat{b}_{I}^{\dagger })$ \textrm{as in eqs(\ref{ap}-\ref{pa}). It
reads explicitly as}
\begin{equation}
\begin{tabular}{lll}
$Q_{\text{{\small ortho}}}$ & $=$ & $\hat{c}_{^{\dot{I}}}^{\dagger }\left(
\mathbf{r}_{i}\right) [\left( \boldsymbol{J}_{1}\right) _{ij}]_{J}^{^{\dot{I}%
}}\hat{b}^{J}\left( \mathbf{r}_{j}\right) +\hat{c}^{\dot{I}}\left( \mathbf{r}%
_{i}\right) [\left( \boldsymbol{J}_{4}\right) _{ij}]_{\dot{I}}^{J}\hat{b}%
_{J}^{\dagger }\left( \mathbf{r}_{j}\right) +$ \\
&  & $\hat{c}^{\dot{I}}\left( \mathbf{r}_{i}\right) [\left( \boldsymbol{J}%
_{3}\right) _{ij}]_{\dot{I}J}\hat{b}^{J}\left( \mathbf{r}_{j}\right) +\hat{c}%
_{^{\dot{I}}}^{\dagger }\left( \mathbf{r}_{i}\right) [\left( \boldsymbol{J}%
_{2}\right) _{ij}]^{^{\dot{I}J}}\hat{b}_{J}^{\dagger }\left( \mathbf{r}%
_{j}\right) $%
\end{tabular}
\label{Q1}
\end{equation}%
\textrm{where }$[\left( \boldsymbol{J}_{1}\right) _{ij}]_{J}^{^{\dot{I}}},$ $%
[\left( \boldsymbol{J}_{2}\right) _{ij}]^{^{\dot{I}J}},$ $[\left(
\boldsymbol{J}_{3}\right) _{ij}]_{\dot{I}J},$ and $[\left( \boldsymbol{J}%
_{4}\right) _{ij}]_{J}^{^{\dot{I}}}$\ \textrm{are coupling tensors.} This
fermionic $Q_{\text{{\small ortho}}}$ which reads also as%
\begin{equation}
Q_{\text{{\small ortho}}}=(\hat{c}_{^{\dot{I}}}^{\dagger }\left( \mathbf{r}%
_{i}\right) ,\hat{c}^{\dot{I}}\left( \mathbf{r}_{i}\right) )\left(
\begin{array}{cc}
\left( \boldsymbol{J}_{1}\right) _{ij} & \left( \boldsymbol{J}_{2}\right)
_{ij} \\
\left( \boldsymbol{J}_{3}\right) _{ij} & \left( \boldsymbol{J}_{4}\right)
_{ij}%
\end{array}%
\right) \left(
\begin{array}{c}
\hat{b}^{I}\left( \mathbf{r}_{j}\right) \\
\hat{b}_{I}^{\dagger }\left( \mathbf{r}_{j}\right)%
\end{array}%
\right)
\end{equation}%
defines the \emph{orthosymplectic}\textrm{\ }TBM; it is characterised by the
translation invariant coupling matrix $\boldsymbol{J}_{ij}=J\left( \mathbf{r}%
_{i}-\mathbf{r}_{j}\right) .$ \textrm{For a hermitian fermionic charge} ($Q_{%
\text{{\small ortho}}}^{\dagger }=Q_{\text{{\small ortho}}})$, the coupling
tensors $\boldsymbol{J}_{1},$ $\boldsymbol{J}_{2},$\ $\boldsymbol{J}_{3}$\
\textrm{and} $\boldsymbol{J}_{4}$ \textrm{are related like }$\boldsymbol{J}%
_{4}^{\dagger }=\boldsymbol{J}_{1}$ and $\boldsymbol{J}_{3}^{\dagger }=%
\boldsymbol{J}_{2}.$ An illustration of this coupling is given by the Figure
\textbf{\ref{DD}}.
By performing the Fourier transform, the (\ref{Q1}) can put $Q_{\text{%
{\small ortho}}}$ into the form $\sum_{\mathbf{k}}Q_{\mathbf{k}}$ with
Fourier modes $Q_{\mathbf{k}}$ constituting the basic object of the
orthosymplectic TBM.

\ \ \ \

$\left( \mathbf{2}\right) $ \textbf{Orthosymplectic} $H_{\text{{\small ortho}%
}}$ and \textbf{supersymmetric }$H_{\text{{\small susy}}}$\newline
The orthosymplectic $H_{\text{{\small ortho}}}$ is related to the fermionic $%
Q_{\text{{\small ortho}}}$\ as follows
\begin{equation}
2H_{\text{{\small ortho}}}=[Q_{\text{{\small ortho}}},Q_{\text{{\small ortho}%
}}^{\dagger }]  \label{H1}
\end{equation}%
\textrm{Notice that for a fermionic charge that is\ nilpotent (}$Q_{\text{%
{\small ortho}}}^{2}=0)$\textrm{\ and} commuting with the orthosymplectic
Hamiltonian $([H_{\text{{\small ortho}}},Q_{\text{{\small ortho}}}]=0),$ eq(%
\ref{H1}) should be read like \textrm{\ }%
\begin{equation}
2H_{\text{{\small susy}}}=[Q_{\text{{\small susy}}},Q_{\text{{\small susy}}%
}^{\dagger }]  \label{hsusy}
\end{equation}%
with%
\begin{equation}
\begin{tabular}{lll}
$\left\{ Q_{\text{{\small susy}}},Q_{\text{{\small susy}}}\right\} $ & $=$ &
$0$ \\
$\left[ H_{\text{{\small susy}}},Q_{\text{{\small susy}}}\right] $ & $=$ & $%
0 $%
\end{tabular}
\label{susy}
\end{equation}%
For $Q_{\text{{\small susy}}}^{\dagger }\neq Q_{\text{{\small susy}}}$, eqs(%
\ref{hsusy}-\ref{susy}) define the $\mathcal{N}=2$ supersymmetric band
theory while for $Q_{\text{{\small susy}}}^{\dagger }=Q_{\text{{\small susy}}%
}$, it defines $\mathcal{N}=1$ theory. Eq(\ref{susy}) define the
supersymmetric constraints that distinguishes $Q_{\text{{\small susy}}}$
from $Q_{\text{{\small orth}}};$ the SUSY models form then a subfamily of
ORTIC models. Moreover, being an even operator and quadratic in the super
oscillator operators; the $H_{\text{{\small ortho}}}$ (resp. $H_{\text{%
{\small susy}}}$) splits as the sum $H_{f}+H_{b}$ with $\left( \mathbf{i}%
\right) $ fermionic contribution having the form $H_{f}=\sum \hat{\lambda}_{%
\mathbf{r}_{i}}\left( h_{f}\right) _{ij}\hat{\lambda}_{\mathbf{r}_{j}}$ that
can be interpreted as in the realisation of the AZ table; and $\left(
\mathbf{ii}\right) $ bosonic contribution like $H_{b}=\sum \hat{\xi}_{%
\mathbf{r}_{i}}[\left( h_{b}\right) _{ij}]\hat{\xi}_{\mathbf{r}_{j}}$.
Furthermore, using translation invariance and the Fourier modes $Q_{\mathbf{k%
}}^{\dagger }=Q_{-\mathbf{k}}$, we have $Q_{\text{{\small ortho}}}^{\dagger
}=\sum_{\mathbf{k}}Q_{-\mathbf{k}}$ which by renaming the variable as $%
\mathbf{p}=-\mathbf{k}$ is equal $Q_{\text{{\small ortho}}}$; then we also
have $H_{\text{{\small ortho}}}=Q_{\text{{\small ortho}}}^{2}$. The same
feature holds for hermitian $Q_{\text{{\small susy}}}$ leading to $H_{\text{%
{\small susy}}}=Q_{\text{{\small susy}}}^{2}$.\ \ \ \newline
To deal with the $Q_{\text{{\small ortho}}}$ (resp. $Q_{\text{{\small susy}}%
} $) and the $H_{\text{{\small ortho}}}$ (resp. $H_{\text{{\small susy}}}$),
we start by the oscillator operators on lattice represented by the Fourier
modes $\hat{\xi}_{\mathbf{k}}^{A}$ and $\hat{\lambda}_{\mathbf{k}\dot{C}}$
with symplectic/orthogonal labels $A/\dot{A}$ ranging from 1 to 2N. They
read in terms of the usual $\hat{b}/\hat{c}$ operators as follows
\begin{equation}
\hat{\xi}_{\mathbf{k}}^{A}=\left(
\begin{array}{c}
\hat{b}_{\mathbf{k}}^{I} \\
\hat{b}_{\mathbf{k}I}^{\dagger }%
\end{array}%
\right) \qquad ,\qquad \hat{\lambda}_{\mathbf{k}\dot{C}}=\left( \hat{c}_{%
\mathbf{k}\dot{K}}^{\dagger },\hat{c}_{\mathbf{k}}^{\dot{K}}\right)
\end{equation}%
with momentum vector $\mathbf{k=(}k_{1},...,k_{d}\mathbf{)}$ parameterising
the Brillouin torus $\mathbb{T}^{d}.$ For later calculations, we use the
adjoint conjugation property $\hat{b}_{\mathbf{k}I}^{\dagger }=\hat{b}_{-%
\mathbf{k}I}$ and $\hat{c}_{\mathbf{k}\dot{I}}^{\dagger }=\hat{c}_{-\mathbf{k%
}\dot{I}}$ relating Fourier modes at $\mathbf{k}$ and $-\mathbf{k}.$ Using
the operators $\hat{\xi}_{\mathbf{k}}^{A}$ and $\hat{\lambda}_{\mathbf{k}%
\dot{C}}$, we calculate the following graded commutators
\begin{equation}
\lbrack \hat{\xi}_{\mathbf{k}}^{A},\hat{\xi}_{-\mathbf{k}}^{B}]=Z^{AB}\qquad
,\qquad \{\hat{\lambda}_{\mathbf{k}\dot{C}},\hat{\lambda}_{-\mathbf{k}\dot{D}%
}\}=G_{\dot{C}\dot{D}}  \label{zg}
\end{equation}%
with $2N\times 2N$ matrices $Z^{AB}$ and $G_{\dot{C}\dot{D}}$ as follows
\begin{equation}
Z^{AB}=\left(
\begin{array}{cc}
\delta ^{IJ} & 0 \\
0 & -\delta _{IJ}%
\end{array}%
\right) \qquad ,\qquad G_{\dot{C}\dot{D}}=\left(
\begin{array}{cc}
\delta _{\dot{K}\dot{L}} & 0 \\
0 & \delta ^{\dot{K}\dot{L}}%
\end{array}%
\right)
\end{equation}%
they read in a condensed form as the tensor products $Z=\sigma _{z}\otimes
I_{N}$ and $G=\sigma _{0}\otimes I_{N}$ with $\sigma _{\mu }$ referring to
the $2\times 2$ Pauli matrices. Notice that to get the relations (\ref{zg}),
we used the properties $[\hat{b}_{\mathbf{k}}^{I},\hat{b}_{-\mathbf{k}%
}^{J}]=\delta ^{IJ}$ and $[\hat{b}_{\mathbf{k}I}^{\dagger },\hat{b}_{-%
\mathbf{k}I}^{\dagger }]=-\delta _{IJ}$ as well $\{\hat{c}_{\mathbf{k}\dot{K}%
}^{\dagger },\hat{c}_{-\mathbf{k}\dot{L}}^{\dagger }\}=\delta _{\dot{K}\dot{L%
}}$ and $\{\hat{c}_{\mathbf{k}}^{\dot{K}},\hat{c}_{-\mathbf{k}}^{\dot{L}%
}\}=\delta ^{\dot{K}\dot{L}}.$ From these expressions, we learn the
interesting relations
\begin{equation}
\hat{\xi}_{\mathbf{k}}^{A}\hat{\xi}_{-\mathbf{k}}^{B}=Z^{AB}+\hat{\xi}_{-%
\mathbf{k}}^{B}\hat{\xi}_{\mathbf{k}}^{A}\qquad ,\qquad \hat{\lambda}_{%
\mathbf{k}\dot{C}}\hat{\lambda}_{-\mathbf{k}\dot{D}}=G_{\dot{C}\dot{D}}-\hat{%
\lambda}_{-\mathbf{k}\dot{D}}\hat{\lambda}_{\mathbf{k}\dot{C}}  \label{xl}
\end{equation}%
Notice that at a given $\mathbf{k}$, we also have%
\begin{equation}
\lbrack \hat{\xi}_{\mathbf{k}}^{A},\hat{\xi}_{\mathbf{k}}^{B}]=Y^{AB}\qquad
,\qquad \{\hat{\lambda}_{\mathbf{k}\dot{C}},\hat{\lambda}_{\mathbf{k}\dot{D}%
}\}=X_{\dot{C}\dot{D}}
\end{equation}%
with
\begin{equation}
Y^{AB}=\left(
\begin{array}{cc}
0 & \delta _{J}^{I} \\
-\delta _{I}^{J} & 0%
\end{array}%
\right) \qquad ,\qquad X_{\dot{C}\dot{D}}=\left(
\begin{array}{cc}
0 & \delta _{\dot{K}}^{\dot{L}} \\
\delta _{\dot{L}}^{\dot{K}} & 0%
\end{array}%
\right)
\end{equation}%
with $YX=Z.$

\subsubsection{ORTIC and SUSY charges on lattice}

The translation invariant supercharge operator $Q$ (\ref{Q1}) on the
hypercubic super lattice leads to the expansion $Q=\sum_{\mathbf{k}}Q_{%
\mathbf{k}}$. The Fourier modes $Q_{\mathbf{k}}$ are complex fermionic
operators expressed in terms of the super oscillator operators $\hat{\xi}_{%
\mathbf{k}}^{A}$ and $\hat{\lambda}_{\mathbf{k}\dot{C}}$ as follows
\begin{equation}
Q_{\mathbf{k}}=\hat{\lambda}_{\mathbf{k}\dot{C}}[\boldsymbol{q}_{\mathbf{k}%
}]_{A}^{\dot{C}}\hat{\xi}_{\mathbf{k}}^{A}\qquad ,\qquad \lbrack \left(
\boldsymbol{q}_{\mathbf{k}}\right) _{A}^{\dot{C}}]^{\dagger }=\left(
\boldsymbol{q}_{\mathbf{k}}^{\ast }\right) _{\dot{C}}^{A}  \label{QK}
\end{equation}%
where the complex coupling tensor
\begin{equation}
\left( \boldsymbol{q}_{\mathbf{k}}\right) _{A}^{\dot{C}}=\left(
\begin{array}{cc}
\left( \boldsymbol{q}_{1}\right) _{I}^{\dot{K}}\left( \mathbf{k}\right) &
\left( \boldsymbol{q}_{2}\right) ^{\dot{K}I}\left( \mathbf{k}\right) \\
\left( \boldsymbol{q}_{3}\right) _{\dot{K}I}\left( \mathbf{k}\right) &
\left( \boldsymbol{q}_{4}\right) _{\dot{K}}^{I}\left( \mathbf{k}\right)%
\end{array}%
\right) =\left(
\begin{array}{cc}
\left( \boldsymbol{q}_{1\mathbf{k}}\right) _{I}^{\dot{K}} & \left(
\boldsymbol{q}_{2\mathbf{k}}\right) ^{\dot{K}I} \\
\left( \boldsymbol{q}_{3\mathbf{k}}\right) _{\dot{K}I} & \left( \boldsymbol{q%
}_{4\mathbf{k}}\right) _{\dot{K}}^{I}%
\end{array}%
\right)  \label{qca}
\end{equation}%
is a $2N\times 2N$ matrix; it is the Fourier transform of $J\left( \mathbf{r}%
_{i}-\mathbf{r}_{j}\right) $. This coupling matrix plays an important role
in our super TBM as it captures the information on the physical properties
of the super bands; it is a complex function of momentum $\mathbf{k}$ and
takes values in the $(2N,2\dot{N}\mathbf{)}$ bi-fundamental representation
of $Sp(2N)\times SO(2N).$ In this regard, recall that $\hat{\xi}_{\mathbf{k}%
}^{A}$ transforms in the $2N$ representation of $Sp(2N)$ while $\hat{\lambda}%
_{\mathbf{k}\dot{C}}$ transforms in the $2\dot{N}$ representation of $Sp(2N)$%
. Below, we give other useful features of this coupling matrix. Notice that
by setting $\hat{\Phi}_{\mathbf{k}}^{\dot{C}}=\left( \boldsymbol{q}_{\mathbf{%
k}}\right) _{A}^{\dot{C}}\hat{\xi}_{\mathbf{k}}^{A},$ the supercharge (\ref%
{QK}) reads simply as
\begin{equation}
Q_{\mathbf{k}}=\hat{\lambda}_{\mathbf{k}\dot{C}}\hat{\Phi}_{\mathbf{k}}^{%
\dot{C}}  \label{FK}
\end{equation}%
Notice also that being a local matrix in the Brillouin torus, the coupling $%
\boldsymbol{q}_{\mathbf{k}}$ satisfies symmetry properties that we present
as a list of \textrm{five} points:

\textbf{1)} \textbf{Periodicity of} $\boldsymbol{q}_{A}^{\dot{C}}\left(
\mathbf{k}\right) $\newline
The $\boldsymbol{q}_{\mathbf{k}}$\ lives on the d-torus $\mathbb{T}^{d}$; it
obeys the periodicity conditions which are given by
\begin{equation}
\boldsymbol{q}_{A}^{\dot{C}}\left( \mathbf{k}+2\pi \mathbf{e}_{j}\right) =%
\boldsymbol{q}_{A}^{\dot{C}}\left( \mathbf{k}\right)  \label{per}
\end{equation}%
with $\mathbf{e}_{j}$ being the unit vector in the $k_{j}$ direction of $%
\mathbf{k}$. Explicit expressions of $\boldsymbol{q}_{\mathbf{k}}$ involve $%
\sin k_{j}$ and $\cos k_{j}$ functions in addition to constant moduli,%
\begin{equation}
\boldsymbol{q}_{\mathbf{k}}=\boldsymbol{q}\left( \sin k_{j},\cos k_{j}\right)
\end{equation}%
For the example of the Brillouin torus $\mathbb{T}^{2}$ parameterised by $%
\left( k_{x},k_{y}\right) ;$ and for the case of two bands, an interesting $%
\boldsymbol{q}_{\mathbf{k}}$ is given by
\begin{equation}
\boldsymbol{q}_{\mathbf{k}}=\left(
\begin{array}{cc}
e^{-i\varphi _{\mathbf{k}}}\sqrt{{\small \varepsilon +}({\small M-}\cos
{\small k}_{x})} & \sqrt{{\small \varepsilon -}(M{\small -}\cos {\small k}%
_{x})} \\
\sqrt{{\small \varepsilon -}({\small M-}\cos {\small k}_{x})} & {\small -}%
e^{+i\varphi _{\mathbf{k}}}\sqrt{{\small \varepsilon +(M-}\cos {\small k}_{x}%
{\small )}}%
\end{array}%
\right)  \label{qq}
\end{equation}%
with ${\small M}$ constant and
\begin{equation}
\begin{tabular}{lll}
$\varepsilon $ & $=$ & $(\sin ^{{\small 2}}{\small k}_{x}{\small +}\sin ^{%
{\small 2}}{\small k}_{y}+({\small M-}\cos {\small k}_{x})^{{\small 2}%
})^{1/2}$ \\
$e^{-i\varphi _{\mathbf{k}}}$ & $=$ & $\frac{\sin k_{x}-i\sin k_{y}}{\sqrt{%
\sin ^{{\small 2}}{\small k}_{x}{\small +}\sin ^{{\small 2}}{\small k}_{y}}}$%
\end{tabular}%
\end{equation}%
The (\ref{qq}) obeys the property (\ref{per}). \textrm{As far as the
coupling matrix (\ref{qq}) is concerned, we give the following comments. }$%
\left( \mathbf{i}\right) $ \textrm{the SUSY models based on eq(\ref{qq})\
are just super extensions of the 2D tight binding hamiltonian }$h_{\mathbf{k}%
}^{fermi}$ \textrm{given by} $d_{x}\sigma ^{x}+d_{y}\sigma ^{y}+d_{z}\sigma
^{z}$ with $d_{x}=\sin k_{x},$ $d_{y}=\sin k_{y}$ \textrm{and} $d_{z}=M-\cos
k_{x};$ it \textrm{is invariant under the symmetry generated by} $P=\sigma
_{y}K.$ $\left( \mathbf{ii}\right) $ \textrm{The structure of this
hamiltonian} $h_{\mathbf{k}}^{fermi}$ \textrm{may be also used to study the
super extension of higher order topological insulators along the
construction of \cite{1DB,TF}.}

\textbf{2)} \textbf{Invariance of the ORTIC charge} \newline
It transforms as $Sp(2N)\times SO(2N)$ vectors with changes given by $%
\Lambda _{\dot{D}}^{\dot{C}}q_{B}^{\dot{D}}S_{A}^{B}$ where $\Lambda _{\dot{D%
}}^{\dot{C}}$ and $S_{A}^{B}$ are orthogonal and symplectic matrices.
Invariance of the ORTIC $Q$ under the orthosymplectic symmetry requires the
constraint relation
\begin{equation}
\boldsymbol{q}_{A}^{\dot{C}}=\Lambda _{\dot{D}}^{\dot{C}}\boldsymbol{q}_{B}^{%
\dot{D}}S_{A}^{B}  \label{413}
\end{equation}%
This condition is too strong to fulfill; so the $Sp(2N)\times SO(2N)$
symmetry will be broken down to subgroups.

\textbf{3)} \textbf{Algebraic structure of} $\boldsymbol{q}_{A}^{\dot{C}%
}\left( \mathbf{k}\right) $ \newline
Under the particular breakings $Sp(2N)\rightarrow SU\left( N\right) $ and $%
SO(2N)\rightarrow SU\left( N\right) ^{\prime }$, we have the following
representations decomposition: \newline
$\left( \mathbf{i}\right) $ Adjoint representations%
\begin{equation}
\begin{tabular}{lll}
$sp(2N)$ & $\rightarrow $ & $u\left( N\right) \oplus \boldsymbol{n}%
_{+}\oplus \boldsymbol{n}_{-}$ \\
$N\left( 2N+1\right) $ & $\rightarrow $ & $N^{2}+\frac{N\left( N+1\right) }{2%
}+\frac{N\left( N+1\right) }{2}$%
\end{tabular}%
\end{equation}%
and%
\begin{equation}
\begin{tabular}{lll}
$so(2N)$ & $\rightarrow $ & $u\left( N\right) ^{\prime }\oplus \boldsymbol{n}%
_{+}^{\prime }\oplus \boldsymbol{n}_{-}^{\prime }$ \\
$N\left( 2N-1\right) $ & $\rightarrow $ & $N^{2}+\frac{N\left( N-1\right) }{2%
}+\frac{N\left( N-1\right) }{2}$%
\end{tabular}%
\end{equation}%
$\left( \mathbf{ii}\right) $ The bi-fundamental $(2N,2\dot{N})$ as direct
sum of four blocs of $U\left( N\right) \times U\left( N\right) ^{\prime }$
as follows%
\begin{equation}
(2N,2\dot{N})=(N_{+},\dot{N}_{-})\oplus (N_{-},\dot{N}_{+})\oplus (N_{+},%
\dot{N}_{+})\oplus (N_{-},\dot{N}_{-})
\end{equation}%
This reduction, corresponds to the decomposition of $\boldsymbol{q}_{A}^{%
\dot{C}}\left( \mathbf{k}\right) $ in terms of four blocs of $N\times N$
matrices like in (\ref{qca}). There, the blocs $\boldsymbol{q}_{2}^{\dot{K}%
I} $ and $\left( \boldsymbol{q}_{3}\right) _{\dot{K}I}$ take values in the $%
(N_{+},\dot{N}_{+})$ and the $(N_{-},\dot{N}_{-})$ representations while the
$\left( \boldsymbol{q}_{1}\right) _{I}^{\dot{K}}$ and $\left( \boldsymbol{q}%
_{4}\right) _{\dot{K}}^{I}$ are respectively valued in $(N_{+},\dot{N}_{-})$
and $(N_{-},\dot{N}_{+})$. The transformation (\ref{413}) splits as follows%
\begin{equation}
\begin{tabular}{lllllll}
$\left( \boldsymbol{q}_{1}\right) _{I}^{\dot{K}}$ & $=$ & $V_{\dot{L}}^{\dot{%
K}}\left( \boldsymbol{q}_{1}\right) _{J}^{\dot{L}}\bar{U}_{I}^{J}$ & \qquad
,\qquad & $\left( \boldsymbol{q}_{2}\right) ^{\dot{K}I}$ & $=$ & $V_{\dot{L}%
}^{\dot{K}}\left( \boldsymbol{q}_{2}\right) ^{\dot{L}J}U_{J}^{I}$ \\
$\left( \boldsymbol{q}_{4}\right) _{\dot{K}}^{I}$ & $=$ & $\bar{V}_{\dot{K}%
}^{\dot{L}}\left( \boldsymbol{q}_{4}\right) _{\dot{L}}^{J}U_{J}^{I}$ &
\qquad ,\qquad & $\left( \boldsymbol{q}_{3}\right) _{\dot{K}I}$ & $=$ & $%
\bar{V}_{\dot{K}}^{\dot{L}}\left( \boldsymbol{q}_{3}\right) _{\dot{L}J}\bar{U%
}_{I}^{J}$%
\end{tabular}%
\end{equation}%
with unitary matrices obeying $U_{J}^{I}\bar{U}_{K}^{J}=\delta _{K}^{I}$ and
$V_{\dot{L}}^{\dot{K}}\bar{V}_{\dot{J}}^{\dot{L}}=\delta _{\dot{J}}^{\dot{K}%
}.$

\textbf{4)} \textbf{TPC transformations of} $\boldsymbol{q}_{A}^{\dot{C}%
}\left( \mathbf{k}\right) $\newline
For the breaking down to $S\left[ U\left( 2\right) \times U\left( N\right) %
\right] $ materialised by the double label notations $A=\left( \alpha
,I\right) $ and $\dot{A}=(\dot{\alpha},\dot{I})$, the splitting (\ref{qca})
is expressed like%
\begin{equation}
\boldsymbol{q}_{A}^{\dot{B}}\left( \mathbf{k}\right) =q_{\alpha I}^{\dot{%
\beta}\dot{J}}\left( \mathbf{k}\right) \qquad ,\qquad \alpha ,\dot{\alpha}%
=1,2  \label{416}
\end{equation}%
This tensor can be remarkably expanded in terms of Pauli matrices as follows%
\begin{equation}
q_{\alpha I}^{\dot{\beta}\dot{J}}\left( \mathbf{k}\right) =\sum_{\mu
=0}^{3}\left( \sigma ^{\mu }\right) _{\alpha }^{\dot{\beta}}[\mathfrak{q}%
_{\mu }\left( \mathbf{k}\right) ]_{I}^{\dot{J}}  \label{q}
\end{equation}%
with $\mathfrak{q}_{\mu }\left( \mathbf{k}\right) $ four $N\times N$
matrices whose topological properties are given by requiring discrete
symmetries \textrm{like} $\mathcal{T}=K,$ $\mathcal{P}=X\circ K$ and $%
\mathcal{C}$ with $X=\sigma _{x}\otimes I_{N}$. In this spinless matter
case, the $q_{A}^{\dot{B}}\left( \mathbf{k}\right) $ must be constrained as%
\begin{equation}
\begin{tabular}{lllll}
$\mathcal{T}$ & : & $q\left( \mathbf{k}\right) ^{\ast }$ & $=$ & $+q\left( -%
\mathbf{k}\right) $ \\
$\mathcal{P}$ & : & $Xq\left( \mathbf{k}\right) ^{\ast }X$ & $=$ & $+q\left(
-\mathbf{k}\right) $ \\
$\mathcal{C}$ & : & $\mathcal{C}q\left( \mathbf{k}\right) \mathcal{C}^{-1}$
& $=$ & $+q\left( \mathbf{k}\right) $%
\end{tabular}
\label{TRS}
\end{equation}%
as it will be checked later on when considering the Hamiltonian language. By
substituting $q\left( \mathbf{k}\right) =\sigma ^{\mu }\mathfrak{q}_{\mu
}\left( \mathbf{k}\right) $, we get the following constraint relations
\begin{equation}
\begin{tabular}{lllll}
$\mathcal{T}$ & : & $\mathfrak{q}_{2}\left( \mathbf{k}\right) ^{\ast }=-%
\mathfrak{q}_{2}\left( -\mathbf{k}\right) $ & $,$ & $\mathfrak{q}%
_{0,1,3}\left( \mathbf{k}\right) ^{\ast }=\mathfrak{q}_{0,1,3}\left( -%
\mathbf{k}\right) $ \\
$\mathcal{P}$ & : & $\mathfrak{q}_{3}\left( \mathbf{k}\right) ^{\ast }=-%
\mathfrak{q}_{3}\left( -\mathbf{k}\right) $ & $,$ & $\mathfrak{q}%
_{0,1,2}\left( \mathbf{k}\right) ^{\ast }=\mathfrak{q}_{0,1,2}\left( -%
\mathbf{k}\right) $ \\
$\mathcal{C}$ & : & $\mathfrak{q}_{2,3}\left( \mathbf{k}\right) =0$ &  &
\end{tabular}%
\end{equation}

\textbf{5)} \textbf{Oscillator realisation of (\ref{FK})} \newline
Using (\ref{qca}) we can express the supercharge (\ref{QK}) in terms the
bosonic $\hat{b}_{\mathbf{k}}$ and the fermionic $\hat{c}_{\mathbf{k}}$
operators. First, we have for $\hat{\Phi}_{\mathbf{k}}^{\dot{C}}=\left(
\boldsymbol{q}_{\mathbf{k}}\right) _{A}^{\dot{C}}\hat{\xi}_{\mathbf{k}}^{A},$
the following%
\begin{equation}
\hat{\Phi}_{\mathbf{k}}^{\dot{C}}=\left(
\begin{array}{c}
\left( \boldsymbol{q}_{1\mathbf{k}}\right) _{I}^{\dot{K}}\hat{b}_{\mathbf{k}%
}^{I}+\hat{b}_{\mathbf{k}I}^{\dagger }\left( \boldsymbol{q}_{2\mathbf{k}%
}\right) ^{I\dot{K}} \\
\left( \boldsymbol{q}_{3\mathbf{k}}\right) _{\dot{K}I}\hat{b}_{\mathbf{k}%
}^{I}+\hat{b}_{\mathbf{k}I}^{\dagger }\left( \boldsymbol{q}_{4\mathbf{k}%
}\right) _{\dot{K}}^{I}%
\end{array}%
\right)
\end{equation}%
By putting into (\ref{QK}), we get the supercharge at $\mathbf{k}$ namely%
\begin{equation}
\begin{tabular}{lll}
$Q_{\mathbf{k}}$ & $=$ & $\hat{c}_{\mathbf{k}\dot{K}}^{\dagger }\left(
\boldsymbol{q}_{1\mathbf{k}}\right) _{I}^{\dot{K}}\hat{b}_{\mathbf{k}}^{I}+%
\hat{c}_{\mathbf{k}}^{\dot{K}}\left( \boldsymbol{q}_{4\mathbf{k}}\right) _{%
\dot{K}}^{I}\hat{b}_{\mathbf{k}I}^{\dagger }+$ \\
&  & $\hat{c}_{\mathbf{k}\dot{K}}^{\dagger }\left( \boldsymbol{q}_{2\mathbf{k%
}}\right) ^{\dot{K}I}\hat{b}_{\mathbf{k}I}^{\dagger }+\hat{c}_{\mathbf{k}}^{%
\dot{K}}\left( \boldsymbol{q}_{3\mathbf{k}}\right) _{\dot{K}I}\hat{b}_{%
\mathbf{k}}^{I}$%
\end{tabular}
\label{kq}
\end{equation}%
It has four block terms generated by the fermionic operators $\hat{c}_{%
\mathbf{k}}^{\dagger }\hat{b}_{\mathbf{k}},$ $\hat{c}_{\mathbf{k}}\hat{b}_{%
\mathbf{k}}^{\dagger },$ $\hat{c}_{\mathbf{k}}^{\dagger }\hat{b}_{\mathbf{k}%
}^{\dagger }$ and $\hat{c}_{\mathbf{k}}\hat{b}_{\mathbf{k}}.$ Notice that by
using the notation (\textbf{\ref{FK}}), the above relation (\ref{kq}) reads
simply as%
\begin{equation}
Q_{\mathbf{k}}=\hat{c}_{\mathbf{k}\dot{A}}^{\dagger }\hat{B}_{\mathbf{k}}^{%
\dot{A}}+\hat{c}_{\mathbf{k}}^{\dot{A}}D_{\mathbf{k}\dot{A}}  \label{KQ}
\end{equation}%
where we have set
\begin{equation}
\begin{tabular}{lll}
$\hat{B}_{\mathbf{k}}^{\dot{A}}$ & $=$ & $\left( \boldsymbol{q}_{1\mathbf{k}%
}\right) _{I}^{\dot{A}}\hat{b}_{\mathbf{k}}^{I}+\left( \boldsymbol{q}_{2%
\mathbf{k}}\right) ^{\dot{A}I}\hat{b}_{\mathbf{k}I}^{\dagger }$ \\
$D_{\mathbf{k}\dot{A}}$ & $=$ & $\left( \boldsymbol{q}_{3\mathbf{k}}\right)
_{\dot{A}I}\hat{b}_{\mathbf{k}}^{I}+\left( \boldsymbol{q}_{4\mathbf{k}%
}\right) _{\dot{A}}^{I}\hat{b}_{\mathbf{k}I}^{\dagger }$%
\end{tabular}
\label{dbd}
\end{equation}%
Eq(\ref{dbd}) is a mapping from the $\hat{b}_{\mathbf{k}}^{I}/\hat{b}_{%
\mathbf{k}I}^{\dagger }$ to linear local combinations $\hat{B}_{\mathbf{k}}^{%
\dot{A}}$ and $D_{\mathbf{k}\dot{A}}.$ These new operators obey the
commutation relations%
\begin{equation}
\begin{tabular}{lll}
$\left[ \hat{B}_{\mathbf{k}}^{\dot{A}},\hat{B}_{\mathbf{k}}^{\dot{D}}\right]
$ & $=$ & $\left( \boldsymbol{q}_{1\mathbf{k}}\boldsymbol{q}_{2\mathbf{k}%
}^{T}\right) ^{\dot{A}\dot{D}}-\left( \boldsymbol{q}_{1\mathbf{k}}%
\boldsymbol{q}_{2\mathbf{k}}^{T}\right) ^{\dot{D}\dot{A}}$ \\
$\left[ D_{\mathbf{k}\dot{A}},D_{\mathbf{k}\dot{B}}\right] $ & $=$ & $\left(
\boldsymbol{q}_{3\mathbf{k}}\boldsymbol{q}_{4\mathbf{k}}^{T}\right) _{\dot{A}%
\dot{B}}-\left( \boldsymbol{q}_{3\mathbf{k}}\boldsymbol{q}_{4\mathbf{k}%
}^{T}\right) _{\dot{B}\dot{A}}$ \\
$\left[ \hat{B}_{\mathbf{k}}^{\dot{A}},D_{\mathbf{k}\dot{C}}\right] $ & $=$
& $\left( \boldsymbol{q}_{1\mathbf{k}}\boldsymbol{q}_{4\mathbf{k}%
}^{T}\right) _{\dot{C}}^{\dot{A}}-\left( \boldsymbol{q}_{3\mathbf{k}}%
\boldsymbol{q}_{2\mathbf{k}}^{T}\right) _{\dot{C}}^{\dot{A}}$%
\end{tabular}
\label{BD}
\end{equation}%
Notice that a necessary condition to go from orthosymplectic (\ref{H1}) to
the supersymmetric (\ref{hsusy}-\ref{susy}) is given by the nilpotency
condition $Q_{\mathbf{k}}^{2}=0$. This demands the vanishing of the
commutators (\ref{BD}) requiring in turns that the $\boldsymbol{q}_{1\mathbf{%
k}}\boldsymbol{q}_{2\mathbf{k}}^{T}$ and $\boldsymbol{q}_{3\mathbf{k}}%
\boldsymbol{q}_{4\mathbf{k}}^{T}$ must be symmetric matrices and $%
\boldsymbol{q}_{1\mathbf{k}}\boldsymbol{q}_{4\mathbf{k}}^{T}=\boldsymbol{q}%
_{3\mathbf{k}}\boldsymbol{q}_{2\mathbf{k}}^{T}.$

\subsubsection{Classification of coupling matrix $q_{\mathbf{k}}$ by solving
(\protect\ref{413})}

The super band models have definite supercharges $Q_{\mathbf{k}}$; as such
they are completely characterised by the coupling matrix $(\boldsymbol{q}_{%
\mathbf{k}})_{A}^{\dot{C}}$; so ORTIC and SUSY TBMs can be classified by
those $\boldsymbol{q}_{\mathbf{k}}$'s solving (\ref{413}) and their
topological properties by the discrete TPC (\ref{TRS}). \newline
In what follows, we focus on the constraint $\boldsymbol{q}_{A}^{\dot{C}%
}=\Lambda _{\dot{D}}^{\dot{C}}\boldsymbol{q}_{B}^{\dot{D}}S_{A}^{B}$ whose
structure indicates that non trivial solutions require relating the
orthogonal $\Lambda _{\dot{D}}^{\dot{C}}$ and the symplectic $S_{A}^{B}$
matrices. This is achieved by breaking
\begin{equation*}
Sp\left( 2N\right) \times SO\left( 2N\right) \rightarrow \mathcal{G}
\end{equation*}%
with maximal $\mathcal{G}$ symmetry given by $U\left( 2\right) \times
U\left( N\right) $ containing the $U\left( N\right) $ describing the complex
structure $J$. Below, we look for solutions of (\ref{413}) by focussing on
the case of unitary groups $\mathcal{G}$ is contained in $U\left( N\right) $
while the general situation will be considered in the discussion section.
For concreteness, we give here after three families of solutions to the
constraint relations.

\ \

\textbf{A) Super Model with }$U\left( N\right) $\textbf{\ symmetry}\newline
One of the interesting coupling matrices $\left( \boldsymbol{q}_{\mathbf{k}%
}\right) _{A}^{\dot{C}}$ solving the constraint (\ref{413}), while using the
splitting (\ref{qca}), is given by
\begin{equation}
\begin{tabular}{lllllll}
$\left( \boldsymbol{q}_{1\mathbf{k}}\right) _{I}^{\dot{K}}$ & $=$ & $\mu _{%
\mathbf{k}}\delta _{I}^{\dot{K}}$ & $,\qquad $ & $\left( \boldsymbol{q}_{2%
\mathbf{k}}\right) ^{\dot{K}I}$ & $=$ & $0$ \\
$\left( \boldsymbol{q}_{4\mathbf{k}}\right) _{\dot{K}}^{I}$ & $=$ & $\nu _{%
\mathbf{k}}\delta _{\dot{K}}^{I}$ & $,\qquad $ & $\left( \boldsymbol{q}_{3%
\mathbf{k}}\right) _{\dot{K}I}$ & $=$ & $0$%
\end{tabular}
\label{chp}
\end{equation}%
It characterised by two complex scalars $\mu _{\mathbf{k}}$ and $\nu _{%
\mathbf{k}}$ which are functions of the momentum $\mathbf{k}.$ For this
model, the fermionic charge (\ref{kq}) reduces to
\begin{equation}
\begin{tabular}{lll}
$Q_{\mathbf{k}}$ & $=$ & $\mu _{\mathbf{k}}(\hat{c}_{\mathbf{k}I}^{\dagger }%
\hat{b}_{\mathbf{k}}^{I})+\nu _{\mathbf{k}}(\hat{b}_{\mathbf{k}I}^{\dagger }%
\hat{c}^{I})$ \\
$Q_{\mathbf{k}}^{\dagger }$ & $=$ & $\bar{\nu}_{\mathbf{k}}(\hat{c}_{\mathbf{%
k}I}^{\dagger }\hat{b}_{\mathbf{k}}^{I})+\bar{\mu}_{\mathbf{k}}(\hat{b}_{%
\mathbf{k}I}^{\dagger }\hat{c}^{I})$%
\end{tabular}
\label{QQ}
\end{equation}%
For this super TBM, the coupling matrix $\boldsymbol{q}_{\mathbf{k}}$ is
proportional to the $N\times N$ identity $\delta _{I}^{\dot{K}}$; so it has
a $U\left( N\right) $ symmetry as manifestly exhibited by (\ref{QQ}). In
this regard, recall that generally speaking, the $\left( \boldsymbol{q}_{%
\mathbf{k}}\right) _{A}^{\dot{C}}$ has $4N^{2}$ complex functions. The
choice (\ref{chp}) corresponds just to the diagonal%
\begin{equation}
\boldsymbol{q}_{\mathbf{k}}=\left(
\begin{array}{cc}
\mu _{\mathbf{k}}I_{N} & 0 \\
0 & \nu _{\mathbf{k}}I_{N}%
\end{array}%
\right) =\left(
\begin{array}{cc}
\mu _{\mathbf{k}} & 0 \\
0 & \nu _{\mathbf{k}}%
\end{array}%
\right) \otimes I_{N}
\end{equation}%
reading also like
\begin{equation}
\boldsymbol{q}_{\mathbf{k}}=\frac{1}{2}\left( \mu _{\mathbf{k}}+\nu _{%
\mathbf{k}}\right) \sigma ^{0}\otimes I_{N}+\frac{1}{2}\left( \mu _{\mathbf{k%
}}-\nu _{\mathbf{k}}\right) \sigma ^{z}\otimes I_{N}  \label{qun}
\end{equation}%
To get more information on the functions $\mu _{\mathbf{k}}$ and $\nu _{%
\mathbf{k}}$, we think about them in terms of $\cos k_{j}$ and $\sin k_{j}$
and impose TPC symmetries acting as
\begin{equation}
\begin{tabular}{lllll}
$\mathcal{T}$ & : & $\left. \mu \left( \mathbf{k}\right) ^{\ast },\nu \left(
\mathbf{k}\right) ^{\ast }\right. $ & $=$ & $\left. \mu \left( -\mathbf{k}%
\right) ,\nu \left( -\mathbf{k}\right) \right. $ \\
$\mathcal{P}$ & : & $\mu \left( \mathbf{k}\right) ^{\ast }$ & $=$ & $\nu
\left( -\mathbf{k}\right) $ \\
$\mathcal{C}$ & : & $\mu \left( \mathbf{k}\right) $ & $=$ & $\nu \left(
\mathbf{k}\right) $%
\end{tabular}%
\end{equation}%
For the case of time reversal invariance, examples of the $\mu _{\mathbf{k}}$
and the $\nu _{\mathbf{k}}$ are given by
\begin{eqnarray}
\mu _{\mathbf{k}} &=&\left( M-\sum_{j=1}^{d}\Delta _{j}\cos k_{j}\right)
+i\sum_{j=1}^{d}t_{j}\sin k_{j}  \notag \\
\nu _{\mathbf{k}} &=&\left( M^{\prime }-\sum_{j=1}^{d}\Delta _{j}^{\prime
}\cos k_{j}\right) +i\sum_{j=1}^{d}t_{j}^{\prime }\sin k_{j}
\end{eqnarray}%
If in addition to time reversal, we demand moreover particle-hole symmetry,
the condition $\mu \left( \mathbf{k}\right) ^{\ast }=\nu \left( -\mathbf{k}%
\right) $ requires the identification of the coupling parameters; that is $%
\left( M,\Delta ,t\right) =\left( M^{\prime },\Delta ^{\prime },t^{\prime
}\right) .$

\ \ \

\textbf{B) Super TBMs}\emph{\ }\textbf{with}\emph{\ }$\prod U\left(
n_{i}\right) $ \textbf{symmetries}\newline
Starting from the above $U\left( N\right) $ super TBM, we can engineer other
super TBMs having symmetries $\mathcal{G}$ contained into it. These
subsymmetries are given by the tensor product group $\prod U\left(
n_{i}\right) $ with the condition $\sum_{i=1}^{\eta }n_{i}=N;$ i.e:%
\begin{equation}
\mathcal{G}=\prod_{i=1}^{\eta }U\left( n_{i}\right) \subset U\left( N\right)
\end{equation}%
A particular super TBM model belonging to this family is given by $U\left(
1\right) ^{N}$ which is the maximal breaking of the $U\left( N\right) $
symmetry. For this super TBM, the coupling matrix $\boldsymbol{q}_{A}^{\dot{C%
}}\left( \mathbf{k}\right) $ of (\ref{qca}) is given by
\begin{equation}
\begin{tabular}{lllllll}
$\left( \boldsymbol{q}_{1\mathbf{k}}\right) _{I}^{\dot{K}}$ & $=$ & $\left(
\mu _{\mathbf{k}}\right) _{I}\delta _{I}^{\dot{K}}$ & $,\qquad $ & $\left(
\boldsymbol{q}_{2\mathbf{k}}\right) ^{\dot{K}I}$ & $=$ & $0$ \\
$\left( \boldsymbol{q}_{4\mathbf{k}}\right) _{\dot{K}}^{I}$ & $=$ & $\left(
\nu _{\mathbf{k}}\right) _{I}\delta _{\dot{K}}^{I}$ & $,\qquad $ & $\left(
\boldsymbol{q}_{3\mathbf{k}}\right) _{\dot{K}I}$ & $=$ & $0$%
\end{tabular}%
\end{equation}%
having 2N parameters given by $\left( \mu _{\mathbf{k}}\right) _{I}$ and $%
\left( \nu _{\mathbf{k}}\right) _{I}$. The supercharges are
\begin{equation}
\begin{tabular}{lll}
$Q_{\mathbf{k}}$ & $=$ & $\dsum\limits_{I}\left( \mu _{\mathbf{k}}\right)
_{I}\left( \hat{c}_{\mathbf{k}I}^{\dagger }\hat{b}_{\mathbf{k}}^{I}\right)
+\left( \nu _{\mathbf{k}}\right) _{I}\left( \hat{b}_{\mathbf{k}I}^{\dagger }%
\hat{c}^{I}\right) $ \\
$Q_{\mathbf{k}}^{\dagger }$ & $=$ & $\dsum\limits_{I}\left( \bar{\nu}_{%
\mathbf{k}}\right) _{I}\left( \hat{c}_{\mathbf{k}I}^{\dagger }\hat{b}_{%
\mathbf{k}}^{I}\right) +\left( \bar{\mu}_{\mathbf{k}}\right) _{I}\left( \hat{%
b}_{\mathbf{k}I}^{\dagger }\hat{c}^{I}\right) $%
\end{tabular}%
\end{equation}%
By equating all the $\left( \mu _{\mathbf{k}}\right) _{I}$'s and equating
all the $\left( \nu _{\mathbf{k}}\right) _{I}$'s, we recover the $U\left(
N\right) $ model described above. For this super TBM, the coupling matrix is
still diagonal as shown below%
\begin{equation}
\boldsymbol{q}_{\mathbf{k}}=\left(
\begin{array}{cccccc}
\left( \mu _{\mathbf{k}}\right) _{1} &  &  &  &  &  \\
& \ddots &  &  &  &  \\
&  & \left( \mu _{\mathbf{k}}\right) _{N} &  &  &  \\
&  &  & \left( \nu _{\mathbf{k}}\right) _{1} &  &  \\
&  &  &  & \ddots &  \\
&  &  &  &  & \left( \nu _{\mathbf{k}}\right) _{N}%
\end{array}%
\right)  \label{qn}
\end{equation}%
Notice that by equating a part of the $\left( \mu _{\mathbf{k}}\right) _{I}$%
's and the $\left( \nu _{\mathbf{k}}\right) _{I}$'s in above (\ref{qn}), we
get new super models with $\prod U_{\left( n_{i}\right) }$ symmetry.
Examples of such symmetries are listed in the following table
\begin{equation}
\begin{tabular}{|c|c|c|c|}
\hline
{\small parameters} & \multicolumn{3}{|c|}{\small unitary symmetry groups}
\\ \hline
${\small 1}$ & ${\small U(N)}$ & - & - \\ \hline
${\small 2}$ & ${\small U(1)\times U(N-1)}$ & - & - \\ \hline
$\left( 3,2\right) $ & ${\small U(1)}^{2}{\small \times U(N-2)}$ & ${\small %
U(2)\times U(N-2)}$ & - \\ \hline
$\left( 4,3,2\right) $ & ${\small U(1)}^{3}{\small \times U(N-3)}$ & $%
{\small U(1)\times U}\left( 2\right) {\small \times U(N-3)}$ & ${\small %
U(3)\times U(N-3)}$ \\ \hline
${\small \ \vdots }$ & ${\small \ \ \ \ \ \ \ \ \ \vdots }$ &  &  \\ \hline
${\small N-1}$ & ${\small U(1)}^{N-2}{\small \times U(2)}$ & - & - \\ \hline
${\small N}$ & ${\small U(1)}^{N}$ & - & - \\ \hline
\end{tabular}
\label{437}
\end{equation}%
\begin{equation*}
\text{ \ \ }
\end{equation*}

\textbf{C) Super models}\emph{\ }\textbf{III with }$U\left( N\right) $
\textbf{symmetry }\newline
Another interesting choice of the coupling $\left( \boldsymbol{q}_{\mathbf{k}%
}\right) _{A}^{\dot{C}}$ is that given by the following non diagonal matrix%
\begin{equation}
\boldsymbol{q}_{\mathbf{k}}=\left(
\begin{array}{cc}
\mu _{\mathbf{k}}I_{N} & \rho _{\mathbf{k}}I_{N} \\
\varsigma _{\mathbf{k}}I_{N} & \nu _{\mathbf{k}}I_{N}%
\end{array}%
\right) =\left(
\begin{array}{cc}
\mu _{\mathbf{k}} & \rho _{\mathbf{k}} \\
\varsigma _{\mathbf{k}} & \nu _{\mathbf{k}}%
\end{array}%
\right) \otimes I_{N}  \label{qqk}
\end{equation}%
It has four complex functions $\mu _{\mathbf{k}},$ $\nu _{\mathbf{k}},$ $%
\rho _{\mathbf{k}}$ and $\varsigma _{\mathbf{k}}$. It describes the
following supercharge%
\begin{equation}
\begin{tabular}{lll}
$Q_{\mathbf{k}}$ & $=$ & $\mu _{\mathbf{k}}(\hat{c}_{\mathbf{k}I}^{\dagger }%
\hat{b}_{\mathbf{k}}^{I})+\nu _{\mathbf{k}}(\hat{b}_{\mathbf{k}I}^{\dagger }%
\hat{c}^{I})+$ \\
&  & $\rho _{\mathbf{k}}(\hat{c}_{\mathbf{k}\dot{K}}^{\dagger }\delta ^{\dot{%
K}I}\hat{b}_{\mathbf{k}I}^{\dagger })+\varsigma _{\mathbf{k}}(\hat{c}_{%
\mathbf{k}}^{\dot{K}}\delta _{\dot{K}I}\hat{b}_{\mathbf{k}}^{I})$%
\end{tabular}%
\end{equation}%
By comparing (\ref{qqk}) with (\ref{qun}), we learn that they have different
numbers of entries. This indicates that (\ref{qun}) has a bigger symmetry
than (\ref{qqk}) which is given by $U\left( 1\right) ^{2}\times U\left(
N\right) $.

\subsection{The super TBM Hamiltonian}

On the lattice, the ORTIC hamiltonian $H_{\mathbf{k}}^{ortic}$ is given by
the anticommutator of the ORTIC supercharge $Q_{\mathbf{k}}^{ortic}$ with
its \textrm{adjoint namely} ($Q_{\mathbf{k}}Q_{\mathbf{k}}^{\dagger }+Q_{%
\mathbf{k}}^{\dagger }Q_{\mathbf{k}})/2$. By using $Q_{\mathbf{k}}^{\dagger
}=Q_{-\mathbf{k}},$ we also have%
\begin{equation}
H_{\mathbf{k}}^{ortic}=\frac{1}{2}\left\{ Q_{\mathbf{k}},Q_{-\mathbf{k}%
}\right\}  \label{hk}
\end{equation}%
The SUSY hamiltonian $H_{\mathbf{k}}^{susy}$ is given by (\ref{hk})
constrained by the nilpotency condition $Q_{\mathbf{k}}^{2}=0$ and the
commutativity $\left[ H_{\mathbf{k}}^{susy},Q_{\pm \mathbf{k}}\right] =0.$
Unitary symmetries of the super $Q_{\pm \mathbf{k}}=UQ_{\pm \mathbf{k}%
}U^{\dagger }$ are also symmetries of the SUSY Hamiltonian $H_{\mathbf{k}%
}=UH_{\mathbf{k}}U^{\dagger }.$

\subsubsection{ORTIC Hamiltonian matrices}

To get the realisation of the $H_{\mathbf{k}}^{ortic}$ in terms of the super
oscillators, we substitute $Q_{\pm \mathbf{k}}$ by their expressions given
by (\ref{QK}). We get a bosonic operator with a \emph{quartic} dependence
into the super oscillators as shown here below,
\begin{equation}
H_{\mathbf{k}}^{ortic}=\frac{1}{2}\left( \boldsymbol{q}_{\mathbf{k}}\right)
_{A}^{\dot{C}}\left( \boldsymbol{q}_{-\mathbf{k}}\right) _{B}^{\dot{D}%
}\left( \hat{\lambda}_{\mathbf{k}\dot{C}}\hat{\lambda}_{-\mathbf{k}\dot{D}}%
\hat{\xi}_{\mathbf{k}}^{A}\hat{\xi}_{-\mathbf{k}}^{B}+\hat{\lambda}_{-%
\mathbf{k}\dot{D}}\hat{\lambda}_{\mathbf{k}\dot{C}}\hat{\xi}_{-\mathbf{k}%
}^{B}\hat{\xi}_{\mathbf{k}}^{A}\right)
\end{equation}%
By using eq(\ref{xl}), we can bring the above relation to the following form%
\begin{equation}
\begin{tabular}{lll}
$H_{\mathbf{k}}^{ortic}$ & $=$ & $\frac{1}{2}\left( \boldsymbol{q}_{\mathbf{k%
}}\right) _{A}^{\dot{C}}\left( \boldsymbol{q}_{-\mathbf{k}}\right) _{B}^{%
\dot{D}}\left[ Z^{AB}\hat{\lambda}_{\mathbf{k}\dot{C}}\hat{\lambda}_{-%
\mathbf{k}\dot{D}}+\hat{\xi}_{-\mathbf{k}}^{B}\hat{\xi}_{\mathbf{k}}^{A}\hat{%
\lambda}_{\mathbf{k}\dot{C}}\hat{\lambda}_{-\mathbf{k}\dot{D}}\right] $ \\
&  & $+\frac{1}{2}\left( \boldsymbol{q}_{\mathbf{k}}\right) _{A}^{\dot{C}%
}\left( \boldsymbol{q}_{-\mathbf{k}}\right) _{B}^{\dot{D}}\left[ G_{\dot{C}%
\dot{D}}\hat{\xi}_{-\mathbf{k}}^{B}\hat{\xi}_{\mathbf{k}}^{A}-\hat{\lambda}_{%
\mathbf{k}\dot{C}}\hat{\lambda}_{-\mathbf{k}\dot{D}}\hat{\xi}_{-\mathbf{k}%
}^{B}\hat{\xi}_{\mathbf{k}}^{A}\right] $%
\end{tabular}
\label{HK}
\end{equation}%
This expression is remarkable and deserves a comment. First it can be
presented like the sum of two contributions $H_{\mathbf{k}}^{(I)}+H_{\mathbf{%
k}}^{(II)}$ with%
\begin{equation}
\begin{tabular}{lll}
$H_{\mathbf{k}}^{(I)}$ & $=$ & $\frac{1}{2}\hat{\lambda}_{\mathbf{k}\dot{C}%
}\left( \mathcal{H}_{\mathbf{k}}^{I}\right) ^{\dot{C}\dot{D}}\hat{\lambda}_{-%
\mathbf{k}\dot{D}}$ \\
$H_{\mathbf{k}}^{(II)}$ & $=$ & $\frac{1}{2}\hat{\xi}_{-\mathbf{k}%
}^{B}\left( \mathcal{H}_{\mathbf{k}}^{II}\right) _{AB}\hat{\xi}_{\mathbf{k}%
}^{A}$%
\end{tabular}%
\end{equation}%
and%
\begin{equation}
\begin{tabular}{lll}
$\left( \mathcal{H}_{\mathbf{k}}^{I}\right) ^{\dot{C}\dot{D}}$ & $=$ & $%
\left( \boldsymbol{q}_{\mathbf{k}}\right) _{A}^{\dot{C}}Z^{AB}\left(
\boldsymbol{q}_{-\mathbf{k}}\right) _{B}^{\dot{D}}+\left( \boldsymbol{q}_{%
\mathbf{k}}\right) _{A}^{\dot{C}}\left[ \hat{\xi}_{-\mathbf{k}}^{B}\hat{\xi}%
_{\mathbf{k}}^{A}\right] \left( \boldsymbol{q}_{-\mathbf{k}}\right) _{B}^{%
\dot{D}}$ \\
$\left( \mathcal{H}_{\mathbf{k}}^{II}\right) _{AB}$ & $=$ & $\left(
\boldsymbol{q}_{\mathbf{k}}\right) _{A}^{\dot{C}}G_{\dot{C}\dot{D}}\left(
\boldsymbol{q}_{-\mathbf{k}}\right) _{B}^{\dot{D}}-\left( \boldsymbol{q}_{%
\mathbf{k}}\right) _{A}^{\dot{C}}\left[ \hat{\lambda}_{\mathbf{k}\dot{C}}%
\hat{\lambda}_{-\mathbf{k}\dot{D}}\right] \left( \boldsymbol{q}_{-\mathbf{k}%
}\right) _{B}^{\dot{D}}$%
\end{tabular}%
\end{equation}%
Second, the coupling matrices $\left( \mathcal{H}_{\mathbf{k}}^{I}\right) ^{%
\dot{C}\dot{D}}$ and $\left( \mathcal{H}_{\mathbf{k}}^{II}\right) _{AB}$
have interesting features that we describe here after.

\begin{description}
\item[$\left( \mathbf{a}\right) $] As required by orthosymplectic invariance
(and then supersymmetry), the coupling matrices $\mathcal{H}_{\mathbf{k}%
}^{I} $ and $\mathcal{H}_{\mathbf{k}}^{II}$ are quadratic in the Fermi-Bose
coupling matrix $\boldsymbol{q}_{\mathbf{k}}.$

\item[$\left( \mathbf{b}\right) $] The $\left( \mathcal{H}_{\mathbf{k}%
}^{I}\right) ^{\dot{C}\dot{D}}$ matrix gives the coupling between $\hat{%
\lambda}_{\mathbf{k}\dot{C}}$ and $\hat{\lambda}_{-\mathbf{k}\dot{D}}$; it
has two terms: $\left( \mathbf{i}\right) $ a field independent term
\begin{equation}
\left( h_{f}\right) ^{\dot{C}\dot{D}}=\left( \boldsymbol{q}_{\mathbf{k}%
}\right) _{A}^{\dot{C}}Z^{AB}\left( \boldsymbol{q}_{-\mathbf{k}}\right)
_{B}^{\dot{D}}
\end{equation}%
with constant $Z^{AB}$. $\left( \mathbf{ii}\right) $ a field dependent term
\begin{equation}
\left( h_{f}^{\prime }\right) ^{\dot{C}\dot{D}}=\left( \boldsymbol{q}_{%
\mathbf{k}}\right) _{A}^{\dot{C}}[\hat{\xi}_{-\mathbf{k}}^{B}\hat{\xi}_{%
\mathbf{k}}^{A}]\left( \boldsymbol{q}_{-\mathbf{k}}\right) _{B}^{\dot{D}}
\end{equation}%
with local field dependence given by $\hat{\xi}_{-\mathbf{k}}^{B}\hat{\xi}_{%
\mathbf{k}}^{A}$.

\item[$\left( \mathbf{c}\right) $] The $\left( \mathcal{H}_{\mathbf{k}%
}^{II}\right) _{AB}$ coupling matrix gives the interaction between $\hat{\xi}%
_{-\mathbf{k}}^{B}$ and $\hat{\xi}_{\mathbf{k}}^{A}$; it also has also two
terms: $\left( \mathbf{i}\right) $ a field independent term
\begin{equation}
\left( h_{b}\right) _{AB}=\left( \boldsymbol{q}_{\mathbf{k}}\right) _{A}^{%
\dot{C}}G_{\dot{C}\dot{D}}\left( \boldsymbol{q}_{-\mathbf{k}}\right) _{B}^{%
\dot{D}}
\end{equation}%
with constant $G_{\dot{C}\dot{D}}.$ $\left( \mathbf{ii}\right) $ a field
dependent term
\begin{equation}
\left( h_{b}^{\prime }\right) _{AB}=\left( \boldsymbol{q}_{\mathbf{k}%
}\right) _{A}^{\dot{C}}[\hat{\lambda}_{\mathbf{k}\dot{C}}\hat{\lambda}_{-%
\mathbf{k}\dot{D}}]\left( \boldsymbol{q}_{-\mathbf{k}}\right) _{B}^{\dot{D}}
\end{equation}%
with local field dependence given by $\hat{\lambda}_{\mathbf{k}\dot{C}}\hat{%
\lambda}_{-\mathbf{k}\dot{D}}$.

\item[$\left( \mathbf{d}\right) $] \emph{Cancellation effect}: Expressing eq(%
\ref{HK}) as%
\begin{equation}
\begin{tabular}{lll}
$H_{\mathbf{k}}^{ortic}$ & $=$ & $\frac{1}{2}\left[ \hat{\lambda}_{\mathbf{k}%
\dot{C}}\left( h_{f}\right) ^{\dot{C}\dot{D}}\hat{\lambda}_{-\mathbf{k}\dot{D%
}}+\hat{\xi}_{-\mathbf{k}}^{B}\left( h_{b}\right) _{AB}\hat{\xi}_{\mathbf{k}%
}^{A}\right] +$ \\
&  & $\frac{1}{2}\left[ \hat{\lambda}_{\mathbf{k}\dot{C}}\left(
h_{f}^{\prime }\right) ^{\dot{C}\dot{D}}\hat{\lambda}_{-\mathbf{k}\dot{D}}-%
\hat{\xi}_{-\mathbf{k}}^{B}\left( h_{b}^{\prime }\right) _{AB}\hat{\xi}_{%
\mathbf{k}}^{A}\right] $%
\end{tabular}%
\end{equation}%
and using the commutation relation $\hat{\xi}_{\mathbf{k}}\hat{\lambda}_{%
\mathbf{k}}=\hat{\lambda}_{\mathbf{k}}\hat{\xi}_{\mathbf{k}},$ we see that
the second line in the above relation vanishes identically due to the
following compensation property
\begin{equation}
\hat{\lambda}_{\mathbf{k}\dot{C}}\left( h_{f}^{\prime }\right) ^{\dot{C}\dot{%
D}}\hat{\lambda}_{-\mathbf{k}\dot{D}}-\hat{\xi}_{-\mathbf{k}}^{B}\left(
h_{b}^{\prime }\right) _{AB}\hat{\xi}_{\mathbf{k}}^{A}=\hat{\xi}_{-\mathbf{k}%
}^{B}\hat{\xi}_{\mathbf{k}}^{A}\hat{\lambda}_{\mathbf{k}\dot{C}}\hat{\lambda}%
_{-\mathbf{k}\dot{D}}-\hat{\lambda}_{\mathbf{k}\dot{C}}\hat{\lambda}_{-%
\mathbf{k}\dot{D}}\hat{\xi}_{-\mathbf{k}}^{B}\hat{\xi}_{\mathbf{k}}^{A}=0
\label{cp}
\end{equation}%
This feature indicates that the presence of bosons in topological
supermatter is not without effect the quantum physics.
\end{description}

By taking into account the compensation property, we find that the
supersymmetric Hamiltonian $H_{\mathbf{k}}^{ortic}$ can be presented like $%
H_{f}+H_{b}$ with fermionic part%
\begin{equation}
\begin{tabular}{lll}
$H_{f}$ & $=$ & $\frac{1}{2}\hat{\lambda}_{\mathbf{k}\dot{C}}\left[
h_{f}\left( \mathbf{k}\right) \right] ^{\dot{C}\dot{D}}\hat{\lambda}_{-%
\mathbf{k}\dot{D}}$ \\
& $=$ & $\frac{1}{2}\mathbf{\hat{\lambda}}_{\mathbf{k}}\left[ \mathbf{h}%
_{f}\left( \mathbf{k}\right) \right] \mathbf{\hat{\lambda}}_{-\mathbf{k}}$%
\end{tabular}
\label{1}
\end{equation}%
and bosonic%
\begin{equation}
\begin{tabular}{lll}
$H_{b}$ & $=$ & $\frac{1}{2}\hat{\xi}_{\mathbf{k}}^{A}\left[ h_{b}\left(
\mathbf{k}\right) \right] _{AB}\hat{\xi}_{-\mathbf{k}}^{B}$ \\
& $=$ & $\frac{1}{2}\mathbf{\hat{\xi}}_{\mathbf{k}}\left[ h_{b}\left(
\mathbf{k}\right) \right] \mathbf{\hat{\xi}}_{-\mathbf{k}}$%
\end{tabular}
\label{2}
\end{equation}%
The coupling matrices $h_{f}\left( \mathbf{k}\right) $ and $h_{b}\left(
\mathbf{k}\right) $ are quadratic in the coupling $q_{\mathbf{k}}$ as given
below%
\begin{equation}
\begin{tabular}{lll}
$\left[ h_{f}\left( \mathbf{k}\right) \right] ^{\dot{C}\dot{D}}$ & $=$ & $%
\left( \boldsymbol{q}_{\mathbf{k}}\right) _{A}^{\dot{C}}Z^{AB}\left(
\boldsymbol{q}_{-\mathbf{k}}\right) _{B}^{\dot{D}}$ \\
$\left[ h_{b}\left( \mathbf{k}\right) \right] _{AB}$ & $=$ & $\left(
\boldsymbol{q}_{\mathbf{k}}\right) _{A}^{\dot{C}}G_{\dot{C}\dot{D}}\left(
\boldsymbol{q}_{-\mathbf{k}}\right) _{B}^{\dot{D}}$%
\end{tabular}%
\end{equation}%
They read shortly as follows%
\begin{equation}
h_{f}\left( \mathbf{k}\right) =\boldsymbol{q}_{\mathbf{k}}Z\boldsymbol{q}_{%
\mathbf{k}}^{\dagger }\qquad ,\qquad h_{b}\left( \mathbf{k}\right) =%
\boldsymbol{q}_{\mathbf{k}}^{\dagger }\boldsymbol{q}_{\mathbf{k}}  \label{fb}
\end{equation}%
with $\det \boldsymbol{h}_{b}=\left\vert \det_{\mathbf{k}}\boldsymbol{q}%
\right\vert ^{2}$ and $\det \boldsymbol{h}_{f}=\left( \det Z\right)
\left\vert \det_{\mathbf{k}}\boldsymbol{q}\right\vert ^{2}$. \newline
By using eq(\ref{qca}), we have the following explicit relations
\begin{equation}
\begin{tabular}{lll}
$h_{f}$ & $=$ & $\left(
\begin{array}{cc}
{\small (q}_{\mathbf{k}}{\small )}_{I}^{\dot{K}}{\small (q}_{-\mathbf{k}}%
{\small )}_{I}^{\dot{L}}-{\small (q}_{\mathbf{k}}{\small )}^{\dot{K}I}%
{\small (q}_{-\mathbf{k}}{\small )}_{I\dot{L}} & {\small (q}_{\mathbf{k}}%
{\small )}_{I}^{\dot{K}}{\small (q}_{-\mathbf{k}}{\small )}^{I\dot{L}}-%
{\small (q}_{\mathbf{k}}{\small )}^{\dot{K}I}{\small (q}_{-\mathbf{k}}%
{\small )}_{\dot{L}}^{I} \\
{\small (q}_{\mathbf{k}}{\small )}_{\dot{K}I}{\small (q}_{-\mathbf{k}}%
{\small )}_{I}^{\dot{L}}-{\small (q}_{\mathbf{k}}{\small )}_{\dot{K}}^{I}%
{\small (q}_{-\mathbf{k}}{\small )}_{I\dot{L}} & {\small (q}_{\mathbf{k}}%
{\small )}_{\dot{K}I}{\small (q}_{-\mathbf{k}}{\small )}^{I\dot{L}}-{\small %
(q}_{\mathbf{k}}{\small )}_{\dot{K}}^{I}{\small (q}_{-\mathbf{k}}{\small )}_{%
\dot{L}}^{I}%
\end{array}%
\right) $ \\
&  &  \\
$h_{b}$ & $=$ & $\left(
\begin{array}{cc}
{\small (q}_{\mathbf{k}}{\small )}_{I}^{\dot{K}}{\small (q}_{-\mathbf{k}}%
{\small )}_{I}^{\dot{L}}+{\small (q}_{\mathbf{k}}{\small )}^{\dot{K}I}%
{\small (q}_{-\mathbf{k}}{\small )}_{I\dot{L}} & {\small (q}_{\mathbf{k}}%
{\small )}_{I}^{\dot{K}}{\small (q}_{-\mathbf{k}}{\small )}^{I\dot{L}}+%
{\small (q}_{\mathbf{k}}{\small )}^{\dot{K}I}{\small (q}_{-\mathbf{k}}%
{\small )}_{\dot{L}}^{I} \\
{\small (q}_{\mathbf{k}}{\small )}_{\dot{K}I}{\small (q}_{-\mathbf{k}}%
{\small )}_{I}^{\dot{L}}+{\small (q}_{\mathbf{k}}{\small )}_{\dot{K}}^{I}%
{\small (q}_{-\mathbf{k}}{\small )}_{I\dot{L}} & {\small (q}_{\mathbf{k}}%
{\small )}_{\dot{K}I}{\small (q}_{-\mathbf{k}}{\small )}^{I\dot{L}}+{\small %
(q}_{\mathbf{k}}{\small )}_{\dot{K}}^{I}{\small (q}_{-\mathbf{k}}{\small )}_{%
\dot{L}}^{I}%
\end{array}%
\right) $%
\end{tabular}%
\end{equation}

\subsubsection{Massless modes from singular couplings}

Here, we give properties of the massless states in ORTIC and SUSY TBMs while
focussing on the coupling matrix $\left( \boldsymbol{q}_{\mathbf{k}}\right)
_{A}^{\dot{C}}$ given by the $U\left( N\right) $ super family. These are the
topological super states that are protected by discrete symmetries.\ \ \

\begin{figure}[tbph]
\begin{center}
\includegraphics[width=10cm]{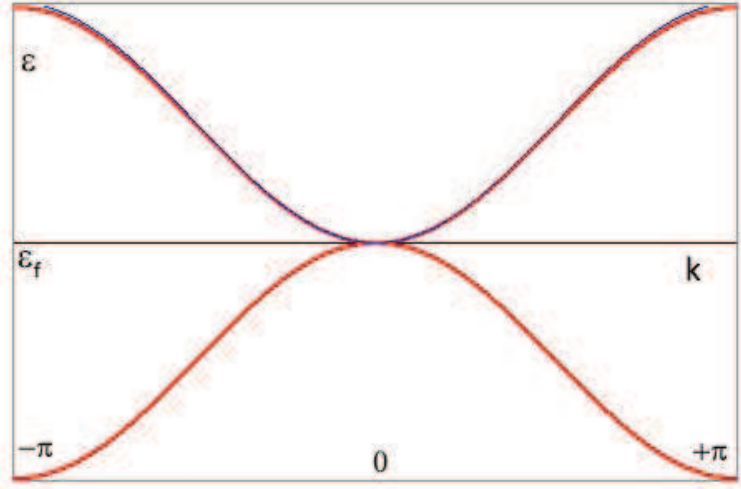}
\end{center}
\par
\vspace{-0.5cm}
\caption{The four bands $\protect\epsilon _{\pm }^{{\protect\small (f)}}=\pm
\left\vert \protect\mu _{\mathbf{k}}\right\vert ^{2}$, $\protect\epsilon %
_{\pm }^{{\protect\small (b)}}=\left\vert \protect\mu _{\mathbf{k}%
}\right\vert ^{2}$ of the $U\left( 1\right) $ super model given by eq(%
\protect\ref{mu}). {In red, the two symmetric bands of the $h_{f}$ with
respect to $\protect\varepsilon _{F}$. In blue, the degenerate bosonic bands
having positive energies. The parameter M is taken around $1$.}}
\label{SB}
\end{figure}

$\bullet $ \textbf{Zeros of the Hamiltonian}\emph{\ }$H_{tot}$\newline
By setting $H_{tot}=H_{b}+H_{f}$ given by eqs(\ref{1}-\ref{2}) and using the
$osp(2N|2N)$ matrix notation with 4N dimensional super vector basis $(%
\mathbf{\hat{\xi}}_{-\mathbf{k}},\mathbf{\hat{\lambda}}_{-\mathbf{k}})$, the
total Hamiltonian can be presented like
\begin{equation}
\left( H_{\mathbf{k}}\right) _{tot}=\frac{1}{2}\left( \mathbf{\hat{\xi}}_{%
\mathbf{k}},\mathbf{\hat{\lambda}}_{\mathbf{k}}\right) \left( \mathbf{h}_{%
\mathbf{k}}\right) _{tot}\left(
\begin{array}{c}
\mathbf{\hat{\xi}}_{-\mathbf{k}} \\
\mathbf{\hat{\lambda}}_{-\mathbf{k}}%
\end{array}%
\right)   \label{451}
\end{equation}%
with $4N\times 4N$ matrix as follows
\begin{equation}
\left( \mathbf{h}_{\mathbf{k}}\right) _{tot}=\left(
\begin{array}{cc}
\left( \mathbf{h}_{\mathbf{k}}\right) _{b} & 0 \\
0 & \left( \mathbf{h}_{\mathbf{k}}\right) _{f}%
\end{array}%
\right)   \label{452}
\end{equation}%
This matrix has $4N$ eigenstates states: 2N of them for the bosonic $\left(
\mathbf{h}_{\mathbf{k}}\right) _{b}$ and the other 2N for the fermion $%
\left( \mathbf{h}_{\mathbf{k}}\right) _{f}$. Because of the decoupling of $%
\left( \mathbf{h}_{\mathbf{k}}\right) _{b}$ and $\left( \mathbf{h}_{\mathbf{k%
}}\right) _{f}$, the determinant $\det \mathbf{h}_{tot}$ is given by the
product $\left( \det \mathbf{h}_{b}\right) .\left( \det \mathbf{h}%
_{f}\right) $. Moreover, substituting (\ref{fb}), we get
\begin{equation}
\det \mathbf{h}_{tot}=\left( -\right) ^{N}\left\vert \det q_{\mathbf{k}%
}\right\vert ^{4}
\end{equation}%
Zero modes of $\mathbf{h}_{tot}$ are given by the zeros of $\det q_{\mathbf{k%
}}$; so massless states of $\mathbf{h}_{tot}$ have multiplicity $4$; two
fermionic modes with vanishing gap and two bosonic partners. For an
illustration, we give in the Figure \textbf{\ref{SB} }the four super band%
\textbf{\ }energies $\epsilon _{\pm }^{{\small (f)}}=\pm \lbrack \left(
1-\cos k\right) ^{2}+\sin ^{2}k]$ and $\epsilon _{\pm }^{{\small (b)}}=\pm
\lbrack \left( 1-\cos k\right) ^{2}+\sin ^{2}k]$.

$\bullet $ \textbf{Supersymmetric bands in the }$U\left( 1\right) $\emph{\ }%
\textbf{super model}\newline
For the super model (\ref{qqk}) with $N=1$, the $\boldsymbol{q}_{\mathbf{k}}$%
-matrix has four entries and the hamiltonian matrices $h_{f}$ and $h_{b}$
read as follows%
\begin{equation}
\begin{tabular}{lll}
$h_{f}$ & $=$ & $\left(
\begin{array}{cc}
\left\vert \mu _{\mathbf{k}}\right\vert ^{2}-\left\vert \rho _{\mathbf{k}%
}\right\vert ^{2} & \mu _{\mathbf{k}}\bar{\varsigma}_{\mathbf{k}}-\rho _{%
\mathbf{k}}\bar{\nu}_{\mathbf{k}} \\
\varsigma _{\mathbf{k}}\bar{\mu}_{\mathbf{k}}-\nu _{\mathbf{k}}\bar{\rho}_{%
\mathbf{k}} & \left\vert \varsigma _{\mathbf{k}}\right\vert ^{2}-\left\vert
\nu _{\mathbf{k}}\right\vert ^{2}%
\end{array}%
\right) $ \\
&  &  \\
$h_{b}$ & $=$ & $\left(
\begin{array}{cc}
\left\vert \mu _{\mathbf{k}}\right\vert ^{2}+\left\vert \varsigma _{\mathbf{k%
}}\right\vert ^{2} & \bar{\mu}_{\mathbf{k}}\rho _{\mathbf{k}}+\bar{\varsigma}%
_{\mathbf{k}}\nu _{\mathbf{k}} \\
\bar{\rho}_{\mathbf{k}}\mu _{\mathbf{k}}+\bar{\nu}_{\mathbf{k}}\varsigma _{%
\mathbf{k}} & \left\vert \rho _{\mathbf{k}}\right\vert ^{2}+\left\vert \nu _{%
\mathbf{k}}\right\vert ^{2}%
\end{array}%
\right) $%
\end{tabular}
\label{hfhb}
\end{equation}%
Referring to these two matrices formally like%
\begin{equation}
h=\left(
\begin{array}{cc}
A & B \\
B^{\ast } & D%
\end{array}%
\right)
\end{equation}%
with entries as (\ref{hfhb}); we can determine their eigenstates and
eigenvalues. The eigenvalues $E_{\pm }$ read like%
\begin{equation}
E_{\pm }=\frac{1}{2}\left( A+D\right) \pm \frac{1}{2}\sqrt{\left( A-D\right)
^{2}+4\left\vert B\right\vert ^{2}}
\end{equation}%
and the corresponding eigenstates $\left\vert \upsilon _{\pm }\right\rangle $
as%
\begin{equation}
\left\vert \upsilon _{\pm }\right\rangle =\frac{1}{\sqrt{\mathcal{N}_{\pm }}}%
\left(
\begin{array}{c}
A-D\pm \sqrt{\left( A-D\right) ^{2}+4\left\vert B\right\vert ^{2}} \\
2B^{\ast }%
\end{array}%
\right)
\end{equation}%
with $\mathcal{N}_{\pm }$ given by the normalisation $\left\langle \upsilon
_{i}|\upsilon _{j}\right\rangle =\delta _{ij}.$ By demanding charge
conjugation invariance ( particle-hole symmetry), the number of functions
gets reduced as $\bar{\mu}_{-\mathbf{k}}=\nu _{\mathbf{k}}$ and $\bar{\rho}%
_{-\mathbf{k}}=\varsigma _{\mathbf{k}}.$ By substituting, we have%
\begin{equation}
\begin{tabular}{lll}
$h_{f}$ & $=$ & $\left(
\begin{array}{cc}
\left\vert \mu _{\mathbf{k}}\right\vert ^{2}-\left\vert \rho _{\mathbf{k}%
}\right\vert ^{2} & \mu _{\mathbf{k}}\rho _{-\mathbf{k}}-\rho _{\mathbf{k}%
}\mu _{-\mathbf{k}} \\
\bar{\rho}_{-\mathbf{k}}\bar{\mu}_{\mathbf{k}}-\bar{\mu}_{-\mathbf{k}}\bar{%
\rho}_{\mathbf{k}} & \left\vert \bar{\rho}_{-\mathbf{k}}\right\vert
^{2}-\left\vert \bar{\mu}_{-\mathbf{k}}\right\vert ^{2}%
\end{array}%
\right) $ \\
&  &  \\
$h_{b}$ & $=$ & $\left(
\begin{array}{cc}
\left\vert \mu _{\mathbf{k}}\right\vert ^{2}+\left\vert \bar{\rho}_{-\mathbf{%
k}}\right\vert ^{2} & \bar{\mu}_{\mathbf{k}}\rho _{\mathbf{k}}+\rho _{-%
\mathbf{k}}\bar{\mu}_{-\mathbf{k}} \\
\bar{\rho}_{\mathbf{k}}\mu _{\mathbf{k}}+\mu _{-\mathbf{k}}\bar{\rho}_{-%
\mathbf{k}} & \left\vert \rho _{\mathbf{k}}\right\vert ^{2}+\left\vert \bar{%
\mu}_{-\mathbf{k}}\right\vert ^{2}%
\end{array}%
\right) $%
\end{tabular}%
\end{equation}%
with traces given by%
\begin{equation}
\begin{tabular}{lll}
$tr\left( h_{f}\right) $ & $=$ & $\left\vert \mu _{\mathbf{k}}\right\vert
^{2}+\left\vert \bar{\rho}_{-\mathbf{k}}\right\vert ^{2}-\left\vert \rho _{%
\mathbf{k}}\right\vert ^{2}-\left\vert \bar{\mu}_{-\mathbf{k}}\right\vert
^{2}\gtrless 0$ \\
$tr\left( h_{b}\right) $ & $=$ & $\left\vert \mu _{\mathbf{k}}\right\vert
^{2}+\left\vert \rho _{\mathbf{k}}\right\vert ^{2}+\left\vert \bar{\rho}_{-%
\mathbf{k}}\right\vert ^{2}+\left\vert \bar{\mu}_{-\mathbf{k}}\right\vert
^{2}\geq 0$%
\end{tabular}%
\end{equation}%
For the case where we set $\rho _{\pm \mathbf{k}}=0$ corresponding to $%
U\left( 1\right) \times U\left( 1\right) $ symmetric model, we have%
\begin{equation}
h_{f}=\left(
\begin{array}{cc}
\left\vert \mu _{\mathbf{k}}\right\vert ^{2} & 0 \\
0 & -\left\vert \bar{\mu}_{-\mathbf{k}}\right\vert ^{2}%
\end{array}%
\right) \qquad ,\qquad h_{b}=\left(
\begin{array}{cc}
\left\vert \mu _{\mathbf{k}}\right\vert ^{2} & 0 \\
0 & \left\vert \bar{\mu}_{-\mathbf{k}}\right\vert ^{2}%
\end{array}%
\right)
\end{equation}%
If moreover, we demand time reversal symmetry, we must have $\bar{\mu}_{%
\mathbf{k}}=\mu _{-\mathbf{k}}$; this leads to the following matrices $%
h_{f}=\left\vert \mu _{\mathbf{k}}\right\vert ^{2}\sigma _{z}$ and $%
h_{b}=\left\vert \mu _{\mathbf{k}}\right\vert ^{2}\sigma _{0}$. For this
model, an interesting function $\mu _{\mathbf{k}}$ is given by
\begin{equation}
\begin{tabular}{lll}
$\mu _{\mathbf{k}}$ & $=$ & $\left( M-\cos k\right) +i\sin k$ \\
$\left\vert \mu _{\mathbf{k}}\right\vert ^{2}$ & $=$ & $\left( M-\cos
k\right) ^{2}+\sin ^{2}k$%
\end{tabular}
\label{mu}
\end{equation}%
showing the existence of four massless states at the fix points $k_{\ast
}=0,\pi $ and $M=\cos k_{\ast }.$ For $\left\vert M\right\vert >1$, we have
massive states.

\section{More on topological supermatter}

In this section, we deepen the investigation of the topological properties
of the super bands while illustrating these features on some classes of the
AZ table \textrm{\cite{1D,1DA}}.

\subsection{Constraints from ORTIC and SUSY}

Topological supermatter is given by ordinary matter constrained by \textrm{%
orthosymmetric invariance or supersymmetry}; as such it constitutes a subset
of the AZ matter; but \textrm{requires bosons}. Here, we focus on spinless
supermatter and develop a method for constructing a family of coupling
matrices $\mathbf{q}_{\mathbf{k}}$ by starting from known $h_{f}$'s. This
proposal of engineering $\mathbf{q}_{\mathbf{k}}$'s out of $h_{f}$ was first
suggested in \textrm{\cite{1M}}; \textrm{it will be given below an
interpretation in terms of symmetries}.

\subsubsection{TPC symmetries of $q_{\mathbf{k}}$}

In the AZ table, the topological classes are modeled by hamiltonians $%
h_{f}\left( \mathbf{k}\right) $ constrained by TPC. One may also demand
other discrete symmetries like crystalline symmetries \textrm{\cite{CR1,CR2}}%
. For the spinless case, we have
\begin{equation}
\mathcal{T}=K,\qquad \mathcal{P}=XK,\qquad \mathcal{C}=X,\qquad X=\sigma
_{x}\otimes I_{{\small N\times N}}
\end{equation}%
The actions of these symmetry generators on $h_{f}\left( \mathbf{k}\right) $
are given by%
\begin{equation}
\begin{tabular}{lllll}
$\mathcal{T}$ & : & $h_{f}\left( \mathbf{k}\right) ^{\ast }$ & $=$ & $%
+h_{f}\left( -\mathbf{k}\right) $ \\
$\mathcal{P}$ & : & $Xh_{f}\left( \mathbf{k}\right) ^{\ast }X$ & $=$ & $%
-h_{f}\left( -\mathbf{k}\right) $ \\
$\mathcal{C}$ & : & $Xh_{f}\left( \mathbf{k}\right) X$ & $=$ & $-h_{f}\left(
\mathbf{k}\right) $%
\end{tabular}
\label{xhf}
\end{equation}%
The topological indices $\mathcal{I}_{m}$ characterising the matter phases
are nicely derived by considering \textrm{involution} hamiltonian matrices $%
\hat{h}_{f}$ on the Brillouin torus $\mathbb{T}^{d}.$ These $\mathcal{I}_{m}$%
's are integers that can be either in $\mathbb{Z}$, $2\mathbb{Z}$ or $%
\mathbb{Z}_{2}$; the expression of $\mathcal{I}_{m}$ in terms of $\hat{h}%
_{f} $ depends on the parity of the spatial dimensions d. For even $d=2m$
for instance, the integer $\mathcal{I}_{m}$ is either given by the m-th
Chern number $Ch_{m}$ or the Fu-Kane index as follows,%
\begin{equation}
Ch_{m}=\frac{1}{m!}\int_{\mathbb{T}^{2m}}Tr\left( \frac{i\mathcal{F}}{2\pi }%
\right) ^{m}
\end{equation}%
and
\begin{eqnarray}
FK_{m} &=&\frac{i^{m}}{\left( 2\pi \right) ^{m}m!}\dint_{\frac{1}{2}\mathbb{T%
}^{2m}}Tr\left( \mathcal{F}^{m}\right)  \notag \\
&&-\frac{i^{m}}{\left( 2\pi \right) ^{m}\left( m-1\right) !}%
\dint\nolimits_{\partial (\frac{1}{2}\mathbb{T}^{2m})}\left[
\dint\nolimits_{0}^{1}dtTr\left( \mathcal{AF}_{t}^{m-1}\right) \right]
\end{eqnarray}%
In these relations, the Berry curvature $\mathcal{F}$ is given by the 2-form
$d\mathcal{A}+\mathcal{A}\wedge \mathcal{A}$ and the parametric $\mathcal{F}%
_{t}$ by $td\mathcal{A}+t^{2}\mathcal{A}^{2}$.\newline
In the super band theory, the hamiltonian matrix $h_{f}\left( \mathbf{k}%
\right) $ is given by $\boldsymbol{q}_{\mathbf{k}}Z\boldsymbol{q}_{\mathbf{k}%
}^{\dagger }$ and the bosonic partner $h_{b}\left( \mathbf{k}\right) $ by $%
\boldsymbol{q}_{\mathbf{k}}\boldsymbol{q}_{\mathbf{k}}^{\dagger }$. \newline
Putting $h_{f}\left( \mathbf{k}\right) =\boldsymbol{q}_{\mathbf{k}}Z%
\boldsymbol{q}_{\mathbf{k}}^{\dagger }$ into (\ref{xhf}), we get the
following transformation of the coupling matrix $q\left( \mathbf{k}\right) $
for spinless fermions%
\begin{equation}
\begin{tabular}{lllll}
$\mathcal{T}$ & : & $\boldsymbol{q}\left( \mathbf{k}\right) ^{\ast }$ & $=$
& $+\boldsymbol{q}\left( -\mathbf{k}\right) $ \\
$\mathcal{P}$ & : & $X\boldsymbol{q}\left( \mathbf{k}\right) ^{\ast }X$ & $=$
& $+\boldsymbol{q}\left( -\mathbf{k}\right) $ \\
$\mathcal{C}$ & : & $\mathcal{C}\boldsymbol{q}\left( \mathbf{k}\right)
\mathcal{C}^{-1}$ & $=$ & $+\boldsymbol{q}\left( \mathbf{k}\right) $%
\end{tabular}
\label{xq}
\end{equation}%
Substituting these transformations into the bosonic $h_{b}\left( \mathbf{k}%
\right) =\boldsymbol{q}_{\mathbf{k}}^{\dagger }\boldsymbol{q}_{\mathbf{k}}$,
we end up with the following transformations%
\begin{equation}
\begin{tabular}{lllll}
$\mathcal{T}$ & : & $h_{b}\left( \mathbf{k}\right) ^{\ast }$ & $=$ & $%
+h_{b}\left( -\mathbf{k}\right) $ \\
$\mathcal{P}$ & : & $Xh_{b}\left( \mathbf{k}\right) ^{\ast }X$ & $=$ & $%
+h_{b}\left( -\mathbf{k}\right) $ \\
$\mathcal{C}$ & : & $\mathcal{C}h_{b}\left( \mathbf{k}\right) \mathcal{C}%
^{-1}$ & $=$ & $+h_{b}\left( \mathbf{k}\right) $%
\end{tabular}%
\end{equation}%
from which we see that $\mathcal{P}$ and $\mathcal{C}$ have different
actions on $h_{f}\left( \mathbf{k}\right) $ and $h_{b}\left( \mathbf{k}%
\right) $. Notice that also that $h_{f}\left( \mathbf{k}\right) $ and $%
Zh_{b}\left( \mathbf{k}\right) $ have similar transformations%
\begin{equation}
\begin{tabular}{lllll}
$\mathcal{T}$ & : & $Zh_{b}\left( \mathbf{k}\right) ^{\ast }$ & $=$ & $%
+Zh_{b}\left( -\mathbf{k}\right) $ \\
$\mathcal{P}$ & : & $X[Zh_{b}\left( \mathbf{k}\right) ^{\ast }]X$ & $=$ & $%
-Zh_{b}\left( -\mathbf{k}\right) $ \\
$\mathcal{C}$ & : & $\mathcal{C}[Zh_{b}\left( \mathbf{k}\right) ^{\ast }]%
\mathcal{C}^{-1}$ & $=$ & $-h_{b}\left( \mathbf{k}\right) $%
\end{tabular}%
\end{equation}%
showing that the TPC symmetries agree with the fact that the matrix $%
Zh_{b}\left( \mathbf{k}\right) =Zq_{\mathbf{k}}^{\dagger }q_{\mathbf{k}}$
has the same spectrum as $h_{f}\left( \mathbf{k}\right) .$ This feature
follows from $\left( i\right) $ the fact that $h_{f}$ and $Zh_{b}$ factorise
like \textsc{AB} and \textsc{BA} with \textsc{A}$=\boldsymbol{q}_{\mathbf{k}%
} $ and \textsc{B}$=Z\boldsymbol{q}_{\mathbf{k}}^{\dagger }$; and $\left(
ii\right) $ the factors \textsc{AB} and \textsc{BA} have the same spectrum.%
\newline
Below, we consider the super TBM family with $U\left( 1\right) ^{N}$
symmetry; they allow to extract straightforwardly information on the
topological phases.

\subsubsection{Coupling $\boldsymbol{q}\left( \mathbf{k}\right) $ of super
models $U\left( 1\right) ^{N}$}

To engineer the $2N\times 2N$ coupling matrix $\boldsymbol{q}\left( \mathbf{k%
}\right) $ for the super models $U\left( 1\right) ^{N},$ we start from an AZ
hamiltonian $h_{f}\left( \mathbf{k}\right) $ with a given TPC symmetry like
in (\ref{xhf}). Because $h_{f}$ is given by $\boldsymbol{q}_{\mathbf{k}}Z%
\boldsymbol{q}_{\mathbf{k}}^{\dagger },$ the fermionic Hamiltonian to start
with must be $2N\times 2N.$\ \ \

$\bullet $ \textbf{Spectrum of} $h_{f}\left( \mathbf{k}\right) $\newline
To get the eigenstates and the eigenvalues for $U\left( 1\right) ^{N}$ super
models, we use the even parity 2N to expand this $h_{f}\left( \mathbf{k}%
\right) $ like
\begin{equation}
h_{f}\left( \mathbf{k}\right) =\sum_{\mu =x,y,z}\sigma ^{\mu }d_{\mu }\left(
\mathbf{k}\right)  \label{dm}
\end{equation}%
where we have taken $h_{f}\left( \mathbf{k}\right) $ traceless. Explicitly,
we have
\begin{equation}
h_{f}=\left(
\begin{array}{cc}
d_{z} & d_{x}-id_{y} \\
d_{x}+id_{y} & -d_{z}%
\end{array}%
\right)  \label{hff}
\end{equation}%
In this relation, the three functions $d_{\mu }=d_{\mu }\left( \mathbf{k}%
\right) $ are hermitian $N\times N$ matrices valued in the $u\left( N\right)
$ Lie algebra with $N^{2}$ dimensions. Denoting by $E_{IJ}=\left\vert
I\right\rangle \left\langle J\right\vert $ the generators of $u\left(
N\right) ,$ we can expand the $d_{\mu }$s like $\sum_{I,J}d_{\mu
}^{IJ}E_{IJ} $ that split as%
\begin{equation}
d_{\mu }=\dsum\limits_{I=1}^{N}d_{\mu
}^{I}E_{I}^{0}+\dsum\limits_{I<J}^{N}d_{\mu
}^{IJ}E_{IJ}^{-}+\dsum\limits_{I>J=1}^{N}d_{\mu }^{IJ}E_{IJ}^{+}
\end{equation}%
with $E_{I}^{0}=\left\vert I\right\rangle \left\langle I\right\vert .$
According to the classification (\ref{437}), we have the following
constraints on the $d_{\mu }^{IJ}$'s%
\begin{equation}
\begin{tabular}{|c|c|c|c|}
\hline
{\small symmetry} & {\small constraints} & {\small values of }$d_{\mu }^{IJ}$
& {\small parameters} \\ \hline
$U\left( N\right) $ & $\left[ E_{IJ},d_{\mu }\right] =0$ & $d_{\mu }=\mathrm{%
d}_{\mu }^{0}I_{_{{\small N\times N}}}$ & $\mathrm{d}_{\mu }^{0}$ \\ \hline
$U\left( 1\right) ^{N}$ & $\left[ E_{I}^{0},d_{\mu }\right] =0$ & $d_{\mu
}=\dsum\limits_{I=1}^{N}d_{\mu }^{I}E_{I}^{0}$ & $d_{\mu }^{1},...,d_{\mu
}^{N}$ \\ \hline
\end{tabular}%
\end{equation}%
Because of the abelian symmetry, the calculations for the $U\left( 1\right)
^{N}$ super models family are similar to the case of one factor $U\left(
1\right) $; as such we will hide the Table l in the $d_{\mu }^{I}$s seen
that this omission does not affect the calculations. \newline
The eigenvalues of $h_{f}$ are $\varepsilon _{\pm }=\pm \varepsilon $ with $%
\varepsilon =\sqrt{d_{x}^{2}+d_{y}^{2}+d_{z}^{2}}$ with eigenvectors $V_{\pm
}$ given by%
\begin{equation}
V_{+}=\left(
\begin{array}{c}
\frac{d_{x}-id_{y}}{d_{x}^{2}+d_{y}^{2}}\left[ d_{z}+\varepsilon \right] \\
1%
\end{array}%
\right) \qquad ,\qquad V_{-}=\left(
\begin{array}{c}
-\frac{d_{x}-id_{y}}{d_{x}^{2}+d_{y}^{2}}\left[ \varepsilon -d_{z}\right] \\
1%
\end{array}%
\right)  \label{ev}
\end{equation}%
By setting $e^{i\phi }=\left( d_{x}-id_{y}\right) /\sqrt{d_{x}^{2}+d_{y}^{2}}
$, the eigenvectors $V_{\pm }$ are normalised as follows%
\begin{equation}
V_{+}=\frac{1}{\sqrt{2\varepsilon }}\left(
\begin{array}{c}
e^{-i\phi }\sqrt{d_{z}+\varepsilon } \\
\sqrt{\varepsilon -d_{z}}%
\end{array}%
\right) \qquad ,\qquad V_{-}=\frac{1}{\sqrt{2\varepsilon }}\left(
\begin{array}{c}
\sqrt{\varepsilon -d_{z}} \\
-e^{+i\phi }\sqrt{\varepsilon +d_{z}}%
\end{array}%
\right)  \label{ew}
\end{equation}

$\bullet $ \textbf{Building}\emph{\ }$q_{\mathbf{k}}$\emph{\ }\textbf{from}%
\emph{\ }$h_{f}$ \newline
Given the above hermitian matrix $h_{f}$, we can diagonalise it by a unitary
transformation $h_{f}=\boldsymbol{V}_{\mathbf{k}}\Delta _{\mathbf{k}}%
\boldsymbol{V}_{\mathbf{k}}^{\dagger }$ where the diagonal matrix $\Delta _{%
\mathbf{k}}$ is given by%
\begin{equation}
\Delta _{\mathbf{k}}=\left(
\begin{array}{cc}
{\small \varepsilon }_{\mathbf{k}{\small (I)}} & 0 \\
0 & -{\small \varepsilon }_{-\mathbf{k}{\small (I)}}%
\end{array}%
\right)  \label{del}
\end{equation}%
and
\begin{equation}
{\small \varepsilon }_{\pm \mathbf{k}{\small (I)}}=\left(
\begin{array}{ccc}
{\small \varepsilon }_{\pm \mathbf{k}{\small 1}} &  &  \\
& {\small \ddots } &  \\
&  & {\small \varepsilon }_{\pm \mathbf{k}{\small N}}%
\end{array}%
\right)
\end{equation}%
with the property ${\small \varepsilon }_{\pm \mathbf{k}{\small i}}\geq 0$
and the ordering ${\small \varepsilon }_{\pm \mathbf{k}\left( {\small i+1}%
\right) }\geq {\small \varepsilon }_{\pm \mathbf{k}{\small i}}.$ For later
use, it is interesting to introduce the two following matrices%
\begin{equation}
D_{\mathbf{k}}=\left(
\begin{array}{cc}
{\small \varepsilon }_{\mathbf{k}{\small (I)}} & 0 \\
0 & {\small \varepsilon }_{-\mathbf{k}{\small (I)}}%
\end{array}%
\right) \qquad ,\qquad D_{\mathbf{k}}^{1/2}=\left(
\begin{array}{cc}
\sqrt{{\small \varepsilon }_{\mathbf{k}{\small (I)}}} & 0 \\
0 & \sqrt{{\small \varepsilon }_{-\mathbf{k}{\small (I)}}}%
\end{array}%
\right)
\end{equation}%
which are related to (\ref{del}) like $\Delta _{\mathbf{k}}=ZD_{\mathbf{k}%
}=D_{\mathbf{k}}Z.$ Notice that the square root $D_{\mathbf{k}}^{1/2}$ is
well defined because of the positivity of the eigenvalues ${\small %
\varepsilon }_{\pm \mathbf{k}{\small i}}.$ Using the relation $h_{f}=%
\boldsymbol{V}_{\mathbf{k}}\Delta _{\mathbf{k}}\boldsymbol{V}_{\mathbf{k}%
}^{\dagger }$ and the factorisation $D_{\mathbf{k}}=D_{\mathbf{k}}^{1/2}D_{%
\mathbf{k}}^{1/2}$ as well as $ZD_{\mathbf{k}}^{1/2}=D_{\mathbf{k}}^{1/2}Z$,
we can first express $h_{f}$ like $\boldsymbol{V}_{\mathbf{k}}\left( ZD_{%
\mathbf{k}}\right) \boldsymbol{V}_{\mathbf{k}}^{\dagger }$ and then as
follows%
\begin{equation}
h_{f}=\left( \boldsymbol{V}_{\mathbf{k}}D_{\mathbf{k}}^{1/2}\right) Z\left(
D_{\mathbf{k}}^{1/2}\boldsymbol{V}_{\mathbf{k}}^{\dagger }\right)
\end{equation}%
Equating with $h_{f}=q_{\mathbf{k}}Zq_{\mathbf{k}}^{\dagger }$, we end up
with%
\begin{equation}
q_{\mathbf{k}}=\boldsymbol{V}_{\mathbf{k}}D_{\mathbf{k}}^{1/2}  \label{qk}
\end{equation}%
Putting this expression of $D_{\mathbf{k}}^{1/2}$ back into the value of the
bosonic $h_{b}=q_{\mathbf{k}}^{\dagger }q_{\mathbf{k}},$ we obtain the
bosonic hamiltonian $h_{b}=D_{\mathbf{k}}^{1/2}V_{\mathbf{k}}^{\dagger }V_{%
\mathbf{k}}D_{\mathbf{k}}^{1/2}$ reading as follows%
\begin{equation}
h_{b}=D_{\mathbf{k}}  \label{kh}
\end{equation}%
with no dependence into $\boldsymbol{V}_{\mathbf{k}}$ indicating that $h_{b}$
is topologically trivial.

\subsection{\textbf{Interpreting the }$\boldsymbol{q}_{\mathbf{k}}$-\textbf{%
\ coupling tensor}}

Here, we give an algebraic interpretation of the 2N$\times $2N Bose-Fermi
coupling tensor ${\small [}\boldsymbol{q}_{\mathbf{k}}{\small ]}_{A}^{\dot{C}%
}$ given by (\ref{qca}) and derive supersymmetric constraints relating the
four N$\times $N block matrices $\boldsymbol{q}_{1\mathbf{k}},$ $\boldsymbol{%
q}_{2\mathbf{k}},$ $\boldsymbol{q}_{3\mathbf{k}},$ $\boldsymbol{q}_{4\mathbf{%
k}}$ making ${\small [}\boldsymbol{q}_{\mathbf{k}}{\small ]}_{A}^{\dot{C}}$.
The basic idea behind this interpretation goes back to eqs(\ref{cB}-\ref{Xb}%
) that we discuss them further in this subsection. For this purpose, we
first study special limits of ${\small [}\boldsymbol{q}_{\mathbf{k}}{\small ]%
}_{A}^{\dot{C}}$ given by the diagonal $z_{\mathbf{k}}^{A}{\small \delta }%
_{A}^{\dot{C}}$ with the $z_{\mathbf{k}}^{A}$ 's 2N complex numbers. Then,
we turn to investigate the deformation away from $z_{\mathbf{k}}^{A}{\small %
\delta }_{A}^{\dot{C}}$.

\subsubsection{The coupling limit $\boldsymbol{q}_{\mathbf{k}}=z_{\mathbf{k}}%
{\protect\small I}_{id}$}

We begin by the fermionic $\hat{c}_{\mathbf{k}}^{\dot{J}}/\hat{c}_{\mathbf{k}%
\dot{J}}^{\dagger }$ and the bosonic $\hat{b}_{\mathbf{k}}^{I}/\hat{b}_{%
\mathbf{k}I}^{\dagger }$ oscillator realisation of the ORTIC charge $Q_{%
\mathbf{k}}$ given by $\hat{\lambda}_{\mathbf{k}\dot{C}}[\boldsymbol{q}_{%
\mathbf{k}}]_{A}^{\dot{C}}\hat{\xi}_{\mathbf{k}}^{A}$ reading explicitly as
follows%
\begin{equation}
\begin{tabular}{lll}
$Q_{\mathbf{k}}$ & $=$ & $\dsum\limits_{I,\dot{J}}\hat{c}_{\mathbf{k}\dot{J}%
}^{\dagger }\left( \boldsymbol{q}_{1\mathbf{k}}\right) _{I}^{\dot{J}}\hat{b}%
_{\mathbf{k}}^{I}+\hat{c}_{\mathbf{k}\dot{J}}^{\dagger }\hat{b}_{\mathbf{k}%
I}^{\dagger }\left( \boldsymbol{q}_{2\mathbf{k}}\right) ^{\dot{J}I}$ \\
&  & $+\dsum\limits_{I,\dot{J}}\left( \boldsymbol{q}_{3\mathbf{k}}\right) _{%
\dot{J}I}\hat{c}_{\mathbf{k}}^{\dot{J}}\hat{b}_{\mathbf{k}}^{I}+\hat{b}_{%
\mathbf{k}I}^{\dagger }\left( \boldsymbol{q}_{4\mathbf{k}}\right) _{\dot{J}%
}^{I}\hat{c}_{\mathbf{k}}^{\dot{J}}$%
\end{tabular}
\label{QB}
\end{equation}%
where the four blocks ($\boldsymbol{q}_{1\mathbf{k}})_{I}^{\dot{J}},$ ($%
\boldsymbol{q}_{2\mathbf{k}})^{\dot{J}I},$ ($\boldsymbol{q}_{3\mathbf{k}})_{%
\dot{J}I}$ and ($\boldsymbol{q}_{4\mathbf{k}})_{\dot{J}}^{I}$ are $N\times N$
coupling matrices as in (\ref{qca}); they are functions of the momentum $%
\mathbf{k}.$\ In the diagonal case where $[\boldsymbol{q}_{\mathbf{k}}]_{A}^{%
\dot{C}}$ is given by%
\begin{equation*}
\lbrack \boldsymbol{q}_{\mathbf{k}}]_{A}^{\dot{C}}=z_{\mathbf{k}}^{A}\delta
_{A}^{\dot{C}}
\end{equation*}%
we have $\left( \boldsymbol{q}_{1\mathbf{k}}\right) _{I}^{\dot{J}}=u_{%
\mathbf{k}}^{I}\delta _{I}^{\dot{J}}$ and $\left( \boldsymbol{q}_{4\mathbf{k}%
}\right) _{\dot{J}}^{I}=v_{\mathbf{k}}^{\dot{J}}\delta _{\dot{J}}^{I}$ while
$\boldsymbol{q}_{2\mathbf{k}}=\boldsymbol{q}_{3\mathbf{k}}=0_{{\small %
N\times N}}$. By substituting, the above ORTIC charge $Q_{\mathbf{k}}$
becomes%
\begin{equation}
Q_{\mathbf{k}}=\sum u_{\mathbf{k}}^{I}\hat{c}_{\mathbf{k}I}^{\dagger }\hat{b}%
_{\mathbf{k}}^{I}+\sum v_{\mathbf{k}}^{\dot{I}}\hat{b}_{\mathbf{k}\dot{I}%
}^{\dagger }\hat{c}_{\mathbf{k}}^{\dot{I}})
\end{equation}%
Here, the $z_{\mathbf{k}}^{A}=(u_{\mathbf{k}}^{I};v_{\mathbf{k}}^{\dot{I}})$
are complex functions of momentum with $\left\vert z_{\mathbf{k}%
}^{A}\right\vert ^{2}=\omega _{\mathbf{k}}^{A}$ thought of in terms of real
frequencies scaling as energy. In the very special limit where $[\boldsymbol{%
q}_{\mathbf{k}}]_{A}^{\dot{C}}$ is proportional to the identity $z_{\mathbf{k%
}}\delta _{A}^{\dot{C}}$, we have $u_{\mathbf{k}}^{I}=v_{\mathbf{k}}^{\dot{I}%
}=z_{\mathbf{k}}$ and $\boldsymbol{q}_{1\mathbf{k}}=\boldsymbol{q}_{4\mathbf{%
k}}=z_{\mathbf{k}}I_{{\small N\times N}}$. In this particular situation, the
above ORTIC charge $Q_{\mathbf{k}}$ reduces further to
\begin{equation}
Q_{\mathbf{k}}=z_{\mathbf{k}}\hat{c}_{\mathbf{k}I}^{\dagger }\hat{b}_{%
\mathbf{k}}^{I}+z_{\mathbf{k}}\hat{b}_{\mathbf{k}I}^{\dagger }\hat{c}_{%
\mathbf{k}}^{I}
\end{equation}%
Up to the scale factor $z_{\mathbf{k}}$, this supercharge is just the sum of
two fermionic operators $\hat{c}_{\mathbf{k}I}^{\dagger }\hat{b}_{\mathbf{k}%
}^{I}$ and its adjoint conjugate $\hat{b}_{\mathbf{k}I}^{\dagger }\hat{c}_{%
\mathbf{k}}^{I}$ which are nothing but the $F_{-}^{+}$ and $\bar{F}_{+}^{-}$
generators of osp(2\TEXTsymbol{\vert}2) within osp(2N\TEXTsymbol{\vert}2N).
So, for the choice $\boldsymbol{q}_{\mathbf{k}}=z_{\mathbf{k}}{\small I}%
_{id} $, the orthosymplectic Hamiltonian is given by
\begin{equation}
H_{\mathbf{k}}^{ortic}=\frac{\left\vert z_{\mathbf{k}}\right\vert ^{2}}{2}%
\left( \hat{b}_{\mathbf{k}\dot{I}}^{\dagger }\hat{b}_{\mathbf{k}}^{\dot{I}}+%
\hat{b}_{\mathbf{k}}^{\dot{I}}\hat{b}_{\mathbf{k}\dot{I}}^{\dagger }\right) +%
\frac{\left\vert z_{\mathbf{k}}\right\vert ^{2}}{2}\left( \hat{c}_{\mathbf{k}%
\dot{I}}^{\dagger }\hat{c}_{\mathbf{k}}^{\dot{I}}-\hat{c}_{\mathbf{k}}^{\dot{%
I}}\hat{c}_{\mathbf{k}\dot{I}}^{\dagger }\right)  \label{hortic}
\end{equation}%
with matrix representation in the basis ($\hat{b}_{\mathbf{k}}^{\dot{I}},%
\hat{b}_{\mathbf{k}\dot{I}}^{\dagger },\hat{c}_{\mathbf{k}}^{\dot{I}},\hat{c}%
_{\mathbf{k}\dot{I}}^{\dagger }$) as follows%
\begin{equation}
\frac{1}{2}\left(
\begin{array}{cccc}
+\omega _{\mathbf{k}} &  &  &  \\
& +\omega _{\mathbf{k}} &  &  \\
&  & +\omega _{\mathbf{k}} &  \\
&  &  & -\omega _{\mathbf{k}}%
\end{array}%
\right)  \label{ome}
\end{equation}%
where we have set $\left\vert z_{\mathbf{k}}\right\vert ^{2}=\omega _{%
\mathbf{k}}$. From this particular limit, one may think about eqs (\ref{QB})
and (\ref{QB}) as follows. $\left( \mathbf{i}\right) $ Eq(\ref{QB}) is given
by the sum of two terms like $Q_{\mathbf{k}}=Q_{\mathbf{k}}^{+}+Q_{\mathbf{k}%
}^{-}$ with
\begin{equation}
\begin{tabular}{lll}
$Q_{\mathbf{k}}^{+}$ & $=$ & $z_{\mathbf{k}}\hat{c}_{\mathbf{k}I}^{\dagger }%
\hat{b}_{\mathbf{k}}^{I}$ \\
$Q_{\mathbf{k}}^{-}$ & $=$ & $z_{\mathbf{k}}\hat{b}_{\mathbf{k}I}^{\dagger }%
\hat{c}_{\mathbf{k}}^{I}$%
\end{tabular}
\label{qp}
\end{equation}%
$\left( \mathbf{ii}\right) $ Eq(\ref{QB}) is a deformation of the above (\ref%
{qp}); and its coupling $\boldsymbol{q}_{\mathbf{k}}$ as a deviation away
from $z_{\mathbf{k}}I_{id}$. Before exploring this deformation, notice that
the orthosymplectic eq(\ref{hortic}) corresponding to $\boldsymbol{q}_{%
\mathbf{k}}=z_{\mathbf{k}}{\small I}_{id}$ coincides with the supersymmetric
Hamiltonian $H_{\mathbf{k}}^{susy}$. This is because the $Q_{\mathbf{k}%
}^{\pm }$ of (\ref{qp}) are nilpotent and the Hamiltonian eq(\ref{hortic})
commutes with $Q_{\mathbf{k}}^{\pm }$; that is
\begin{equation}
\left[ H_{\mathbf{k}}^{ortic},Q_{\mathbf{k}}^{\pm }\right] =0\quad ,\quad
\left( Q_{\mathbf{k}}^{-}\right) ^{2}=\left( Q_{\mathbf{k}}^{+}\right) ^{2}=0
\end{equation}%
Extending the fermionic charges (\ref{qp}) for couplings $[\boldsymbol{q}_{%
\mathbf{k}}]_{A}^{\dot{C}}$ beyond the diagonal $z_{\mathbf{k}}\delta _{A}^{%
\dot{C}}$, we can present the $Q_{\mathbf{k}}^{+}$ and the $Q_{\mathbf{k}%
}^{-}$ as well as their adjoint conjugates $\bar{Q}_{\mathbf{k}}^{-}$ and $%
\bar{Q}_{\mathbf{k}}^{+}$ as follows,%
\begin{equation}
\begin{tabular}{lll}
$Q_{\mathbf{k}}^{+}$ & $=$ & $\hat{c}_{\mathbf{k}\dot{J}}^{\dagger }\left(
\boldsymbol{q}_{1\mathbf{k}}\right) _{I}^{\dot{J}}\hat{b}_{\mathbf{k}}^{I}+%
\hat{c}_{\mathbf{k}\dot{J}}^{\dagger }\hat{b}_{\mathbf{k}I}^{\dagger }\left(
\boldsymbol{q}_{2\mathbf{k}}\right) ^{\dot{J}I}$ \\
$Q_{\mathbf{k}}^{-}$ & $=$ & $\left( \boldsymbol{q}_{3\mathbf{k}}\right) _{%
\dot{J}I}\hat{c}_{\mathbf{k}}^{\dot{J}}\hat{b}_{\mathbf{k}}^{I}+\hat{b}_{%
\mathbf{k}I}^{\dagger }\left( \boldsymbol{q}_{4\mathbf{k}}\right) _{\dot{J}%
}^{I}\hat{c}_{\mathbf{k}}^{\dot{J}}$ \\
&  &  \\
$\bar{Q}_{\mathbf{k}}^{-}$ & $=$ & $\hat{b}_{\mathbf{k}I}^{\dagger }\left(
\boldsymbol{q}_{1\mathbf{k}}^{\dagger }\right) _{\dot{J}}^{I}\hat{c}_{%
\mathbf{k}}^{\dot{J}}+\left( \boldsymbol{q}_{2\mathbf{k}}^{\dagger }\right)
_{I\dot{J}}\hat{b}_{\mathbf{k}}^{I}\hat{c}_{\mathbf{k}}^{\dot{J}}$ \\
$\bar{Q}_{\mathbf{k}}^{+}$ & $=$ & $\hat{b}_{I\mathbf{k}}^{\dagger }\hat{c}_{%
\dot{J}\mathbf{k}}^{\dagger }\left( \boldsymbol{q}_{3\mathbf{k}}^{\dagger
}\right) ^{I\dot{J}}+\hat{c}_{\dot{J}\mathbf{k}}^{\dagger }\left(
\boldsymbol{q}_{4\mathbf{k}}^{\dagger }\right) _{I}^{\dot{J}}\hat{b}_{%
\mathbf{k}}^{I}$%
\end{tabular}
\label{qm}
\end{equation}%
By mimicking eqs(\ref{cB}-\ref{Xb}), we see that the above supercharges (\ref%
{qm}) can be handled in two ways as given here after:

\begin{itemize}
\item \emph{First way: }The $Q_{\mathbf{k}}^{\pm }$ and $\bar{Q}_{\mathbf{k}%
}^{\pm }$ in (\ref{qm}) are imagined in a condensed form as follows
\begin{equation}
\begin{tabular}{lll}
$Q_{\mathbf{k}}^{+}$ & $=$ & $\hat{c}_{\mathbf{k}\dot{J}}^{\dagger }\hat{B}_{%
\mathbf{k}}^{\dot{J}}$ \\
$Q_{\mathbf{k}}^{-}$ & $=$ & $\hat{D}_{\mathbf{k}\dot{J}}^{\dagger }\hat{c}_{%
\mathbf{k}}^{\dot{J}}$%
\end{tabular}%
\qquad ,\qquad
\begin{tabular}{lll}
$\bar{Q}_{\mathbf{k}}^{-}$ & $=$ & $\hat{B}_{\mathbf{k}\dot{J}}^{\dagger }%
\hat{c}_{\mathbf{k}}^{\dot{J}}$ \\
$\bar{Q}_{\mathbf{k}}^{+}$ & $=$ & $\hat{c}_{\mathbf{k}\dot{J}}^{\dagger }%
\hat{D}_{\mathbf{k}}^{\dot{J}}$%
\end{tabular}
\label{w1}
\end{equation}%
where the $\hat{B}_{\mathbf{k}}^{\dot{J}}$ and $\hat{D}_{\mathbf{k}}^{\dot{J}%
}$ are linear functions of the bosonic $\hat{b}_{\mathbf{k}}^{I}$ and $\hat{b%
}_{\mathbf{k}I}^{\dagger }$.

\item \emph{Second way: }The $Q_{\mathbf{k}}^{\pm }$ and $\bar{Q}_{\mathbf{k}%
}^{\pm }$ are thought of like%
\begin{equation}
\begin{tabular}{lll}
$Q_{\mathbf{k}}^{+}$ & $=$ & $\hat{C}_{\mathbf{k}\dot{J}}^{\dagger }\hat{b}_{%
\mathbf{k}}^{\dot{J}}$ \\
$Q_{\mathbf{k}}^{-}$ & $=$ & $\hat{b}_{\mathbf{k}\dot{J}}^{\dagger }\hat{E}_{%
\mathbf{k}}^{\dot{J}}$%
\end{tabular}%
\qquad ,\qquad
\begin{tabular}{lll}
$\bar{Q}_{\mathbf{k}}^{-}$ & $=$ & $\hat{b}_{\mathbf{k}\dot{J}}^{\dagger }%
\hat{C}_{\mathbf{k}}^{\dot{J}}$ \\
$\bar{Q}_{\mathbf{k}}^{+}$ & $=$ & $\hat{E}_{\mathbf{k}\dot{J}}^{\dagger }%
\hat{b}_{\mathbf{k}}^{\dot{J}}$%
\end{tabular}
\label{w2}
\end{equation}%
where now $\hat{C}_{\mathbf{k}}^{\dot{J}}$ and $\hat{E}_{\mathbf{k}}^{\dot{J}%
}$ are linear functions of the fermionic $\hat{c}_{\mathbf{k}}^{I}$ and $%
\hat{c}_{\mathbf{k}I}^{\dagger }$.
\end{itemize}

In what follows, we develop the picture given by eq(\ref{w1}); a similar
analysis can be done for (\ref{w2}).

\subsubsection{Coupling $\boldsymbol{q}_{\mathbf{k}}$ as a deviation away
from $\boldsymbol{q}_{\mathbf{k}}=z_{\mathbf{k}}{\protect\small I}_{id}$}

By singling out the fermionic operators $\hat{c}_{\mathbf{k}}^{\dot{J}}$ and
$\hat{c}_{\mathbf{k}\dot{J}}^{\dagger },$ one can put the ORTIC charge (\ref%
{QB}) like $Q_{\mathbf{k}}=Q_{\mathbf{k}}^{+}+Q_{\mathbf{k}}^{-}$ with $Q_{%
\mathbf{k}}^{\pm }$ as in (\ref{w1}) and the new bosonic operators $\hat{B}_{%
\mathbf{k}}^{\dot{J}}$ and $\hat{D}_{\mathbf{k}}^{\dot{J}}$ given by%
\begin{equation}
\begin{tabular}{lll}
$\hat{B}_{\mathbf{k}}^{\dot{J}}$ & $=$ & $\left( \boldsymbol{q}_{1\mathbf{k}%
}\right) _{I}^{\dot{J}}\hat{b}_{\mathbf{k}}^{I}+\hat{b}_{\mathbf{k}%
I}^{\dagger }\left( \boldsymbol{q}_{2\mathbf{k}}\right) ^{I\dot{J}}$ \\
$\hat{B}_{\mathbf{k}\dot{J}}^{\dagger }$ & $=$ & $\left( \boldsymbol{\bar{q}}%
_{2\mathbf{k}}\right) _{\dot{J}I}\hat{b}_{\mathbf{k}}^{I}+\hat{b}_{\mathbf{k}%
I}^{\dagger }\left( \bar{q}_{1\mathbf{k}}\right) _{\dot{J}}^{I}$%
\end{tabular}%
\quad ,\quad
\begin{tabular}{lll}
$\hat{D}_{\mathbf{k}\dot{J}}^{\dagger }$ & $=$ & $\left( \boldsymbol{q}_{3%
\mathbf{k}}\right) _{\dot{J}I}\hat{b}_{\mathbf{k}}^{I}+\hat{b}_{\mathbf{k}%
I}^{\dagger }\left( \boldsymbol{q}_{4\mathbf{k}}\right) _{\dot{J}}^{I}$ \\
$\hat{D}_{\mathbf{k}}^{\dot{J}}$ & $=$ & $\hat{b}_{\mathbf{k}I}^{\dagger
}\left( \boldsymbol{\bar{q}}_{3\mathbf{k}}\right) ^{I\dot{J}}+\left(
\boldsymbol{\bar{q}}_{4\mathbf{k}}\right) _{I}^{\dot{J}}\hat{b}_{\mathbf{k}%
}^{I}$%
\end{tabular}
\label{2BD}
\end{equation}%
For the case $\boldsymbol{q}_{1\mathbf{k}}=\boldsymbol{q}_{4\mathbf{k}}=z_{%
\mathbf{k}}I_{{\small N\times N}}$ and $\boldsymbol{q}_{2\mathbf{k}}=%
\boldsymbol{q}_{3\mathbf{k}}=0_{{\small N\times N}},$ one has $\hat{B}_{%
\mathbf{k}}^{I}=z\hat{b}_{\mathbf{k}}^{I}$ and $\hat{D}_{\mathbf{k}}^{I}=%
\bar{z}\hat{b}_{\mathbf{k}}^{I}$.

\ \ \

\textbf{A) Orthosymplectic hamiltonian}\emph{\ }\newline
The ORTIC hamiltonian $H_{ortic}=\sum_{_{\mathbf{k}}}H_{\mathbf{k}}^{ortic}$
is defined by the anticommutators $\{Q_{\mathbf{k}},Q_{\mathbf{k}}^{\dagger
}\}=H_{\mathbf{k}}^{ortic}$; it reads in terms of $Q_{\mathbf{k}}^{\pm }$
and their adjoint conjugates $\bar{Q}_{\mathbf{k}}^{\mp }$ as follows
\begin{equation}
H_{\mathbf{k}}^{ortic}=\left\{ Q_{\mathbf{k}}^{+},\bar{Q}_{\mathbf{k}%
}^{-}\right\} +\left\{ Q_{\mathbf{k}}^{-},\bar{Q}_{\mathbf{k}}^{+}\right\}
+\left\{ Q_{\mathbf{k}}^{-},\bar{Q}_{\mathbf{k}}^{-}\right\} +\left\{ Q_{%
\mathbf{k}}^{+},\bar{Q}_{\mathbf{k}}^{+}\right\}  \label{HOR}
\end{equation}%
As for the ORTIC supercharge $Q_{\mathbf{k}}$, the bosonic $H_{\mathbf{k}%
}^{ortic}$ is valued in the osp(2N\TEXTsymbol{\vert}2N) Lie superalgebra;
and has four anticommutators blocks $H_{1\mathbf{k}}+H_{2\mathbf{k}}+Z_{1%
\mathbf{k}}+\bar{Z}_{1\mathbf{k}}$ given by
\begin{equation}
\begin{tabular}{lllllll}
$\left\{ Q_{\mathbf{k}}^{+},\bar{Q}_{\mathbf{k}}^{-}\right\} $ & $=$ & $H_{1%
\mathbf{k}}$ & $\qquad ,\qquad $ & $\left\{ Q_{\mathbf{k}}^{-},\bar{Q}_{%
\mathbf{k}}^{-}\right\} $ & $=$ & $Z_{1\mathbf{k}}$ \\
$\left\{ Q_{\mathbf{k}}^{-},\bar{Q}_{\mathbf{k}}^{+}\right\} $ & $=$ & $H_{2%
\mathbf{k}}$ & $\qquad ,\qquad $ & $\left\{ Q_{\mathbf{k}}^{+},\bar{Q}_{%
\mathbf{k}}^{+}\right\} $ & $=$ & $\bar{Z}_{1\mathbf{k}}$%
\end{tabular}%
\end{equation}%
where $H_{1\mathbf{k}}$ and $H_{2\mathbf{k}}$ are hermitian while $Z_{1%
\mathbf{k}}$ and $\bar{Z}_{1\mathbf{k}}$ are exchanged under adjoint
conjugation. They read in terms of the oscillators $\hat{c}_{\mathbf{k}}/%
\hat{c}_{\mathbf{k}}^{\dagger }$ and the new $\hat{B}_{\mathbf{k}}/\hat{B}_{%
\mathbf{k}}^{\dagger }$ as well as $\hat{D}_{\mathbf{k}}/\hat{D}_{\mathbf{k}%
}^{\dagger }$ as follows%
\begin{equation}
\begin{tabular}{lllllll}
$H_{1\mathbf{k}}$ & $=$ & $\hat{B}_{\mathbf{k}\dot{I}}^{\dagger }\hat{B}_{%
\mathbf{k}}^{\dot{I}}+\hat{c}_{\mathbf{k}\dot{I}}^{\dagger }\mathbb{B}_{\dot{%
J}}^{\dot{I}}\hat{c}_{\mathbf{k}}^{\dot{J}}$ & $\qquad ,\qquad $ & $Z_{1%
\mathbf{k}}$ & $=$ & $\hat{c}_{\mathbf{k}}^{\dot{I}}\mathbb{\bar{F}}_{\dot{I}%
\dot{J}}\hat{c}_{\mathbf{k}}^{\dot{J}}$ \\
$H_{2\mathbf{k}}$ & $=$ & $\hat{D}_{\mathbf{k}\dot{I}}^{\dagger }\hat{D}_{%
\mathbf{k}}^{\dot{I}}-\hat{c}_{\mathbf{k}\dot{I}}^{\dagger }\mathbb{D}_{\dot{%
J}}^{\dot{I}}\hat{c}_{\mathbf{k}}^{\dot{J}}$ & $\qquad ,\qquad $ & $\bar{Z}%
_{1\mathbf{k}}$ & $=$ & $\hat{c}_{\mathbf{k}\dot{I}}^{\dagger }\mathbb{F}^{%
\dot{I}\dot{J}}\hat{c}_{\mathbf{k}\dot{J}}^{\dagger }$%
\end{tabular}%
\end{equation}%
with matrices $\mathbb{B}_{\dot{J}}^{\dot{I}},$ $\mathbb{D}_{\dot{J}}^{\dot{I%
}}$ and so on given by the following commutators%
\begin{equation}
\begin{tabular}{lllllll}
$\mathbb{B}_{\dot{J}}^{\dot{I}}$ & $=$ & $[\hat{B}_{\mathbf{k}}^{\dot{I}},%
\hat{B}_{\mathbf{k}\dot{J}}^{\dagger }]$ & $\qquad ,\qquad $ & $\mathbb{D}_{%
\dot{J}}^{\dot{I}}$ & $=$ & $[\hat{D}_{\mathbf{k}}^{\dot{I}},\hat{D}_{%
\mathbf{k}\dot{J}}^{\dagger }]$ \\
$\mathbb{X}^{IJ}$ & $=$ & $[\hat{B}_{\mathbf{k}}^{I},\hat{B}_{\mathbf{k}%
}^{J}]$ & $\qquad ,\qquad $ & $\mathbb{Y}^{IJ}$ & $=$ & $[\hat{D}_{\mathbf{k}%
}^{I},\hat{D}_{\mathbf{k}}^{J}]$ \\
$\mathbb{F}^{\dot{I}\dot{J}}$ & $=$ & $[\hat{B}_{\mathbf{k}}^{\dot{I}},\hat{D%
}_{\mathbf{k}}^{\dot{J}}]$ & $\qquad ,\qquad $ & $\mathbb{\bar{F}}_{\dot{I}%
\dot{J}}$ & $=$ & $[\hat{D}_{\mathbf{k}\dot{I}}^{\dagger },\hat{B}_{\mathbf{k%
}\dot{J}}^{\dagger }]$ \\
$\mathbb{G}_{\dot{I}}^{\dot{J}}$ & $=$ & $[\hat{B}_{\mathbf{k}}^{\dot{J}},%
\hat{D}_{\mathbf{k}\dot{I}}^{\dagger }]$ & $\qquad ,\qquad $ & $\mathbb{\bar{%
G}}_{\dot{J}}^{\dot{I}}$ & $=$ & $[\hat{D}_{\mathbf{k}}^{\dot{I}},\hat{B}_{%
\mathbf{k}\dot{J}}^{\dagger }]$%
\end{tabular}%
\end{equation}%
Using (\ref{2BD}), these quantities read in terms of the coupling tensors as
follows%
\begin{equation}
\begin{tabular}{lll}
$\mathbb{B}$ & $=$ & $\boldsymbol{q}_{1\mathbf{k}}\boldsymbol{q}_{1\mathbf{k}%
}^{\dagger }-\boldsymbol{q}_{2\mathbf{k}}^{\dagger }\boldsymbol{q}_{2\mathbf{%
k}}$ \\
$\mathbb{D}$ & $=$ & $\boldsymbol{q}_{4\mathbf{k}}^{\dagger }\boldsymbol{q}%
_{4\mathbf{k}}-\boldsymbol{q}_{3\mathbf{k}}\boldsymbol{q}_{3\mathbf{k}%
}^{\dagger }$ \\
$\mathbb{X}$ & $=$ & $\boldsymbol{q}_{1\mathbf{k}}\boldsymbol{q}_{2\mathbf{k}%
}-\boldsymbol{q}_{1\mathbf{k}}\boldsymbol{q}_{2\mathbf{k}}$%
\end{tabular}%
\qquad ,\qquad
\begin{tabular}{lll}
$\mathbb{F}$ & $=$ & $\boldsymbol{q}_{1\mathbf{k}}\boldsymbol{q}_{3\mathbf{k}%
}^{\dagger }-\boldsymbol{q}_{4\mathbf{k}}^{\dagger }\boldsymbol{q}_{2\mathbf{%
k}}$ \\
$\mathbb{G}$ & $=$ & $\boldsymbol{q}_{1\mathbf{k}}\boldsymbol{q}_{4\mathbf{k}%
}-\boldsymbol{q}_{3\mathbf{k}}\boldsymbol{q}_{2\mathbf{k}}$ \\
$\mathbb{Y}$ & $=$ & $\boldsymbol{\bar{q}}_{4\mathbf{k}}\boldsymbol{\bar{q}}%
_{3\mathbf{k}}-\boldsymbol{\bar{q}}_{4\mathbf{k}}\boldsymbol{\bar{q}}_{3%
\mathbf{k}}$%
\end{tabular}
\label{BDF}
\end{equation}%
where the diagonal blocks are hermitian; that is $\mathbb{B}^{\dagger }=%
\mathbb{B}$ and $\mathbb{D}^{\dagger }=\mathbb{D}$. Putting these relations
into (\ref{HOR}), we end up with%
\begin{equation}
\begin{tabular}{lll}
$H_{\mathbf{k}}^{ortic}$ & $=$ & $\hat{B}_{\mathbf{k}\dot{I}}^{\dagger }\hat{%
B}_{\mathbf{k}}^{\dot{I}}+\hat{D}_{\mathbf{k}\dot{I}}^{\dagger }\hat{D}_{%
\mathbf{k}}^{\dot{I}}+\hat{c}_{\mathbf{k}\dot{I}}^{\dagger }\mathbb{M}_{\dot{%
J}}^{\dot{I}}\hat{c}_{\mathbf{k}}^{\dot{J}}$ \\
&  & $+\mathbb{\bar{F}}_{\dot{I}\dot{J}}\hat{c}_{\mathbf{k}}^{\dot{I}}\hat{c}%
_{\mathbf{k}}^{\dot{J}}+\hat{c}_{\mathbf{k}\dot{I}}^{\dagger }\hat{c}_{%
\mathbf{k}\dot{J}}^{\dagger }\mathbb{F}^{\dot{I}\dot{J}}$%
\end{tabular}
\label{OR}
\end{equation}%
where $\mathbb{M}=\mathbb{B}-\mathbb{D}$. From this hamiltonian, we learn
the bosonic $H_{\mathbf{k}}^{bose}$ and the fermionic $H_{\mathbf{k}%
}^{fermi} $ contributions namely%
\begin{equation}
\begin{tabular}{lll}
$H_{\mathbf{k}}^{bose}$ & $=$ & $\hat{B}_{\mathbf{k}\dot{I}}^{\dagger }\hat{B%
}_{\mathbf{k}}^{\dot{I}}+\hat{D}_{\mathbf{k}\dot{I}}^{\dagger }\hat{D}_{%
\mathbf{k}}^{\dot{I}}$ \\
$H_{\mathbf{k}}^{fermi}$ & $=$ & $\hat{c}_{\mathbf{k}\dot{I}}^{\dagger }%
\mathbb{M}_{\dot{J}}^{\dot{I}}\hat{c}_{\mathbf{k}}^{\dot{J}}+\mathbb{\bar{F}}%
_{\dot{I}\dot{J}}\hat{c}_{\mathbf{k}}^{\dot{I}}\hat{c}_{\mathbf{k}}^{\dot{J}%
}+\hat{c}_{\mathbf{k}\dot{I}}^{\dagger }\hat{c}_{\mathbf{k}\dot{J}}^{\dagger
}\mathbb{F}^{\dot{I}\dot{J}}$%
\end{tabular}
\label{RO}
\end{equation}

\textbf{B) Supersymmetric Hamiltonian}\newline
In the absence of supersymmetric central charges, the hamiltonian $H_{%
\mathbf{k}}^{susy}$ is obtained from the ORTIC $H_{\mathbf{k}}^{ortic}$ by
imposing the supersymmetric conditions required by the supersymmetric
algebra on world line. First, we impose the nilpotency of the
anticommutators $\left\{ Q_{\mathbf{k}}^{+},\bar{Q}_{\mathbf{k}}^{+}\right\}
$ and $\left\{ Q_{\mathbf{k}}^{-},\bar{Q}_{\mathbf{k}}^{-}\right\} ;$ that
is $Z_{1\mathbf{k}}=\bar{Z}_{1\mathbf{k}}=0.$ These nilpotencies reduce the
ORTIC $H_{\mathbf{k}}^{ortic}$ to the sum of two terms namely $H_{\mathbf{k}%
}^{susy}=H_{1\mathbf{k}}+H_{2\mathbf{k}}$. Second, we require the following
commutation relations to hold
\begin{equation}
\lbrack H_{\mathbf{k}}^{susy},Q_{\mathbf{k}}^{+}]=H_{\mathbf{k}}^{susy},Q_{%
\mathbf{k}}^{-}]=0
\end{equation}%
Clearly, this supersymmetric invariance put constraints on the coupling
tensors $\boldsymbol{q}_{1\mathbf{k}},$ $\boldsymbol{q}_{2\mathbf{k}},$ $%
\boldsymbol{q}_{3\mathbf{k}}$ and $\boldsymbol{q}_{4\mathbf{k}}$; they are
no longuer free tensors; candidate of such $\boldsymbol{q}_{n\mathbf{k}}$'s
can be determined by using (\ref{BDF}). Here, we omit the details; but to
fix the ideas, we describe below some steps of the calculations regarding
the resolution of the constraints $\left\{ Q_{\mathbf{k}}^{+},\bar{Q}_{%
\mathbf{k}}^{+}\right\} =\left\{ Q_{\mathbf{k}}^{-},\bar{Q}_{\mathbf{k}%
}^{-}\right\} =0$. \newline
These constraints correspond to imposing $Z_{1\mathbf{k}}=0$ in (\ref{HOR})
and (\ref{OR}); as such they require $\hat{c}_{\mathbf{k}\dot{I}}^{\dagger }%
\hat{c}_{\mathbf{k}\dot{J}}^{\dagger }\mathbb{F}^{\dot{I}\dot{J}}$ $=0$ with
tensor $\mathbb{F}^{\dot{I}\dot{J}}$ as in eq(\ref{BDF}). This condition
gives a matrix relation between the four coupling tensors $\boldsymbol{q}_{1%
\mathbf{k}},$ $\boldsymbol{q}_{2\mathbf{k}},$ $\boldsymbol{q}_{3\mathbf{k}}$
and $\boldsymbol{q}_{4\mathbf{k}}$ namely%
\begin{equation}
\left( \boldsymbol{q}_{1\mathbf{k}}\boldsymbol{q}_{3\mathbf{k}}^{\dagger
}\right) ^{\dot{I}\dot{J}}-\left( \boldsymbol{q}_{4\mathbf{k}}^{\dagger }%
\boldsymbol{q}_{2\mathbf{k}}\right) ^{\dot{I}\dot{J}}=\eta _{q}\Sigma _{%
\mathbf{k}}^{\dot{I}\dot{J}}  \label{sig}
\end{equation}%
where $\Sigma _{\mathbf{k}}^{\dot{I}\dot{J}}$ is a symmetric matrix ($\Sigma
_{\mathbf{k}}^{\dot{I}\dot{J}}=\Sigma _{\mathbf{k}}^{\dot{J}\dot{I}}$) \
because $\hat{c}_{\mathbf{k}\dot{I}}^{\dagger }\hat{c}_{\mathbf{k}\dot{J}%
}^{\dagger }\Sigma _{\mathbf{k}}^{\dot{I}\dot{J}}=0$. As far as the
constraint (\ref{sig}) is concerned, let us give \textrm{some} comments
regarding its solutions. \newline
$\left( \mathbf{1}\right) $ The simplest solution is given by the diagonal
coupling $\boldsymbol{q}_{\mathbf{k}}=z_{\mathbf{k}}{\small I}_{{\small %
2N\times 2N}}$ considered previously with sub-blocks $\boldsymbol{q}_{1%
\mathbf{k}}=\boldsymbol{q}_{4\mathbf{k}}=z_{\mathbf{k}}{\small I}_{{\small %
N\times N}}$ and $\boldsymbol{q}_{2\mathbf{k}}=\boldsymbol{q}_{3\mathbf{k}%
}=0_{{\small N\times N}}$. It corresponds just to the trivial case $\eta
_{q}=0.$ This requires%
\begin{equation}
\boldsymbol{q}_{1\mathbf{k}}\boldsymbol{q}_{3\mathbf{k}}^{\dagger }=%
\boldsymbol{q}_{4\mathbf{k}}^{\dagger }\boldsymbol{q}_{2\mathbf{k}}
\end{equation}%
$\left( \mathbf{2}\right) $ A second set of solutions of (\ref{sig}) is
given by the remarkable case where $\Sigma _{\mathbf{k}}^{\dot{I}\dot{J}%
}=\delta ^{\dot{I}\dot{J}}$ with $\eta _{q}\neq 0.$ In this case, a solution
of (\ref{sig}) is given by%
\begin{equation}
\boldsymbol{q}_{1\mathbf{k}}\boldsymbol{q}_{3\mathbf{k}}^{\dagger }=\varrho
_{\mathbf{k}}I_{{\small N\times N}}\quad ,\quad \boldsymbol{q}_{4\mathbf{k}%
}^{\dagger }\boldsymbol{q}_{2\mathbf{k}}=\tilde{\varrho}_{\mathbf{k}}I_{%
{\small N\times N}}
\end{equation}%
with the relation $\varrho _{\mathbf{k}}-\tilde{\varrho}_{\mathbf{k}}=\zeta
_{\mathbf{k}}.$\newline
$\left( \mathbf{3}\right) $ Other solutions of the constraint (\ref{sig})
can be also written down; for instance by taking $\boldsymbol{q}_{3\mathbf{k}%
}^{\dagger }=\boldsymbol{q}_{1\mathbf{k}}^{T}$ and $\boldsymbol{q}_{4\mathbf{%
k}}^{\dagger }=\boldsymbol{q}_{2\mathbf{k}}^{T}.$

\subsection{Topological super model with two bands}

Here, we consider a 2D Brillouin $\mathbb{T}^{2}$ parameterised by $%
(k_{x},k_{y})$ with $0\leq k_{x},k_{y}<2\pi ;$ and we apply the above
construction to the hamiltonian (\ref{hff}). Because of the particle-hole
symmetry, we must have $\sigma _{x}h_{f}\left( \mathbf{k}\right) ^{\ast
}\sigma _{x}=-h_{f}\left( -\mathbf{k}\right) ;$ thus requiring conditions on
the $d_{x,y,z}$ functions namely
\begin{equation}
\begin{tabular}{lll}
$d_{x,y}\left( -\mathbf{k}\right) $ & $=$ & $-d_{x,y}\left( \mathbf{k}%
\right) $ \\
$d_{z}\left( -\mathbf{k}\right) $ & $=$ & $d_{z}\left( \mathbf{k}\right) $%
\end{tabular}
\label{ct}
\end{equation}%
We solve these constraint relations as
\begin{equation}
d_{x}=t_{z}\sin k_{x},\qquad d_{y}=t_{t}\sin k_{y},\qquad d_{z}=M-\cos
k_{x}-\cos k_{y}
\end{equation}

\subsubsection{Constructing the coupling matrix $\boldsymbol{q}_{\mathbf{k}}$%
}

For this supersymmetric two bands model, the diagonal matrices $\mathcal{D}_{%
\mathbf{k}}$ and $\Delta _{\mathbf{k}}$ are given by
\begin{equation}
\mathcal{D}_{\mathbf{k}}=\left(
\begin{array}{cc}
{\small \varepsilon } & 0 \\
0 & {\small \varepsilon }%
\end{array}%
\right) \qquad ,\qquad \Delta _{\mathbf{k}}=\left(
\begin{array}{cc}
{\small \varepsilon } & 0 \\
0 & -{\small \varepsilon }%
\end{array}%
\right)  \label{DO}
\end{equation}%
where $\varepsilon =\sqrt{d_{x}^{2}+d_{y}^{2}+d_{z}^{2}}$ with the
remarkable property $\mathcal{D}_{\mathbf{k}}=\varepsilon \sigma _{0}$. We
also have $\Delta _{\mathbf{k}}=\varepsilon \sigma _{z}$ with gap energy $%
2\varepsilon $. The above diagonal (\ref{DO}) can be compared to (\ref{ome})
with $\varepsilon $ given by $\omega _{\mathbf{k}}$. For fermionic gapless
states, the hamiltonians $h_{f}$ and $h_{b}$ have zero modes ($\det
h_{f}=\det h_{b}=0$).

\begin{figure}[tbph]
\begin{center}
\includegraphics[width=12cm]{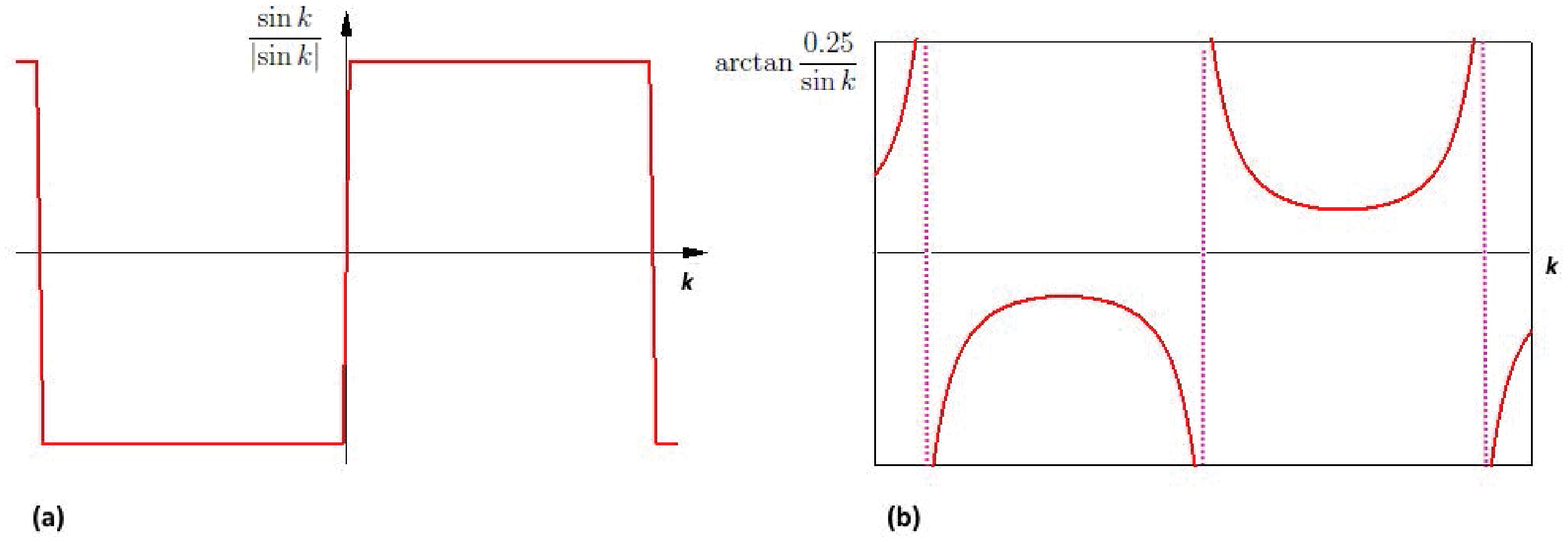}
\end{center}
\par
\vspace{-0.5cm}
\caption{Variation of factor the phase $\left( d_{x}+id_{y}\right) /\protect%
\sqrt{d_{x}^{2}+d_{y}^{2}}$ with $d_{i}=\sin k_{i}$ for $\sin k_{y}=0$ and $%
\sin k_{y}\neq 0.$ \textbf{a}) The plot is given for $\sin k_{y}=0$; thus
reducing to $\cos \protect\phi =\sin k/\left\vert \sin k\right\vert $
showing a discontinuity at $\sin k_{x}=0.$ \textbf{b}) $\sin k_{y}\neq 0$
taken as $0.25$ having a discontinuity at $\sin k_{x}=0$.}
\label{fi}
\end{figure}

$\bullet $ \emph{Building} \emph{the passage matrix} $\boldsymbol{V}_{%
\mathbf{k}}$\newline
The unitary matrix $\boldsymbol{V}_{\mathbf{k}}$ involved in the
construction of (\ref{qk}) is given by $\left( V_{+},V_{-}\right) $ where $%
V_{\pm }$ are the normalisation of the eigenvectors (\ref{ew}) of $h_{f}$.
Substituting, we obtain
\begin{equation}
\boldsymbol{V}_{\mathbf{k}}=\frac{1}{\sqrt{2\varepsilon }}\left(
\begin{array}{cc}
e^{-i\phi }\sqrt{\varepsilon +d_{z}} & \sqrt{\varepsilon -d_{z}} \\
\sqrt{\varepsilon -d_{z}} & -e^{+i\phi }\sqrt{\varepsilon +d_{z}}%
\end{array}%
\right)
\end{equation}%
Putting this expression back into $\boldsymbol{q}_{\mathbf{k}}=\boldsymbol{V}%
_{\mathbf{k}}D_{\mathbf{k}}^{1/2}$, we end up with the coupling matrix
\begin{equation}
\boldsymbol{q}_{\mathbf{k}}=\frac{1}{\sqrt{2}}\left(
\begin{array}{cc}
e^{-i\phi }\sqrt{\varepsilon +d_{z}} & \sqrt{\varepsilon -d_{z}} \\
\sqrt{\varepsilon -d_{z}} & -e^{+i\phi }\sqrt{\varepsilon +d_{z}}%
\end{array}%
\right)  \label{qkk}
\end{equation}%
with $\det \boldsymbol{q}=-\varepsilon $ and
\begin{equation}
e^{i\phi }=\frac{d_{x}+id_{y}}{\sqrt{d_{x}^{2}+d_{y}^{2}}}  \label{phi}
\end{equation}%
Comparing (\ref{qkk}) to (\ref{qca}), we learn the expressions of $%
\boldsymbol{q}_{1\mathbf{k}},$ $\boldsymbol{q}_{2\mathbf{k}},$ $\boldsymbol{q%
}_{3\mathbf{k}}$ and $\boldsymbol{q}_{4\mathbf{k}}$ namely%
\begin{equation}
\boldsymbol{q}_{1\mathbf{k}}=\frac{\sqrt{\varepsilon +d_{z}}}{\sqrt{2}}%
e^{-i\phi },\quad \boldsymbol{q}_{2\mathbf{k}}=\boldsymbol{q}_{3\mathbf{k}}=%
\frac{\sqrt{\varepsilon -d_{z}}}{\sqrt{2}},\quad \boldsymbol{q}_{4\mathbf{k}%
}=-\frac{\sqrt{\varepsilon +d_{z}}}{\sqrt{2}}e^{+i\phi }
\end{equation}%
In the Figure \textbf{\ref{fi}}, we plot the real part $e^{i\phi }$ where a
distortion lives at the high symmetry points $k=0,\pi .$%

$\bullet $ \emph{Topological distortion}\newline
The above coupling $\boldsymbol{q}_{\mathbf{k}}$ matrix is a function of $%
\left( k_{x},k_{y}\right) $; it obeys the PH symmetry (\ref{qq}). At the
high symmetry point, the matrix coupling has distortion manifested by an
ill-definiteness. The fix points of PH are given by the solution of the
vanishing $d_{x,y}\left( \mathbf{k}\right) =0.$ For the model where $%
d_{x}=t_{x}\sin k_{x},$ $d_{y}=t_{t}\sin k_{y},$ the points are given by $%
\left( k_{x}^{\ast },k_{y}^{\ast }\right) =\left( n_{x}\pi ,n_{y}\pi \right)
$ reading explicitly as%
\begin{equation}
\left( k_{x}^{\ast },k_{y}^{\ast }\right) =\left( 0,0\right) ,\qquad \left(
\pi ,0\right) ,\qquad \left( 0,\pi \right) ,\qquad \left( \pi ,\pi \right)
\end{equation}%
At these points $\mathbf{k}_{\ast }$, the phase $e^{i\phi }$ shows an
obstruction; it behaves like $\frac{0}{0^{+}}+i\frac{0}{0^{+}}$ as it can be
checked on eq(\ref{phi}). This singularity survives in the limit $%
d_{z}\rightarrow 0$ where $\det q\rightarrow 0$ and where live fermionic
gapless states and bosonic partners.

\subsubsection{Time reversal symmetry}

Under time reversal symmetry, the Hamiltonian for spinless fermions is
constrained by $h_{f}\left( \mathbf{k}\right) ^{\ast }=h_{f}\left( -\mathbf{k%
}\right) .$ For the two band model with hamiltonian $h_{f}=\sum d_{\mu
}\left( \mathbf{k}\right) \sigma ^{\mu }$ the three $d_{\mu }\left( \mathbf{k%
}\right) $ functions are constrained like%
\begin{equation}
\begin{tabular}{lll}
$d_{x,z}\left( \mathbf{k}\right) ^{\ast }$ & $=$ & $d_{x,z}\left( -\mathbf{k}%
\right) $ \\
$d_{y}\left( -\mathbf{k}\right) $ & $=$ & $-d_{y}\left( -\mathbf{k}\right) $%
\end{tabular}
\label{trs}
\end{equation}%
We solve these conditions as
\begin{equation}
d_{x}=M-t\cos k_{x},\qquad d_{y}=t\sin k_{y},\qquad d_{z}=0
\end{equation}
leading to%
\begin{equation}
h_{f}=\left(
\begin{array}{cc}
0 & d_{x}-id_{y} \\
d_{x}+id_{y} & 0%
\end{array}%
\right)
\end{equation}%
Its eigenvalues $\varepsilon _{\pm }$ are given by $\pm \varepsilon =\pm
\sqrt{d_{x}^{2}+d_{y}^{2}}$ with gap $E_{g}=2\varepsilon .$ The vanishing
condition for this gap corresponds to $d_{x}=d_{y}=0.$ The normalised
eigenvectors $V_{\pm }$ are given by%
\begin{equation}
V_{+}=\frac{1}{\sqrt{2}}\left(
\begin{array}{c}
e^{-i\phi } \\
1%
\end{array}%
\right) \qquad ,\qquad V_{-}=\frac{1}{\sqrt{2}}\left(
\begin{array}{c}
1 \\
-e^{+i\phi }%
\end{array}%
\right)
\end{equation}%
with $e^{-i\phi }=\left( d_{x}-id_{y}\right) /\sqrt{d_{x}^{2}+d_{y}^{2}}$.
Using the above eigenvectors, we obtain the unitary matrix $\boldsymbol{V}_{%
\mathbf{k}}$ diagonalising $h_{f}$ namely
\begin{equation}
\boldsymbol{V}_{\mathbf{k}}=\frac{1}{\sqrt{2}}\left(
\begin{array}{cc}
e^{-i\phi } & 1 \\
1 & -e^{+i\phi }%
\end{array}%
\right)
\end{equation}%
Putting this expression into $q_{\mathbf{k}}=\boldsymbol{V}_{\mathbf{k}}D_{%
\mathbf{k}}^{1/2}$, we end up with the coupling matrix
\begin{equation}
q_{\mathbf{k}}=\frac{\sqrt{\varepsilon }}{\sqrt{2}}\left(
\begin{array}{cc}
e^{-i\phi } & 1 \\
1 & -e^{+i\phi }%
\end{array}%
\right)
\end{equation}%
with $\det q=-\varepsilon $ showing that $q_{\mathbf{k}}$ has a singularity
for gapless states. Here also the bosonic $h_{b}$ is given by $\varepsilon
\sigma _{0}$ and $h_{f}=\varepsilon \sigma _{z}.$

\section{Conclusion and Discussions}

In this paper, we used the orthosymplectic structure of the quantized graded
phase space of quantum super oscillators $\hat{c}^{i}/\hat{c}_{i}^{\dagger }$
and $\hat{b}^{\alpha }/\hat{b}_{\alpha }^{\dagger }$ to develop tight
binding models for super bands and use this construction to investigate
topological phases of supermatter. We distinguished two families of
supermatter: supersymmetric and orthosymplectic. \newline
$\left( \mathbf{1}\right) $ the supersymmetric family, termed as SUSY
matter; is based on two fermionic charges $Q_{{\small susy}}^{\pm }=$ $%
Q_{1}\pm iQ_{2}$ constrained by the usual $\mathcal{N}=2$ \emph{%
supersymmetric} algebra of QM. This is a four dimensional graded Lie algebra
generated by the hermitian $Q_{1},Q_{2}$, interpreted as $\mathcal{N}=2$
supercharge operators, and the bosonic $H_{{\small susy}}^{\mathcal{N}=2}$
defining the $\mathcal{N}=2$ SUSY Hamiltonian as well as a bosonic charge
operator $T_{0}$ generating the SO(2) R-symmetry rotating the two $Q_{i}$'s.
Given $Q_{{\small susy}}^{\pm },$ we have the anticommutator $\{Q_{{\small %
susy}}^{+},Q_{{\small susy}}^{-}\}$ allowing to construct $H_{{\small susy}%
}^{\mathcal{N}=2}$ and the $\{Q_{{\small susy}}^{\pm },Q_{{\small susy}%
}^{\pm }\}=0$ as well as the $[Q_{{\small susy}}^{\pm },H_{{\small susy}}^{%
\mathcal{N}=2}]=0$ giving the supersymmetric constraints. The nilpotencies $%
(Q_{{\small susy}}^{\pm })^{2}=0$ lead to finite dimensional
supermultiplets, and the commutativities $[Q_{{\small susy}}^{\pm },H_{%
{\small susy}}^{\mathcal{N}=2}]=0$ give superstates with same energy. Notice
that in absence of central charges, this $\mathcal{N}=2$ system is just the
union of two isomorphic $\mathcal{N}=1$ SUSY systems; one generated by $%
Q_{1} $ and the other by $Q_{2}$. For the $\mathcal{N}=1$ SUSY family
generated by $Q_{1}$, we have one anticommutator $\{Q_{1},Q_{1}\}$ giving
the SUSY Hamiltonian $H_{{\small susy}}^{\mathcal{N}=1}$ while the previous
supersymmetric constraints get restricted to the commutativity $[Q_{{\small 1%
}},H_{{\small susy}}^{\mathcal{N}=1}]=0$.\newline
$\left( \mathbf{2}\right) $ the orthosymplectic family, termed as ORTIC
matter, is based on the \emph{orthosymplectic} osp(2N\TEXTsymbol{\vert}2M)
Lie superalgebra with $M\geq N\geq 1$. For the particular case $M=N$, the
osp(2N\TEXTsymbol{\vert}2N) superalgebra is generated by $\left( i\right) $ $%
4N^{2}$ bosonic operators given by the $N\left( 2N-1\right) $ generators of
so(2N) and the $N\left( 2N+1\right) $ ones of sp(2N); and $\left( ii\right) $
$4N^{2}$ fermionic operators realised as given by
\begin{equation}
F_{i}^{\alpha }=\hat{c}_{i}^{\dagger }\hat{b}^{\alpha },\qquad \bar{F}%
_{\alpha }^{i}=\hat{b}_{\alpha }^{\dagger }\hat{c}^{i},\qquad \bar{F}%
_{\alpha i}=\hat{c}_{i}^{\dagger }\hat{b}_{\alpha }^{\dagger },\qquad
F^{\alpha i}=\hat{b}^{\alpha }\hat{c}^{i}  \label{lcm}
\end{equation}%
Using this orthosymplectic structure, we can construct four hermitian
fermionic operators $Q_{1},Q_{2},Q_{3},Q_{4}$ that characterise ORTIC
matter. These odd operators can be combined like $Q^{\pm }=Q_{1}\pm iQ_{2}$
and $\tilde{Q}^{\pm }=Q_{3}\pm iQ_{4}$, and are realised by linear
combinations of the fermionic generators (\ref{lcm}) line $Q^{+}=\sum
R_{\alpha }^{i}F_{i}^{\alpha }$ and so on. In terms of the oscillators, we
have
\begin{equation}
Q^{+}=\hat{c}_{i}^{\dagger }R_{\alpha }^{i}\hat{b}^{\alpha },\qquad Q^{-}=%
\hat{b}_{\alpha }^{\dagger }\bar{R}_{i}^{\alpha }\hat{c}^{i},\qquad \tilde{Q}%
^{+}=\hat{c}_{i}^{\dagger }T^{i\alpha }\hat{b}_{\alpha }^{\dagger },\qquad
\tilde{Q}^{-}=\hat{b}^{\alpha }\bar{T}_{\alpha i}\hat{c}^{i}  \label{71}
\end{equation}%
where the $R_{\alpha }^{i}$ and $T^{i\alpha }$ are complex coupling tensors.
With these four fermionic charges (\ref{71}), we can engineer the basic
observables for the tight binding modeling of ORTIC matter as follows.%
\newline
$\left( \mathbf{i}\right) $ Two ORTIC charges $Q_{{\small ortic}}^{\pm }$
given by the most general linear combination of (\ref{71}); the $Q_{{\small %
ortic}}^{+}$ is given by $z_{+}Q^{+}+z_{-}Q^{-}+w_{+}\tilde{Q}^{+}+w_{-}%
\tilde{Q}^{-}$ with complex $z_{\pm }$ and $w_{\pm };$ and its adjoint
conjugate $Q_{{\small ortic}}^{-}$ by $\bar{z}_{-}Q^{+}+\bar{z}_{+}Q^{-}+%
\bar{w}_{-}\tilde{Q}^{+}+\bar{w}_{+}\tilde{Q}^{-}.$ Using (\ref{71}), the $%
Q_{{\small ortic}}^{\pm }$ can be also presented like%
\begin{equation}
\begin{tabular}{lll}
${\small Q}_{{\small ortic}}^{{\small +}}$ & ${\small =}$ & ${\small (\hat{c}%
}_{i}^{{\small \dagger }}{\small ,\hat{c}}^{i}{\small )}\left(
\begin{array}{cc}
{\small z}_{+}{\small R}_{\alpha }^{i} & {\small w}_{+}{\small T}^{i\alpha }
\\
{\small w}_{-}{\small \bar{T}}_{\alpha i} & {\small z}_{-}{\small \bar{R}}%
_{i}^{\alpha }%
\end{array}%
\right) \left(
\begin{array}{c}
{\small \hat{b}}^{{\small \alpha }} \\
{\small \hat{b}}_{{\small \alpha }}^{{\small \dagger }}%
\end{array}%
\right) $ \\
${\small Q}_{{\small ortic}}^{{\small -}}$ & ${\small =}$ & ${\small (\hat{b}%
_{\alpha }^{\dagger },\hat{b}^{\alpha })}\left(
\begin{array}{cc}
{\small \bar{z}}_{+}{\small \bar{R}}_{i}^{\alpha } & {\small w}_{-}{\small T}%
^{i\alpha } \\
{\small \bar{w}}_{+}{\small \bar{T}}_{\alpha i} & {\small \bar{z}}_{-}%
{\small R}_{\alpha }^{i}%
\end{array}%
\right) \left(
\begin{array}{c}
{\small \hat{c}}^{i} \\
{\small \hat{c}}_{i}^{{\small \dagger }}%
\end{array}%
\right) $%
\end{tabular}
\label{72}
\end{equation}%
they have been denoted in the main text like $\hat{\lambda}_{\dot{A}}[%
\boldsymbol{q}_{\mathbf{k}}]_{A}^{\dot{A}}\hat{\xi}^{A}$ and $\hat{\xi}_{A}[%
\boldsymbol{q}_{\mathbf{k}}^{\dagger }]_{\dot{A}}^{A}\hat{\lambda}^{\dot{A}%
}; $ see also (\ref{qqpm}) in appendix B. Notice that a hermitian ORTIC
charge $Q_{{\small ortic}}$ requires the identification $Q_{{\small ortic}%
}^{+}=Q_{{\small ortic}}^{-}$ which is solved by setting $\bar{z}_{\mp
}=z_{\pm }$ and $\bar{w}_{\mp }=w_{\pm }.$\newline
$\left( \mathbf{ii}\right) $ Three anticommutators; the first $\{Q_{{\small %
ortic}}^{+},Q_{{\small ortic}}^{-}\}$ defines the ORTIC hamiltonian $H_{%
{\small ortic}}.$ The $\{Q_{{\small ortic}}^{+},Q_{{\small ortic}}^{+}\}$
and its adjoint $\{Q_{{\small ortic}}^{-},Q_{{\small ortic}}^{-}\}$ give
bosonic charge operators $Z_{{\small ortic}}^{++}$ and $Z_{{\small ortic}%
}^{--}$. The $H_{{\small ortic}}$ and $Z_{{\small ortic}}^{++}$ and $Z_{%
{\small ortic}}^{--}$ are valued in the bosonic sector of osp(2N\TEXTsymbol{%
\vert}2N). By imposing the constraints%
\begin{equation}
\lbrack H_{{\small ortic}},Q_{{\small ortic}}^{\pm }]=0\qquad ,\qquad Z_{%
{\small ortic}}^{++}=Z_{{\small ortic}}^{--}=0
\end{equation}%
one obtains the SUSY matter. These conditions are non trivial as they are
mapped into constraint relations on the coupling tensor $[\boldsymbol{q}_{%
\mathbf{k}}]_{A}^{\dot{A}}$ that has been studied in this investigation; see
also appendix B for other details. \newline
To deal with the super TBM and the coupling matrix $\boldsymbol{q}_{\mathbf{k%
}}$ of the supercharge $Q_{{\small ortic}}^{\pm }$ (\ref{QK}), we revisited
properties of the quantum super oscillators in connection with: $\left(
\mathbf{i}\right) $ The orthosymplectic structure given by the triplet $%
\left( \Omega ,G,J\right) $ and described by the supergroup $OSP(2N|2M).$ $%
\left( \mathbf{ii}\right) $ The construction of models for super topological
matter on hypercubic super lattice in relation with the periodic AZ table.
\textrm{For the triplet} $\left( \Omega ,G,J\right) ,$ we found that the
three structures are remarkably accommodated in the $OSP(2N|2M)$ supergroup.
For application of our study, we have considered \textrm{two} typical
topological models from the AZ table; one having charge conjugation
(particle-hole) symmetry and the other has time reversal invariance; but the
method extends straightforwardly to other discrete symmetries like those of
\textrm{\cite{1D,1DA}} and also to spinfull matter. As an important fact of
supermatter resulting from this study, is that the topological obstructions
is captured by $h_{f}$ while $h_{b}$ is some how topologically trivial in
agreement with literature results.\newline
The general set up of the super TBM has been given in the core of the paper
and in the appendices A and B. In the remainder of this discussion section,
we want to give additional comments regarding the classification of the
symmetries of the supercharge
\begin{equation}
Q_{\mathbf{k}}^{+}=\hat{\lambda}_{\mathbf{k}\dot{C}}\left[ \left(
\boldsymbol{q}_{\mathbf{k}}\right) _{A}^{\dot{C}}\right] \hat{\xi}_{\mathbf{k%
}}^{A},\qquad Q_{\mathbf{k}}^{-}=\hat{\xi}_{\mathbf{k}A}\left[ \left(
\boldsymbol{q}_{\mathbf{k}}^{\dagger }\right) _{\dot{C}}^{A}\right] \hat{%
\lambda}_{\mathbf{k}}^{\dot{C}}  \label{61}
\end{equation}%
We show below that the constraint eq(\ref{413}) giving $(\boldsymbol{q}_{%
\mathbf{k}})_{A}^{\dot{B}}$ has $2\times P_{\left( N\right) }$ solutions
with $P_{\left( N\right) }$ being the number of partitions of $N.$ To that
purpose, we start by recalling that our super TBM lives on hypercubic super
lattice and is based on a supercharge $\boldsymbol{Q}$ with Fourier modes $%
Q_{\mathbf{k}}^{\pm }$ as in (\ref{61}). From the view of the $\mathcal{O}%
_{\left( \text{\textsc{N,M}}\right) }$ observables of eqs(\ref{ob}-\ref{bo}%
), this is the simplest fermionic operator living on the Brillouin torus $%
\mathbb{T}^{d}$ combining bosons $\hat{\xi}_{\mathbf{k}}^{A}$ and fermions $%
\hat{\lambda}_{\mathbf{k}\dot{C}}$. The $(q_{\mathbf{k}})_{A}^{\dot{B}}$ is
a $2N\times 2N$ matrix coupling fermions and bosons; it plays a basic role
in the super modeling. It belongs to the bi-fundamental of $Sp\left(
2N\right) \times SO\left( 2N\right) $ which is just the even part of the
orthosymplectic $osp(2N|2N)$. The novelty of the super TBM is that the
square of the supercharge generates a Hamiltonian operator $H_{ortic}$
(resp. $H_{susy}$) which $(i)$ splits as the sum $%
H_{f}^{ortic}+H_{b}^{ortic} $ (resp. $H_{f}^{susy}+H_{b}^{susy}$) describing
fermionic and bosonic contributions (\ref{1}-\ref{2}); and $(ii)$ hides a
remarkable cancellation property (\ref{cp}) indicating that the effect of
bosons in topological supermatter is not as trivial as one might think. The
coupling matrices $h_{f}\left( \mathbf{k}\right) $ and $h_{b}\left( \mathbf{k%
}\right) $ associated with $H_{f}$ and $H_{b}$ are respectively given by $%
\boldsymbol{q}_{\mathbf{k}}Z\boldsymbol{q}_{\mathbf{k}}^{\dagger }$ and $%
h_{b}\left( \mathbf{k}\right) =\boldsymbol{q}_{\mathbf{k}}^{\dagger }%
\boldsymbol{q}_{\mathbf{k}}$ with $Z=\sigma _{z}\otimes I_{N}.$ While $%
h_{b}\left( \mathbf{k}\right) $ is a positive definite operator, the $%
h_{f}\left( \mathbf{k}\right) $ has an indefinite sign because of the $\pm 1$
eigenvalues of $\sigma _{z}$ in agreement with valence and conduction bands.
As additional comments on the results obtained in this study, we cite the
following:

\begin{description}
\item[1)] The set of topological supermatter is a subset of ordinary
topological matter; but with the presence of bosonic matter described by $%
h_{b}\left( \mathbf{k}\right) $. It is classified by the periodic AZ table
with hamiltonian $h_{f}\left( \mathbf{k}\right) $ factorised like
\begin{equation}
h_{f}\left( \mathbf{k}\right) =\boldsymbol{q}_{\mathbf{k}}Z\boldsymbol{q}_{%
\mathbf{k}}^{\dagger }
\end{equation}%
For this subset of AZ matter, "gapless states" are given by \textrm{SUSY
(ORTIC)} multiplets having massless fermions and massless bosons. These
massless super states constraint $h_{f}\left( \mathbf{k}\right) $ and $%
h_{b}\left( \mathbf{k}\right) $ to have zero modes; thus requiring $\det
h_{f}\left( \mathbf{k}\right) =0$ and $\det h_{b}\left( \mathbf{k}\right)
=0. $ These two requirements are solved by $\det \boldsymbol{q}\left(
\mathbf{k}\right) =0$; and so is the condition for massless super states.
Another consequence of \textrm{SUSY (ORTIC) }is that super bands come in
representation multiplets. For instance, in the case of $\mathcal{N}=2$
supersymmetry, the SUSY multiplets have \textrm{four modes}: 2 fermionic
with energies $\pm \epsilon $ (conduction ane valence) and 2 bosonic
positive energy $+\epsilon $. So, $\mathcal{N}=2$ multiplets in super TBM (%
\ref{SQ}, \ref{Q1}) have 3 states with positive $+\epsilon $\ and one with
negative $-\epsilon $. This feature is illustrated by the Figure \textbf{\ref%
{DD}}.

\item[2)] Besides TPC discrete symmetries (\ref{xq}) and crystalline ones
like mirrors, a classification of the super TBM charges is given by global
symmetries of $q_{A}^{\dot{C}}.$ As investigated in the main text and as it
will be described below, it turns out that the classes of $\boldsymbol{q}%
_{A}^{\dot{C}}$ are given by subgroups $Sp\left( 2N\right) \times SO\left(
2N\right) $ by help of the splitting (\ref{416})
\begin{equation}
\begin{tabular}{lll}
$SO\left( 2N\right) $ & $:$ & $S\left[ U\left( 2\right) \times U\left(
N\right) \right] $ \\
$Sp\left( 2N\right) $ & $:$ & $S\left[ U\left( 2\right) ^{\prime }\times
SU\left( N\right) ^{\prime }\right] $%
\end{tabular}
\label{br}
\end{equation}%
The invariance of the supercharge (\ref{QK}) requires the condition (\ref%
{413}). By using (\ref{416}) and (\ref{br}), this condition reads as follows
\begin{equation}
q_{\alpha I}^{\dot{\gamma}\dot{K}}=\left[ \left( U_{1}\right) _{\dot{\delta}%
}^{\dot{\gamma}}\left( U_{2}\right) _{\dot{L}}^{\dot{K}}\right] q_{\beta J}^{%
\dot{\delta}\dot{L}}\left[ \left( U_{1}^{\prime }\right) _{\alpha }^{\beta
}\left( U_{2}^{\prime }\right) _{I}^{J}\right]  \label{qa}
\end{equation}%
where the $2\times 2$ matrices $U_{1}$ and $U_{1}^{\prime }$ are elements of
the $U\left( 2\right) \times U\left( 2\right) ^{\prime }$ in (\ref{br}) and
the $N\times N$ matrices $U_{2}$ and $U_{2}^{\prime }$ are elements of $%
U\left( N\right) \times U\left( N\right) ^{\prime }$. The condition (\ref{qa}%
) has several solutions classified by subgroups of (\ref{br}). We comment
here after on this classification while illustrating the constructions on
particular examples.

\begin{description}
\item[a)] \emph{\ }$U\left( 2\right) \times U\left( N\right) $\emph{\
symmetry:} The simplest solution of the condition (\ref{qa}) is given by the
following one parameter family
\begin{equation}
q_{\beta J}^{\dot{\delta}\dot{L}}\left( \mathbf{k}\right) =\mu \left(
\mathbf{k}\right) \delta _{\beta }^{\dot{\delta}}\delta _{J}^{\dot{L}}
\end{equation}%
where the complex $\mu \left( \mathbf{k}\right) $ lives on $\mathbb{T}^{d}.$
This particular solution has a strong symmetry given by the $U\left(
2\right) \times U\left( N\right) $ group. This invariance is derived as
follows: First, substitute the above realisation in the condition; so the
right hand side of eq(\ref{qa}) becomes
\begin{equation}
q_{\alpha I}^{\dot{\gamma}\dot{K}}=\mu \left( \mathbf{k}\right) \left[
\left( U_{1}^{\prime }U_{1}\right) _{\alpha }^{\dot{\gamma}}\right] \left[
\left( U_{2}^{\prime }U_{2}\right) _{I}^{\dot{K}}\right]
\end{equation}%
To insure invariance, we must have $U_{1}^{\prime }U_{1}=I_{2}$ and $%
U_{2}^{\prime }U_{2}=I_{N}$ which are respectively solved by requiring $%
U_{1}^{\prime }=U_{1}^{\dagger }$ and $U_{2}^{\prime }=U_{2}^{\dagger }$
generating the group $U\left( 2\right) \times U\left( N\right) .$

\item[b)] \emph{\ }$U\left( 1\right) ^{2}\times U\left( N\right) $\emph{\
symmetry:} A second solution of the constraint eq(\ref{qa}) is given by%
\begin{equation}
q_{\beta J}^{\dot{\delta}\dot{L}}\left( \mathbf{k}\right) =\left(
\begin{array}{cc}
\mu \left( \mathbf{k}\right) \delta _{J}^{\dot{L}} & 0 \\
0 & \nu \left( \mathbf{k}\right) \delta _{J}^{\dot{L}}%
\end{array}%
\right)
\end{equation}%
It has two complex parameters $\mu \left( \mathbf{k}\right) $ and $\nu
\left( \mathbf{k}\right) $\ living on $\mathbb{T}^{d}.$ This solution is
invariant under $U\left( 1\right) ^{2}\times U\left( N\right) $ which is a
subgroup of $U\left( 2\right) \times U\left( N\right) .$ By setting $\mu
\left( \mathbf{k}\right) =\nu \left( \mathbf{k}\right) ,$ we recover the
previous class.

\item[c)] \emph{\ }$U\left( N\right) $\emph{\ symmetry:} A third solution of
eq(\ref{qa}) is given by%
\begin{equation}
q_{\beta J}^{\dot{\delta}\dot{L}}\left( \mathbf{k}\right) =\left(
\begin{array}{cc}
\mu \left( \mathbf{k}\right) & \rho \left( \mathbf{k}\right) \\
\varsigma \left( \mathbf{k}\right) & \nu \left( \mathbf{k}\right)%
\end{array}%
\right) \otimes \delta _{J}^{\dot{L}}
\end{equation}%
It has four complex parameters $\mu \left( \mathbf{k}\right) ,$ $\nu \left(
\mathbf{k}\right) ,$ $\rho \left( \mathbf{k}\right) $\ and $\varsigma \left(
\mathbf{k}\right) $\ living on $\mathbb{T}^{d}.$ This solution is invariant
under $U\left( N\right) $ which is a subgroup of $U\left( 2\right) \times
U\left( N\right) .$
\end{description}
\end{description}

\ \ \newline
Following this method, one can construct several solutions of eq(\ref{qa});
they are classified by the subgroups of $U\left( 2\right) \times U\left(
N\right) $; the classes follow from the factorisation
\begin{equation}
q_{\beta J}^{\dot{\delta}\dot{L}}\left( \mathbf{k}\right) =\mathcal{A}%
_{\beta }^{\dot{\delta}}\left( \mathbf{k}\right) \times \mathcal{B}_{J}^{%
\dot{L}}\left( \mathbf{k}\right)
\end{equation}%
where in general $\mathcal{A}_{\beta }^{\dot{\delta}}\left( \mathbf{k}%
\right) $ has four complex functions and $\mathcal{B}_{J}^{\dot{L}}\left(
\mathbf{k}\right) $ has $N^{2}$ complex functions. In this regard, we recall
that unitary subgroups of $U\left( N\right) $ are given by%
\begin{equation}
\prod\limits_{i=1}^{l}U\left( n_{i}\right) \qquad ,\qquad
\sum_{i=1}^{l}n_{i}=N
\end{equation}%
including the maximal abelian $U\left( 1\right) ^{N}$ corresponding to the
choice $\mathcal{B}_{J}^{\dot{L}}=\mathcal{B}_{J}\delta _{J}^{\dot{L}}$
where $\mathcal{B}_{J}$ are $N$ complex functions. Notice that the partition
$P\left( 2\right) =2$ because there are two ways to decompose the integer $2$%
: either just as $2$ of like $1+1.$ Notice also that the number of
possibilities of decomposing a positive integer like $\sum_{i=1}^{l}n_{i}=N$
is given by the partition $P\left( N\right) .$ This number is obtained by
expanding the Mac-Mahon partition function in a series as follows \textrm{%
\cite{Jeh}},%
\begin{equation}
\prod\limits_{l=1}^{\infty }\left( 1-X^{l}\right) ^{-1}=\sum_{n=1}^{\infty
}P_{\left( n\right) }X^{n}
\end{equation}%
So the number of classes of is given by $P_{\left( 2\right) }\times
P_{\left( N\right) }=2P_{\left( N\right) }$.

\section{Two appendices}

In this section, we give two appendices A and B where we collect useful
ingredients and complete some results given in the main text. In appendix A,
we give useful tools on $\left( 1\right) $ the supersymmetric algebras in 2D
and 1D having $n$ supersymmetric charges ($n\leq 4$); $\left( 2\right) $
their embedding in the orthosymplectic osp(2\TEXTsymbol{\vert}2) and $\left(
3\right) $ the embedding of osp(2\TEXTsymbol{\vert}2) in the $\mathcal{N}=2$
super conformal invariance in the Neveu-Schwaz (NS) sector. In appendix B,
we revisit basic algebraic aspects of the orthosymplectic osp(2N\TEXTsymbol{%
\vert}2M) underlying results obtained in this paper and also in the study of
the supersymmetric fivefold ways of \cite{1D2}. We also complete partial
results given in section 4 and subsection 6.2 of our investigation.

\subsection{Appendix A: fermions' algebra in 2D and 1D}

We begin by recalling useful features on fermions in 2D world sheet and in
1D world line. The two worlds are somehow related due to the abelian
property of the SO(2) rotation group of the real plane $\mathbb{R}^{2}$
(SO(1,1) for $\mathbb{R}^{1,1}$). Indeed 2D fermions are described by $%
SO\left( 2\right) $ spinors $\mathbf{\psi }_{\alpha }$ having two components
$\left( \psi _{+1/2},\psi _{-1/2}\right) $. They can be either hermitian
with $\psi _{\pm 1/2}$ real (Majorana) or complex (Dirac). Because of the
abelian property of $SO\left( 2\right) $, these fermionic components $\psi
_{\pm 1/2}$ can be handled separately (as Weyl or Majorana-Weyl spinors).
So; one real fermion $\mathbf{\psi }_{\alpha }$ in 2D may be imagined in
terms of two real 1D fermions $\lambda _{1}$ and $\lambda _{2}$ respectively
associated with $\psi _{+1/2}$ and $\psi _{-1/2}$. A complex $\mathbf{\psi }%
_{\alpha }$ in 2D splitting as $\mathbf{\xi }_{\alpha }+i\mathbf{\chi }%
_{\alpha }$ can thought of in terms of four real fermions in 1D denoted as $%
\lambda _{1},$ $\lambda _{2},$ $\lambda _{3},$ $\lambda _{4}$ and
respectively given by $\xi _{+1/2},$ $\xi _{-1/2},$ $\chi _{+1/2},$ $\chi
_{-1/2}.$ In sum, we have the following 2D/1D correspondence
\begin{equation}
n\text{ fermions }\mathbf{\psi }_{\alpha }^{i}\text{ in 2D \qquad }%
\rightarrow \text{\qquad }2n\text{ fermions }\lambda _{A}\text{ in 2D}
\label{vi}
\end{equation}%
This correspondence applies also to the conserved supersymmetric charges in
2D and 1D (super QM). In 2D $\mathcal{N}=1$ supersymmetric theory, the
fermionic generator $Q_{\alpha }$ is a Majorana spinor; it has two conserved
super charges: a left supercharge $Q_{-1/2}$ and a right one $Q_{+1/2}$ that
can realised independently due to the reducibility of SO(2) representations.
This 2D $\mathcal{N}=1$ theory can be put in correspondence with $\mathcal{N}%
=2$ supersymmetry in 1D. So, by using (\ref{vi}), we have the following
dictionary: \newline
$\left( i\right) $ $\mathcal{N}=1$ supersymmetry in 2D generated by a
Majorana spinor operator $Q_{\alpha }=Q_{\pm 1/2}$ is generally termed as 2D
$\mathcal{N}=\left( 1,1\right) $. In this case, we have two real
supersymmetric charges: $Q_{+1/2}$ and $Q_{-1/2}$. From the 1D view, the
super QM has two supercharges given by $Q_{1}$ and $Q_{2}$. \newline
$\left( ii\right) $ $\mathcal{N}=2$ supersymmetry in 2D has two Majorana
supercharges $Q_{\alpha }^{1},$ $Q_{\alpha }^{2}$ often denoted like $Q_{\pm
1/2}^{\pm }$ with the upper $\pm $ charges referring to an extra SO$\left(
2\right) _{R}$ symmetry rotating $Q_{\alpha }^{1}$ and $Q_{\alpha }^{2}$.
This structure is generally termed as 2D $\mathcal{N}=\left( 2,2\right) .$
Here, we have four real supersymmetric charges given by the following
Majorana-Weyl fermions,
\begin{equation}
Q_{+1/2}^{1},\qquad Q_{-1/2}^{1},\qquad Q_{+1/2}^{2},\qquad Q_{-1/2}^{2}
\end{equation}%
From the 1D view, the associated theory has four supercharges $Q_{1},$ $%
Q_{2},$ $Q_{3}$ and $Q_{4}$.\newline
$\left( iii\right) $ In the case of $n+m$ Majorana-Weyl spinors; we use the
terminology $\mathcal{N}=\left( n,m\right) $ supersymmetry having n right
charges $Q_{+1/2}^{1},...Q_{+1/2}^{n}$; and m left charges $%
Q_{-1/2}^{1},...Q_{-1/2}^{m}$. In this case, we have $n+m$ real
supersymmetric charges. So, for $n\neq m,$ the some how apparent classical
left-right symmetry is violated. \newline
We end this intro hat by noticing that our interest in going to 1D $\mathcal{%
N}=2$ super QM through 2D theory is motivated by its embedding in osp(2%
\TEXTsymbol{\vert}2).

\subsubsection{Supersymmetry in 2D and 1D}

Focusing on the interesting case of a supersymmetric theory with two real
supercharges $Q_{1}$ and $Q_{2}$ combined into a complex $Q=Q_{1}+iQ_{2}$
and its adjoint $\bar{Q}=Q_{1}-iQ_{2}$, the underlying Lie superalgebra is,
generally speaking, defined by
\begin{equation}
\begin{tabular}{lll}
$QQ^{\dagger }+Q^{\dagger }Q$ & $=$ & $2P$ \\
$\left\{ Q,Q\right\} $ & $=$ & $2Z$ \\
$\left\{ Q^{\dagger },Q^{\dagger }\right\} $ & $=$ & $2Z^{\dagger }$%
\end{tabular}
\label{ap1}
\end{equation}%
In these relations, the P is a bosonic hermitian operator (the Hamiltonian
in $\mathcal{N}=2$ super QM) and the complex $Z$ is the central charge of
the superalgebra. They obey the commutations%
\begin{eqnarray}
\left[ P,Q\right] &=&\left[ P,Q^{\dagger }\right] =0  \notag \\
\left[ Z,Q\right] &=&\left[ Z,Q^{\dagger }\right] =0  \label{ap2} \\
\left[ P,Z\right] &=&0  \notag
\end{eqnarray}%
As far as this 1D $\mathcal{N}=2$ superalgebra is concerned, notice that the
\textrm{four} following features:\newline
$\left( \mathbf{1}\right) $ the central charge Z scales as energy, the same
as P; it plays an important role in the study of BPS states. These states
are beyond the the scope of the present paper; and so we will disregard Z
here. So, the relation $\left\{ Q,Q\right\} =Z$ becomes a nilpotency
condition of the fermionic charge; that is
\begin{equation}
Q^{2}=\bar{Q}^{2}=0
\end{equation}%
$\left( \mathbf{2}\right) $ The highest weight representations $\mathcal{R}%
_{susy}$ of the superalgebra (\ref{ap1}-\ref{ap2}) are two complex
dimensional (even complex dimensional in general) due to $Q^{2}=0$; they
contain as complex bosonic state degrees $\left\vert b\right\rangle $ as
complex fermionic ones $\left\vert f\right\rangle ;$ and they have the same
energy $\varepsilon $. For the example of the 1D $\mathcal{N}=2$
supersymmetric scalar representation with ground state as%
\begin{equation}
Q\left\vert b\right\rangle =0\qquad ,\qquad P\left\vert b\right\rangle
=\varepsilon \left\vert b\right\rangle
\end{equation}%
the fermionic partner state is given by $\left\vert f\right\rangle =\bar{Q}%
\left\vert b\right\rangle .$\newline
$\left( \mathbf{3}\right) $ The energy of this state is determined by
computing $P\left\vert f\right\rangle =P\bar{Q}\left\vert b\right\rangle $;
it equal to $\varepsilon \left\vert f\right\rangle ;$ thanks to the
commutativity $P\bar{Q}=\bar{Q}P$. The feature $Q^{2}=\bar{Q}^{2}=0$ and $%
\left[ P,\bar{Q}\right] =0$ are present in supersymmetric algebra (\ref{ap1}-%
\ref{ap2}); but are violated for osp(2\TEXTsymbol{\vert}2) as its
supercharges $Q_{ortic}$ have the typical property%
\begin{equation}
\left[ P_{ortic},Q_{ortic}\right] \neq 0
\end{equation}%
This orthosymplectic feature will be described in the next sub-subsection.%
\newline
$\left( \mathbf{4}\right) $ In the case where the supercharge is hermitian $%
Q^{\dagger }=Q$, the eqs(\ref{ap1}-\ref{ap2}) reduce to%
\begin{equation}
Q^{2}=2P\qquad ,\qquad \left[ P,Q\right] =0
\end{equation}%
They define the 1D $\mathcal{N}=1$ supersymmetric algebra underlying the $%
\mathcal{N}=1$\ super QM.

\subsubsection{The orthosymplectic osp(2\TEXTsymbol{\vert}2)}

The osp(2\TEXTsymbol{\vert}2) is eight dimensional; it has a bosonic sector $%
\boldsymbol{g}_{\bar{0}}$ given by $so(2,\mathbb{R})\oplus sp(2,\mathbb{R});$
and a fermionic sector $\boldsymbol{g}_{\bar{1}}$ given by a 4-dimensional
module of $\boldsymbol{g}_{\bar{0}}.$ Their graded commutations are as
follows:

\paragraph{I. \textbf{The bosonic sector} $so(2)\oplus sp(2):$\newline
}

The even part $\boldsymbol{g}_{\bar{0}}$ has four bosonic generators; the
hermitian $J_{0}$ generating the abelian $so(2);$ and the three $S_{0,\pm }$
generating the $sp(2)$ obeying $S_{0}^{\dagger }=S_{0}$ and $S_{\pm
}^{\dagger }=S_{\mp }$. The commutation relations defining $\boldsymbol{g}_{%
\bar{0}}$ are given by%
\begin{equation}
\begin{tabular}{lll}
$\left[ J_{0},S_{0,\pm }\right] $ & $=$ & $0$ \\
$\left[ S_{0},S_{\pm }\right] $ & $=$ & $\pm S_{\pm }$ \\
$\left[ S_{-},S_{+}\right] $ & $=$ & $2S_{0}$%
\end{tabular}
\label{sp1}
\end{equation}%
As far as these relations are concerned, notice the following interesting
features.\newline
$\left( \mathbf{1}\right) $ the two hermitian $J_{0}$ and $S_{0}$ are the
commuting generators of $so(2)\oplus sp(2)$; they can be diagonalised
simultaneously in the same basis. Their quantum charges $\left( q,p\right) $
label the irreducible representations $\mathcal{R}_{\left( q,p\right)
}^{ortic}$ of osp(2\TEXTsymbol{\vert}2). The quantum number $p=s_{z},$ it is
the usual spin projection ranging as $-s\leq s_{z}\leq s.$ \newline
$\left( \mathbf{2}\right) $ The two $J_{0}$ and $S_{0}$ appear in the graded
commutations of the osp(2\TEXTsymbol{\vert}2) superalgebra through the
linear combinations $S_{0}\pm J_{0};$ see eq(\ref{lc}). They can be imagined
as candidates for the $H_{susy}$ of eq(\ref{ap1}) and a charge operator
generating a U$\left( 1\right) _{R}$ symmetry; see below for further
details. \newline
In terms of the supersymmetric oscillators $\hat{b}/\hat{b}^{\dagger }$ and $%
\hat{c}/\hat{c}^{\dagger }$, we have the following oscillator realisation of
the bosonic operators $J_{0},S_{0}$ and $S_{\pm },$%
\begin{eqnarray}
J_{0} &=&\frac{1}{4}\left( \hat{c}^{\dagger }\hat{c}-\hat{c}\hat{c}^{\dagger
}\right)  \label{J0} \\
S_{0} &=&\frac{1}{4}\left( \hat{b}^{\dagger }\hat{b}+\hat{b}\hat{b}^{\dagger
}\right) ,\qquad S_{-}=\frac{1}{2}\hat{b}\hat{b},\qquad S_{+}=\frac{1}{2}%
\hat{b}^{\dagger }\hat{b}^{\dagger }  \label{S0}
\end{eqnarray}%
From this realisation, we can check that we have
\begin{equation}
\begin{tabular}{lllllll}
$\left[ J_{0},\hat{c}\right] $ & $=$ & $-\frac{1}{2}\hat{c}$ & $,\qquad $ & $%
\left[ J_{0},\hat{c}^{\dagger }\right] $ & $=$ & $\frac{1}{2}\hat{c}%
^{\dagger }$ \\
$\left[ S_{0},\hat{b}\right] $ & $=$ & $-\frac{1}{2}\hat{b}$ & $,\qquad $ & $%
\left[ S_{0},\hat{b}^{\dagger }\right] $ & $=$ & $\frac{1}{2}\hat{b}%
^{\dagger }$%
\end{tabular}
\label{cr1}
\end{equation}%
indicating that $\left( \mathbf{i}\right) $ the fermionic $\hat{c}/\hat{c}%
^{\dagger }$ carry a half charge of $so\left( 2\right) $ ($q=\pm 1/2$) but
no charge of $S_{0}$ ($p=0$)$;$ $\left( \mathbf{ii}\right) $ the bosonic $%
\hat{b}/\hat{b}^{\dagger }$ carry a half charge of $sp\left( 2\right) $ ($%
p=\pm 1/2$) but no $so\left( 2\right) $ charge. \newline
Notice that in the $\mathcal{R}_{\left( q,p\right) }$ representation
language, we can can think of $\hat{c}/\hat{c}^{\dagger }$ in terms of an
so(2) doublet $\hat{c}^{q}=(\hat{c},\hat{c}^{\dagger })$ and about $\hat{b}/%
\hat{b}^{\dagger }$ in terms of an sp(2) doublet $\hat{b}^{p}=(\hat{b},\hat{b%
}^{\dagger })$.

\paragraph{II. \textbf{The fermionic sector:}\newline
}

The four fermionic operators generating $\boldsymbol{g}_{\bar{1}}$ are
denoted like $\boldsymbol{F}_{p}^{q}$ with $q=\pm $ (short of $\pm 1/2$)
labelling the charges of $so(2)$ and $p=\pm $ indexing the charges of $%
sp(2). $ They behave as doublets under $so\left( 2\right) $ and under $%
sp\left( 2\right) .$ These $\boldsymbol{F}_{p}^{q}$'s are realised in terms
of the super oscillators as follows%
\begin{equation}
\begin{tabular}{lll}
$\boldsymbol{F}_{-}^{-}=\hat{b}\hat{c}$ & $,\qquad $ & $\boldsymbol{F}%
_{+}^{+}=\hat{b}^{\dagger }\hat{c}^{\dagger }$ \\
$\boldsymbol{F}_{-}^{+}=\hat{b}\hat{c}^{\dagger }$ & $,\qquad $ & $%
\boldsymbol{F}_{+}^{-}=\hat{b}^{\dagger }\hat{c}$%
\end{tabular}
\label{fbc}
\end{equation}%
with $\left( \mathbf{i}\right) $ the adjoint conjugations $\left(
\boldsymbol{F}_{-}^{-}\right) ^{\dagger }=\boldsymbol{F}_{+}^{+}$ and $%
\left( \boldsymbol{F}_{-}^{+}\right) ^{\dagger }=\boldsymbol{F}_{+}^{-};$
and $\left( \mathbf{ii}\right) \ $the nilpotency $\left( \boldsymbol{F}%
_{p}^{q}\right) ^{2}=0$. By using the notation $\hat{c}^{q}$ and $\hat{b}%
_{p},$ the above fermionic generators can be presented collectively like $%
F_{p}^{q}=\hat{c}^{q}\hat{b}_{p}.$The graded commutation relations between
these fermionic operators are given by%
\begin{equation}
\begin{tabular}{lll}
$\left\{ \boldsymbol{F}_{-}^{-},\boldsymbol{F}_{+}^{+}\right\} $ & $=$ & $%
2S_{0}-2J_{0}$ \\
$\left\{ \boldsymbol{F}_{-}^{-},\boldsymbol{F}_{-}^{+}\right\} $ & $=$ & $%
2S_{-}$ \\
$\left\{ \boldsymbol{F}_{-}^{-},\boldsymbol{F}_{+}^{-}\right\} $ & $=$ & $0$%
\end{tabular}%
,\qquad
\begin{tabular}{lll}
$\left\{ \boldsymbol{F}_{-}^{+},\boldsymbol{F}_{+}^{-}\right\} $ & $=$ & $%
2S_{0}+2J_{0}$ \\
$\left\{ \boldsymbol{F}_{+}^{-},\boldsymbol{F}_{+}^{+}\right\} $ & $=$ & $%
2S_{+}$ \\
$\left\{ \boldsymbol{F}_{-}^{+},\boldsymbol{F}_{+}^{+}\right\} $ & $=$ & $0$%
\end{tabular}
\label{lc}
\end{equation}%
and%
\begin{equation}
\begin{tabular}{lll}
$\left[ J_{0},\boldsymbol{F}_{\pm }^{-}\right] $ & $=$ & $-\frac{1}{2}%
\boldsymbol{F}_{\pm }^{-}$ \\
$\left[ J_{0},\boldsymbol{F}_{\pm }^{+}\right] $ & $=$ & $+\frac{1}{2}%
\boldsymbol{F}_{\pm }^{+}$%
\end{tabular}%
,\qquad
\begin{tabular}{lll}
$\left[ S_{0},\boldsymbol{F}_{-}^{\pm }\right] $ & $=$ & $-\frac{1}{2}%
\boldsymbol{F}_{-}^{\pm }$ \\
$\left[ S_{0},\boldsymbol{F}_{+}^{\pm }\right] $ & $=$ & $+\frac{1}{2}%
\boldsymbol{F}_{+}^{\pm }$%
\end{tabular}%
\end{equation}%
as well as
\begin{equation}
\begin{tabular}{lll}
$\left[ S_{-},\boldsymbol{F}_{-}^{\pm }\right] $ & $=$ & $0$ \\
$\left[ S_{+},\boldsymbol{F}_{+}^{\pm }\right] $ & $=$ & $0$%
\end{tabular}%
,\qquad
\begin{tabular}{lll}
$\left[ S_{+},\boldsymbol{F}_{-}^{\pm }\right] $ & $=$ & $-\boldsymbol{F}%
_{+}^{\pm }$ \\
$\left[ S_{-},\boldsymbol{F}_{+}^{\pm }\right] $ & $=$ & $+\boldsymbol{F}%
_{-}^{+}$%
\end{tabular}
\label{crf}
\end{equation}%
The quadratic Casimir $C_{2}$ commuting with the generators of osp(2%
\TEXTsymbol{\vert}2) and characterising the osp(2\TEXTsymbol{\vert}2)
representations is given by \textrm{\cite{osp22}}
\begin{equation}
C_{2}=S_{0}\left( 2S_{0}+1\right) -J_{0}\left( 2J_{0}+1\right) +2S_{+}S_{-}+2%
\boldsymbol{F}_{-}^{-}\boldsymbol{F}_{-}^{+}-2\boldsymbol{F}_{+}^{+}%
\boldsymbol{F}_{+}^{-}
\end{equation}

\subsubsection{Embedding $\mathcal{N}=2$ SUSY in osp(2\TEXTsymbol{\vert}2)
and in $\mathcal{N}=2$ CFT$_{2}$}

From the above graded commutation relations of osp(2\TEXTsymbol{\vert}2), we
can deduce a set of interesting properties; those useful features for us are
as listed below:

\paragraph{I. \textbf{Two complex generators}\ $\boldsymbol{F}_{-}^{-}$
\textbf{and} $\boldsymbol{F}_{+}^{-}:$\newline
}

The osp(2\TEXTsymbol{\vert}2) has two complex supercharges $\boldsymbol{F}%
_{-}^{-}$ and $\boldsymbol{F}_{+}^{-}$ (four real ones $F_{1},$ $F_{2},$ $%
F_{3},$ $F_{4}$). Using eqs(\ref{cr1}-\ref{crf}), we draw the following
relations%
\begin{equation}
\begin{tabular}{lll}
$\left[ 2S_{0}-2J_{0},\boldsymbol{F}_{p}^{q}\right] $ & $=$ & $\left(
p-q\right) \boldsymbol{F}_{p}^{q}$ \\
$\left[ 2S_{0}+2J_{0},\boldsymbol{F}_{p}^{q}\right] $ & $=$ & $\left(
p+q\right) \boldsymbol{F}_{p}^{q}$%
\end{tabular}%
\end{equation}%
So for $p=+q,$ we have the following%
\begin{equation}
\begin{tabular}{lll}
$\left[ 2S_{0}-2J_{0},\boldsymbol{F}_{-}^{-}\right] $ & $=$ & $0$ \\
$\left[ 2S_{0}-2J_{0},\boldsymbol{F}_{+}^{+}\right] $ & $=$ & $0$%
\end{tabular}%
,\qquad
\begin{tabular}{lll}
$\left[ 2S_{0}-2J_{0},\boldsymbol{F}_{+}^{-}\right] $ & $=$ & $+2\boldsymbol{%
F}_{+}^{-}$ \\
$\left[ 2S_{0}-2J_{0},\boldsymbol{F}_{-}^{+}\right] $ & $=$ & $-2\boldsymbol{%
F}_{-}^{+}$%
\end{tabular}%
\end{equation}%
and for $p=-q,$ we have%
\begin{equation}
\begin{tabular}{lll}
$\left[ 2S_{0}+2J_{0},\boldsymbol{F}_{-}^{-}\right] $ & $=$ & $-2\boldsymbol{%
F}_{-}^{-}$ \\
$\left[ 2S_{0}+2J_{0},\boldsymbol{F}_{+}^{+}\right] $ & $=$ & $+2\boldsymbol{%
F}_{+}^{+}$%
\end{tabular}%
,\qquad
\begin{tabular}{lll}
$\left[ 2S_{0}+2J_{0},\boldsymbol{F}_{+}^{-}\right] $ & $=$ & $0$ \\
$\left[ 2S_{0}+2J_{0},\boldsymbol{F}_{-}^{+}\right] $ & $=$ & $0$%
\end{tabular}%
\end{equation}

\paragraph{II. $\mathcal{N}=2$ \textbf{SUSY as a subalgebra of osp(2%
\TEXTsymbol{\vert}2):}\newline
}

If restricting the four bosonic generators of osp(2\TEXTsymbol{\vert}2) down
to the combination $P=\omega \left( S_{0}+J_{0}\right) $ and $T=\upsilon
\left( S_{0}-J_{0}\right) $ with no $S_{\pm };$ and the four fermionic ones
down to the two $\boldsymbol{F}_{-}^{+}\sim Q/\sqrt{\omega }$ and $%
\boldsymbol{F}_{+}^{-}\sim \bar{Q}/\sqrt{\omega }$ with no $\boldsymbol{F}%
_{+}^{+}$ nor $\boldsymbol{F}_{-}^{-}$, the above graded commutations reduce
to%
\begin{equation}
\begin{tabular}{lll}
$\left\{ Q,\bar{Q}\right\} $ & $=$ & $P$ \\
$\left[ P,Q\right] $ & $=$ & $0$%
\end{tabular}%
\qquad ,\qquad
\begin{tabular}{lll}
$\left\{ Q,Q\right\} $ & $=$ & $0$ \\
$\left\{ \bar{Q},\bar{Q}\right\} $ & $=$ & $0$%
\end{tabular}%
\end{equation}%
These relations should be compared with eqs(\ref{ap1}); they define a
supersymmetric algebra with two supercharges $Q$ and $\bar{Q}$. Notice the
two following: $\left( \mathbf{i}\right) $ in terms of the harmonic
oscillators, the supercharges are realised as $Q=\sqrt{\omega }\hat{c}%
^{\dagger }\hat{b}$ and $\bar{Q}=\sqrt{\omega }\hat{b}^{\dagger }\hat{c}$
while the bosonic operator P is given by
\begin{equation}
P=\omega (\hat{b}^{\dagger }\hat{b}+\hat{c}^{\dagger }\hat{c})
\end{equation}%
$\left( \mathbf{ii}\right) $ Because of supersymmetry, the $\mathcal{N}=2$
superalgebra has another bosonic charge namely the charge operator which we
denoted as T. It generates the $U\left( 1\right) _{R}$ symmetry of the $%
\mathcal{N}=2$ $U\left( 1\right) $ superalgebra and it acts as $\left[ T,Q%
\right] =Q$. Its oscillator realisation is given by $(\hat{c}^{\dagger }\hat{%
c}-\hat{b}^{\dagger }\hat{b})/2.$

\paragraph{III. \textbf{OSp(2\TEXTsymbol{\vert}2) as subinvariance of} $%
\mathcal{N}=2$ \textbf{CFT}$_{2}:$\newline
}

From the view of the 2D $\mathcal{N}=2$ superconformal field theory (CFT$%
_{2} $), the osp(2\TEXTsymbol{\vert}2) is a subalgebra of the $\mathcal{N}=2$
super CFT$_{2}$ algebra in the NS sector. In terms of the generators of the $%
\mathcal{N}=2$ super Virasoro algebra \textrm{\cite{scft}} given by: $\left(
\mathbf{i}\right) $ the usual bosonic $L_{n}$ and $J_{n}$ with n integer,
respectively referring the Virasoro and the $U\left( 1\right) $ Kac-Moody
generators, and $\left( \mathbf{ii}\right) $ the fermionic partners $%
G_{n+1/2}^{+}$ and $G_{n+1/2}^{-}$, we have
\begin{equation}
\begin{tabular}{lll}
$L_{0}=-S_{0}$ & $,\qquad $ & $G_{\pm 1/2}^{-}=F_{\pm }^{-}$ \\
$L_{\pm }=\pm S_{\pm }$ & $,\qquad $ & $G_{\pm 1/2}^{+}=F_{\pm }^{+}$%
\end{tabular}
\label{LG}
\end{equation}%
Recall that the osp(2\TEXTsymbol{\vert}2) Lie superalgebra corresponds just
to the anomaly free sub- superalgebra of the $\mathcal{N}=2$ super Virasoro
in NS sector defined as%
\begin{equation}
\begin{tabular}{lll}
$\left[ L_{m},L_{n}\right] $ & $=$ & $\left( m-n\right) L_{m+n}+\frac{c}{12}%
\left( m^{3}-m\right) \delta _{m+n}$ \\
$\left[ J_{m},J_{n}\right] $ & $=$ & $\frac{c}{3}m\delta _{m+n}$ \\
$\left\{ G_{r}^{+},G_{s}^{-}\right\} $ & $=$ & $2L_{r+s}+\left( r-s\right)
J_{r+s}+\frac{c}{12}\left( 4r^{2}-1\right) \delta _{r+s}$ \\
$\left\{ G_{r}^{\pm },G_{s}^{\pm }\right\} $ & $=$ & $0$ \\
$\left[ L_{m},G_{r}^{\pm }\right] $ & $=$ & $\left( \frac{m}{2}-r\right)
G_{m+r}^{\pm }$ \\
$\left[ J_{m},G_{r}^{\pm }\right] $ & $=$ & $\pm G_{m+r}^{\pm }$%
\end{tabular}%
\end{equation}%
with $r,s\in \mathbb{Z}+1/2.$ The anomalous terms $\frac{c}{12}\left(
m^{3}-m\right) \delta _{m+n}$ and $\frac{c}{12}\left( 4r^{2}-1\right) \delta
_{r+s}$ disappear for $m=0,\pm 1$ and $r=\pm 1/2.$ This anomaly free
condition reduces the infinite set $L_{m},J_{m},G_{r}^{\pm }$ down to the
four bosonic $L_{0},L_{\pm },J_{0}$ and the four fermionic $G_{\pm
1/2}^{+},G_{\pm 1/2}^{-}$ as in (\ref{LG}).

\subsection{Appendix B: From osp(2N\TEXTsymbol{\vert}2M) to SUSY}

In this appendix, we use the tools developed in the present study to shed
some light on the graded algebraic structure underlying the section 4 and
subsection 6.2 as well as the supersymmetric fivefold ways of \textrm{\cite%
{1D2}}.

\subsubsection{General on oscillator realisation of fermionic charges}

For simplicity of the presentation, we use short cuts to reach some results
of this construction while keeping quite similar notations as in \cite{1D2}.
For the details regarding other interesting results in particular the
intrinsic aspects of the fivefold way classes and the physical applications,
we report the reader to the above mentioned reference. To make an idea on
the types of fermionic charges we will consider in this appendix, we
anticipate this construction by noticing that we will consider three
families (I, II, III) of fermionic charges as summarised in the following
table%
\begin{equation}
\begin{tabular}{|c|c|c|c|c|}
\hline
{\small families} & {\small supercharges} & {\small realisations} & {\small %
couplings} & {\small number of Q's} \\ \hline
{\small I} & $\left. Q^{+}\right. $ & $\hat{c}_{i}^{\dagger }R_{\alpha }^{i}%
\hat{b}^{\alpha }$ & $\left. R_{\alpha }^{i}\right. $ & {\small 1 complex}
\\ \hline
{\small II} & $\left.
\begin{array}{c}
Q_{1}^{+} \\
Q_{2}^{+}%
\end{array}%
\right. $ & $\left.
\begin{array}{c}
\hat{c}_{i}^{\dagger }R_{\alpha }^{i}\hat{b}^{\alpha } \\
\hat{c}_{i}^{\dagger }\hat{b}_{\alpha }^{\dagger }T^{\alpha i}%
\end{array}%
\right. $ & $\left.
\begin{array}{c}
R_{\alpha }^{i} \\
T^{\alpha i}%
\end{array}%
\right. $ & {\small 2 complex} \\ \hline
{\small III} & $\left.
\begin{array}{c}
Q_{1}^{+} \\
Q_{2}^{+} \\
Q_{3}^{-} \\
Q_{4}^{-}%
\end{array}%
\right. $ & $\left.
\begin{array}{c}
\hat{c}_{i}^{\dagger }R_{\alpha }^{i}\hat{b}^{\alpha } \\
\hat{c}_{i}^{\dagger }\hat{b}_{\alpha }^{\dagger }T^{\alpha i} \\
S_{\alpha i}\hat{c}^{i}\hat{b}^{\alpha } \\
\hat{b}_{\alpha }^{\dagger }W_{i}^{\alpha }\hat{c}^{i}%
\end{array}%
\right. $ & $\left.
\begin{array}{c}
R_{\alpha }^{i} \\
T^{\alpha i} \\
S_{\alpha i} \\
W_{i}^{\alpha }%
\end{array}%
\right. $ & {\small 4 complex} \\ \hline
\end{tabular}
\label{t}
\end{equation}%
\begin{equation*}
\end{equation*}%
From these supercharges one can construct several anticommutators
(observables). For example, from the $Q^{+}$ in the first row of the above
table (family I) and its adjoint ($Q^{+}$)$^{\dagger }=Q^{-}$, one can build
$\{Q^{+},Q^{-}\}=H$ and $\{Q^{+},Q^{+}\}=Z^{++}$\ as well as $%
\{Q^{-},Q^{-}\}=Z^{--}.$ From the $Q_{1}^{+},Q_{2}^{+}$ of the family II and
their adjoints $Q_{1}^{-},Q_{2}^{-},$ we can build 10 anticommutators.
Special supersymmetric Hamiltonians $H_{susy}$ using fermionic charges as in
the two first rows of this table (families I and II) were considered in
\textrm{\cite{1D2}}. General ones will be given here.

\subsubsection{ Fermionic charge operators: examples}

Following \textrm{\cite{1D2}}, the fermionic charge operator $Q$ can be
realised in terms of tensor products of N+M supersymmetric quantum
oscillators with the following features; see also\textrm{\ subsection 4.2 of
our present study.}

\begin{itemize}
\item N free fermionic quantum oscillators $\hat{c}^{i}$ and their adjoint $%
\hat{c}_{i}^{\dagger }$ labeled by $i=1,...,N$ and obeying the usual
anticommutation relations. These $\hat{c}^{i}$ and $\hat{c}_{i}^{\dagger }$
transform in the fundamental representations of $U\left( N\right) $. This
symmetry group $U\left( N\right) $ is the maximal unitary part of the SO(2N)
orthogonal group rotating the underlying 2N Majorana fermions $\hat{\gamma}%
_{2l-1}$ and $\hat{\gamma}_{2l}$ making $\hat{c}^{l}=(\hat{\gamma}_{2l-1}+i%
\hat{\gamma}_{2l-1})/2$ and $\hat{c}_{l}^{\dagger }=(\hat{\gamma}_{2l-1}+i%
\hat{\gamma}_{2l-1})/2$. The $N^{2}$ generators of $U\left( N\right) $ are
given by%
\begin{equation}
\mathcal{O}_{i}^{j}=\frac{1}{4}\left( \hat{c}_{i}^{\dagger }\hat{c}^{j}-\hat{%
c}^{j}\hat{c}_{i}^{\dagger }\right)
\end{equation}%
including the commuting Cartan charge operators as $J_{i}=\mathcal{O}%
_{i}^{i} $ namely%
\begin{equation}
J_{i}=\frac{1}{4}\left( \hat{c}_{i}^{\dagger }\hat{c}^{i}-\hat{c}^{i}\hat{c}%
_{i}^{\dagger }\right) ,\qquad i=1,...,N
\end{equation}%
From these abelian $J_{i}$'s; the previous $J_{0}$ corresponds to $%
\sum_{i}J_{i}$. We also have $\mathcal{O}^{\left[ ij\right] }=\hat{c}^{i}%
\hat{c}^{j}/2$ and $\mathcal{O}_{\left[ ij\right] }=\hat{c}_{i}^{\dagger }%
\hat{c}_{j}^{\dagger }/2$ transforming under the antisymmetric
representations of $U(N).$ In the tight binding modeling, the above
fermionic oscillator operators are fibered over the Brillouin zone as $\hat{c%
}_{\mathbf{k}}^{i}$ and $\hat{c}_{\mathbf{k}i}^{\dagger }$ with momentum $%
\mathbf{k}$.

\item M free bosonic operators $\hat{b}^{\alpha }$ and their adjoint $\hat{b}%
_{\alpha }^{\dagger }$ with label $i=1,...,M$. The $\hat{b}^{\alpha }$ and $%
\hat{b}_{\alpha }^{\dagger }$ transform in the fundamental representations
of $U\left( M\right) $. This unitary $U\left( M\right) $ is the maximal
unitary subgroup within the usual $Sp(2M)$ phase space symplectic symmetry.
The $M^{2}$ generators of $U\left( M\right) $ are given by%
\begin{equation}
\mathcal{S}_{\alpha }^{\beta }=\frac{1}{4}\left( \hat{b}_{\alpha }^{\dagger }%
\hat{b}^{\beta }+\hat{b}^{\beta }\hat{b}_{\alpha }^{\dagger }\right)
\end{equation}%
including the commuting Cartan charge operators%
\begin{equation}
S_{\alpha }=\frac{1}{4}\left( \hat{b}_{\alpha }^{\dagger }\hat{b}^{\alpha }+%
\hat{b}^{\beta }\hat{b}_{\alpha }^{\dagger }\right) ,\qquad \alpha =1,...,M
\end{equation}%
with $S_{0}=\sum_{\alpha }S_{\alpha }.$ We also have $\mathcal{S}^{\left(
\alpha \beta \right) }=\hat{b}^{\alpha }\hat{b}^{\beta }/2$ and $\mathcal{S}%
_{\left( \alpha \beta \right) }=\hat{b}_{\alpha }^{\dagger }\hat{b}_{\beta
}^{\dagger }/2$ transforming under the symmetric representations of $U(N).$%
\newline
Here also the tight binding model operators are fibered over the Brillouin
zone as $\hat{b}_{\mathbf{k}}^{\alpha }$ and $\hat{b}_{\mathbf{k}\alpha
}^{\dagger }$.
\end{itemize}

\paragraph{\textbf{I. Oscillator realisation of} $Q_{susy}$:\newline
}

In terms of the fermionic and bosonic quantum oscillators, a particular
realisation of the supersymmetric charge $Q$ reads as follows
\begin{equation}
Q=\dsum\limits_{i,\alpha }\hat{c}_{i}^{\dagger }R_{\alpha }^{i}\hat{b}%
^{\alpha },\qquad Q^{\dagger }=\dsum\limits_{j,\beta }\hat{b}_{\beta
}^{\dagger }\bar{R}_{j}^{\beta }\hat{c}^{j}  \label{R}
\end{equation}%
with coupling tensors given by the $N\times M$ rectangular matrix $R_{\alpha
}^{i}$ and its adjoint conjugate $\left( R^{\dagger }\right) _{j}^{\beta
}\equiv \bar{R}_{j}^{\beta }.$ This $R_{\alpha }^{i}$ carries $NM$ complex
degrees of freedom. A far this complex supercharge is concerned, notice the
following features.

\ \ \

\textbf{(1) fermionic generators of osp(2N\TEXTsymbol{\vert}2M)}\newline
The Lie superalgebras has 4NM fermionic generators ($F^{i\alpha },\bar{F}%
_{i\alpha },G_{i}^{\alpha },\bar{G}_{\alpha }^{i}$); they extend the ones
given by (\ref{fbc}) and are realised in terms of the fermionic $\hat{c}%
_{i}^{\dagger }/\hat{c}^{i}$ and bosonic $\hat{b}_{\alpha }^{\dagger }/\hat{b%
}^{\alpha }$ operators as follows%
\begin{equation}
\begin{tabular}{lll}
$\boldsymbol{F}_{i\alpha }=\hat{c}_{i}^{\dagger }\hat{b}_{\alpha }^{\dagger
} $ & $,\qquad $ & $\boldsymbol{\bar{F}}^{\alpha i}=\hat{b}^{\alpha }\hat{c}%
^{i}$ \\
$\boldsymbol{G}_{i}^{\alpha }=\hat{c}_{i}^{\dagger }\hat{b}^{\alpha }$ & $%
,\qquad $ & $\boldsymbol{\bar{G}}_{\alpha }^{i}=\hat{b}_{\alpha }^{\dagger }%
\hat{c}^{i}$%
\end{tabular}
\label{GF}
\end{equation}%
obeying amongst others the following anticommutations
\begin{equation}
\begin{tabular}{lll}
$\left\{ \boldsymbol{F}^{i\alpha },\boldsymbol{\bar{F}}_{j\beta }\right\} $
& $=$ & $2\delta _{j}^{i}\mathcal{S}_{\alpha }^{\beta }-2\delta _{\beta
}^{\alpha }\mathcal{O}_{i}^{j}$ \\
$\left\{ \boldsymbol{G}_{i}^{\alpha },\boldsymbol{\bar{G}}_{\beta
}^{j}\right\} $ & $=$ & $2\delta _{j}^{i}\mathcal{S}_{\alpha }^{\beta
}+2\delta _{\beta }^{\alpha }\mathcal{O}_{i}^{j}$%
\end{tabular}%
\qquad ,\qquad
\begin{tabular}{lll}
$\left\{ \boldsymbol{F}^{i\alpha },\boldsymbol{F}^{j\beta }\right\} $ & $=$
& $0$ \\
$\left\{ \boldsymbol{G}_{i}^{\alpha },\boldsymbol{G}_{j}^{\beta }\right\} $
& $=$ & $0$%
\end{tabular}
\label{GO}
\end{equation}%
and%
\begin{equation}
\begin{tabular}{lllllll}
$\left\{ \boldsymbol{\bar{F}}^{\alpha i},\boldsymbol{G}_{j}^{\beta }\right\}
$ & $=$ & $2\delta _{j}^{i}\mathcal{S}^{\left( \alpha \beta \right) }$ & $%
\qquad ,\qquad $ & $\{\boldsymbol{F}_{i\alpha },\boldsymbol{G}_{j}^{\beta
}\} $ & $=$ & $2\delta _{\beta }^{\alpha }\overline{\mathcal{O}}_{\left[ ij%
\right] }$ \\
$\left\{ \boldsymbol{\bar{F}}^{\alpha i},\boldsymbol{\bar{G}}_{\beta
}^{j}\right\} $ & $=$ & $2\delta _{\beta }^{\alpha }\mathcal{O}^{\left[ ij%
\right] }$ & $\qquad ,\qquad $ & $\{\boldsymbol{F}_{i\alpha },\boldsymbol{%
\bar{G}}_{\beta }^{j}\}$ & $=$ & $2\delta _{i}^{j}\mathcal{\bar{S}}_{\left(
\alpha \beta \right) }$%
\end{tabular}
\label{FG}
\end{equation}%
In terms of these fermionic operators, we learn that the supercharge (\ref{R}%
) and its adjoint conjugate $Q^{\dagger }$ are given by the linear
combinations
\begin{equation}
Q=\dsum\limits_{i,\alpha }R_{\alpha }^{i}\boldsymbol{G}_{i}^{\alpha }\qquad
,\qquad Q^{\dagger }=\dsum\limits_{i,\alpha }\bar{R}_{i}^{\alpha }%
\boldsymbol{\bar{G}}_{\alpha }^{i}  \label{QR}
\end{equation}%
These are complex fermionic charges carrying charges $q=\pm $ of the
orthogonal $J_{0}$ and charges $p=\pm $ under the symplectic $S_{0}$; they
correspond to the $Q$ and its adjoint $\bar{Q}$ in eqs(\ref{ap1}-\ref{ap2}).
Notice that $Q$ is valued in the bifundamental $\left( \bar{N},M\right) $ of
the group $U\left( N\right) \times U\left( M\right) ;$ this feature will
have important consequences on supersymmetry.

\ \ \

\textbf{(2) the superalgebra of }$Q$\emph{\ }\textbf{and}\emph{\ }$%
Q^{\dagger }$ \textbf{of (\ref{QR})}\newline
Here we show that eqs(\ref{QR}) generate a supersymmetric algebra with two
fermionic charges $Q$ and $\bar{Q}$. To that purpose, we first check that $%
Q^{2}=0$; then we calculate the anticommutator $QQ^{\dagger }+Q^{\dagger
}Q=H_{susy}$ and after that we verify that we have $\left[ H_{susy},Q\right]
=0.$\emph{\ }

\emph{a) Calculating }$Q^{2}$\newline
By substituting (\ref{QR}) into $Q^{2};$ we find that it reads as $\mathcal{O%
}_{\left[ ij\right] }T_{\alpha \beta }^{ij}\mathcal{S}^{\left( \alpha \beta
\right) }$ with complex coupling $T_{\alpha \beta }^{ij}$ quadratic into $%
R_{\alpha }^{i}$\ namely
\begin{equation}
T_{\alpha \beta }^{ij}=4R_{\alpha }^{i}R_{\beta }^{j}
\end{equation}%
and $\mathcal{O}_{\left[ ij\right] }=\hat{c}_{i}^{\dagger }\hat{c}%
_{j}^{\dagger }/2$ as well as $\mathcal{S}^{\left( \alpha \beta \right) }=%
\hat{b}^{\alpha }\hat{b}^{\beta }/2$. Clearly the contraction of this tensor
$T_{\alpha \beta }^{ij}$ vanishes identically due to $T_{\alpha \beta
}^{ij}=T_{\beta \alpha }^{ji}$ and because of the properties $\mathcal{O}_{%
\left[ ij\right] }=-\mathcal{O}_{\left[ ji\right] }$ and $\mathcal{S}%
^{\left( \alpha \beta \right) }=\mathcal{S}^{\left( \beta \alpha \right) }.$
This nilpotency feature can be explicitly exhibited. First, by using the
symmetry of $\mathcal{S}^{\left( \alpha \beta \right) },$ we have%
\begin{equation}
Q^{2}=2\mathcal{O}_{ij}\left( R_{\alpha }^{i}R_{\beta }^{j}+R_{\alpha
}^{j}R_{\beta }^{i}\right) \mathcal{S}^{\left( \alpha \beta \right) }
\end{equation}%
then using the antisymmetry $\mathcal{O}_{\left[ ij\right] },$ we end up with%
\begin{equation}
Q^{2}=\mathcal{O}_{ij}\left( R_{\alpha }^{i}R_{\beta }^{j}-R_{\alpha
}^{j}R_{\beta }^{i}+R_{\alpha }^{j}R_{\beta }^{i}-R_{\alpha }^{i}R_{\beta
}^{j}\right) \mathcal{S}^{\left( \alpha \beta \right) }
\end{equation}%
indicating that $Q^{2}$ vanishes identically with no constraint on $%
R_{\alpha }^{i}$.

\emph{b) Computing }$H_{susy}$\newline
For the calculation of the anticommutator $QQ^{\dagger }+Q^{\dagger
}Q=H_{susy},$ we substitute $Q=\hat{c}_{i}^{\dagger }R_{\alpha }^{i}\hat{b}%
^{\alpha }$ and $Q^{\dagger }=\hat{b}_{\beta }^{\dagger }\bar{R}_{j}^{\beta }%
\hat{c}^{j}$, we obtain%
\begin{equation}
H_{susy}=\hat{c}_{i}^{\dagger }[h_{f}]_{j}^{i}\hat{c}^{j}+\hat{b}_{\alpha
}^{\dagger }[h_{b}]_{\beta }^{\alpha }\hat{b}^{\beta }  \label{hs}
\end{equation}%
with hermitian coupling tensors $[h_{f}]_{j}^{i}$ and $[h_{b}]_{\beta
}^{\alpha }$ respectively given by the $N\times N$ square matrix $%
[RR^{\dagger }]_{j}^{j}$ and the $M\times M$ square matrix $[R^{\dagger
}R]_{\beta }^{\alpha }.$ So, we have
\begin{equation}
h_{f}=RR^{\dagger },\qquad h_{b}=R^{\dagger }R
\end{equation}%
with $H_{b}^{susy}=\hat{b}_{\alpha }^{\dagger }[h_{b}]_{\beta }^{\alpha }%
\hat{b}^{\beta }$ and $H_{f}^{susy}=\hat{c}_{i}^{\dagger }[h_{f}]_{j}^{i}%
\hat{c}^{j}$ reading also as%
\begin{equation}
\begin{tabular}{lll}
$H_{b}^{susy}$ & $=$ & $2[h_{b}]_{\beta }^{\alpha }\mathcal{S}_{\alpha
}^{\beta }-\frac{1}{2}tr\left( h_{b}\right) $ \\
$H_{f}^{susy}$ & $=$ & $2[h_{f}]_{j}^{i}\mathcal{O}_{j}^{i}+\frac{1}{2}%
tr\left( h_{f}\right) $%
\end{tabular}
\label{sh}
\end{equation}%
indicating that $H^{susy}=2[h_{b}]_{\beta }^{\alpha }\mathcal{S}_{\alpha
}^{\beta }+2[h_{f}]_{j}^{i}\mathcal{O}_{j}^{i}$ is valued into $u\left(
N\right) \oplus u\left( M\right) $. Notice that we have%
\begin{equation}
h_{f}R-Rh_{b}=0  \label{HR}
\end{equation}

\emph{c) Checking the commutativity }$H_{susy},Q=QH_{susy},$\newline
Regarding the calculation of $\left[ H_{susy},Q\right] ,$ we compute the
commutators $[H_{b}^{susy},Q]$ and $[H_{f}^{susy},Q];$ we find%
\begin{equation}
\begin{tabular}{lll}
$\left[ H_{b}^{susy},\hat{c}_{i}^{\dagger }R_{\gamma }^{i}\hat{b}^{\gamma }%
\right] $ & $=$ & $-\hat{c}_{i}^{\dagger }[RR^{\dagger }R]_{\beta }^{i}\hat{b%
}^{\beta }$ \\
$\left[ H_{f}^{susy},\hat{c}_{i}^{\dagger }R_{\gamma }^{i}\hat{b}^{\gamma }%
\right] $ & $=$ & $+\hat{c}_{i}^{\dagger }[RR^{\dagger }R]_{\beta }^{i}\hat{b%
}^{\beta }$%
\end{tabular}%
\end{equation}%
whose sum vanishes identically. So, eqs(\ref{R}\textbf{-}\ref{QR}) realise a
SUSY charge.

\paragraph{\textbf{II. Beyond the realisation (\protect\ref{R}-\protect\ref%
{QR}):}\newline
}

Clearly, eq(\ref{QR}) is a special realisation of the fermionic charge $Q$;
a general oscillator realisation can be written down; it involves the
coupling tensors $T^{i\alpha }$ and $\bar{T}_{i\alpha }$ in addition $%
R_{\alpha }^{i}$ and $\bar{R}_{i}^{\alpha }.$ We refer to these
orthosymplectic supercharges like $Q_{ortic}\equiv Q^{+}$ and $\bar{Q}%
_{ortic}\equiv Q^{-}$\ with realisations given by
\begin{equation}
\begin{tabular}{lll}
$Q^{+}$ & $=$ & $\hat{c}_{i}^{\dagger }R_{\alpha }^{i}\hat{b}^{\alpha }+\hat{%
c}_{i}^{\dagger }\hat{b}_{\alpha }^{\dagger }T^{\alpha i}$ \\
$Q^{-}$ & $=$ & $\hat{b}_{\alpha }^{\dagger }\bar{R}_{i}^{\alpha }\hat{c}%
^{i}+\bar{T}_{i\alpha }\hat{b}^{\alpha }\hat{c}^{i}$%
\end{tabular}
\label{rs}
\end{equation}%
These supercharges goes beyond (\ref{R}-\ref{QR}) corresponding to $%
T^{\alpha i}=0$; the above $Q^{\pm }$ have some specific properties as
listed below:

\ \ \

\textbf{(1) Chiral odd spaces}\newline
The supercharges $Q^{\pm }$ belong to two adjoint conjugate subspaces in the
odd sector of the osp(2N\TEXTsymbol{\vert}2M). Indeed, using the fermionic
operators generating the odd sector of osp(2N\TEXTsymbol{\vert}2M) namely $%
\boldsymbol{G}_{i}^{\alpha },$ $\boldsymbol{F}_{i\alpha },$ $\boldsymbol{%
\bar{G}}_{\alpha }^{i},$ $\boldsymbol{\bar{F}}^{\alpha i}$ given eq(\ref{GF}%
), then, we have
\begin{equation}
\begin{tabular}{lll}
$Q^{+}$ & $=$ & $R_{\alpha }^{i}\boldsymbol{G}_{i}^{\alpha }+T^{\alpha i}%
\boldsymbol{F}_{i\alpha }$ \\
$Q^{-}$ & $=$ & $\bar{R}_{i}^{\alpha }\boldsymbol{\bar{G}}_{\alpha }^{i}+%
\bar{T}_{i\alpha }\boldsymbol{\bar{F}}^{\alpha i}$%
\end{tabular}
\label{pm}
\end{equation}%
with $Q^{+}$ sitting in the complex directions $\boldsymbol{G}_{i}^{\alpha },%
\boldsymbol{F}_{i\alpha };$ and the $Q^{-}=(Q^{+})^{\dagger }$ in their
images $\boldsymbol{\bar{G}}_{\alpha }^{i},\boldsymbol{\bar{F}}^{\alpha i}$
under conjugation. From these $Q^{\pm }$, one can also construct a hermitian
supercharge $\mathfrak{Q}$ and an antihermitian $\mathfrak{P}$ as usual like%
\begin{equation}
\mathfrak{Q}=\frac{Q^{+}+Q^{-}}{\sqrt{2}}\qquad ,\qquad i\mathfrak{P}=\frac{%
Q^{+}-Q^{-}}{\sqrt{2}}
\end{equation}%
Notice that the two following features of eq(\ref{pm}). First, it carries $%
2NM$ complex degrees of freedom; $NM$ coming from $R_{\alpha }^{i}$\ and $NM$
from $T^{i\alpha }$. Second, the $Q^{\pm }$ are valued in reducible
representations of $U\left( N\right) \times U\left( M\right) $. This is
because, the $\boldsymbol{G}_{i}^{\alpha }$ sits in the bifundamental $%
\left( \bar{N},M\right) $ while the $\boldsymbol{F}_{i\alpha }$ sits in the
antisymmetric $\bar{N}\wedge \bar{M}.$ This feature deforms the expressions
of the supersymmetric constraints satisfied by $\{Q^{\pm },Q^{\pm }\}=0$ and
$[H,Q^{\pm }]=0.$

\ \

\textbf{(2) Anticommutators }$\left\{ Q^{\pm },Q^{\pm }\right\} $\newline
From the orthosymplectic supercharges (\ref{pm}), we can define three
anticommutators: $\left( i\right) $ The anticommutator \{$Q^{+},Q^{-}$\}
defining the orthosymplectic Hamiltonian $H_{ortic}$ which is hermitian. $%
\left( ii\right) $ The anticommutator \{$Q^{+},Q^{+}$\}, which in general is
non vanishing, defines a complex operator $Z_{ortic}$ scaling as $H_{ortic}$%
. The third anticommutator is given by $Z_{ortic}^{\dagger
}=\{Q^{-},Q^{-}\}; $ it is the adjoint conjugate of $Z_{ortic}$. \newline
These three bosonic operators $H_{ortic},$ $Z_{ortic}$ and $%
Z_{ortic}^{\dagger }$ are valued in the bosonic subalgebra $so(2N)\oplus
sp(2M)$ generated by $\mathcal{O}_{i}^{j},$ $\mathcal{O}^{\left[ ij\right]
}, $ $\overline{\mathcal{O}}_{\left[ ij\right] }$ (for $so_{2N}$) and $%
\mathcal{S}_{\beta }^{\alpha },$ $\mathcal{S}^{\left( \alpha \beta \right)
}, $ $\mathcal{\bar{S}}_{\left( \alpha \beta \right) }$ (for $sp_{2M}$). To
exhibit this feature, we calculate these anticommutators.

\ \ \

\textbf{a) the Hamiltonian}\emph{\ }$H_{ortic}$\newline
Substituting (\ref{pm}) into $\{Q^{+},Q^{-}\}$, we get%
\begin{equation}
\begin{tabular}{lll}
$H_{ortic}$ & $=$ & $R_{\alpha }^{i}\bar{R}_{j}^{\beta }\{\boldsymbol{G}%
_{i}^{\alpha },\boldsymbol{\bar{G}}_{\beta }^{j}\}+T^{\alpha i}\bar{R}%
_{j}^{\beta }\{\boldsymbol{F}_{i\alpha },\boldsymbol{\bar{G}}_{\beta }^{j}\}$
\\
&  & $R_{\alpha }^{i}\bar{T}_{j\beta }\{\boldsymbol{G}_{i}^{\alpha },%
\boldsymbol{\bar{F}}^{\beta j}\}+T^{\alpha i}\bar{T}_{j\beta }\{\boldsymbol{F%
}_{i\alpha },\boldsymbol{\bar{F}}^{\beta j}\}$%
\end{tabular}%
\end{equation}%
By using (\ref{GO}-\ref{FG}), this $H_{ortic}$ reads as a linear combination
of the above mentioned generators namely%
\begin{equation}
\begin{tabular}{lll}
$H_{ortic}$ & $=$ & $2\left( R_{\alpha }^{i}\bar{R}_{i}^{\beta }+T^{\alpha i}%
\bar{T}_{i\beta }\right) \mathcal{S}_{\beta }^{\alpha }+$ \\
&  & $2\left( R_{\alpha }^{i}\bar{R}_{j}^{\alpha }-\bar{T}_{j\alpha
}T^{\alpha i}\right) \mathcal{O}_{i}^{j}+$ \\
&  & $2\left( R_{\alpha }^{i}\bar{T}_{i\beta }\right) \mathcal{S}^{\left(
\alpha \beta \right) }+2(T^{\alpha i}\bar{R}_{i}^{\beta })\mathcal{\bar{S}}%
_{\left( \alpha \beta \right) }$%
\end{tabular}
\label{orti}
\end{equation}%
As it has no $\mathcal{O}^{\left[ ij\right] },$ $\overline{\mathcal{O}}_{%
\left[ ij\right] },$ itt is valued in $u_{N}\oplus sp_{2M}.$ Interesting
families of such Hamiltonians are given by the two following constraints on
the coupling tensors: $\left( \mathbf{i}\right) $ The $R_{\alpha }^{i}$ and $%
T^{\alpha i}$ couplings are constrained like%
\begin{equation}
\begin{tabular}{lll}
$R_{\alpha }^{i}\bar{T}_{i\beta }$ & $=$ & $\eta _{RT}\Omega _{\alpha \beta
} $ \\
$T^{\beta i}\bar{R}_{i}^{\alpha }$ & $=$ & $\bar{\eta}_{RT}\Omega ^{\beta
\alpha }$%
\end{tabular}%
\qquad ,\qquad
\begin{tabular}{lll}
$\Omega _{\alpha \beta }$ & $=$ & $-\Omega _{\beta \alpha }$ \\
$\Omega ^{\beta \alpha }$ & $=$ & $-\Omega ^{\alpha \beta }$%
\end{tabular}
\label{RT}
\end{equation}%
where $\eta _{RT}$ scaling as energy. For simple calculations, we will set $%
\eta _{RT}=0$. For this choice, the $\mathcal{S}^{\left( \alpha \beta
\right) }$ and $\mathcal{\bar{S}}_{\left( \alpha \beta \right) }$ terms
disappear and the Hamiltonian (\ref{orti}) reduces to a form similar to (\ref%
{sh}); it reads as follows%
\begin{equation}
\begin{tabular}{lll}
$H_{ortic}$ & $=$ & $2\left( \bar{R}_{i}^{\beta }R_{\alpha }^{i}+T^{\beta i}%
\bar{T}_{i\alpha }\right) \mathcal{S}_{\beta }^{\alpha }+$ \\
&  & $2\left( R_{\alpha }^{i}\bar{R}_{j}^{\alpha }-\bar{T}_{j\alpha
}T^{\alpha i}\right) \mathcal{O}_{i}^{j}$%
\end{tabular}%
\end{equation}%
it is valued into $u\left( N\right) \oplus u\left( M\right) .$ In this case,
we have the following bosonic and fermionic contributions%
\begin{equation}
\begin{tabular}{lll}
$\left( h_{b}\right) _{\gamma }^{\beta }$ & $=$ & $\bar{R}_{l}^{\beta
}R_{\gamma }^{l}+T^{\beta l}\bar{T}_{l\gamma }$ \\
$\left( h_{f}\right) _{j}^{i}$ & $=$ & $R_{\alpha }^{i}\bar{R}_{j}^{\alpha }-%
\bar{T}_{j\alpha }T^{\alpha i}$%
\end{tabular}
\label{BFC}
\end{equation}
$\left( \mathbf{ii}\right) $ If in addition to (\ref{RT}) with $\eta _{RT}=0$%
, the $R_{\alpha }^{i}$ and $T^{\alpha i}$ couplings are restricted to the
following diagonal choices
\begin{equation}
\begin{tabular}{lllllll}
$R_{\alpha }^{i}\bar{R}_{i}^{\beta }$ & $=$ & $\lambda _{R}^{\alpha }\delta
_{\alpha }^{\beta }$ & $\qquad ,\qquad $ & $R_{\alpha }^{i}\bar{R}%
_{j}^{\alpha }$ & $=$ & $\mu _{R}^{i}\delta _{j}^{i}$ \\
$T^{\alpha i}\bar{T}_{i\beta }$ & $=$ & $\lambda _{T}^{\alpha }\delta
_{\alpha }^{\beta }$ & $\qquad ,\qquad $ & $T^{\alpha i}\bar{T}_{j\alpha }$
& $=$ & $\mu _{T}^{i}\delta _{j}^{i}$%
\end{tabular}%
\end{equation}%
where $\lambda _{R}^{\alpha },$ $\lambda _{T}^{\alpha }$\ and $\mu _{R}^{i},$
$\mu _{T}^{i}$ are $M+N$ real numbers, the Hamiltonian takes an interesting
form. For this choice, the resulting $H_{ortic}$ is valued in the Cartan
subalgebra of $so(2N)\oplus sp(2M)$ as given here after%
\begin{equation}
H_{ortic}=2\sum_{\alpha =1}^{M}\left( \lambda _{R}^{\alpha }+\lambda
_{T}^{\alpha }\right) \mathcal{S}_{\alpha }+2\sum_{i=1}^{N}\left( \mu
_{R}^{i}-\mu _{T}^{i}\right) \mathcal{J}_{i}
\end{equation}

\textbf{b) the anticommutator} \{$Q^{+},Q^{+}\}$\newline
This anticommutator defines the observable $Z_{ortic}$; we find after
substituting $Q^{+}=R_{\alpha }^{i}\boldsymbol{G}_{i}^{\alpha }+T^{\alpha i}%
\boldsymbol{F}_{i\alpha }$ and using$\{\boldsymbol{G}_{i}^{\alpha },%
\boldsymbol{G}_{j}^{\beta }\}=\{\boldsymbol{F}_{i\alpha },\boldsymbol{F}%
_{j\beta }\}=0$, the following expression%
\begin{equation}
Z_{ortic}=R_{\alpha }^{i}T^{\beta j}\left\{ \boldsymbol{G}_{i}^{\alpha },%
\boldsymbol{F}_{j\beta }\right\} +T^{\beta j}R_{\alpha }^{i}\left\{
\boldsymbol{F}_{j\beta },\boldsymbol{G}_{i}^{\alpha }\right\}
\end{equation}%
Using (\ref{GO}-\ref{FG}), we can put this relation as follows%
\begin{equation}
Z_{ortic}=-4\left( R_{\alpha }^{i}T^{\alpha j}\right) \overline{\mathcal{O}}%
_{\left[ ij\right] }
\end{equation}%
showing that $Z_{ortic}$ is valued in the antisymmetric representation of
the $u\left( N\right) $ subalgebra of so$\left( 2N\right) $. It vanishes for
the case
\begin{equation}
R_{\alpha }^{i}T^{\alpha j}=\zeta _{RT}G^{ij}\qquad ,\qquad G^{ij}=G^{ji}
\end{equation}

\textbf{c) the commutator}\emph{\ }[$H,Q^{+}]$\newline
Substituting the bosonic $H_{ortic}=\left( h_{b}\right) _{\alpha }^{\beta }%
\mathcal{S}_{\beta }^{\alpha }+\left( h_{f}\right) _{j}^{i}\mathcal{O}%
_{i}^{j}$ and the fermionic $Q^{+}=R_{\alpha }^{i}\boldsymbol{G}_{i}^{\alpha
}+T^{\alpha i}\boldsymbol{F}_{i\alpha }$, then using%
\begin{equation}
\begin{tabular}{lll}
$\left[ \mathcal{S}_{\beta }^{\alpha },\boldsymbol{G}_{l}^{\gamma }\right] $
& $=$ & $-\frac{1}{2}\delta _{\beta }^{\gamma }\boldsymbol{G}_{l}^{\alpha }$
\\
$\left[ \mathcal{S}_{\beta }^{\alpha },\boldsymbol{F}_{l\gamma }\right] $ & $%
=$ & $+\frac{1}{2}\delta _{\gamma }^{\alpha }\boldsymbol{F}_{l\beta }$%
\end{tabular}%
\qquad ,\qquad
\begin{tabular}{lll}
$\left[ \mathcal{O}_{i}^{j},\boldsymbol{G}_{l}^{\gamma }\right] $ & $=$ & $%
\frac{1}{2}\delta _{l}^{j}\boldsymbol{G}_{i}^{\gamma }$ \\
$\left[ \mathcal{O}_{i}^{j},\boldsymbol{F}_{l\gamma }\right] $ & $=$ & $%
\frac{1}{2}\delta _{l}^{j}\boldsymbol{F}_{i\gamma }$%
\end{tabular}%
\end{equation}%
we find after some algebra, the following%
\begin{equation}
\begin{tabular}{lll}
$\lbrack H,Q^{+}]$ & $=$ & $\frac{1}{2}\{\left( h_{f}\right)
_{j}^{i}R_{\gamma }^{j}-\left( h_{b}\right) _{\gamma }^{\beta }R_{\beta
}^{i}\}\boldsymbol{G}_{i}^{\gamma }+$ \\
&  & $\frac{1}{2}\{T^{\gamma j}\left( h_{f}\right) _{j}^{i}+\left(
h_{b}\right) _{\beta }^{\gamma }T^{\beta i}\}\boldsymbol{F}_{i\gamma }$%
\end{tabular}%
\end{equation}%
The vanishing condition of this commutator requires%
\begin{equation}
\begin{tabular}{lll}
$\left( h_{f}\right) _{j}^{i}R_{\gamma }^{j}-\left( h_{b}\right) _{\gamma
}^{\beta }R_{\beta }^{i}$ & $=$ & $0$ \\
$T^{\gamma j}\left( h_{f}\right) _{j}^{i}+\left( h_{b}\right) _{\beta
}^{\gamma }T^{\beta i}$ & $=$ & $0$%
\end{tabular}
\label{hfb}
\end{equation}%
By replacing $\left( h_{f}\right) _{j}^{i}$ and $\left( h_{b}\right)
_{\gamma }^{\beta }$ by eq(\ref{BFC}), we obtain%
\begin{equation}
\begin{tabular}{lll}
$\left( h_{b}\right) _{\gamma }^{\beta }R_{\beta }^{i}$ & $=$ & $R_{\beta
}^{i}\bar{R}_{l}^{\beta }R_{\gamma }^{l}+R_{\beta }^{i}T^{\beta l}\bar{T}%
_{l\gamma }$ \\
$\left( h_{f}\right) _{j}^{i}R_{\gamma }^{j}$ & $=$ & $R_{\beta }^{i}\bar{R}%
_{l}^{\beta }R_{\gamma }^{l}-R_{\gamma }^{j}\bar{T}_{j\alpha }T^{\alpha i}$%
\end{tabular}%
,\qquad
\begin{tabular}{lll}
$\left( h_{b}\right) _{\gamma }^{\beta }T^{\gamma i}$ & $=$ & $\bar{R}%
_{l}^{\beta }R_{\gamma }^{l}T^{\gamma i}+T^{\beta l}\bar{T}_{l\gamma
}T^{\gamma i}$ \\
$\left( h_{f}\right) _{j}^{i}T^{\beta j}$ & $=$ & $T^{\beta j}\bar{R}%
_{j}^{\gamma }R_{\gamma }^{i}-T^{\beta l}\bar{T}_{l\gamma }T^{\gamma i}$%
\end{tabular}%
\end{equation}%
which by substituting into (\ref{hfb}), we get the following constraint
relations%
\begin{equation}
\begin{tabular}{lll}
$R_{\gamma }^{j}\bar{T}_{j\alpha }T^{\alpha i}+R_{\alpha }^{i}T^{\alpha j}%
\bar{T}_{j\gamma }$ & $=$ & $0$ \\
$\bar{R}_{j}^{\beta }R_{\gamma }^{j}T^{\gamma i}+T^{\beta j}\bar{R}%
_{j}^{\gamma }R_{\gamma }^{i}$ & $=$ & $0$%
\end{tabular}%
\end{equation}%
These constraints are naturally solved by the orthogonalites $R_{\alpha
}^{i}T^{\alpha j}=R_{\alpha }^{i}\bar{T}_{i\beta }=0.$ If substituting $%
R_{\alpha }^{i}T^{\alpha j}=\zeta _{RT}G^{ij}$ and $R_{\alpha }^{i}\bar{T}%
_{i\beta }=\eta _{RT}\Omega _{\alpha \beta }$, these constraints read as%
\begin{equation}
\begin{tabular}{lll}
$\eta _{RT}\Omega _{\gamma \alpha }T^{\alpha i}+\zeta _{RT}G^{ij}\bar{T}%
_{j\gamma }$ & $=$ & $0$ \\
$\zeta _{RT}G^{ji}\bar{R}_{j}^{\beta }+\bar{\eta}_{RT}\Omega ^{\beta \gamma
}R_{\gamma }^{i}$ & $=$ & $0$%
\end{tabular}%
\end{equation}

\paragraph{III. Fermionic charges beyond (\protect\ref{pm})\newline
}

In eq(\ref{pm}) the two fermionic charge $Q^{+}$\ and $Q^{-}$ are related by
adjoint conjugation; the combination $\sqrt{2}\mathfrak{Q}=Q^{+}+Q^{-}$ is
hermitian and reads like%
\begin{equation}
\sqrt{2}\mathfrak{Q}=R_{\alpha }^{i}\boldsymbol{G}_{i}^{\alpha }+T^{\alpha i}%
\boldsymbol{F}_{i\alpha }+\bar{R}_{i}^{\alpha }\boldsymbol{\bar{G}}_{\alpha
}^{i}+\bar{T}_{i\alpha }\boldsymbol{\bar{F}}^{\alpha i}
\end{equation}%
having two complex coupling tensors $R_{\alpha }^{i}$ and $T^{\alpha i}.$
However, this is still a particular case because one may relax this
hermiticity constraint by thinking about the above fermionic charge $\sqrt{2}%
\mathfrak{Q}$ as follows%
\begin{equation}
\mathfrak{Q}^{+}=R_{\alpha }^{i}\boldsymbol{G}_{i}^{\alpha }+T^{\alpha i}%
\boldsymbol{F}_{i\alpha }+W_{i}^{\alpha }\boldsymbol{\bar{G}}_{\alpha
}^{i}+S_{i\alpha }\boldsymbol{\bar{F}}^{\alpha i}
\end{equation}%
where now we have four complex coupling tensors namely the old complex $%
R_{\alpha }^{i}$ and $T^{\alpha i}$ and the two new $W_{i}^{\alpha }$ and $%
S_{i\alpha }$. This $\mathfrak{Q}^{+}$ involves $4NM$ complex degrees of
freedom; it is the general form of the fermionic generator one can build out
of the fermionic $\hat{c}_{i}^{\dagger }/\hat{c}^{i}$ and bosonic $\hat{b}%
_{\alpha }^{\dagger }/\hat{b}^{\alpha }$ operators. The $\mathfrak{Q}^{+}$
is valued in the complexified osp(2N\TEXTsymbol{\vert}2M) and can be viewed
as the given by sum $\mathfrak{Q}^{+}=Q_{1}^{+}+Q_{2}^{+}$ with
\begin{equation}
\begin{tabular}{lll}
$Q_{1}^{+}$ & $=$ & $R_{\alpha }^{i}\boldsymbol{G}_{i}^{\alpha }+T^{\alpha i}%
\boldsymbol{F}_{i\alpha }$ \\
$Q_{2}^{+}$ & $=$ & $W_{i}^{\alpha }\boldsymbol{\bar{G}}_{\alpha
}^{i}+S_{i\alpha }\boldsymbol{\bar{F}}^{\alpha i}$%
\end{tabular}%
\end{equation}%
with $Q_{1}^{+}$ involving the $R_{\alpha }^{i},$ $T^{\alpha i}$ and the $%
Q_{2}^{+}$ using the new coupling tensors\ $S_{i\alpha }$\ and\ $%
W_{i}^{\alpha }.$ The adjoint conjugate of the $\mathfrak{Q}^{+}$ is given
by $\mathfrak{Q}^{-}=Q_{1}^{-}+Q_{2}^{-}$ with%
\begin{equation}
\begin{tabular}{lll}
$Q_{1}^{-}$ & $=$ & $\bar{R}_{i}^{\alpha }\boldsymbol{\bar{G}}_{\alpha }^{i}+%
\bar{T}_{i\alpha }\boldsymbol{\bar{F}}^{\alpha i}$ \\
$Q_{2}^{-}$ & $=$ & $\bar{W}_{\alpha }^{i}\boldsymbol{G}_{i}^{\alpha }+\bar{S%
}^{\alpha i}\boldsymbol{F}_{i\alpha }$%
\end{tabular}%
\end{equation}%
An interesting way to deal with these $\mathfrak{Q}^{\pm }$ is to use the
notations $\boldsymbol{\hat{\lambda}}_{\dot{A}}^{\dagger }=(\boldsymbol{\hat{%
c}}_{\dot{I}}^{\dagger },\boldsymbol{\hat{c}}^{\dot{I}})$ and $\hat{\xi}%
_{A}^{\dagger }=(\boldsymbol{\hat{b}}_{I}^{\dagger },\boldsymbol{\hat{b}}%
^{I})$\ as well as $\boldsymbol{\hat{\lambda}}^{\dot{A}}=(\boldsymbol{\hat{c}%
}^{\dot{I}},\boldsymbol{\hat{c}}_{\dot{I}}^{\dagger })^{T}$ and $\hat{\xi}%
^{A}=(\boldsymbol{\hat{b}}^{I},\boldsymbol{\hat{b}}_{I}^{\dagger })$ in
terms of which the $\mathfrak{Q}^{+}$ and $\mathfrak{Q}^{-}$ read in a
condensed way as follows
\begin{equation}
\mathfrak{Q}^{-}=\boldsymbol{\hat{\lambda}}_{\dot{A}}^{\dagger }\left[
\boldsymbol{q}\right] _{B}^{\dot{A}}\hat{\xi}^{B}\qquad ,\qquad \mathfrak{Q}%
^{+}=\hat{\xi}_{A}^{\dagger }\left[ \boldsymbol{q}^{\dagger }\right] _{\dot{B%
}}^{A}\boldsymbol{\hat{\lambda}}^{\dot{B}}  \label{qqpm}
\end{equation}%
Here, the $\left[ \boldsymbol{q}\right] _{B}^{\dot{A}}$ coupling tensor$\ $%
and adjoint $\left[ \boldsymbol{q}^{\dagger }\right] _{\dot{B}}^{A}$ are
respectively given by $2N\times 2M$ and $2M\times 2N$ rectangular matrices
reading in $N\times M$ and $M\times N$ sub-block matrices as follows
\begin{equation}
\mathfrak{Q}^{-}=\left( \boldsymbol{\hat{c}}^{\dagger },\boldsymbol{\hat{c}}%
\right) \left(
\begin{array}{cc}
R & T \\
S & W%
\end{array}%
\right) \left(
\begin{array}{c}
\boldsymbol{\hat{b}} \\
\boldsymbol{\hat{b}}^{\dagger }%
\end{array}%
\right) \quad ,\quad \mathfrak{Q}^{+}=\left( \boldsymbol{\hat{b}}^{\dagger },%
\boldsymbol{\hat{b}}\right) \left(
\begin{array}{cc}
R^{\dagger } & S^{\dagger } \\
T^{\dagger } & W^{\dagger }%
\end{array}%
\right) \left(
\begin{array}{c}
\boldsymbol{\hat{c}} \\
\boldsymbol{\hat{c}}^{\dagger }%
\end{array}%
\right)
\end{equation}%
From these fermionic charges, one can calculate three anticommutators: the
hermitian $\{\mathfrak{Q}^{-},\mathfrak{Q}^{+}\}$ and the complex $\{%
\mathfrak{Q}^{+},\mathfrak{Q}^{+}\}$\ and the adjoint conjugate $\{\mathfrak{%
Q}^{-},\mathfrak{Q}^{-}\}$. The calculations of these anticommutators go in
the same manner as we have done before. The novelty is that in this general
case the graded canonical commutation relations are some how unusual: $%
\left( \mathbf{i}\right) $ For the case of fermions, the anticommutation
relations have the form
\begin{equation}
\{\boldsymbol{\hat{\lambda}}_{\dot{A}}^{\dagger },\boldsymbol{\hat{\lambda}}%
^{\dot{B}}\}=\delta _{\dot{A}}^{\dot{B}},\quad \{\boldsymbol{\hat{\lambda}}_{%
\dot{A}}^{\dagger },\boldsymbol{\hat{\lambda}}_{\dot{B}}^{\dagger }\}=\dot{%
\Sigma}_{\dot{A}\dot{B}}^{x},\quad \{\boldsymbol{\hat{\lambda}}^{\dot{A}},%
\boldsymbol{\hat{\lambda}}^{\dot{B}}\}=\dot{\Sigma}_{x}^{\dot{A}\dot{B}}
\end{equation}%
where $\{\boldsymbol{\hat{\lambda}}_{\dot{A}}^{\dagger },\boldsymbol{\hat{%
\lambda}}_{\dot{B}}^{\dagger }\}$ and $\{\boldsymbol{\hat{\lambda}}^{\dot{A}%
},\boldsymbol{\hat{\lambda}}^{\dot{B}}\}$ are non vanishing. This is because%
\begin{equation}
\{\boldsymbol{\hat{\lambda}}_{\dot{A}}^{\dagger },\boldsymbol{\hat{\lambda}}%
_{\dot{B}}^{\dagger }\}=\left(
\begin{array}{cc}
\{\boldsymbol{\hat{c}}_{\dot{I}}^{\dagger },\boldsymbol{\hat{c}}^{\dot{J}}\}
& \{\boldsymbol{\hat{c}}_{\dot{I}}^{\dagger },\boldsymbol{\hat{c}}_{\dot{J}%
}^{\dagger }\} \\
\{\boldsymbol{\hat{c}}^{\dot{I}},\boldsymbol{\hat{c}}^{\dot{J}}\} & \{%
\boldsymbol{\hat{c}}^{\dot{I}},\boldsymbol{\hat{c}}_{\dot{J}}^{\dagger }\}%
\end{array}%
\right)
\end{equation}%
In these relations, the $\dot{\Sigma}_{\dot{A}\dot{B}}^{x}$ and $\dot{\Sigma}%
_{x}^{\dot{A}\dot{B}}$\ are given by
\begin{equation}
\delta _{\dot{A}}^{\dot{B}}=\left(
\begin{array}{cc}
\delta _{\dot{I}}^{\dot{J}} & 0 \\
0 & \delta _{\dot{J}}^{\dot{I}}%
\end{array}%
\right) ,\quad \dot{\Sigma}_{\dot{A}\dot{B}}^{x}=\left(
\begin{array}{cc}
0 & \delta _{\dot{I}}^{\dot{J}} \\
\delta _{\dot{J}}^{\dot{I}} & 0%
\end{array}%
\right) ,\quad \dot{\Sigma}_{x}^{\dot{A}\dot{B}}=\left(
\begin{array}{cc}
0 & \delta _{\dot{J}}^{\dot{I}} \\
\delta _{\dot{I}}^{\dot{J}} & 0%
\end{array}%
\right)
\end{equation}%
$\left( \mathbf{ii}\right) $ For the bosons, we also have%
\begin{equation}
\left[ \hat{\xi}^{A},\hat{\xi}_{B}^{\dagger }\right] =\Sigma _{B}^{zA},\quad %
\left[ \hat{\xi}^{A},\hat{\xi}^{B}\right] =i\Sigma _{y}^{AB},\quad \left[
\hat{\xi}_{A}^{\dagger },\hat{\xi}_{B}^{\dagger }\right] =-i\Sigma _{AB}^{y}
\end{equation}%
where here also the $[\hat{\xi}^{A},\hat{\xi}^{B}]$ and $[\hat{\xi}%
_{A}^{\dagger },\hat{\xi}_{B}^{\dagger }]$ are non vanishing. In these
relations, the $\Sigma _{B}^{zA}$ and $\Sigma _{y}^{AB}$ are given by%
\begin{equation}
\left( \Sigma ^{z}\right) _{B}^{A}=\left(
\begin{array}{cc}
\delta _{J}^{I} & 0 \\
0 & -\delta _{J}^{I}%
\end{array}%
\right) ,\quad i\Sigma _{y}^{AB}=\left(
\begin{array}{cc}
0 & \delta _{J}^{I} \\
-\delta _{I}^{J} & 0%
\end{array}%
\right) ,\quad -i\Sigma _{AB}^{y}=\left(
\begin{array}{cc}
0 & -\delta _{I}^{J} \\
\delta _{J}^{I} & 0%
\end{array}%
\right)
\end{equation}%
\begin{equation*}
\end{equation*}


\begin{thebibliography}{99}
\bibitem{1A} J.Wess and J. Bagger, Supersymmetry and Supergravity, Princeton
Series in Physics (1983).

\bibitem{1B} Michael B. Green, John H. Schwarz, Edward Witten,, Superstring
Theory, Cambridge University Press (2012).

\bibitem{1C} Gerard Queralto, Mark Kremer, Lukas J. Maczewsky, Matthias
Heinrich, Jordi Mompart, Veronica Ahufinger, Alexander Szameit,
Communications Physics volume 3, Article number: 49 (2020),
arXiv:1911.01160v1 [physics.optics].

\bibitem{1C0} P. V. Buividovich, Phys. Rev. D 106 (2022) 046001,
arXiv:2205.09704 [hep-th].

\bibitem{2C} Ken K. W. Ma, Ruojun Wang, Kun Yang, Phys. Rev. Lett. 126,
206801 (2021), arXiv:2101.05448v2 [cond-mat.mes-hall].

\bibitem{3C} Tianlin Li, Junyu Liu, Yuan Xin, Yehao Zhou, JHEP 1706 (2017)
111, arXiv:1702.01738v5 [hep-th] .

\bibitem{1CA} Zongping Gong, Robert H. Jonsson, and Daniel Malz,
Supersymmetric Free Fermions and Bosons: Locality, Symmetry and Topology,
Phys. Rev. B 105, 085423 (2022), arXiv:2112.07527v3 [quant-ph].

\bibitem{1CB} G. Junker, Supersymmetric Methods in Quantum and Statistical
Physics (Springer Berlin Heidelberg, Berlin, Heidelberg, 1996).

\bibitem{1CC} Sung-Sik Lee, Phys.Rev.B76:075103,2007,
arXiv:cond-mat/0611658v5 [cond-mat.str-el].

\bibitem{1CD} Natalia Chepiga, Ji\v{r}\'{\i} Min\'{a}\v{r}, Kareljan
Schoutens, SciPost Phys. 11, 059 (2021), arXiv:2105.04359v3
[cond-mat.str-el].

\bibitem{2C1} Sourlas, N. Introduction to supersymmetry in condensed matter
physics. Physica D 15, 115-122 (1985).

\bibitem{2CA} Junker, G. Supersymmetric Methods in Quantum and Statistical
Physics (Springer,1996).

\bibitem{2CB} Cooper, F., Khare, A., Sukhatme, U. Supersymmetry and quantum
mechanics. Phys.Rep. 251, 267-385 (1995).

\bibitem{2CC} Chumakov, S. M., Wolf, K. B. Supersymmetry in Helmholtz
Optics. Phys. Lett. A 193,51-52 (1994).

\bibitem{2CD} Miri, M. A., Heinrich, M., El-Ganainy, R., Christodoulides, D.
N. Supersymmetric optical structures. Phys. Rev. Lett. 110, 233902 (2013).

\bibitem{2CE} Chris Elliott \& Pavel Safronov, Communications in
Mathematical Physics volume 371, pages727--786 (2019).

\bibitem{2CF} Dine, M. Supersymmetry and string theory: beyond the standard
model (Cambridge University Press, 2007).

\bibitem{1D} A. Altland and M. R. Zirnbauer, Phys. Rev. B 55, 1142 (1997).

\bibitem{1D1} Michael J. Lawler, Phys. Rev. B 94, 165101 (2016),
arXiv:1510.03697 [cond-mat.str-el].

\bibitem{1D2} Krishanu Roychowdhury, Jan Attig, Simon Trebst, Michael J.
Lawler, Supersymmetry on the lattice: Geometry, Topology, and Spin Liquids,
arXiv:2207.09475 [cond-mat.str-el].

\bibitem{1DA} S. Ryu, A. P. Schnyder, A. Furusaki, and A. W. W. Ludwig, New
J. Phys. 12, 065010 (2010).

\bibitem{1DB} L. B Drissi, E. H Saidi, Journal Phys Condensed Matter, 32
(36) (2020), arXiv:2207.02901v1 [cond-mat.mes-hall].

\bibitem{TF} T. Fukui, Phys. Rev. B 99, 165129 (2019).

\bibitem{1DC} Lalla Btissam Drissi, El Hassan Saidi, European Physical
Journal Plus 136, (68) (2021), arXiv:2206.11984v1 [cond-mat.mtrl-sci].

\bibitem{1DD} L. B. Drissi, S. Lounis and E. H. Saidi, Eur. Phys. J. Plus,
137 7 (2022) 796.

\bibitem{1E} Robert H. Jonsson, Lucas Hackl, Krishanu Roychowdhury,
Entanglement dualities in supersymmetry, Phys. Rev. Research 3, 023213
(2021), arXiv:2103.09657v2 [quant-ph].

\bibitem{1E1} Amit Giveon, David Kutasov, Journal of High Energy Physics 42
(2016), \qquad arXiv:1510.08872 [hep-th].

\bibitem{1EA} Pramod Padmanabhan, Fumihiko Sugino, Diego Trancanelli, Quant.
Inf. and Comp., Vol 20, No 1-2 (2020) 37-64, arXiv:1911.02577v2 [quant-ph].

\bibitem{1EB} Eugenio Bianchi, Lucas Hackl, and Nelson Yokomizo,
\textquotedblleft Entanglement entropy of squeezed vacua on a
lattice,\textquotedblright \ Phys. Rev. D 92, 085045 (2015),
arXiv:1507.01567 [hepth].

\bibitem{1EC} Eugenio Bianchi, Lucas Hackl, and Nelson Yokomizo,
\textquotedblleft Linear growth of the entanglement entropy and the
kolmogorov-sinai rate,\textquotedblright \ Journal of High Energy Physics
2018, 25 (2018).

\bibitem{1ED} G. Junker, Eur. Phys. J. Plus 2020, 135, 464 (13pp),
arXiv:1911.00454v2 [math-ph].\newline
Georg Junker, Symmetry 2021, 13(5), 835, arXiv:2105.03240v1 [math-ph].

\bibitem{1EE} N. Srinivas, A. Shukla, R. P. Malik, Int. J. Mod. Phys. A 30:
1550166 [p01--p12], 2015, arXiv:1410.2486v3 [hep-th].

\bibitem{1F} E. H Saidi, L. B Drissi, Nucl.Phys.B 974 (2022) 115632,
arXiv:2112.04695 [hep-th].

\bibitem{2F} Maurice de Gosson, Symplectic Geometry and Quantum Mechanics
(Springer, 2006).

\bibitem{1G} E H Saidi, M B Sedra and J Zerouaoui, 1995 Class. Quantum Grav.
12 2705.

\bibitem{1GA} Lucas Hackl, Tommaso Guaita, Tao Shi, Jutho Haegeman, Eugene
Demler, and Ignacio Cirac, SciPost Phys. 9, 048 (2020), arXiv:2004.01015.

\bibitem{1GB} Alexandru Macridin, Panagiotis Spentzouris, James Amundson,
Roni Harnik, Phys. Rev. A 98, 042312 (2018), arXiv:1805.09928v2 [quant-ph].%
\newline
Gavriel Segre, Phase operator of the quantum supersymmetric harmonic
oscillator, arXiv:0710.3138v1 [quant-ph].

\bibitem{1GC} Eyal Cornfeld, Shachar Carmeli, Phys. Rev. Research 3, 013052
(2021), arXiv:2009.04486v3 [cond-mat.mes-hall].

\bibitem{1H0} Gabriel Lopes Cardoso, Thomas Mohaupt, Physics Reports (2020),
arXiv:1909.06240v4 [hep-th].

\bibitem{1H} Hackl, Lucas; Bianchi, Eugenio, Bosonic and fermionic Gaussian
states from K\"{a}hler structures, SciPost Phys. Core 4, 025 (2021),
arXiv:2010.15518.

\bibitem{1HA} Angel Garcia-Chung, Phys. Rev. D 101, 106004 (2020),
arXiv:2003.00388v1 [gr-qc].

\bibitem{YS} Youssra Boujakhrout, El Hassan Saidi, Nucl.Phys.B 981 (2022)
115859, arXiv:2207.14777v1 [hep-th].

\bibitem{sp} Sigiswald Barbier, Jan Frahm, Int. Math. Res. Not. IMRN 2017
(2017), no. 21, 16357-16420, arXiv:1710.07271v2 [math.RT].

\bibitem{SA} Vincent X. Genest, Jean-Michel Lemay, Luc Vinet, Alexei
Zhedanov, J. Phys. A: Math. Theor. 46 (2013) 505204, arXiv:1309.1701v2
[math-ph].

\bibitem{CA1} L. Fu, C. L. Kane, Topological insulators with inversion
symmetry, Phys. Rev. B 76 (2007) 045302.

\bibitem{CA2} C. W. Peterson, W. A. Benalcazar, T. L. Hughes, and G. Bahl,
Nature 555, 346-350 (2018)

\bibitem{CA3} M. Ragragui, L. B. Drissi, E. H. Saidi, Evidence of
topological surface states, Materials Science \& Engineering B (2022).

\bibitem{CA4} Zongping Gong and Tommaso Guaita, Topology of Quantum Gaussian
States and Operations, arXiv:2106.05044v2 [quant-ph].

\bibitem{1M} Zongping Gong, Robert H. Jonsson, and Daniel Malz,
Supersymmetric Free Fermions and Bosons: Locality, Symmetry and Topology,
Phys. Rev. B 105, 085423 (2022), arXiv:2112.07527v3 [quant-ph].

\bibitem{CR1} Ken Shiozaki and Masatoshi Sato, Phys. Rev. B 90, 165114
(2014).

\bibitem{CR2} Yoichi Ando, Liang Fu, Annual Review of Condensed Matter
Physics 6, 361-381 (2015).

\bibitem{Jeh} Lalla Btissam Drissi, Houda Jehjouh, El Hassan Saidi,
Nucl.Phys.B801:316-345,2008, arXiv:0801.2661v2 [hep-th].

\bibitem{APP} \textrm{Tetsuo Deguchi Akira Fujii Katsushi Ito, Volume 238,
Issues 2--4, 5 April 1990, Pages 242-246}

\bibitem{scft} E.H.Saidi, Chiral rings in the N=4 SU(2) conformal theory,
Physics Letters B, Volume 300, Issues 1--2, 4 February 1993, Pages 84-91.%
\newline
E.H.Saidi, M.B.Sedra, A.Serhani, Physics Letters B Volume 353, Issues 2--3,
29 June 1995, Pages 209-212.

\bibitem{osp22} Xiang-Mao Ding, Mark D. Gould, Courtney J. Mewton, Yao-Zhong
Zhang, On osp(2\TEXTsymbol{\vert}2) Conformal Field Theories,

\bibitem{sqm} Monique Combescure, Fran\c{c}ois Gieres, Maurice Kibler,%
\textrm{\ 2004 J. Phys. A: Math. Gen. 37 10385}

\bibitem{witten} E. Witten, Constraints on supersymmetry breaking, Nuclear
Physics B 202, 253, (1982).
\end{thebibliography}
\end{document}